\documentclass[11pt,a4paper]{article}
\pdfoutput=1
\usepackage{amssymb,amsmath,amsfonts, mathtools, mathrsfs}
\usepackage[utf8]{inputenc} 
\usepackage[dvipsnames]{xcolor}

\newif\ifnatbibsort\natbibsorttrue

\relax
\ifnatbibsort\RequirePackage[numbers,sort&compress]{natbib}\else\RequirePackage[numbers,compress]{natbib}\fi
\RequirePackage[colorlinks=true
,urlcolor=blue
,anchorcolor=blue
,citecolor=blue
,filecolor=blue
,linkcolor=blue
,menucolor=blue
,pagecolor=blue
,linktocpage=true
,pdfproducer=medialab
,pdfa=true
]{hyperref}

\usepackage{graphicx}
\usepackage{caption}
\usepackage{tensor}
\usepackage{subfigure}
\usepackage{enumerate}
\usepackage{dsfont}
\usepackage{wrapfig}
\usepackage{verbatim}
\usepackage{amsfonts,euscript,amssymb,stmaryrd,braket}
\usepackage{tikz}
\usetikzlibrary{arrows,decorations.markings,patterns}
\usepackage{slashed}

\def\clock{{\count0=\time
		\divide\count0 60
		\ifnum\count0<10 0\fi\the\count0
		\multiply\count0 -60 \advance\count0 \time
		:\ifnum\count0<10 0\fi \the\count0
}}
\newcommand{\timestamp}{{\small\vbox{\hbox{\tt\jobname.tex}
			\hbox{\the\day/\the\month/\the\year, \clock}}}}

\DeclareMathOperator{\arcsinh}{arcsinh}
\DeclareMathOperator{\arcoth}{arcoth}

\newcommand{\nn}{\nonumber}
\newcommand{\bea}{\begin{eqnarray}}
	\newcommand{\eea}{\end{eqnarray}}

\newcommand{\be}{\begin{equation}}
	\newcommand{\ee}{\end{equation}}

\allowdisplaybreaks

\makeatletter
\let\old@startsection=\@startsection
\let\oldl@section=\l@section
\renewcommand{\@startsection}[6]{\old@startsection{#1}{#2}{#3}{#4}{#5}{#6\mathversion{bold}}}
\renewcommand{\l@section}[2]{\oldl@section{\mathversion{bold}#1}{#2}}
\makeatother

\numberwithin{equation}{section}

\usepackage{color}

\def \RR {{\mathbb R}}

\def\ri {{\rm i}}
\def\rd {{\rm d}}
\def\e {{\rm e}}

\begin{document}
	\renewcommand{\thefootnote}{\arabic{footnote}}

	\overfullrule=0pt
	\parskip=2pt
	\parindent=12pt
	\headheight=0in \headsep=0in \topmargin=0in \oddsidemargin=0in

	\vspace{ -3cm} \thispagestyle{empty} \vspace{-1cm}
	\begin{flushright} 
		\footnotesize
		\textcolor{red}{\phantom{print-report}}
	\end{flushright}

	\definecolor{mygreen}{RGB}{54,115,44}

	\begin{center}
		\vspace{.0cm}


				{\Large\bf \mathversion{bold}
			Holographic thermal entropy from geodesic bit threads
		}

		\vspace{0.8cm} {
			Stefania Caggioli$^{\,a}$,
			Francesco Gentile$^{\,b}$,
			Domenico Seminara$^{\,a}$
			and Erik Tonni$^{\,b}$
		}
		\vskip  0.7cm
		
		\small
		{\em
			$^{a}\,$Dipartimento di Fisica e Astronomia and INFN Sezione di Firenze,  \\ via G. Sansone 1, 50019, Sesto Fiorentino, Italy \\
			\vskip 0.05cm
			$^{b}\,$SISSA and INFN Sezione di Trieste, via Bonomea 265, 34136, Trieste, Italy 
		}
		\normalsize

	\end{center}

	\vspace{0.3cm}
	\begin{abstract} 
		
		The holographic bit threads are an insightful tool to investigate
		the holographic entanglement entropy and other quantities 
		related to the bipartite entanglement in AdS/CFT.
		We mainly explore the geodesic bit threads 
		in various static backgrounds, 
		for the bipartitions characterized by 
		either a sphere or an infinite strip. 
		In pure AdS and for the sphere,
		the geodesic bit threads provide a gravitational dual 
		of the map implementing the geometric action of the modular conjugation
		in the dual CFT.
		In Schwarzschild AdS black brane and for the sphere, 
		our numerical analysis shows that
		the flux of the geodesic bit threads through the horizon 
		gives the holographic thermal entropy of the sphere. 
		This feature is not observed when the subsystem is an infinite strip,
		whenever we can construct the corresponding bit threads.
		The bit threads are also determined by
		the global structure of the gravitational background;
		indeed, for instance, 
		we show that the geodesic bit threads of an arc in the BTZ black hole cannot be constructed. 
	\end{abstract}

	\newpage

	\tableofcontents

	\section{Introduction}
	\label{sec-intro}

	
	In the gauge/gravity correspondence \cite{Maldacena:1997re, Aharony:1999ti},
	the gravitational prescription to evaluate the entanglement entropy of a spatial region $A$ 
	in the dual $(d+1)$-dimensional conformal field theory (CFT$_{d+1}$) 
	for static asymptotically AdS$_{d+2}$ backgrounds,
	proposed by Ryu and Takayanagi (RT) \cite{Ryu:2006bv, Ryu:2006ef},
	and its covariant extension to time-dependent geometries,
	formulated by Hubeny, Rangamani and Takayanagi \cite{Hubeny:2007xt},
	have triggered a wide interdisciplinary research activity 
	with the aim of finding insightful connections between quantum gravity, 
	quantum field theory and quantum information theory
	(see e.g. \cite{Rangamani:2016dms, Headrick:2019eth} for recent reviews).
	%

	
	Freedman and Headrick \cite{Freedman:2016zud} have reformulated the RT proposal
	for static backgrounds through a specific convex optimization problem.
	This alternative prescription requires to consider 
	the set $\mathcal{V} $ of the divergenceless vector fields in the bulk
	that are bounded by a constant,
	finding among them the ones providing the maximum flux through $A$, 
	whose value multiplied by $1/(4 G_{\textrm{\tiny N}})$ gives the holographic entanglement entropy;
	namely
	\be
	\label{HEE-BT-intro}
	S_A = \frac{1}{4 G_{\textrm{\tiny N}} } \;
	\underset{\boldsymbol{V} \in \mathcal{V}}{\textrm{max}}
	\int_A 
	\boldsymbol{V} \cdot \boldsymbol{n} \, \sqrt{h}\; \rd^d s
	\;\;\;\;\qquad\;\;\;
	\mathcal{V} \equiv
	\Big\{
	\boldsymbol{V}\; \big|\; \nabla_\mu V^\mu = 0 \;\textrm{ and }\; | \boldsymbol{V} | \leqslant 1
	\Big\}\,.
	\ee
	The equivalence of this proposal with the one by Ryu and Takayanagi is based on the
	Riemannian geometry version of the max-flow min-cut theorem \cite{Fdrer1974RealFC,Strang1983MaximalFT,Nozawa1990MaxflowMT}.
	The flows occurring in (\ref{HEE-BT-intro}) have been called bit threads 
	and their integral lines naturally provide a link representation 
	of the bipartite entanglement between the boundary region $A$ and its complement $B$. 
	We remark that the prescription (\ref{HEE-BT-intro})
	does not lead to a unique vector field configuration.
	Explicit examples of holographic bit threads in some simple static gravitational backgrounds have been constructed in \cite{Agon:2018lwq}.
	The covariant extension of (\ref{HEE-BT-intro}) has been discussed in \cite{Headrick:2022nbe}.

	
	Various properties of the bit threads have been studied,
	including the ones capturing aspects of the holographic bipartite  entanglement 
	different from the holographic entanglement entropy.
	We find it worth mentioning 
	the proofs based on bit threads of the strong subadditivity \cite{Freedman:2016zud} 
	and of the monogamy \cite{Hubeny:2018bri,Cui:2018dyq},
	the extension of the bit thread constructions to Lorentzian backgrounds \cite{Headrick:2017ucz}
	and the bit thread formulation of the holographic entanglement entropy
	in higher curvature gravity \cite{Harper:2018sdd}.
	The bit threads have been employed to explore also
	the quantum correction to the holographic entanglement entropy \cite{Agon:2021tia, Rolph:2021hgz},
	the entanglement of purification \cite{Harper:2019lff},
	the first law of entanglement in relation to Einstein's equations \cite{Agon:2020mvu},
	the holographic complexity \cite{Pedraza:2021mkh,Pedraza:2021fgp,Caceres:2023ziv},
	the multipartite holographic entanglement \cite{Harper:2021uuq}
	and some
	minimal area problems in string field theory \cite{Headrick:2018dlw, Headrick:2018ncs}.

	A feature of the Freedman-Headrick prescription in (\ref{HEE-BT-intro})
	heavily exploited in this manuscript 
	is based on the simple observation \cite{Tonni18:bariloche-talk} that
	the integrand in (\ref{HEE-BT-intro}) associated with a holographic bit thread configuration
	provides a gravitational dual of a contour function for the entanglement entropy in the dual CFT
	\cite{Chen_2014,Cardy:2016fqc,Coser:2017dtb} living on the boundary, 
	i.e. a specific density of the holographic entanglement entropy in $A$;
	hence the non-uniqueness of the bit thread configuration in (\ref{HEE-BT-intro})
	can also be understood within the dual CFT on the boundary.

	In this manuscript, we mainly investigate the geodesic bit threads introduced in \cite{Agon:2018lwq},
	i.e. the bit thread configurations constructed through the geodesics 
	of a generic time slice of the static asymptotically AdS$_{d+2}$ gravitational background.
	Their relation with the contour function of the entanglement entropy coming from the 
	entanglement Hamiltonian \cite{Haag:1992hx,Hislop:1981uh,Casini:2011kv, Wong:2013gua, Cardy:2016fqc}
	has been explored in \cite{Kudler-Flam:2019oru}.

	In some cases,
	the geodesic bit threads display an interesting relation with the holographic thermal entropy of the region $A$.

	It is expected that
	the entanglement entropy $S_A$ becomes the thermal entropy of $A$,
	denoted by $S_{A,\textrm{\tiny th}}$ hereafter,
	when the size of $A$ is large with respect to the inverse temperature $\beta$ of the CFT.
	For instance, 
	in a CFT$_2$ at finite temperature $1/\beta$ and on the line,
	the entanglement entropy of an interval $A$ of length $\ell$ is 
	$S_A = \tfrac{c}{3} \log\! \big( \tfrac{\beta}{\pi \epsilon} \sinh(\pi \ell /\beta)\big)$
	and, in the high temperature regime where $\ell / \beta \gg \beta/\epsilon \gg 1$,
	it becomes the thermal entropy of $A$ given by $S_{A,\textrm{\tiny th}} =\tfrac{\pi c}{3\beta}  \,\ell$  \cite{Calabrese:2004eu},
	where $\tfrac{\pi c}{3\beta} $ is the thermal entropy density 
	obtained from the Stefan-Boltzmann law for a CFT$_2$ on the line \cite{Cardy:2010fa}.
	This feature occurs also for the 
	expansion of the holographic entanglement entropy in the UV cutoff
	given by the RT prescription,
	whose finite term grows like the thermal entropy of $A$ 
	when $A$ is large with respect to the position of the horizon
	(see e.g. \cite{Tonni:2010pv, Liu:2013una}).

	In a holographic CFT$_2$ at finite temperature on the line
	and for the bipartition given by an interval $A$,
	it has been found \cite{Mintchev:2022fcp}
	that the corresponding geodesic bit threads in the 
	constant time slice of the BTZ black brane
	give the holographic thermal entropy of $A$
	for any $\ell/\beta$,
	not only in the high temperature regime. 
	Any holographic bit thread configuration provides a bijective map 
	between $A$ and $B$ through the bulk. 
	However, in a holographic CFT$_2$,
	for the setups and the bipartitions whose entanglement entropy is known in a universal way \cite{Calabrese:2004eu},
	it has been observed \cite{Mintchev:2022fcp}
	that the corresponding geodesic bit threads implement 
	the geometric action of the modular conjugation \cite{Hislop:1981uh,Haag:1992hx}
	on the gravitational side of the holographic correspondence.

	
	In this work, we study various aspects of the geodesic bit threads 
	in simple static asymptotically AdS$_{d+2}$ backgrounds,
	when the region $A$ is either a sphere (i.e. a ball) or an infinite strip,
	mainly focussing on their relation with the thermal entropy of $A$,
	in order to extend to higher dimensions
	some of the results reported in \cite{Mintchev:2022fcp} for $d=1$.

	
	The paper is organized as follows. 
	In Sec.\,\ref{sec-AdS3} and Sec.\,\ref{sec-BTZ} we revisit the geodesic bit threads for an interval 
	when the gravitational background is the constant time slice of 
	either Poincar\'e AdS$_3$ or BTZ black brane, respectively. 
In Sec.\,\ref{HigherdimAdS} we consider different types of holographic bit threads 
in the constant time slice of AdS$_{d+2}$,
when $A$ is either a sphere or an infinite strip.
In Sec.\,\ref{sec-hyp-bh} we explore the geodesic bit threads for a sphere
in a specific hyperbolic black hole in any dimension,
where analytic results can be obtained. 
Our main results are numerical and are reported in Sec.\,\ref{Schwarzschild AdS black brane},
where we investigated the geodesic bit threads and also another type of bit threads
in the constant time slice of the Schwarzschild AdS$_{d+2}$ black brane.
	In Sec.\,\ref{sec-BTZ-black-hole} we study the possibility of constructing the geodesic bit threads 
	for a circular arc in the constant time slice of a BTZ black hole,
	whose dual CFT$_2$ on its boundary is at finite temperature and finite volume.
	We close in Sec.\,\ref{sec-conclusions} by summarizing our results and mentioning potential future directions. 
	Some details supporting the analyses of the main text
	and further extensions
	are reported in  Appendices\, 
	\ref{app-modulus},  \ref{app-entropy-complementary-btz}, \ref{GeodAdS4},
	\ref{Nestingtotal}, \ref{app-traslatipedraza}, 
	\ref{app-BTZ-global-CTmap} and \ref{app-LongGeodesics-GlobalBTZ}.

	\section{AdS$_3$}
	\label{sec-AdS3}

	
	In this section, we study the geodesic bit threads in a constant time slice of Poincar\'e AdS$_3$ (Sec.\,\ref{subsec-gbt-ads3})
	and the corresponding flow through the boundary (Sec.\,\ref{subsec-flow-bdy-ads3}), 
	which provides a contour function for the holographic entanglement entropy.

	
	In the context of the gauge/gravity correspondence,
	consider a three-dimensional gravity model in AdS$_3$ described by the Poincar\'e coordinates $(t,w,y)$,
	where $w>0$ is the holographic coordinate. 
	The spacetime of the dual CFT$_2$ on the boundary at $w \to 0^+$ is described by the coordinates $(t,y)$.
	On a constant time (i.e. $t=\textrm{const}$) slice of AdS$_3$ the induced metric is
	\be
	\label{H2-metric}
	ds^2 = \frac{L^2_{\textrm{\tiny AdS}}}{w^2} \,\big(  \rd w^2 + \rd y^2 \big)
	\ee
	which characterizes the Euclidean hyperbolic upper half plane $\mathbb{H}_2$
	parameterized by the coordinates $(w,y)$, where $y\in \RR$.

	
	In the CFT$_2$ on the boundary of AdS$_3$, 
	we consider the spatial bipartition of the real line at $t=\textrm{const}$ 
	given by an interval $A$ and its complement $B \equiv \RR \setminus A$. 
	The spatial translation invariance allows us to choose $A\equiv [-b, b]$ with $b>0$
	without loss of generality. 
	The RT curve $\gamma_A$ \cite{Ryu:2006bv, Ryu:2006ef}, 
	whose regularized length provides the holographic entanglement entropy $S_A$ for this setup,
	is given by the following half circle in $\mathbb{H}_2$, namely
	\be
	\label{RT-curve-ads3}
	\gamma_A\; :
	\qquad
	w_m^2 + y_m^2 = b^2
	\ee
	(see the red curve in the top panel of Fig.\,\ref{fig:ads3-main}),
	where $y_m \in [-b,b]$ and $P_m =(w_m, y_m)$ is the generic point of $\gamma_A$.
	Denoting by $\gamma_B$ the RT curve corresponding to $B$,
	in this setup we have $\gamma_A = \gamma_B$, 
	implying that $S_A = S_B$, 
	in agreement with the fact that 
	the dual CFT$_2$ in the boundary is in its ground state, which is a pure state.

	\subsection{Geodesic bit threads}
	\label{subsec-gbt-ads3}

	
	In order to determine the geodesic bit threads of $A \subset \RR$ \cite{Agon:2018lwq},
	let us consider the generic geodesic in a constant time slice of AdS$_3$ (\ref{H2-metric}) 
	with both the endpoints on the boundary, i.e. 
	\be
	\label{gen-geo-ads3}
	w = \sqrt{ b_0^2-( y - c_0)^2}
	\ee
	where $(w,y)=(0,c_0 \pm b_0)$ are the coordinates of the endpoints.
	The RT curve $\gamma_A$ in (\ref{RT-curve-ads3}) corresponds to the special case of (\ref{gen-geo-ads3})
	where $c_0=0$ and $b_0=b$.

\begin{figure}[t!]
		\vspace{-.5cm}
		\hspace{-1.1cm}
		\includegraphics[width=1.15\textwidth]{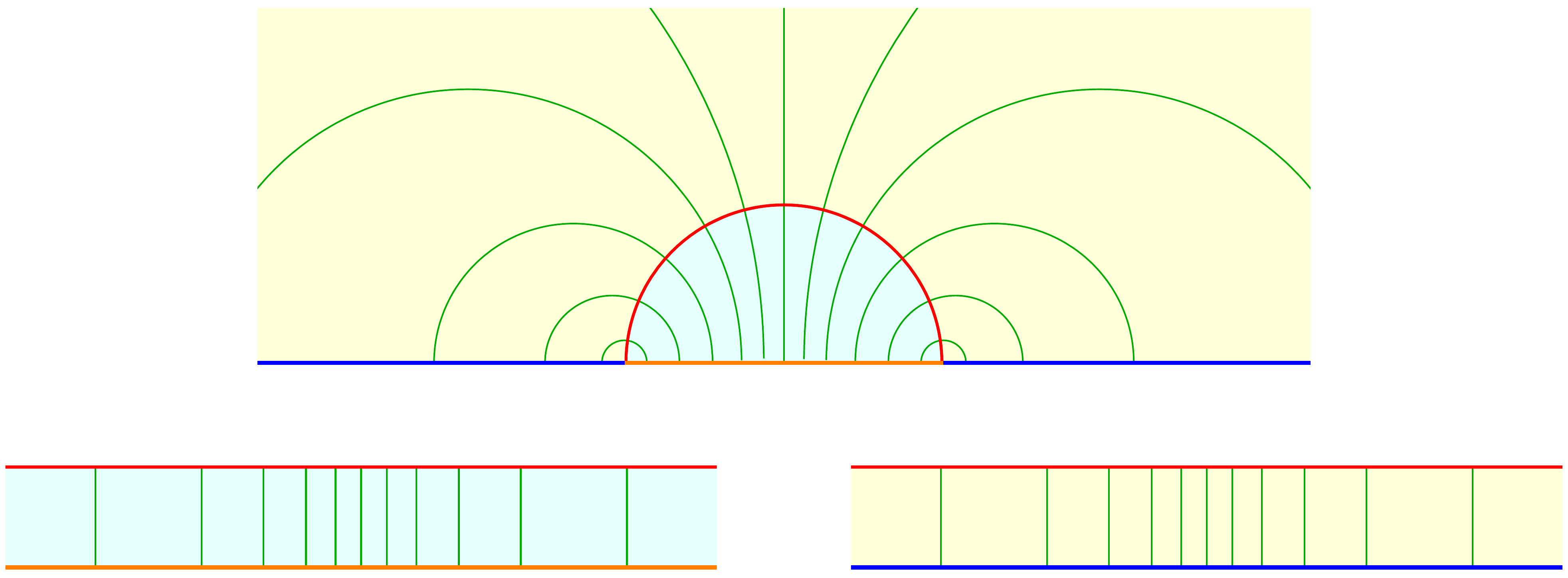}
		\vspace{-.2cm}
		\caption{\small Geodesic bit threads for an interval in the line, in the constant time slice of AdS$_3$ (top panel).
		The cyan and yellow regions in the top panel are mapped through (\ref{chm-ads-inside-inverse})
		onto the region outside the horizon in the BTZ black branes 
		(see (\ref{btz-brane-metric-CHM})) in the bottom left and bottom right panels, respectively.}
		\label{fig:ads3-main}
	\end{figure}

	
	The geodesic bit threads define a flow characterized by a specific vector field $\boldsymbol{V} = (V^w, V^y)$
	whose integral lines are
	the geodesics $w(y)$ of the form (\ref{gen-geo-ads3}) 
	that intersect $\gamma_A$ orthogonally;
	hence they satisfy
	\be
	\label{intersect-ortho-condition-ads}
	\left\{ \begin{array}{l}
		w(y_m) = w_{m}(y_m)
		\\
		\rule{0pt}{.5cm}
		\big[ \,g_{yy} + g_{ww} \, w'(y) \,w'_{m}(y) \,\big] \big|_{(w,y)=(w_m(y_m),y_m)} = 0
	\end{array}\right.
	\ee
	where $g_{yy} = g_{ww} = L^2_{\textrm{\tiny AdS}} / w^2$ are the diagonal components of  (\ref{H2-metric})
	and $( w_m(y_m),y_m)$
	is the generic point of $\gamma_A$ where the intersection occurs. 
	Solving (\ref{intersect-ortho-condition-ads}), one finds
	\be
	\label{c0-b-gbt-ads3}
	c_0 = \frac{b^2}{y_m}
	\;\;\;\qquad\;\;\;
	b_0 = \frac{\sqrt{b^2 - y_m^2}}{|y_m|} \; b\,.
	\ee
	Plugging (\ref{c0-b-gbt-ads3}) into (\ref{gen-geo-ads3}),
	we obtain the integral curves of the geodesic bit threads,
	which are represented by the green curves in the top panel of Fig.\,\ref{fig:ads3-main}.

	The endpoints of these integral curves provide the map implementing
	the geometric action of the modular conjugation in CFT$_2$ for this setup \cite{Mintchev:2022fcp};
	hence, the geodesic bit threads identify the holographic gravitational counterpart
	of this map in the dual CFT$_2$.
	
	
	To fix the vector field $\boldsymbol{V}$ characterizing the geodesic bit threads, 
	whose integral curves are given by (\ref{gen-geo-ads3}) and  (\ref{c0-b-gbt-ads3}),
	we determine their parameterisation by first imposing $|V| =1$ on the RT curve (\ref{RT-curve-ads3}) 
	and then checking that $|V| < 1$ everywhere else in the 
	constant time slice of AdS$_3$ \cite{Freedman:2016zud}.
	This can be achieved by following the procedure discussed in \cite{Agon:2018lwq} 
	and reviewed in Appendix \ref{app-modulus}, 
	which leads to
	\be
	\label{vector-field-ads3}
	\boldsymbol{V} 
	=
	\big| \boldsymbol{V} \big|\, \boldsymbol{\tau}
	=
	\big( \, V^w\, , \, V^y \,\big) 
	=  
	\frac{1}{L_{\textrm{\tiny AdS}}}
	\Bigg( \frac{2b\,w}{\sqrt{(b^2-y^2-w^2)^2+4b^2\, w^2}} \Bigg)^2 \;\bigg( \frac{b^2 -y^2 + w^2}{2b}  \, ,\,  \frac{y \, w}{b} \bigg)
	\ee
	being $\boldsymbol{\tau} = (\tau^w, \tau^y)$ defined as the unit norm vector tangent to the generic geodesic bit thread,
	whose components are
	\be
	\boldsymbol{\tau} = 
	\big( \,\tau^w\, , \, \tau^y \,\big) =  
	\frac{2b\,w}{L_{\textrm{\tiny AdS}}\,\sqrt{(b^2-y^2-w^2)^2+4b^2\, w^2}} \;\bigg( \frac{b^2 -y^2 + w^2}{2b}  \, ,\, \frac{y \,w}{b} \bigg) \,.
	\ee

	
	We find it insightful to bipartite the time slice parameterized by the coordinates $(w,y)$ 
	into the two complementary regions corresponding to the light blue and the yellow domain in the top panel of Fig.\,\ref{fig:ads3-main},
	which are described respectively by $(w_{+},y_{+})$, satisfying $y_+^2 +w_+^2 < b^2$,
	and by $(w_{-}, y_{-})$, satisfying $y_-^2 +w_-^2 > b^2$.
	Now, let us consider the following changes of coordinates 
	\cite{Casini:2011kv, Espindola:2018ozt}\footnote{In \cite{Casini:2011kv, Espindola:2018ozt} only the mapping corresponding to $(y_+, w_+)$ has been considered.
		}
	\be
	\label{chm-ads-in-out}
	y_\pm  \equiv   \frac{b\, \sinh(x_\pm /b)}{\cosh(x_\pm /b) \pm \sqrt{1-(z_\pm /b)^2}  }
	\;\;\;\qquad\;\;\;
	w_\pm \equiv  \frac{z_\pm}{\cosh(x_\pm /b) \pm \sqrt{1-(z_\pm /b)^2}  }
	\ee
	where $x_\pm \in \RR$ and $0<z_\pm  \leqslant b$,
	that satisfy
	\be
	y_{\mp} (x,z) = y_{\pm} (x +\ri \pi b ,-z)
	\;\;\qquad\;\;
	w_{\mp} (x,z) = w_{\pm} (x+\ri \pi b,-z)
	\ee
	and whose inverse maps read
	\be
	\label{chm-ads-inside-inverse}
	x_\pm = \frac{b}{2}\, \log\! \left(  \frac{(b + y_\pm)^2 + w_\pm^2}{(b - y_\pm)^2 + w_\pm^2}  \right)
	\;\;\;\qquad\;\;\;
	z_\pm = \frac{2b^2\, w_\pm}{\sqrt{ \big(b^2-y_\pm^2-w_\pm^2\big)^2+4 b^2 w_\pm^2 }} \,.
	\ee
	The interesting feature of the two pairs of coordinates $(z_\pm,x_\pm )$ defined through (\ref{chm-ads-in-out})
	is that both of them parameterize the region between the horizon and the boundary  
	of a constant time slice of the BTZ black brane whose horizon is placed at $z=b$;
	indeed
	\be
	\label{btz-brane-metric-CHM}
	ds^2 
	= \frac{L^2_{\textrm{\tiny AdS}}}{w_\pm^2} \Big( \rd w_\pm^2 + \rd y_\pm^2  \Big)
	= \frac{L^2_{\textrm{\tiny AdS}}}{z_\pm ^2} \left( \,
	\frac{\rd z_\pm^2}{1-(z_\pm /b)^2} 
	+ \rd x_\pm^2 \right) .
	\ee
	From the second expression in (\ref{chm-ads-inside-inverse})
	one realizes that the RT curve given by $w_\pm^2 + y_\pm^2 = b^2$ (see (\ref{RT-curve-ads3})) 
	is mapped onto the planar horizon of the metric (\ref{btz-brane-metric-CHM}), i.e. $z_\pm |_{\gamma_A} = b$.

	The domains described by the coordinates $(z_{+},x_{+})$ and $(z_{-},x_{-} )$
	are shown in the bottom left and bottom right panel of Fig.\,\ref{fig:ads3-main} respectively,
	where the horizontal red lines represent the horizons,
	that correspond to the RT curve in the top panel of the same figure. 
	In the bottom left and bottom right panel of Fig.\,\ref{fig:ads3-main} 
	we show 
	the images through (\ref{chm-ads-inside-inverse}) of the arcs of the geodesic bit threads 
	contained in the region $y_+^2 +w_+^2 < b^2$ and $y_-^2 +w_-^2 > b^2$ respectively
	(see the vertical green lines connecting each horizon and the corresponding boundary).
	For the region $y_+^2 +w_+^2 < b^2$ and the bottom left panel of Fig.\,\ref{fig:ads3-main},
	see also Fig.\,7 of \cite{Han:2019scu}.
	The definition of the two BTZ black brane geometries parameterized by $(z_{+},x_{+} )$ and $(z_{-},x_{-} )$
	will be used also to explore the geodesic bit threads of an interval 
	in the constant time slice of the BTZ black brane (see Sec.\,\ref{sec-BTZ}).

	\subsection{Fluxes through the boundary}
	\label{subsec-flow-bdy-ads3}

	It is worth investigating the flux of the vector field $\boldsymbol{V}$ characterizing the geodesic bit threads 
	(see (\ref{vector-field-ads3})) through a generic region in the boundary at $w=0$.
	For instance, 
	the flux of $\boldsymbol{V}$ through the interval $A$ provides its holographic entanglement entropy,
	according to the Freedman-Headrick prescription  \cite{Freedman:2016zud} in (\ref{HEE-BT-intro}).


	Let us consider the integrand occurring in the flux of the geodesic bit threads
	corresponding to $A$ through a generic domain $R$ in the boundary
	(see e.g. (\ref{HEE-BT-intro}), where $R=A$). 
	The unit vector $\boldsymbol{n}$ normal to the boundary 
	is $\boldsymbol{n}=(n^w, n^y )=\tfrac{w}{L_{\textrm{\tiny AdS}}} \big( 1 , 0 \big)$, where $w\to 0^+$.
	%
	Thus, in the flux of the geodesic bit threads through $R$,
	the scalar product in the integrand is $g_{ww} V^w n^w$ as $w\to 0^+$, 
	where $\lim_{w\to 0} V^w > 0$ for $y\in A$, while $\lim_{w\to 0} V^w < 0$ for $y \notin A$.
	In such integrand also the square root of the determinant of the metric induced on the $w=\textrm{const}$ slice 
	(i.e. $ds^2 \big|_{w=\textrm{const}} = \tfrac{L^2_{\textrm{\tiny AdS}}}{w^2} \, \rd y^2$) must be taken into account. 
	Thus, for the setup that we are investigating, 
	the integrand occurring in the flux of the geodesic bit threads 
	through a domain $R$ on the boundary reads
	\be
	\label{contour-function-vacuum-chi}
	\lim_{w \to 0^+} \left(  \frac{1}{4 G_{\textrm{\tiny N}} } \; g_{ww} \,V^w n^w \, \frac{L_{\textrm{\tiny AdS}}}{w}  \right) 
	\equiv 
	\, \chi_A(y)\,\mathcal{C}(y)
	\ee
	where $y \in \RR$ corresponds to the generic point on the line, 
	the step function $\chi_A(y)$ is the characteristic function of $A$
	(which is equal to $+1$ for $y\in A$ and $-1$ for $y\in B$) 
	and we have introduced 
	\be
	\label{contour-function-vacuum}
	\mathcal{C}(y)
	\equiv 
	\frac{L_{\textrm{\tiny AdS}} }{4 G_{\textrm{\tiny N}}} 
	\; \frac{2b}{|b^2 - y^2|}
	\,=\,
	\frac{ c_{\textrm{\tiny BH}} }{6} 
	\; \frac{2b}{|b^2 - y^2|}
	\ee
	being $c_{\textrm{\tiny BH}} $ defined as the Brown-Henneaux central charge \cite{Brown:1986nw}
	\be
	\label{BH-central-charge}
	c_{\textrm{\tiny BH}} \equiv \frac{3 L_{\textrm{\tiny AdS}}}{2 G_{\textrm{\tiny N}}} 
	\ee
	which is $c_{\textrm{\tiny BH}} \gg 1$ in the classical regime that we are considering.
	%

%

	The contour functions for the entanglement entropy of a subregion $A$ 
	in the dual CFT on the boundary
	could be interpreted as the holographic duals 
	of the integrands 
	occurring in the fluxes through $A$ of the corresponding gravitational bit threads 
	\cite{Tonni18:bariloche-talk}.
	For a given spatial bipartition of the boundary CFT in a given state, 
	both these quantities are highly non-unique.

	In the setup that we are considering, 
	i.e. for a CFT$_2$ on a line, in its ground state and the bipartition given by an interval,
	it has been observed \cite{Kudler-Flam:2019oru}
	that the function $\mathcal{C}(y)$ in (\ref{contour-function-vacuum}) 
	for $|y| < b$ and with $c_{\textrm{\tiny BH}} $ replaced by a generic central charge $c$ 
	provides  the specific contour function for the entanglement entropies
	proposed in \cite{Chen_2014,Cardy:2016fqc,Coser:2017dtb},
	which has been obtained from the inverse of the weight function occurring in the corresponding modular Hamiltonian 
	\cite{Hislop:1981uh, Casini:2011kv, Cardy:2016fqc}.
	This contour function for the entanglement entropies has also been checked through numerical computations 
	in free lattice models whose continuum limit are CFT$_2$ with $c=1$ \cite{Chen_2014,Coser:2017dtb}.
	Hence, from (\ref{contour-function-vacuum}) one observes that,  
	in the setup we are investigating, 
	the proposal of \cite{Tonni18:bariloche-talk} is realized
	in the specific example given by the geodesic bit threads
	and the contour function for the entanglement entropy 
	provided by the weight function of the entanglement Hamiltonian,
	proposed in \cite{Chen_2014,Cardy:2016fqc,Coser:2017dtb}.
	Thus, in the setup explored throughout this section,
	(\ref{contour-function-vacuum}) can be interpreted 
	as the holographic contour function associated with the geodesic bit threads.


	To evaluate the holographic entanglement entropy of the interval $A$ and of its complement $B$
	through the contour function (\ref{contour-function-vacuum}),
	we adopt the following UV regularisation procedure, 
	called entanglement wedge cross-section regularisation
	\cite{Dutta:2019gen,Han:2019scu,Headrick:2022nbe, Nguyen:2017yqw,Takayanagi:2017knl}.
	Given the holographic cutoff $w \geqslant \varepsilon_{\textrm{\tiny AdS}}$ in the bulk, 
	where $\varepsilon_{\textrm{\tiny AdS}} \ll 1$,
	which corresponds to the UV cutoff $\epsilon$ in the dual CFT employed in Sec.\,\ref{sec-intro}
	according to the AdS/CFT dictionary,
	the intersections of the straight line $w = \varepsilon_{\textrm{\tiny AdS}}$ with the RT curve (\ref{RT-curve-ads3})
	are the points $P^\pm_{\varepsilon_{\textrm{\tiny AdS}}}$ having $ y_{m} =  \pm \sqrt{ b^2-\varepsilon_{\textrm{\tiny AdS}}^2} $.
	The endpoints of the geodesic bit threads (see \eqref{gen-geo-ads3} and \eqref{c0-b-gbt-ads3})
	intersecting the RT curve in $P^{+}_{\varepsilon_{\textrm{\tiny AdS}}}$ and $P^{-}_{\varepsilon_{\textrm{\tiny AdS}}}$
	provide a natural UV regularisation for the dual CFT$_2$ on the boundary of AdS$_3$.
	The $y$-coordinates of these four endpoints are given by 
	$y \in \big\{ b - \varepsilon_{\textrm{\tiny bdy}}^{\textrm{\tiny $A$}} \,, \, b +  \varepsilon_{\textrm{\tiny bdy}}^{\textrm{\tiny $B$}}  \big\}$ 
	for the geodesic bit thread passing through $P^{+}_{\varepsilon_{\textrm{\tiny AdS}}}$
	and by 
	$y \in \big\{ \! - b - \varepsilon_{\textrm{\tiny bdy}}^{\textrm{\tiny $B$}} \,, \,- \,b +  \varepsilon_{\textrm{\tiny bdy}}^{\textrm{\tiny $A$}}  \big\}$ 
	for the geodesic bit thread passing through $P^{-}_{\varepsilon_{\textrm{\tiny AdS}}}$, 
	where $\varepsilon_{\textrm{\tiny bdy}}^{\textrm{\tiny $A$}}   $ and $\varepsilon_{\textrm{\tiny bdy}}^{\textrm{\tiny $B$}}   $ 
	are defined respectively using
	\be 
	\label{ads-EWCS-regEndp}
	b - \varepsilon_{\textrm{\tiny bdy}}^{\textrm{\tiny $A$}}   
	\equiv 
	b \ \sqrt{  \frac{b + \varepsilon_{\textrm{\tiny AdS}} }{  b - \varepsilon_{\textrm{\tiny AdS}}  } }
	\;\;\; \qquad \;\;\;
	b +  \varepsilon_{\textrm{\tiny bdy}}^{\textrm{\tiny $B$}} 
	\equiv 
	b \ \sqrt{  \frac{b - \varepsilon_{\textrm{\tiny AdS}} }{  b + \varepsilon_{\textrm{\tiny AdS}}  } } \,.
	\ee 
	The endpoints of the geodesic bit threads intersecting the RT curve in $P^{\pm}_{\varepsilon_{\textrm{\tiny AdS}}}$
	identify the interval 
	$A_\varepsilon \equiv \big[ \!- b + \varepsilon_{\textrm{\tiny bdy}}^{\textrm{\tiny $A$}} \,  , b - \varepsilon_{\textrm{\tiny bdy}}^{\textrm{\tiny $A$}} \big] \subsetneq A$
	and the region
	$B_\varepsilon \equiv \big(\!- \!\infty \, , - b  - \varepsilon_{\textrm{\tiny bdy}}^{\textrm{\tiny $B$}}   \big]  \cup \big[ b +  \varepsilon_{\textrm{\tiny bdy}}^{\textrm{\tiny $B$}} \, , +\infty \big) \subsetneq B$,
	which provide the integration domains 
	to determine $S_A$ and $S_B$ as the fluxes of the geodesic bit threads, 
	through the contour function for the holographic entanglement entropy in (\ref{contour-function-vacuum}).
	In particular, for $S_A$ we have
	\be
	\label{HEE-SA-full}
	S_A
	\,=
	\int_{- b + \varepsilon_{\textrm{\tiny bdy}}^{\textrm{\tiny $A$}} }^{ b - \varepsilon_{\textrm{\tiny bdy}}^{\textrm{\tiny $A$}}} \mathcal{C}(y)\, \rd y
	\,=\,
	\left[ 
	\frac{c_{\textrm{\tiny BH}} }{6}
	\log \! \left( \frac{b+x}{ b-x} \right) 
	\right]_{- b + \varepsilon_{\textrm{\tiny bdy}}^{\textrm{\tiny $A$}} }^{  b - \varepsilon_{\textrm{\tiny bdy}}^{\textrm{\tiny $A$}}  }
	\! =\,
	\frac{c_{\textrm{\tiny BH}} }{3}  \log \! \left( \frac{2 b}{\varepsilon_{\textrm{\tiny AdS}}} \right)  
	+ 
	R(\varepsilon_{\textrm{\tiny AdS}})
	\ee
	where 
	\be 
	R(\varepsilon_{\textrm{\tiny AdS}}) 
	\equiv
	\frac{c_{\textrm{\tiny BH}} }{6}  \log \! \left( \frac{ b + \sqrt{b^2-\varepsilon_{\textrm{\tiny AdS}}^2}}{2b} \right) 
	\ee 
	which is $O(\varepsilon_{\textrm{\tiny AdS}}^2)$ as $\varepsilon_{\textrm{\tiny AdS}} \to 0$.
	Similarly, the holographic entanglement entropy $S_B$ can be evaluated as the following flux
	\be
	\label{HEE-SB-full}
	S_B
	\,=
	\int_{-\infty}^{- b -  \varepsilon_{\textrm{\tiny bdy}}^{\textrm{\tiny $B$}}  } \mathcal{C}(y)\, \rd y \,
	+ 
	\int^{+\infty}_{ b +  \varepsilon_{\textrm{\tiny bdy}}^{\textrm{\tiny $B$}}  } \mathcal{C}(y)\, \rd y
	\,=\,
	2
	\left[ 
	\frac{c_{\textrm{\tiny BH}} }{6}
	\log \! \left( \frac{x-b}{ b+x} \right) 
	\right]_{ b+ \varepsilon_{\textrm{\tiny bdy}}^{\textrm{\tiny $B$}}  }^{  +\infty   }
	\!\! = \,
	S_A
	\ee
	which turns out to be equal to $S_A$ in (\ref{HEE-SA-full})
	to all orders in $\varepsilon_{\textrm{\tiny AdS}}$,
	as expected from the purity of the ground state
	and as obtained also through the standard RT prescription
	because $\gamma_A = \gamma_B$
	in the setup, we are exploring 
	(see also the text below (\ref{RT-curve-ads3})).

	
	We find it instructive to consider the images $\boldsymbol{V}_{\!\!\pm}$ 
	of the vector field $\boldsymbol{V}$ in (\ref{vector-field-ads3}) 
	through (\ref{chm-ads-in-out}). 
	As anticipated in the final paragraph of Sec.\,\ref{subsec-gbt-ads3},
	the integral curves of $\boldsymbol{V}$ are mapped by (\ref{chm-ads-inside-inverse})
	into straight lines
	connecting the horizon and the corresponding boundary in the constant time slice BTZ black brane geometries
	described by the coordinates $(z_{\pm},x_{\pm} )$
	(see the vertical green lines in the bottom panels of Fig.\,\ref{fig:ads3-main}).
	The modulus $\big| \boldsymbol{V} \big|$ of the vector field $\boldsymbol{V}$ in (\ref{vector-field-ads3}) is mapped to
	\be
	\big| \boldsymbol{V}_{\!\!\pm} \big|
	= \frac{z_\pm}{b}
	\label{eq:CHMmodulusADS}
	\ee 
	where $0 < z_\pm \leqslant b$.
	The expression (\ref{eq:CHMmodulusADS})
	is equal to $1$ only for $z_\pm =b$, that correspond to the images of the RT curve (\ref{RT-curve-ads3})
	(see also the text below (\ref{btz-brane-metric-CHM})),
	as expected from the fact that (\ref{vector-field-ads3}) is the vector field describing the geodesic bit threads.


	As for the images of the  holographic contour function (\ref{contour-function-vacuum}) through (\ref{chm-ads-in-out}),
	we first write $y_\pm |_{z_\pm=0}$ from the first expression in (\ref{chm-ads-in-out}) 
	and then plug the result into (\ref{contour-function-vacuum}) to get $\mathcal{C}\big(y_\pm \big|_{z_\pm=0}\big)$.
	By considering also the jacobian of the map, we find
	\be
	\label{contour-ads3-after-CHM}
	\mathcal{C}^{(\pm)}_{\textrm{\tiny BTZ}} (x_\pm)
	\,\equiv \,
	\mathcal{C}\big(y_\pm \big|_{z_\pm=0}\big) \, \partial_{x_\pm} \big( y_\pm \big|_{z_\pm =0} \big) 
	\,=\, 
	\frac{L_{\textrm{\tiny AdS}} }{4 G_{\textrm{\tiny N}}} \; \frac{1}{b} 
	\,=\,  
	\frac{L_{\textrm{\tiny AdS}} }{4 G_{\textrm{\tiny N}}} \; \frac{2\pi}{\beta_0}
	\,=\, 
	\frac{\pi c_{\textrm{\tiny BH}} }{3\beta_0}
	\;\;\;\qquad\;\;
	\beta_0 \equiv 2\pi b
	\ee
	where $\beta_0 $ is the inverse temperature of a thermal CFT$_2$
	dual to a BTZ black brane with horizon at $z=b$ according to the standard AdS/CFT dictionary
	and also the Brown-Henneaux central charge (\ref{BH-central-charge}) has been employed in the last step. 
	%
	

	In a CFT$_2$ on the line at finite temperature $1/\beta$ with central charge $c$,
	the spatial density of the free energy is $f_{\textrm{\tiny th}} =  - \pi c/(6\beta^2)$ 
	\cite{Cardy:2010fa,Affleck:1986bv,Bloete:1986qm}.
	By applying a standard thermodynamic relation \cite{Huangbook}, one finds the corresponding entropy density
	\be
	\label{SBthermalentropy}
	s_{\textrm{\tiny th}} = - \frac{ \partial f_{\textrm{\tiny th}}  }{\partial (1/\beta) } = \frac{\pi c}{3\beta} \,.
	\ee 
	By employing these expressions into the definition of the spatial density of the free energy 
	$f_{\textrm{\tiny th}} = u_{\textrm{\tiny th}} - s_{\textrm{\tiny th}} /\beta$,
	we obtain the energy density $u_{\textrm{\tiny th}} = \pi c/(6 \beta^2)$, 
	i.e. the Stefan-Boltzmann law for a CFT$_2$ on the line at finite temperature \cite{Cardy:2010fa}.

	Comparing (\ref{SBthermalentropy}) with the last expression in (\ref{contour-ads3-after-CHM}),
	one concludes that both $\mathcal{C}^{(\pm)}_{\textrm{\tiny BTZ}} (x_\pm)$ 
	are equal to the thermal entropy density $s_{\textrm{\tiny th}}$ of a CFT$_2$ in (\ref{SBthermalentropy})
	specialized to $c= c_{\textrm{\tiny BH}}$ and $\beta = \beta_0$,
	which correspond to the holographic setup whose gravitational background is 
	the last expression in (\ref{btz-brane-metric-CHM}).


	The expressions for the holographic entanglement entropy in (\ref{HEE-SA-full}) and (\ref{HEE-SB-full})
	can be found also from (\ref{contour-ads3-after-CHM}) as follows. 
	First one observes that the transformation $x_{\pm} |_{w_\pm =0}$ 
	from a time slice of the boundary of AdS$_3$ to a time slice of the boundary of a BTZ black brane,
	obtained from the first equation in (\ref{chm-ads-inside-inverse}),
	is such that $x_+ =0$ is the image of $y_+=0$, while both $y_- \to \pm \infty$ are sent into $x_- =0$. 
	Hence, $x_{+} |_{w_+ =0}$  maps $ \big[\! - b + \varepsilon_{\textrm{\tiny bdy}}^{\textrm{\tiny $A$}}  \, ,\, b - \varepsilon_{\textrm{\tiny bdy}}^{\textrm{\tiny $A$}} \big]  $ 
	onto $\big[ - x_{0,+} \, , x_{0,+} \big]$
	while $x_{-} |_{w_- =0}$ sends $\big( \!- \!\infty, - b  - \varepsilon_{\textrm{\tiny bdy}}^{\textrm{\tiny $B$}}   \big]  
	\cup \big[ b +  \varepsilon_{\textrm{\tiny bdy}}^{\textrm{\tiny $B$}}  , +\infty \big)$ 
	onto $\big[  -x_{0,-} \, , x_{0,-} \big]$,
	where $x_{0,+} = x_{0,-} $ are given by 
	\be 
	\label{x0-from-CHM}
	x_{0,\pm} \, \equiv  \,
	b \, \log \!\left(\frac{ b + \sqrt{b^2-\varepsilon_{\textrm{\tiny AdS}}^2} }{\varepsilon_{\textrm{\tiny AdS}} }\right)
	\ee 
	that label the points in the boundary of the two BTZ backgrounds identified by \eqref{chm-ads-in-out} and \eqref{chm-ads-inside-inverse}. 
	From this observation and the holographic contour function (\ref{contour-ads3-after-CHM}), one finds
	\be 
	\label{SA-SB-from-C}
	S_A 
	= \int_{-x_{0,+}}^{x_{0,+} } \! \! \mathcal{C}^{(+)}_{\textrm{\tiny BTZ}} (x)\, \rd x
	= 
	\frac{\pi c_{\textrm{\tiny BH}} }{3\beta_0}\;2 x_{0,+}
	\;\;\qquad\;\;
	S_B 
	= \int_{-x_{0,-}}^{x_{0,-} } \! \! \mathcal{C}^{(-)}_{\textrm{\tiny BTZ}} (x)\, \rd x
	= 
	\frac{\pi c_{\textrm{\tiny BH}}}{3\beta_0}\;2 x_{0,-}
	\ee 
	which are equal because of (\ref{x0-from-CHM})
	and coincide with (\ref{HEE-SA-full}) and (\ref{HEE-SB-full}) respectively. 
	
	These considerations tell us that 
	the holographic entanglement entropy can be interpreted as a thermal entropy 
	in the setup explored throughout this section, as first realized in \cite{Casini:2011kv}.

	\section{BTZ black brane}
	\label{sec-BTZ}

	In this section, we study the geodesic bit threads for an interval in the constant time slice of the BTZ black brane. 
	In Sec.\,\ref{subsec-BTZ-plane-GBT} the vector field characterizing these geodesic bit threads 
	and the corresponding auxiliary geodesics are constructed,
	revisiting the corresponding analyses reported in \cite{Agon:2018lwq} and in \cite{Mintchev:2022fcp}.
	We explore their fluxes through some relevant domains in Sec.\,\ref{subsec-flows-btz-planar}.

	
	The metric on a constant time slice of the BTZ black brane reads
	\be
	\label{btz-brane-metric}
	ds^2 = \frac{L^2_{\textrm{\tiny AdS}}}{z^2} 
	\left( 
	\frac{\rd z^2}{1-(z/z_h)^2} + \rd x^2  \right)
	\ee
	where $x\in \RR$,  the holographic coordinate is $z>0$
	and $z=z_h$ gives the position of the planar horizon.
	According to the AdS/CFT dictionary, 
	the dual CFT$_2$ on the boundary of the BTZ black brane (i.e. at $z \to 0^+$)
	is defined in the two-dimensional Minkowski space described by the coordinates $(t, x)$
	and has finite temperature $1/\beta$,
	which is related to the position of the horizon by $z_h = \beta/(2\pi)$.
	In the vanishing temperature limit,
	we have $z_h \to +\infty$ and the holographic setup considered in Sec.\,\ref{sec-AdS3} is recovered.
	
%

	On the boundary of the constant time slice of the BTZ black brane geometry
	described by (\ref{btz-brane-metric}), which is a line parameterized by $x$,
	let us introduce a bipartition given by an interval $A$ and its complement $B =\RR\setminus A$. 
	The spatial translation invariance of the CFT$_2$ on the boundary
	allows us to choose $A\equiv [-b, b]$ with $b>0$
	without loss of generality, like in Sec.\,\ref{sec-AdS3}.
	The RT curve $\gamma_A$ for the interval $A$,
	whose regularized length provides the holographic entanglement entropy 
	of $A$ in the CFT$_2$ at finite temperature, is given by 
	\be
	\label{RT-curve-btz-brane}
	\gamma_A\; :
	\qquad
	z_m(x_m) \,=\, z_h\, \frac{\sqrt{ \cosh(2b/z_h) - \cosh(2x_m/z_h) }}{\sqrt{2} \,\cosh(b/z_h)}
	\ee
	where $x_m\in A$ and $P_m =(z_m, x_m)$ corresponds to the generic point of $\gamma_A$.
	The maximum value reached by $z_m$ on $\gamma_A$
	is $ z_\ast  \equiv z_h \tanh(b/z_h)$.

	\subsection{Geodesic bit threads}
	\label{subsec-BTZ-plane-GBT}


	In the constant time slice of the BTZ black brane,
	whose metric is (\ref{btz-brane-metric}),
	the geodesics are solutions of the following differential equation
	\cite{Cruz:1994ir,Arefeva:2017pho,Balasubramanian:2012tu}
	\be
	\label{ode-geodesic-btz-brane}
	z \,\sqrt{  1+ \frac{(z')^2}{1-(z/z_h)^2} } \,=\, C
	\ee
	where $C$ is a constant.
	The integral lines of the geodesic bit threads associated with the spatial bipartition 
	of the boundary given by the interval $A$
	belong to the class of the geodesics of (\ref{btz-brane-metric})
	which have one endpoint on the boundary at $z=0$
	and the other endpoint either on the boundary (type I) 
	or on the horizon (type II).
	%

	
	Let us consider first the type I geodesics, described by $z=z_{\textrm{\tiny I}}(x)$,
	with the endpoints on the boundary at $x=p$ and $x=q$, with $p<q$.
	The separation of these two endpoints along the boundary is 
	$2 b_0 \equiv q-p$ and their midpoint on the boundary is located at $c_0 \equiv (p+q)/2$.
	The maximum value $\tilde{z}_\ast < z_h$ for $z=z_{\textrm{\tiny I}}(x)$  is reached at $x=c_0$,
	where we have  $\tilde{z}_\ast \equiv z_{\textrm{\tiny I}}(c_0) $ and $z'_{\textrm{\tiny I}}(c_0) = 0$.
	The geodesic of (\ref{btz-brane-metric}) having 
	both the endpoints on the boundary, at $x=c_0\pm b_0$, is given by
	\be
	\label{geod-btz-gen-2bdy}
	z_{\textrm{\tiny I}}(x) 
	=
	z_h\, \frac{\sqrt{ \cosh(2b_0/z_h) - \cosh(2(x-c_0)/z_h) }}{\sqrt{2} \,\cosh(b_0/z_h)}
	\ee
	which satisfies (\ref{ode-geodesic-btz-brane}) with $C = C_{\textrm{\tiny I}} \equiv z_h \tanh(b_0/z_h)$,
	that also provides the maximum value of $z$ along the geodesic, i.e. $\tilde{z}_\ast = C_{\textrm{\tiny I}}$.
	Notice that $C_{\textrm{\tiny I}} \to z_h^-$ as $b_0 \to +\infty$.
	The RT curve (\ref{RT-curve-btz-brane})
	corresponds to (\ref{geod-btz-gen-2bdy}) in the special case where $c_0 = 0$ and $b_0=b$.


	The geodesics of (\ref{btz-brane-metric}) 
	having one endpoint on the boundary and the other one on the horizon
	are described by $z=z_{\textrm{\tiny II}}(x)$, where 
	\be
	\label{geod-btz-gen-bdy-hor}
	z_{\textrm{\tiny II}}(x) 
	=
	z_h\, \frac{\sqrt{ \cosh(2b_0/z_h) - \cosh(2(x-c_0)/z_h) }}{\sqrt{2} \,\sinh(b_0/z_h)}
	\ee
	which satisfies (\ref{ode-geodesic-btz-brane}) with $C = C_{\textrm{\tiny II}} \equiv z_h \coth(b_0/z_h) > z_h$.
	In this case $C_{\textrm{\tiny II}} \to z_h^+$ as $b_0 \to +\infty$.
	The maximum value $\tilde{z}_\ast$ for $z=z_{\textrm{\tiny II}}(x)$ is reached at $x=c_0$
	and corresponds to the horizon.
	Indeed, from (\ref{geod-btz-gen-bdy-hor}) one finds
	\be
	z_{\textrm{\tiny II}}(c_0) = z_h = \tilde{z}_\ast 
	\;\;\;\;\qquad\;\;\;\;
	z'_{\textrm{\tiny II}}(c_0) = 0\,.
	\ee
	For these geodesics, $b_0 >0 $ gives
	the separation between the values of the $x$-coordinate of the two endpoints.
	Notice  that (\ref{geod-btz-gen-bdy-hor})  
	can be obtained also by replacing $b_0 \mapsto b_0+ \ri \, \pi z_h/2$ 
	and $c_0 \mapsto c_0 + \ri \, \pi z_h/2$ in (\ref{geod-btz-gen-2bdy}).
	Given $b_0>0$ and $c_0 \in \RR$,
	the profile  \eqref{geod-btz-gen-bdy-hor} provides 
	two different geodesics reaching the horizon at $x=c_0$ and with the other endpoint on the boundary,
	depending on whether the endpoint on the boundary has either $x=c_0+b_0$ or $x=c_0-b_0$.

	
	Another type of geodesics of \eqref{btz-brane-metric} occurring in our analysis
	is obtained as a limiting case of either (\ref{geod-btz-gen-2bdy}) or (\ref{geod-btz-gen-bdy-hor}).
	It corresponds to the geodesics with one endpoint on the boundary 
	at a finite $x$-coordinate $x=s$, while the $x$-coordinate of the other endpoint diverges,
	either at $x \to +\infty$ or at $x \to -\infty$.
	These geodesics are given by 
	\be
	\label{geod-btz-gen-bdy-inf}
	z_{\textrm{\tiny I/II}}^\pm (x) 
	= 
	z_h \sqrt{1-e^{ \pm 2(s-x)/z_h}}
	\ee 
	where $\pm$ corresponds to the sign of the divergence in the $x$-coordinate of the second endpoint. 
	The expression (\ref{geod-btz-gen-bdy-inf}) can be obtained by taking the limit 
	$ c_0 \pm  b_0 \to \pm \infty $ with $c_0 \mp  b_0 = s$ kept fixed
	in either \eqref{geod-btz-gen-2bdy} or \eqref{geod-btz-gen-bdy-hor}.

	
	The geodesic bit threads associated with the interval $A$ 
	in the constant time slice of the BTZ black brane (\ref{btz-brane-metric})
	are geodesics belonging to the union of the classes of geodesics described above
	(see (\ref{geod-btz-gen-2bdy}), (\ref{geod-btz-gen-bdy-hor}) and (\ref{geod-btz-gen-bdy-inf}))
	which intersect $\gamma_A$ orthogonally. 
	Thus, the profile $z=z(x)$ of 
	the integral lines of the geodesic bit threads must satisfy
	\be
	\label{intersect-ortho-condition}
	\left\{ \begin{array}{l}
		z(x_m) = z_{m}(x_m)
		\\
		\rule{0pt}{.5cm}
		\big[ \,g_{xx} + g_{zz} \, z'(x) \,z'_{m}(x) \,\big] \big|_{(z,x)=(z_m(x_m),x_m)} = 0
	\end{array}\right.
	\ee
	where $g_{xx} = 1/z^2$ and $g_{zz} = 1/\big[1-(z/z_h)^2\big]$ are the diagonal components of the metric (\ref{btz-brane-metric})
	and $(z_m(x_m), x_m)$ is the generic point of $\gamma_A$  where the intersection occurs.

	The parameters $c_0$ and $b_0$ providing the integral lines of the geodesic bit threads are obtained by solving (\ref{intersect-ortho-condition}).
	Independently of whether such integral lines are described 
	by either (\ref{geod-btz-gen-2bdy}) or (\ref{geod-btz-gen-bdy-hor}),
	for $c_0$ and $b_0$ we find respectively
	\bea
	\label{c0-btz-brane}
	c_0  
	&=&
	x_m
	+ 
	\frac{z_h}{4} \, \log \!
	\left(
	\frac{ \e^{2 b/z_h}  + \e^{-2 b/z_h}  - 2\, \e^{-2 x_m/z_h}  }{ \e^{2 b/z_h}  + \e^{-2 b/z_h}  - 2\, \e^{2 x_m/z_h}} 
	\right)^2
	\\
	\nonumber
	&=&
	x_m
	+ 
	\frac{z_h}{4} \, \log \!
	\left(
	\frac{ \cosh(2b/z_h) - \cosh(2x_m/z_h) + \sinh(2x_m/z_h)  }{  \cosh(2b/z_h) - \cosh(2x_m/z_h) - \sinh(2x_m/z_h)  } 
	\right)^2
	\\
	&=&
	x_m
	+ 
	\frac{z_h}{4} \, \log \!
	\left(
	\frac{  z_m(x_m)^2\cosh(b/z_h)^2 + z_h^2 \cosh(x_m/z_h) \sinh(x_m/z_h)  }{   z_m(x_m)^2 \cosh(b/z_h)^2 - z_h^2\cosh(x_m/z_h) \sinh(x_m/z_h)   } 
	\right)^2
	\nonumber
	\eea
	where the last expression is written in terms of (\ref{RT-curve-btz-brane}),
	and
	\be
	\label{b0-btz-brane}
	b_0  =  \frac{z_h}{4}\, \left|\,
	\log \!
	\left(
	\frac{
		z_m(x_m)\,\sinh (b/z_h)   + z_h\, \sinh (x_m/z_h)
	}{
		z_m(x_m)\,\sinh (b/z_h)   - z_h\, \sinh (x_m/z_h)
	}
	\right)^2 
	\; \right| \,.
	\ee
	As a consistency check, we notice that 
	the limit $z_h \to \infty$ of (\ref{c0-btz-brane}) and (\ref{b0-btz-brane})
	gives the corresponding expressions in (\ref{c0-b-gbt-ads3}) with $y_m$ replaced by $x_m$, as expected.

\begin{figure}[t!]
		\vspace{-.5cm}
		\hspace{-1.1cm}
		\includegraphics[width=1.15\textwidth]{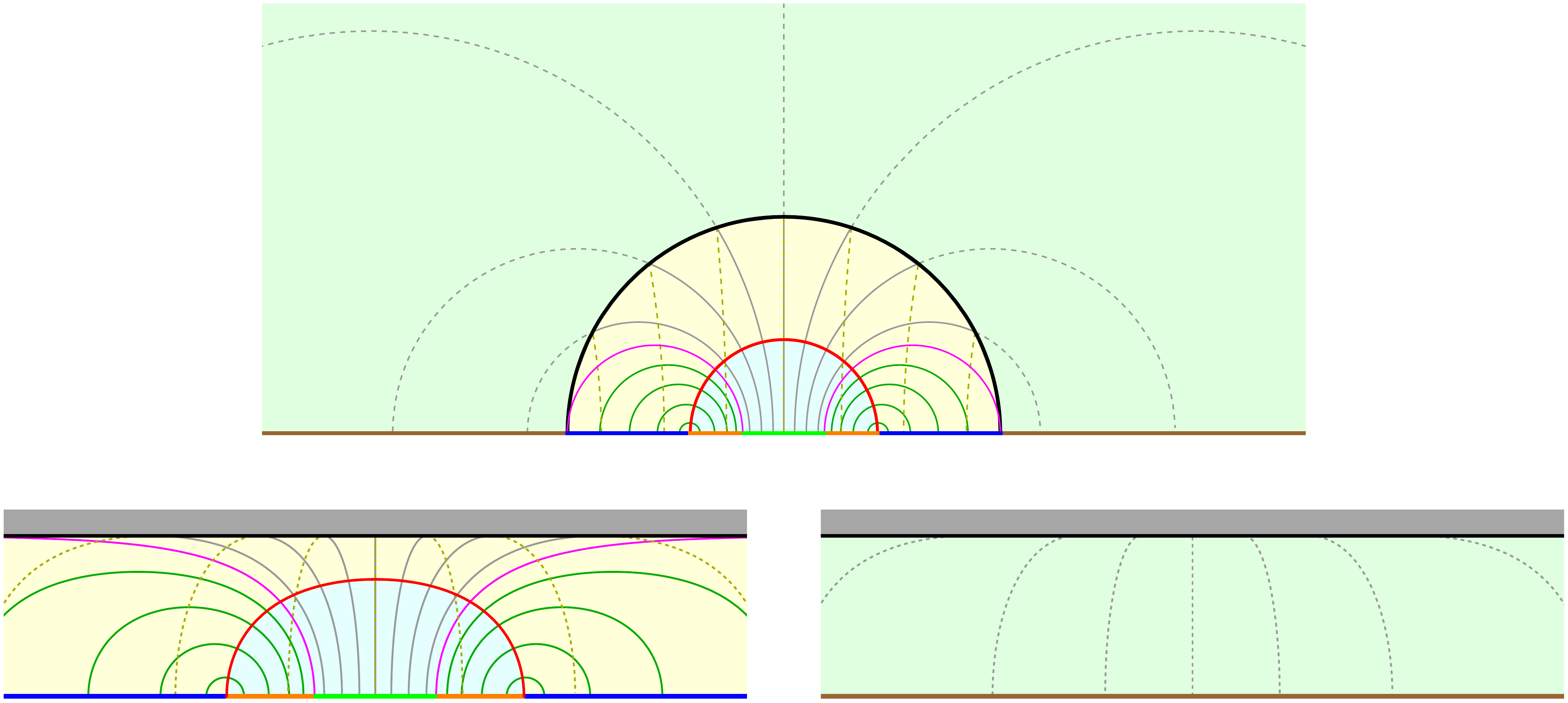}
		\vspace{-.2cm}
		\caption{\small Geodesic bit threads for an interval in the line, in the constant time slice of  the 
		BTZ black brane (bottom left panel).
		The bottom left and bottom right panels display BTZ$_+$ and BTZ$_-$ respectively,
		while the top panel shows the constant time slice of Poincar\'e AdS$_3$,
		which are related through (\ref{chm-btz-inside})-(\ref{chm-btz-inside-inverse}) (see also Fig.\,\ref{fig:ads3-main}).}
		\label{fig:BTZ-brane-main}
	\end{figure}

	
	In the bottom left panel of Fig.\,\ref{fig:BTZ-brane-main}, 
	the solid red curve corresponds to $\gamma_A$,
	while the integral lines of the corresponding geodesic bit threads,
	obtained from either (\ref{geod-btz-gen-2bdy}) or (\ref{geod-btz-gen-bdy-hor}) or (\ref{geod-btz-gen-bdy-inf})
	with the parameters $c_0$ and $b_0$ given by (\ref{c0-btz-brane}) and (\ref{b0-btz-brane}),
	are denoted by the solid green curves, the solid grey curves and the solid magenta curves respectively. 
	These integral lines have been found in \cite{Agon:2018lwq} and further discussed in \cite{Mintchev:2022fcp}.

	Let us consider the two curves of the form $z_{\textrm{\tiny I/II}}^\pm(x)$ defined in  (\ref{geod-btz-gen-bdy-inf}), 
	that separate the geodesic bit threads described by $z_{\textrm{\tiny I}}(x)$ 
	from the ones whose profile is given by $z_{\textrm{\tiny II}}(x)$ 
	(see (\ref{geod-btz-gen-2bdy}) and (\ref{geod-btz-gen-bdy-hor}) respectively)
	and which correspond to the solid magenta curves in the bottom left panel of Fig.\,\ref{fig:BTZ-brane-main}
	(see also Fig.\,4 of \cite{Agon:2018lwq}).
	These two special geodesic bit threads are described 
	by $z_{\textrm{\tiny I/II}}^+(x)$ with $s = b_\beta $
	and by $z_{\textrm{\tiny I/II}}^-(x)$ with $s = -b_\beta $,
	where \cite{Mintchev:2022fcp}
	\be
	b_\beta 
	=
	z_h  \log \! \big[\! \cosh(b/z_h) \big] 
	\label{bbeta}
	\ee
	that naturally identifies the interval $A_\beta \equiv [-b_\beta , b_\beta ] \subsetneq A$
	(see the green interval in the bottom left panel of Fig.\,\ref{fig:BTZ-brane-main}).
	The intersection between these two limiting geodesic bit threads
	and the RT curve (\ref{RT-curve-btz-brane}) is given by $(z_{m,\beta} , \pm \,x_{m,\beta} )$, where
	\cite{Agon:2018lwq}
	\be
	\label{x-z-beta-coord-def}
	x_{m,\beta} 
	\,\equiv \,
	z_h \,\textrm{arccoth}\big(z_h^2/z_\ast^2\big) > 0
	\;\;\; \qquad \;\;\;
	z_{m,\beta} 
	\,\equiv \,
	\frac{z_h \, z_\ast}{ \sqrt{z^2_h+ z^2_\ast}}
	\ee
	in terms of $z_\ast$ introduced in the text below (\ref{RT-curve-btz-brane}).
	Notice that $\pm x_{m,\beta} $ correspond to the values of $x_m$ 
	where the arguments of the logarithm in $c_0$ and $b_0$ (see (\ref{c0-btz-brane}) and (\ref{b0-btz-brane})) either vanish or diverge;
	hence both $c_0$ and $b_0$ diverge as $x_m \to \pm \,x_{m,\beta} $. 
	However, by considering the combinations $c_0 +b_0 $ and $c_0 - b_0 $  in this limit,
	we find that one diverges while the other becomes either $+b_\beta$ or $- b_\beta$.
	This is consistent with the fact that these special geodesic bit threads
	are limiting cases
	for the geodesic bit threads having both their endpoints on the boundary 
	and also for the geodesic bit threads with one endpoint on the boundary and the other on the horizon.
	Notice that $x_{m,\beta} /z_h$ and $z_{m,\beta} /z_h$ in (\ref{x-z-beta-coord-def}) are functions of $z_\ast/z_h$;
	hence, 
	since $z_\ast / z_h  = \tanh(b/z_h)$ (see the text below (\ref{RT-curve-btz-brane})),
	they can be written in terms of $b/z_h$ as follows
	\be
	\label{x-z-beta-coord-from-b}
	\frac{x_{m,\beta} }{z_h}
	=
	\frac{1}{2} \log \! \big[\!  \cosh(2 b/z_h)\big]
	\;\;\; \qquad \;\;\;
	\frac{z_{m,\beta} }{z_h}
	=
	\frac{\sinh(b/z_h)}{\sqrt{\cosh(2 b/z_h)}}
	\ee
	whose zero temperature limit give $x_{m,\beta} \to 0^+$ and $z_{m,\beta} \to b$, as expected.
	The expressions in (\ref{x-z-beta-coord-from-b})
	will be employed in Sec.\,\ref{sec-Sch-AdS-sphere} (see the solid lines in Fig.\,\ref{fig:bbetasphere}).

	Considering the $x$-coordinate $x_m$ of the intersection point 
	between a geodesic bit thread and its RT curve (\ref{RT-curve-btz-brane}),
	the geodesic bit thread is either of the form (\ref{geod-btz-gen-2bdy}) or of the form (\ref{geod-btz-gen-bdy-hor})
	depending on whether $|x_m| \in [ 0, x_{m,\beta} )$ or $|x_m| \in (x_{m,\beta} , b)$ respectively. 
	Given a geodesic bit thread having both its endpoints on the boundary,
	the maximum value reached by the $z$-coordinate along this curve
	can be obtained from (\ref{geod-btz-gen-2bdy}), \eqref{c0-btz-brane} and \eqref{b0-btz-brane},
	finding \cite{Agon:2018lwq}
	\be 
	\tilde{z}_\ast = z_{\textrm{\tiny I}}(c_0) = z_h \tanh(b_0/z_h)  = 
	\frac{z_\ast \,z_m }{\sqrt{z_\ast^2-z_m^2}} 
	\ee 
	(see also the text below (\ref{geod-btz-gen-2bdy})).
	As a consistency check of this result, by employing (\ref{x-z-beta-coord-def}), 
	we observe that $\tilde{z}_\ast \big|_{z_m = z_{m, \beta}} = z_h$, as expected.

	The changes of coordinates introduced in (\ref{chm-ads-in-out}) and (\ref{chm-ads-inside-inverse}) \cite{Casini:2011kv, Espindola:2018ozt},
	which provide the different panels in Fig.\,\ref{fig:ads3-main},
	can also be employed to explore the geodesics bit threads in the BTZ black brane we are investigating. 
	In particular, by renaming the coordinates of the BTZ black brane as $(z_+, x_+)=(z,x)$ (see (\ref{btz-brane-metric})),
	we can introduce
	\be
	\label{chm-btz-inside}
	y_\pm  =   \frac{z_h\, \sinh(x_\pm /z_h)}{ \cosh(x_\pm /z_h) \pm \sqrt{1-(z_\pm/z_h)^2} }
	\;\;\;\qquad\;\;\;
	w_\pm =  \frac{z_\pm}{ \cosh(x_\pm /z_h) \pm \sqrt{1-(z_\pm /z_h)^2} }
	\ee
	where $x_\pm\in \RR$ and $z_\pm >0$, whose inverse maps read
	\be
	\label{chm-btz-inside-inverse}
	x_\pm = \frac{z_h}{2}\, \log\! \left(  \frac{(z_h + y_\pm)^2 + w_\pm^2}{(z_h - y_\pm)^2 + w_\pm^2}  \right)
	\;\;\;\qquad\;\;\;
	z_\pm = \frac{2z_h^2\, w_\pm}{\sqrt{ z_h^4+2 z_h^2\big(w_\pm^2 - y_\pm^2\big) +\big(w_\pm^2 + y_\pm^2\big)^2 }} \,.
	\ee
	The changes of coordinates in (\ref{chm-btz-inside}) and (\ref{chm-btz-inside-inverse})
	are obtained by replacing $b$ with $z_h$
	in (\ref{chm-ads-in-out}) and (\ref{chm-ads-inside-inverse}) respectively;
	hence they satisfy (see (\ref{btz-brane-metric-CHM})) 
	\be
	\label{btz-brane-metric-CHM-bis}
	ds^2 
	= \frac{L^2_{\textrm{\tiny AdS}}}{w_\pm^2} \Big( \rd y_\pm^2 + \rd w_\pm^2 \Big)
	= \frac{L^2_{\textrm{\tiny AdS}}}{z^2} \left( 
	\rd x_\pm^2 + \frac{\rd z_\pm^2}{1-(z_\pm/z_h)^2} \,\right)
	\ee
	where $y_+^2 + w_+^2 < z_h^2$ and $y_-^2 + w_-^2 > z_h^2$.
	This leads us to consider two different BTZ black brane geometries 
	having the same horizon 
	parameterized by the coordinates $(z_+, x_+)$ and $(z_-, x_-)$,
	that we will be denoted by BTZ$_+$ and BTZ$_-$ respectively
	(see the bottom left and bottom right panel 
	of Fig.\,\ref{fig:BTZ-brane-main} respectively).
	The interval $A$ belongs to the boundary of BTZ$_+$;
	hence the corresponding RT curve (\ref{RT-curve-btz-brane})
	is embedded in the bulk of BTZ$_+$
	(see the red curve in the bottom left panel of Fig.\,\ref{fig:BTZ-brane-main}).
	The maps (\ref{chm-btz-inside}) send BTZ$_+$ and BTZ$_-$
	onto the two complementary regions $y_+^2 + w_+^2 < z_h^2$ 
	(see the union of the yellow and light blue domains in the top panel of Fig.\,\ref{fig:BTZ-brane-main})
	and $y_-^2 + w_-^2 > z_h^2$ (see the light green domain in the top panel of Fig.\,\ref{fig:BTZ-brane-main})
	respectively,
	whose union provides a bipartition of 
	a time slice of the Poincar\'e AdS$_3$ parameterized by $(w_\pm, y_\pm)$
	and shown in the top panel of Fig.\,\ref{fig:BTZ-brane-main}.
	The two domains of this bipartition share the black curve in the top panel of Fig.\,\ref{fig:BTZ-brane-main},
	which is mapped onto the horizons of BTZ$_+$ and BTZ$_-$
	in the bottom left and bottom right panels of the same figure. 
	Furthermore, from (\ref{chm-btz-inside}) we observe that
	\be
	y_{\mp} (x_\pm,z_\pm) = y_{\pm} (x_\pm +\ri \pi z_h ,-z_\pm)
	\;\;\qquad\;\;
	w_{\mp} (x_\pm,z_\pm) = w_{\pm} (x_\pm +\ri \pi z_h ,-z_\pm) \,.
	\ee

	The geodesic bit threads discussed in \cite{Agon:2018lwq} belong to BTZ$_+$.
	We find it worth considering also the auxiliary geodesics 
	associated with the geodesic bit threads that reach the horizon,
	which have been introduced in \cite{Mintchev:2022fcp}
	to provide a holographic interpretation of the 
	geometric action of the modular conjugation in CFT$_2$.
	Focussing on a geodesic bit thread reaching the horizon
	(see a solid grey curve in the bottom left panel of Fig.\,\ref{fig:BTZ-brane-main}), 
	i.e. in the form (\ref{geod-btz-gen-bdy-hor}),
	whose $x$-coordinate of the endpoints are  $x=c_0$ and $x=c_0 \mp b_0$
	with $c_0$ and $b_0$ given by  (\ref{c0-btz-brane}) and (\ref{b0-btz-brane}),
	the corresponding auxiliary geodesic is also described by (\ref{geod-btz-gen-bdy-hor})
	but the $x$-coordinates of  its endpoints are $x=c_0$ and $x=c_0 \pm b_0$.
	Hence, an auxiliary geodesic and the corresponding geodesic bit thread share their endpoint on the horizon,
	where they meet smoothly. 
	These auxiliary geodesics are denoted by the dashed dark yellow curves
	in the bottom left panel of Fig.\,\ref{fig:BTZ-brane-main}.
	
	The mappings (\ref{chm-btz-inside}) and their inverse (\ref{chm-btz-inside-inverse})
	provide an alternative interpretation of these auxiliary geodesics. 
	Recalling that the coordinates of the BTZ black brane (\ref{btz-brane-metric}) have been renamed as $(z_+, x_+)=(z,x)$,
	we first observe that \eqref{chm-btz-inside} send
	the interval $A$ (i.e. $|x_+| < b$) and the corresponding RT curve \eqref{RT-curve-btz-brane} 
	respectively 
	onto the interval $|y_+| < \tilde{b} $ on the boundary of Poincar\'e AdS$_3$
	shown in the top panel of Fig.\,\ref{fig:BTZ-brane-main}, where
	\be
	\tilde{b} \equiv   z_h \tanh (b/z_h) 
	\label{interval-before-chm}
	\ee 
	(see the union of the green and the orange segments in the top panel of Fig.\,\ref{fig:BTZ-brane-main})
	and onto its RT curve in the bulk of the Poincar\'e AdS$_3$ parameterized by $(y_\pm, w_\pm)$,
	i.e. the half circle \eqref{RT-curve-ads3} with $b$ replaced by $\tilde{b}$
	(see the red curve in the top panel of Fig.\,\ref{fig:BTZ-brane-main}).
	In this Poincar\'e AdS$_3$,
	the geodesic bit threads provided by the RT curve associated with the interval $|y_+| < \tilde{b} $
	(given by \eqref{gen-geo-ads3} and \eqref{c0-b-gbt-ads3} with $b$ replaced by $\tilde{b}$)
	within the region $y_+^2 + w_+^2 < z_h^2$ 
	are mapped through (\ref{chm-btz-inside-inverse})
	onto all the geodesic bit threads in BTZ$_+$,
	whose integral lines are obtained by plugging the parameters \eqref{c0-btz-brane} and \eqref{b0-btz-brane}
	into either \eqref{geod-btz-gen-2bdy} or \eqref{geod-btz-gen-bdy-hor}
	(see the solid green, magenta and grey curves in the bottom left panel and top panel of Fig.\,\ref{fig:BTZ-brane-main}).
	As for the arcs of the geodesic bit threads of the RT curve associated with the interval $|y_+| < \tilde{b} $
	within the region $y_-^2+w_-^2 >  z_h^2 $ of the Poincar\'e AdS$_3$ parameterized by $(w_\pm, y_\pm)$
	(see the dashed grey curves in the top panel of Fig.\,\ref{fig:BTZ-brane-main}),
	the maps \eqref{chm-btz-inside-inverse} send them 
	onto the auxiliary geodesics introduced above in BTZ$_-$.
	The endpoint on the horizon of an auxiliary geodesic in BTZ$_-$
	and the endpoint on the horizon of the corresponding geodesic bit thread in BTZ$_+$
	have the same value of the $x$-coordinate. 
	Thus, while 
	in \cite{Mintchev:2022fcp} the auxiliary geodesics are introduced as geodesics in BTZ$_+$
	(see the dashed dark yellow curves in Fig.\,\ref{fig:BTZ-brane-main}),
	in this alternative approach they belong to BTZ$_-$ 
	(see the dashed grey curves in Fig.\,\ref{fig:BTZ-brane-main}).

	The maps \eqref{chm-btz-inside}-\eqref{chm-btz-inside-inverse}
	allow us to write analytic expressions for the vector fields $\boldsymbol{V}_{\!\! \pm}$ in BTZ$_\pm$ 
	characterizing the geodesic bit threads and the corresponding auxiliary geodesics, 
	which can be written as
	\be
	\label{vector-field-btz}
	\boldsymbol{V}_{\!\! \pm} 
	=  
	\big( \, V_\pm^{z_\pm} , \, V_\pm^{x_\pm} \,\big) 
	=  
	|\boldsymbol{V}_{\!\! \pm}| \, \boldsymbol{\tau}^\mu_\pm   
	=
	|\boldsymbol{V}_{\!\! \pm}|
	\, \big( \,\tau^{z_\pm}_\pm  , \, \tau^{x_\pm}_\pm \,\big) 
	\ee
	where $|\boldsymbol{V}_{\!\! \pm}|$ and  $\boldsymbol{\tau}_\pm$ are 
	the modulus of $\boldsymbol{V}_{\!\! \pm}$ and its unit tangent vector respectively.
	Considering the vector field $\widetilde{\boldsymbol{V}} $
	of the geodesic bit threads in AdS$_3$ (see the top panel of Fig.\,\ref{fig:BTZ-brane-main})
	defined as \eqref{vector-field-ads3} with $b$ replaced by $\tilde{b}$ (see \eqref{interval-before-chm})
	and then applying the coordinate changes 	\eqref{chm-btz-inside}-\eqref{chm-btz-inside-inverse}
	to the resulting expression, 
	we find 
	\be 
	\label{mod-gbt-btz}
	|\boldsymbol{V}_{\!\! \pm}|  =  
	\frac{ (z_\pm /z_h) \sinh (b/z_h )
	}{ 
		\left\{  \left[\, \cosh ( (b-x_\pm )/z_h )  \mp \sqrt{1-(z_\pm /z_h)^2} \; \right] 
			\left[\, \cosh ((b+x_\pm )/z_h ) \mp \sqrt{1-(z_\pm /z_h)^2} \;\right] \right\}^{1/2}
		}
	\ee 
	and 
	\bea
	\label{tau-gbt-btz-components-2}
	\tau_\pm^{z_\pm}
	&=&  
	\frac{ |\boldsymbol{V}_{\!\! \pm} | }{L_{\textrm{\tiny AdS}} }  \;
	\frac{ \sqrt{z_h^2-z_\pm^2} }{\sinh (b/z_h) }
	\left[\,
	\pm \cosh (b/z_h )  - \sqrt{1-(z_\pm/z_h)^2} \, \cosh (x_\pm/z_h) 
	\,  \right]
	\\
	\rule{0pt}{1cm}
	\label{tau-gbt-btz-components-1}
	\tau_\pm^{x_\pm}
	&=& 
	\frac{ |\boldsymbol{V}_{\!\! \pm} | }{L_{\textrm{\tiny AdS}} } \;
	\frac{ z_\pm \sinh (x_\pm/z_h) }{ \sinh (b/z_h)}\,.
	\eea

	We remark that $|\boldsymbol{V}_+| =1 $ on $\gamma_A$,
	$|\boldsymbol{V}_+| < 1 $ elsewhere in BTZ$_+$ and 
	$|\boldsymbol{V}_-|<1$ everywhere in BTZ$_-$.
	This can be shown by observing that,
	denoting by $N_\pm$ and $D_\pm$ the numerator and the denominator in (\ref{mod-gbt-btz}) respectively, 
	we have that 
	\be
	\label{id-sq-D-N}
	D_\pm^2 - N_\pm^2 = 
	\left[ \,
	\cosh(x_\pm /z_h) \mp \sqrt{1- ( z_\pm/z_h)^2} \; \cosh(b/z_h)
	\,\right]^2
	\ee
	which implies 
	\be
	|\boldsymbol{V}_{\!\! \pm}|^2  =  1 - \frac{D_\pm^2 - N_\pm^2}{D_\pm^2} \,\leqslant\,1 \,.
	\ee
	Indeed, from (\ref{id-sq-D-N}) 
	it is straightforward to realize that $D_{-}^2 - N_{-}^2 >0 $ 
	and that $D_{+}^2 - N_{+}^2  = 0 $ on the RT curve \eqref{RT-curve-btz-brane}.
		On the horizon, i.e. in the limit $z_\pm \to z_h$,
		from (\ref{vector-field-btz})-(\ref{tau-gbt-btz-components-1}) 
		we observe that 
		both $\boldsymbol{V}_+$ and $\boldsymbol{V}_-$ 
		become the same expression,
		with vanishing $\tau_\pm^z$.

	An implicit expression for $\boldsymbol{V}_{\!\! +} $ in \eqref{vector-field-btz} 
	has been written in Eqs.\,(2.61) and (2.64) of \cite{Agon:2018lwq},
	while in \eqref{mod-gbt-btz}-\eqref{tau-gbt-btz-components-1}
	the dependence of the vector field on the spacetime point $(z,x)$ is explicit. 
	We have checked numerically the agreement between these two expressions.

	\subsection{Fluxes through the boundary and the horizon}
	\label{subsec-flows-btz-planar}

	In the following, we investigate the fluxes of $\boldsymbol{V}_{\!\! \pm} $ 
	given by (\ref{vector-field-btz})-(\ref{tau-gbt-btz-components-1})
	either through the boundary or through the horizon of BTZ$_\pm$.
	%
	
	
	From (\ref{btz-brane-metric}) and (\ref{vector-field-btz})-(\ref{tau-gbt-btz-components-1}), 
	we find that the flux $\Phi_\pm (x; z_{\pm,0}) $ of $\boldsymbol{V}_{\!\! \pm} $ 
	through a generic slice at constant $z_\pm = z_{\pm,0}$ reads
	\bea
	\label{btz-flow-zfixed}
	\Phi_\pm (x_\pm ; z_{\pm,0})  
	& \equiv &
	\frac{1}{4 G_{\textrm{\tiny N}}} \; g_{z_\pm z_\pm}  V^{z_\pm}_\pm  \, n^{z_\pm}  \sqrt{g_{x_\pm x_\pm}} \, \Big|_{z_\pm = z_{\pm,0}}
	\\
	\rule{0pt}{.9cm}
	& & 
	\hspace{-2.6cm}
	=
	\frac{L_{\textrm{\tiny AdS}} }{4 G_{\textrm{\tiny N}}}\;
	\frac{
		\big[ 	
		\pm \cosh (b/z_h )  - \sqrt{1-(z_{\pm,0}/z_h)^2}\, \cosh (x_\pm /z_h) 
		\, \big]
		\sinh(b/z_h)   
	}{ 
		z_h \! \left[\, \cosh ( (b-x_\pm)/z_h )  \mp \sqrt{1-(z_{\pm,0}/z_h)^2} \; \right] 
		\left[ \, \cosh ((b+x_\pm)/z_h) \mp \sqrt{1-(z_{\pm,0}/z_h)^2} \; \right]
	}
	\nn
	\eea 
	where $g_{zz}$ and $g_{xx}$ are the non vanishing elements of the metric \eqref{btz-brane-metric},
	$ \boldsymbol{n}  = (n^{z_\pm}, n^{x_\pm} ) $ is the unit normal vector to the constant $z_\pm$ slice
	and  $\sqrt{g_{x_\pm x_\pm}}$ comes from the volume element on the slice
	at constant $z_\pm = z_{\pm,0}$.

	As for the limit $z_{\pm,0} \,\to\, 0$ of \eqref{btz-flow-zfixed},
	by employing  the Brown-Henneaux central charge (\ref{BH-central-charge}) 
	and the relation $\beta = 2 \pi z_h $ (see the text below (\ref{btz-brane-metric})),
	for $x_\pm \in \RR$ we find 
	\be
	\label{contour-bdy-btz}
	\mathcal{C}_A^\pm (x_\pm)\,
	\equiv
	\lim_{z_{\pm,0} \,\to\, 0} 
	\big| \Phi_\pm (x_\pm ; z_{\pm,0})   \big| \,
	=\,
	\frac{\pi  \,c_{\textrm{\tiny BH}} }{3 \beta }\;
	\frac{  \sinh (2 \pi  b / \beta )}{ \big| \cosh (2 \pi  b / \beta ) \mp \cosh (2 \pi  x_\pm  /  \beta  ) \, \big|} \,.
	\ee 
	The expression of $\mathcal{C}_A^+ (x_+  )$ in (\ref{contour-bdy-btz}) for $x_+\in A$
	agrees with the contour function for the entanglement entropy of an interval
	in a line for a CFT$_2$ at finite temperature discussed in \cite{Cardy:2016fqc, Coser:2017dtb},
	specialized to the Brown-Henneaux central charge (\ref{BH-central-charge}).
	The zero temperature limit $\beta \to +\infty$ of (\ref{contour-bdy-btz}) gives
	\be
	\mathcal{C}_A^+ (x_+  ) \to \frac{c_{\textrm{\tiny BH}} }{6}\; \frac{2b}{\big| b^2 - x_+^2 \big|}
	\;\;\;\;\;\qquad\;\;\;\;\;
	\mathcal{C}_A^- (x_-  ) \to \,0
	\ee
	hence $\mathcal{C}_A^+ (x_+  ) $ provides (\ref{contour-function-vacuum}) in this limit, as expected. 
	
	We find it worth considering also the limit $z_{\pm ,0} \,\to\, z_h$,
	where the flux density \eqref{btz-flow-zfixed} becomes
	\be 
	\label{contour-horizon-btz}
	\mathcal{C}_h (x_\pm  )\,
	\equiv
	\lim_{z_{\pm,0} \,\to\,  z_h} 
	\! \big| \Phi_\pm (x_\pm ; z_{\pm,0})   \big| \,
	= 
	\frac{ \pi \,c_{\textrm{\tiny BH}} }{3 \beta}\;
	\frac{ \sinh (4 \pi  b/ \beta )}{ \cosh (4 \pi  b/ \beta )+\cosh (4 \pi  x_\pm /\beta )} \,.
	\ee

	To evaluate the holographic entanglement entropy of the interval $A$ 
	by using the flux $\mathcal{C}_A^+ (x_+  )$ through the boundary of BTZ$_+$ in (\ref{contour-bdy-btz}),
	a UV regularisation must be introduced.
	We adopt again the entanglement wedge cross-section regularization 
	\cite{Dutta:2019gen,Han:2019scu,Headrick:2022nbe, Nguyen:2017yqw,Takayanagi:2017knl};
	i.e. we adapt to the BTZ black brane background
	the procedure described in Sec.\,\ref{subsec-flow-bdy-ads3} to obtain  \eqref{ads-EWCS-regEndp}.
	After introducing the cutoff in the bulk at $z=\varepsilon_{\textrm{\tiny BTZ}} $, 
	with $\varepsilon_{\textrm{\tiny BTZ}} \ll 1$,
	the entanglement wedge cross-section regularization prescription provides
	the interval
	$A_\varepsilon \equiv \big[\! -b + \varepsilon_{\textrm{\tiny bdy}}^{\textrm{\tiny $A$}} \, , \, b - \varepsilon_{\textrm{\tiny bdy}}^{\textrm{\tiny $A$}}  \big] \subsetneq A$ 
	and the domain 
	$B_\varepsilon \equiv \big( \!  -\infty \,,\, -b - \varepsilon_{\textrm{\tiny bdy}}^{\textrm{\tiny $B$}}  \big]
	\cup \big[  b + \varepsilon_{\textrm{\tiny bdy}}^{\textrm{\tiny $B$}}  \,, +\infty \big) \subsetneq B$,
	where $\varepsilon_{\textrm{\tiny bdy}}^{\textrm{\tiny $A$}}   $ and $\varepsilon_{\textrm{\tiny bdy}}^{\textrm{\tiny $B$}}   $ are given by 
	\be
	\label{pBTZ1-eps-AB}
	b - \varepsilon_{\textrm{\tiny bdy}}^{\textrm{\tiny $A$}}   
	= 
	z_h \, \log\!  \left( \frac{ \mathcal{P}_+^{1/2} + \mathcal{P}_-^{1/2} \tanh (b/z_h)  }{ \mathcal{P}_+^{1/2} - \mathcal{P}_-^{1/2} \tanh (b/z_h)  } \right)
	\qquad
	b + \varepsilon_{\textrm{\tiny bdy}}^{\textrm{\tiny $B$}}   
	= 
	z_h \, \log\!  \left( \frac{ \mathcal{P}_-^{1/2} + \mathcal{P}_+^{1/2} \tanh (b/z_h)  }{ \mathcal{P}_-^{1/2} - \mathcal{P}_+^{1/2} \tanh (b/z_h)  } \right)
	\ee
	in terms of 
	\be
	\label{calP-pm-def}
	\mathcal{P}_\pm \equiv \sqrt{z_h^2-\varepsilon_{\textrm{\tiny BTZ}}^2} \, \sinh (b/z_h) \pm \varepsilon_{\textrm{\tiny BTZ}} \,.
	\ee
	The limit $z_h \to +\infty$ of (\ref{pBTZ1-eps-AB}) gives the corresponding expressions in (\ref{ads-EWCS-regEndp}), as expected.

	The flux $\mathcal{C}_A^+ (x_+  )$ in \eqref{contour-bdy-btz} through  $A_\varepsilon$
	provides the holographic entanglement entropy of $A$
	\bea
	\label{hee-thermal-from-C}
	S_A 
	\,= 
	\int_{- b + \varepsilon_{\textrm{\tiny bdy}}^{\textrm{\tiny $A$}}   }^{b - \varepsilon_{\textrm{\tiny bdy}}^{\textrm{\tiny $A$}}  } 
	\! \mathcal{C}^+_A (x_+  )\, \rd x_+
	&=&
	\left[ \,
	\frac{c_{\textrm{\tiny BH}}}{6}\, 
	\log \!
	\left(
	\frac{
		\sinh ( \pi(x_+ + b)/ \beta )
	}{
		\sinh ( \pi(b-x_+ )/ \beta )
	}
	\right) 
	\, \right]_{- b + \varepsilon_{\textrm{\tiny bdy}}^{\textrm{\tiny $A$}}   }^{b - \varepsilon_{\textrm{\tiny bdy}}^{\textrm{\tiny $A$}}  }  
	\nn
	\\ 
	\rule{0pt}{.7cm}
	&=&
	\frac{c_{\textrm{\tiny BH}}}{3} \,
	\log \! \left( \frac{\beta}{\pi \varepsilon_{\textrm{\tiny BTZ}}} \sinh( 2 \pi b / \beta ) \right) 
	+ 
	R_A(\varepsilon_{\textrm{\tiny BTZ}})
	\eea 
	where 
	\be 
	R_A(\varepsilon_{\textrm{\tiny BTZ}}) 
	\equiv
	\frac{c_{\textrm{\tiny BH}} }{3}  \,
	\log\! \left[\,
	\frac{\pi}{\beta} 
	\left(
	\sqrt{ \big[ ( \beta/2 \pi)^2 -\varepsilon_{\textrm{\tiny BTZ}}^2 \big]  -\big[\varepsilon_{\textrm{\tiny BTZ}} /\sinh (2 \pi  b/\beta  ) \big]^2   }  
	+\sqrt{ ( \beta/2 \pi)^2 -\varepsilon_{\textrm{\tiny BTZ}}^2  } \,\right)
	\right]
	\ee 
	which vanishes as $\varepsilon_{\textrm{\tiny BTZ}} \to 0$.
	The leading terms in (\ref{hee-thermal-from-C}) agree with the corresponding expression in CFT$_2$  \cite{Calabrese:2004eu}
	specialized to the Brown-Henneaux central charge (\ref{BH-central-charge}).

	The flux of $\mathcal{C}_A^+ (x_+  )$ in \eqref{contour-bdy-btz} through  $A_\beta $ 
	(see \eqref{bbeta} and the green interval in the boundary in the bottom left panel of Fig.\,\ref{fig:BTZ-brane-main})
	provides the holographic thermal entropy of $A$
	\cite{Mintchev:2022fcp}
	\be 
	\label{Gibbs-ent-definition}
	S_{A, \textrm{\tiny th}}
	\equiv 
	\int_{-b_\beta}^{b_\beta} 
	\mathcal{C}^+_A (x_+  )\, \rd x_+
	\,=\,
	2 b \,s_{\textrm{\tiny th}} 
	=\,
	\frac{\pi c_{\textrm{\tiny BH}} }{3 \beta }\, 2b
	\ee 
	in terms of the Stefan-Boltzmann entropy density 
	$s_{\textrm{\tiny th}}$ for a CFT$_2$ (see \eqref{SBthermalentropy})
	specialized to the Brown-Henneaux central charge (\ref{BH-central-charge}).
	Notice that $S_{A, \textrm{\tiny th}}$ in (\ref{Gibbs-ent-definition}) is UV finite.

	We find it instructive to consider also 
	the flux $\mathcal{C}_A^+ (x_+  )$ in \eqref{contour-bdy-btz} through  $B_\varepsilon$, namely
	\bea
	\label{SB-tilde}
	\widetilde{S}_B 
	&\equiv &
	\int_{- \infty }^{-b - \varepsilon_{\textrm{\tiny bdy}}^{\textrm{\tiny $B$}}  } \!
	\mathcal{C}^+_A (x_+  )\, \rd x_+
	+ 
	\int_{ b + \varepsilon_{\textrm{\tiny bdy}}^{\textrm{\tiny $B$}}  }^{\infty } \!\!
	\mathcal{C}^+_A (x_+ )\, \rd x_+
	\nn
	\\ 
	\rule{0pt}{.7cm}
	& = &
	2
	\left[ \,
	\frac{c_{\textrm{\tiny BH}}}{6} \,
	\log \!
	\left(
	\frac{
		\sinh ( \pi(x_+-b)/ \beta )
	}{
		\sinh ( \pi( x_+ + b )/ \beta )
	}
	\right) 
	\,\right]_{ b + \varepsilon_{\textrm{\tiny bdy}}^{\textrm{\tiny $B$}}   }^{ +\infty  }
	\! =  \,	
	S_A - S_{A, \textrm{\tiny th}}
	\eea  
	where  $S_A$ and $S_{A, \textrm{\tiny th}}$ are given in 
	(\ref{hee-thermal-from-C})) and (\ref{Gibbs-ent-definition}) respectively. 
	Notice that (\ref{SB-tilde}) is different from the entanglement entropy $S_B$ of the complement domain $B = \mathbb{R} \setminus A$,
	which is discussed in Appendix\;\ref{app-entropy-complementary-btz} for completeness.

	All the geodesic bit threads that start in $A_\beta$ arrive at the horizon of BTZ$_+$
	(see the solid grey geodesics in the bottom left panel of Fig.\,\ref{fig:BTZ-brane-main}),
	providing a bijective correspondence between these two domains.
	Similarly, 
	in BTZ$_-$ the corresponding auxiliary geodesics connect the entire horizon with the whole boundary
	in a bijective way (see the dashed grey geodesics in the bottom right panel of Fig.\,\ref{fig:BTZ-brane-main}).
	These observations imply 
		\be
	\label{eq:PhiHorizonIntegral}
	S_{A, \textrm{\tiny th}}
	= \int_{-\infty}^{\infty} \mathcal{C}_h (x_\pm  ) \, \rd x_\pm
	= \int_{-\infty}^{\infty} \mathcal{C}^-_A (x_-  )\, \rd x_-
	= \frac{\pi \,c_{\textrm{\tiny BH}} }{3 \beta } \; 2b
	\ee 
	meaning that 
	another way to obtain 
	the holographic thermal entropy $S_{A, \textrm{\tiny th}}$ in \eqref{Gibbs-ent-definition} 
	is	
	either as the flux of (\ref{contour-horizon-btz}) through the horizon 
	or as the flux of $\mathcal{C}_A^- (x_-  )$ (see \eqref{contour-bdy-btz})  through the boundary of BTZ$_-$.


	\section{AdS$_{d+2}$}
	\label{HigherdimAdS}

	
	In this section we extend the analysis discussed in Sec.\,\ref{sec-AdS3} to higher dimensions
	by considering a gravity model in AdS$_{d+2}$.
	This geometry is described by Poincaré coordinates $(t,w, \boldsymbol{y})$,
	where $w>0$ serves as the holographic coordinate
	and $\boldsymbol{y}\in \RR^d$.
	The dual CFT$_{d+1}$ lives in the flat $(d+1)$-dimensional Minkowski spacetime
	located at the boundary $w \to 0^+$, which is parameterized by the coordinates $(t,\boldsymbol{y})$. 
	
	On a constant time slice of AdS$_{d+2}$ the induced metric is
	\be
	\label{fAdS-metric}
	ds^2 = \frac{L^2_{\textrm{\tiny AdS}}}{w^2} \,\big( \rd w^2 + \rd \boldsymbol{y}^2  \big) 
	\ee
	which characterizes the $(d+1)$-dimensional Euclidean hyperbolic space $\mathbb{H}_{d+1}$.
	When our geometrical configuration exhibits an obvious spherical symmetry, it is convenient to represent $\boldsymbol{y}$ in polar coordinates $\boldsymbol{r} \equiv (r, \boldsymbol{\Omega})$, where $r>0$ is the radial coordinate and $\boldsymbol{\Omega}$ is a set of angular coordinates describing the $(d-1)$-dimensional unit sphere. 
	Thus, the metric \eqref{fAdS-metric} becomes
	\be
	\label{Hd-metric}
	ds^2 =
	\frac{L^2_{\textrm{\tiny AdS}}}{w^2} \,\big( \rd w^2 + \rd \boldsymbol{r}^2 \big)
	\ee
	where the line element $\rd \boldsymbol{y}^2$ in \eqref{fAdS-metric} 
	has been replaced by $\rd \boldsymbol{r}^2 \equiv \rd r^2 + r^2 \rd \boldsymbol{\Omega}^2$, 
	being $\rd \boldsymbol{\Omega}^2$ defined as the metric on the unit $(d-1)$-dimensional sphere.

	\subsection {Sphere}
	\label{sec-ads-gbt-sphere}


	In the constant time slice of the dual CFT$_{d+1}$ living on the boundary of AdS$_{d+2}$,
	let us consider the spatial bipartition of $\mathbb{R}^d$ 
	provided  by a sphere $A$  centered at the origin with radius $b>0$  and its complement  $B \equiv \RR^d \setminus A$.  
	The spherical symmetry of this configuration induces to adopt the polar coordinates in $\RR^d$;
	hence, a point in $\mathbb{H}_{d+1}$ (the constant time slice) is identified by the coordinates $(w, \boldsymbol{r} )$.
	%
	The corresponding RT hypersurface $\gamma_A$,
	whose regularized area gives the holographic entanglement entropy for this setup,
	is the following hemisphere in $\mathbb{H}_{d+1}$  \cite{Ryu:2006bv, Ryu:2006ef} 
	\be
	\label{RT-curve-adsd}
	\gamma_A\; :
	\qquad
	w_m^2 + r_m^2 = b^2
	\ee
	where $r_m \in [0,b]$ and $P_m \equiv(w_m,\boldsymbol{r}_m)=
	(w_m,r_m,\boldsymbol{\Omega}_m)$ identifies a generic point of $\gamma_A$.

\begin{figure}[t!]
		\vspace{-.5cm}
		\hspace{-0.6cm}
		\includegraphics[width=0.5\textwidth]{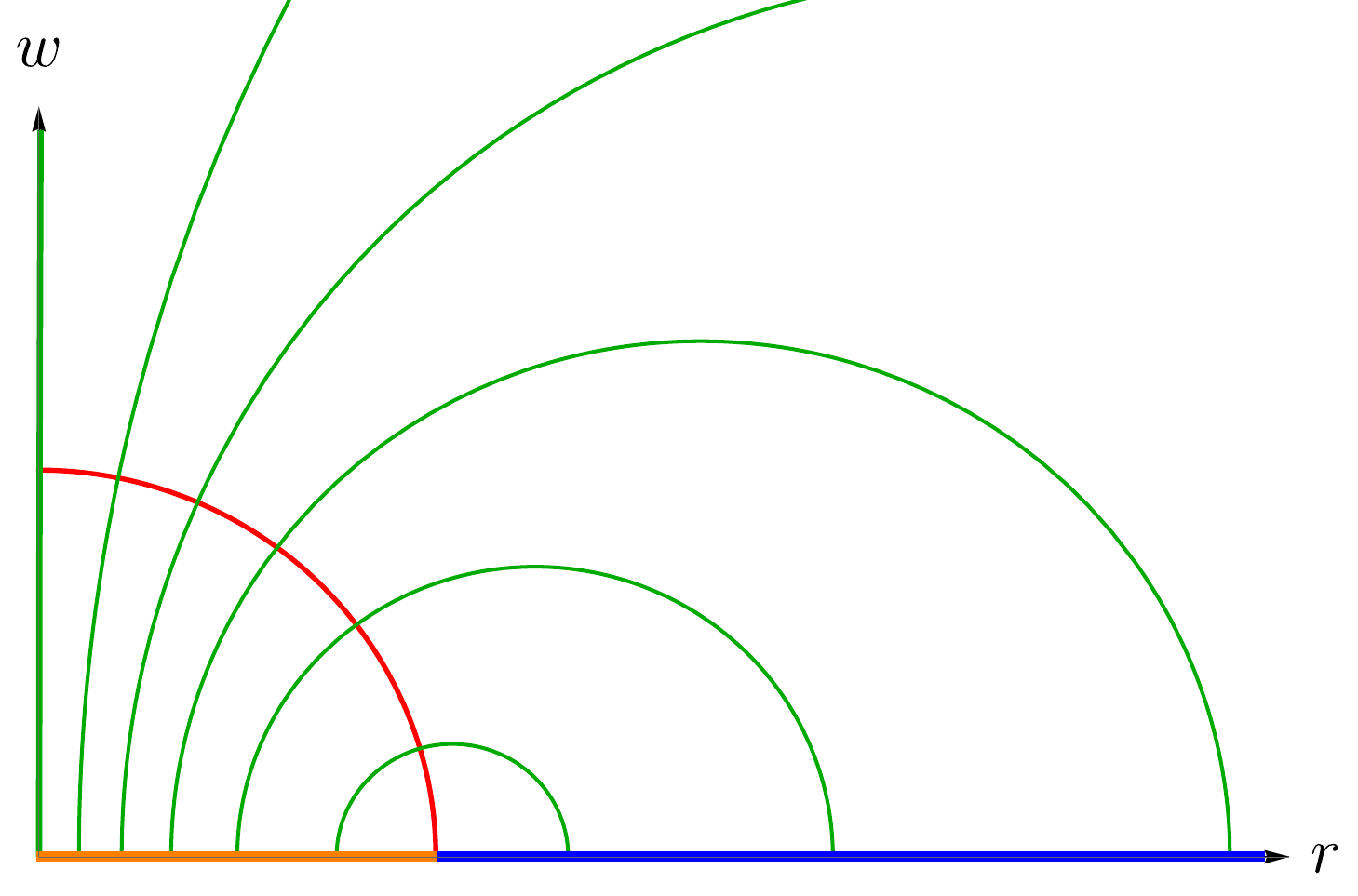}
		\hspace{0.4cm}
		\includegraphics[width=0.55\textwidth]{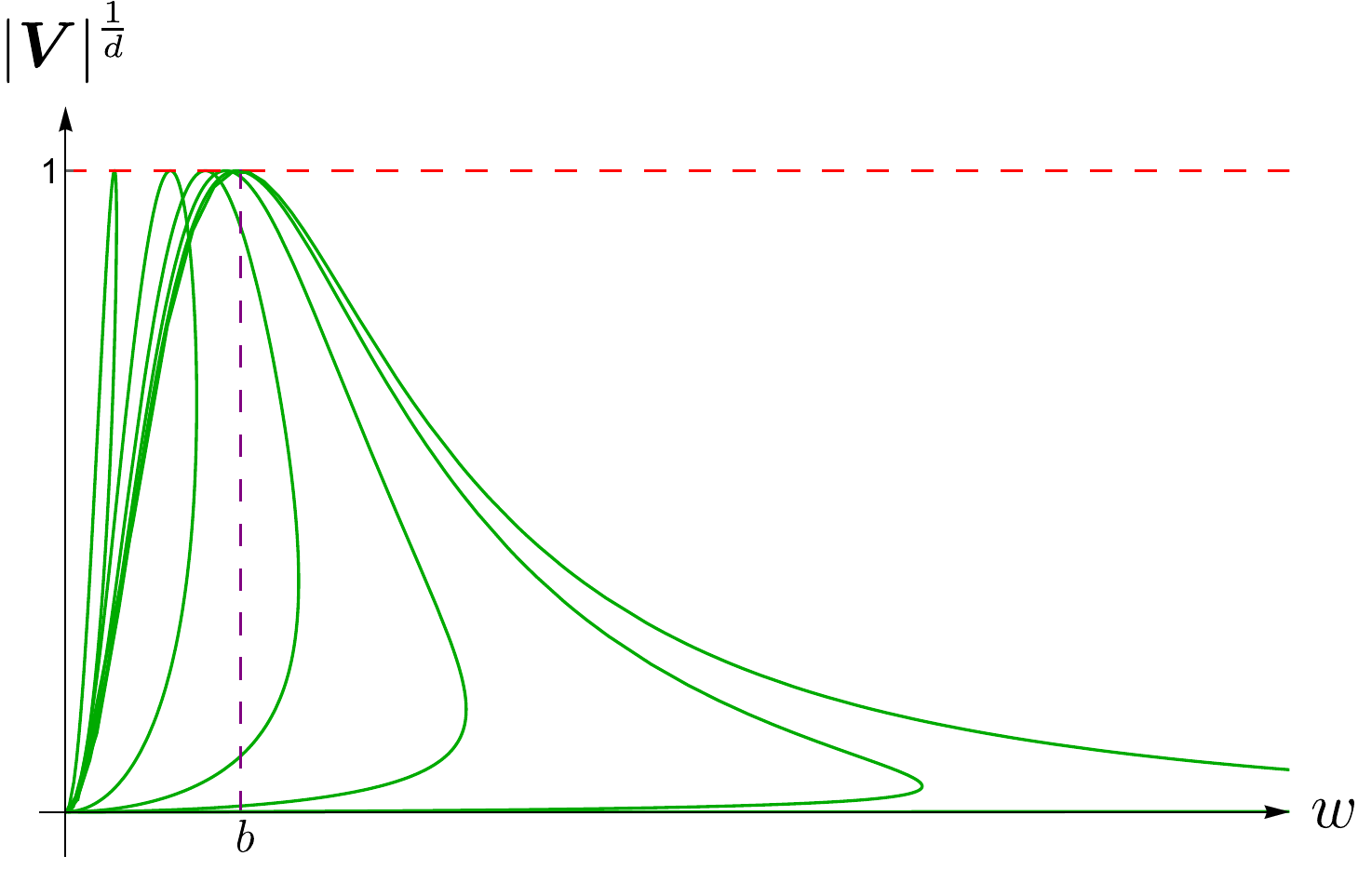}
		\vspace{-.4cm}
		\caption{
			\small
			Geodesic bit threads for a sphere in $\RR^d$, in the constant time slice of AdS$_{d+2}$.
			Left: Integral curves (solid green lines, see (\ref{geodesicAdSd})-(\ref{c0-b-gbt-adsd})).
			Right: $\big|\boldsymbol{V} \big|^{1/d}$ along the geodesic bit threads shown in the left panel 
			(see (\ref{mod-V-AdS-ddim})).
		}
		\label{fig:HigherDspherebitthreads}
	\end{figure}

	The construction of the geodesics bit threads for this configuration \cite{Agon:2018lwq} closely mimics the analysis of Sec.\,\ref{subsec-gbt-ads3}
	because the spherical symmetry allows us to suppress the angular dependence,
	focussing solely on geodesics lying in the $(w,r)$ plane. 
	In this plane, the integral lines of the geodesic bit threads are given by the following half circumferences
	\be
	\label{geodesicAdSd}
	w^2+(r-c_0)^2=b^2_0
	\ee
	where the parameters $c_0$ and $b_0$ are determined by imposing that the geodesic \eqref{geodesicAdSd} 
	intersects orthogonally $\gamma_A$  at  $(w_m, r_m)$.
	This requirement translates into two conditions identical to \eqref{intersect-ortho-condition-ads} with $y$ replaced by $r$. Solving these two conditions  leads to 
	\begin{equation}
		\label{c0-b-gbt-adsd}
		c_0= r_m+\frac{w_m^2}{r_m}= \frac{b^2}{r_m} 
		\;\;\;\qquad\;\;\;
		b_0= \frac{w_m}{r_m}\, b\,.
	\end{equation}
	The result of this construction is illustrated in the left panel of Fig.\,\ref{fig:HigherDspherebitthreads}, 
	where $A$ corresponds to the orange segment in the radial coordinate
	and the integral lines of its geodesic bit threads are depicted as green semicircles with their endpoints on the boundary. 
	One of these endpoints lies within $A$, while the other is situated in the complementary region $B$, marked by the blue half line. 
	The RT hypersurface (\ref{RT-curve-adsd}) corresponds to the red curve.

	Considering the two endpoints in $A$ and $B$ of the geodesic bit thread intersecting $\gamma_A$ at $(w_m, r_m)$,
	their radial coordinates are $r_{\textrm{\tiny $A$}} = {b(b-w_m)}/{r_m}$ and $r_{\textrm{\tiny $B$}} = {b(b+w_m)}/{r_m}$ respectively, 
	which satisfy
	\be
		\label{modconjsphere}
		r_{\textrm{\tiny $B$}}=
		\frac{b^2}{r_{\textrm{\tiny $A$}}}\;. 
	\ee
	
	By extending to arbitrary $d$ the observation made in \cite{Mintchev:2022fcp} for $d=1$ in a straightforward way,
	here we notice that (\ref{modconjsphere}) coincides 
	with the map implementing the geometric action of the modular conjugation in this setup
	\cite{Hislop:1981uh, Haag:1992hx};
	i.e. for a CFT$_{d+1}$ in the ground state and the bipartition given by a sphere of radius $b$.

	In this setup, let us consider
	the modular trajectory in the domain of dependence $\mathcal{D}_A$ of the sphere $A$
	generated by the modular evolution (i.e. through the modular Hamiltonian)
	of a point with spacetime coordinates $(t_0, r_0) \in \mathcal{D}_A$.
	In terms of the null coordinates $r_\pm \equiv r \pm t$,
	the generic point of this modular trajectory
	is given by \cite{Hislop:1981uh, Casini:2011kv}
	\be
	\label{mod-traj-diamond}
	r_\pm(\tau) = b\; 
	\frac{
	\big(b+r_{\pm,0} \big) - \e^{2\pi \tau} \big(b -r_{\pm,0}\big)
	}{
	\big(b+r_{\pm,0}\big) + \e^{2\pi \tau} \big(b -r_{\pm,0}\big)
	}
	\ee
	where $\tau \in \RR$ corresponds to the evolution parameter
	and $r_{\pm,0} \equiv r_0 \pm t_0$ are the null coordinates for the initial point;
	indeed (\ref{mod-traj-diamond}) 
	satisfy the required initial condition $r_\pm(\tau=0)=r_{\pm,0}$.
	The map $(t_0, r_0) \mapsto \big(\,\tilde{t}_0 , \tilde{r}_0 \big)$ 
	implementing the geometric action of the modular conjugation 
	can be obtained by analytically continuing $\tau$ in (\ref{mod-traj-diamond}) to $\tau =\ri/2$.
	Performing this substitution in (\ref{mod-traj-diamond}), one obtains 
	\be
	\tilde{t}_0 =  - \frac{b^2}{r_0^2 - t_0^2} \; t_0
	\;\;\;\;\qquad\;\;\;\;
	\tilde{r}_0 =  \frac{b^2}{r_0^2 - t_0^2} \; r_0\,.
	\ee
	On the constant time slice that we are considering, $t_0=0$;
	hence this map simplifies to $\tilde{t}_0 = 0$, as expected, 
	and $\tilde{r}_0 = b^2/r_0$, which coincides with the 
	relation (\ref{modconjsphere}) between the endpoints of the generic geodesic bit threads. 
	Thus, for  this case where $d >1$,
	the geodesic bit threads provide a gravitational dual of the 
	map in the dual CFT$_{d+1}$
	implementing the geometric action of the modular conjugation,
	as discussed in \cite{Mintchev:2022fcp} in various examples in $d=1$.

	
	The geodesic bit threads are characterized by the divergenceless vector field $\boldsymbol{V} $
	whose integral lines are given by \eqref{geodesicAdSd}-(\ref{c0-b-gbt-adsd})
	and whose normalization can be established by following 
	the procedure developed in  \cite{Agon:2018lwq}, which is briefly reviewed in Appendix\;\ref{app-modulus}. 
	This leads to the following expression \cite{Agon:2018lwq}
	\begin{equation}
		\label{Vd}
		\boldsymbol{V} =\big( V^w,V^r ,\boldsymbol{0}\big) =
		\frac{1}{L_{\textrm{\tiny AdS}}}
		\Bigg( \frac{2b w}{\sqrt{(b^2-r^2-w^2)^2+4b^2 w^2}} \Bigg)^{d+1} \bigg(   \frac{b^2 -r^2 + w^2}{2b}, \frac{r w}{b},  \boldsymbol{0} \bigg)
	\end{equation}
		whose modulus is given by 
		\be
		\label{mod-V-AdS-ddim}
		|\boldsymbol{V}|=\Bigg( \frac{2b w}{\sqrt{(b^2-r^2-w^2)^2+4b^2 w^2}} \Bigg)^{ d}
		\ee
		that manifestly satisfies the bound specified in \eqref{HEE-BT-intro}, 
		which is saturated exclusively at the RT hypersurface (\ref{RT-curve-adsd}).

		For future convenience,  it is worth writing the value of $\boldsymbol{V}$ along a generic geodesic bit thread. 
		This leads us to introduce some notations that will prove effective when explicit analytic solutions 
		are unattainable.  
		Solving  \eqref{geodesicAdSd} with respect to the radial coordinate $r$ gives
		\be
		\label{r-mag-min}
		r_{\textrm{\tiny$\lessgtr$}}=c_0\mp\sqrt{b_0^2-w^2} \,.
		\ee
		The first branch, denoted by  $r_{\textrm{\tiny$<$}}$,  
		represents the part of the thread that runs from the boundary to its maximum with $r'(w)\geqslant 0$
		and corresponds to the minus sign in (\ref{r-mag-min}).  
		The second branch, given by $r_{\textrm{\tiny $>$}}$, 
		describes the thread decreasing from its maximum back to the boundary. 
		Consequently, the vector field on these two branches can be expressed as follows
		\begin{equation}
			\label{Vd2}
			\boldsymbol{V}_{\!\!\textrm{\tiny$\lessgtr$}}
			=
			\frac{1}{L_{\textrm{\tiny AdS}}}
			\left(\frac{r_m w}{w_m r_{\textrm{\tiny$\lessgtr$}}}\right)^{d+1} \bigg(  \frac{b^2 -r_{\textrm{\tiny$\lessgtr$}}^2 + w^2}{2b}, \frac{r_{\textrm{\tiny$\lessgtr$}} \, w}{b},  \boldsymbol{0} \bigg)
		\end{equation}
		where different choices of the point $(w_m,r_m)$ on $\gamma_A$ provide different bit threads. 
		In this notation, the expression of the modulus is compact, and it is given by $|\boldsymbol{V}|=\left(\frac{r_m w}{w_m r_{\textrm{\tiny$\lessgtr$}}}\right)^d$. 
		In the right panel of Fig.\,\ref{fig:HigherDspherebitthreads} 
		we show $|\boldsymbol{V}|^{1/d}$ along the different bit threads. 
		This quantity is independent of $d$ in this setup, and its graph clearly illustrates that $|\boldsymbol{V}|\leqslant 1$. 
		In fact, the green curve corresponding to the geodesic bit thread intersecting $\gamma_A$ at $w_m$
		remains below the dashed red line 
		and reaches its maximum at $1$ only when $w=w_m$.
		Consequently, this maximum can only occur before the dashed vertical purple line, which corresponds to $w=b.$

		It is instructive to investigate the flux of the vector field in  \eqref{Vd} through the two regions of our bipartition: the sphere $A$ and its complement $B \equiv \mathbb{R}^d \setminus A$. 
		Three ingredients are needed for this analysis: 
		the unit vector $\boldsymbol{n}$ orthogonal to the surfaces of constant $w$, 
		the projection of $\boldsymbol{V}$ along $\boldsymbol{n}$, 
		and the volume form induced on the boundary. 
		The components of $\boldsymbol{n}$ are $(n^w, n^r, \boldsymbol{n}^\Omega) = \tfrac{w}{L_{\textrm{\tiny AdS}}} \big(1, 0, \boldsymbol{0}\big)$. 
		The scalar product $\boldsymbol{n}\cdot\boldsymbol{V}$ reduces to $g_{ww} \,n^w \,V^w$,
		which is positive  in $A$ and negative in its complement $B$. 
		Finally, the induced volume form is given by $\left(L_{\textrm{\tiny AdS}} / w\right)^{d}  \rd^d \boldsymbol{r} $,
		where $\rd^d \boldsymbol{r} \equiv r^{d-1} \, \rd r\, \rd \boldsymbol{\Omega}$ stands for the usual Euclidean volume form in polar coordinates. 
		Thus, the flux density through the boundary reads
		\be
		\label{integrandind}
		\lim_{w \to 0^+} \left[\, \frac{1}{4 G_{\textrm{\tiny N}}} \, g_{ww} \, V^w n^w \left(\frac{L_{\textrm{\tiny AdS}}}{w}\right)^{d} \;\right]
		\equiv \chi_A(\boldsymbol{r})\,\mathcal{C}(\boldsymbol{r})
		\ee
		being $\chi_A(\boldsymbol{r})$ defined as the step function equal to $+1$ for $r < b$ and to $-1$ for $r > b$. 
		In (\ref{integrandind}), we have understood the integrand as the object that is naturally integrated over the flat measure of $\mathbb{R}^d$.
		The finite result for the limit in \eqref{integrandind} arises from a compensation among the vanishing of $V^{w}$ as $w^{d+1}$, 
		the vanishing of $n^w$ as $w$
		and the divergent contributions coming from the volume form and the metric component $g_{ww}$.
		This leads to \cite{Kudler-Flam:2019oru,Han:2019scu}
		\be
		\label{contour-function-vacuum-ddim}
		\mathcal{C}(\boldsymbol{r})
		\equiv 
		\frac{L^d_{\textrm{\tiny AdS}} }{4 G_{\textrm{\tiny N}}} 
		\; \frac{(2b)^d }{\big| b^2 - r^2\big|^d} \, .
		\ee
		
		Considering the entanglement Hamiltonian of $A$
		for a CFT$_{d+1}$ in $\RR^d$ and in its ground state \cite{Hislop:1981uh, Casini:2011kv},
		it is straightforward to observe that the r.h.s. of (\ref{contour-function-vacuum-ddim})
		is proportional to 
		the $d$-th power of the reciprocal of the weight function multiplying the energy density 
		in the entanglement Hamiltonian of $A$.
		The coincidence of these two quantities for $d=1$ occurs simply for dimensional reasons.

		The holographic contour function \eqref{contour-function-vacuum-ddim} allows us to evaluate 
		the holographic entanglement entropy of $A$. 
		In this evaluation, it is crucial to consider the nontrivial relation between the field theory cutoff $\varepsilon_{\textrm{\tiny bdy}}$ 
		and the holographic cutoff $\varepsilon_{\textrm{\tiny AdS}}$, 
		as discussed e.g.  in \cite{Headrick:2022nbe,Han:2019scu,Nguyen:2017yqw,Takayanagi:2017knl}.		%
		This mapping can be found by adapting to this case the analysis 
		reported in Sec.\,\ref{subsec-gbt-ads3} for AdS$_3$
		and, as a result,  $\varepsilon^{\textrm{\tiny $A$}}_{\textrm{\tiny bdy}}$ and $\varepsilon^{\textrm{\tiny $B$}}_{\textrm{\tiny bdy}}$  are still given by \eqref{ads-EWCS-regEndp} for  any $d$.
		Then, by introducing the domain 
		$A_\varepsilon \equiv \big\{  (r, \boldsymbol{\Omega}) \,; \, 0\leqslant r\leqslant b-\varepsilon_{\textrm{\tiny bdy}}^{\textrm{\tiny $A$}}  \big\} \subsetneq A$,
		the holographic entanglement entropy of $A$ is obtained 
		from the contour function \eqref{contour-function-vacuum-ddim} as follows
		\bea
		\label{HEE-SA-full-d}
		S_A
		&=&
		\int_{A_\varepsilon }
			\! \mathcal{C}(\boldsymbol{r})\; \rd ^d \boldsymbol{r}
			\, = \,
			 \frac{L^d_{\textrm{\tiny AdS}} \, \Omega_{d-1}}{4 G_{\textrm{\tiny N}}} \int_0^{b-\varepsilon_{\textrm{\tiny bdy}}^{\textrm{\tiny $A$}}}
		\! \frac{(2b)^d }{\big| b^2 - r^2\big|^d} \; r^{d-1} \,\rd r\nonumber
		\\
		\rule{0pt}{.8cm}
		&=&
		\frac{ L^d_{\textrm{\tiny AdS}} \, \Omega_{d-1}}{4 d G_{\textrm{\tiny N}}}\;
		2^d \left(\frac{b-\varepsilon_{\textrm{\tiny AdS}} }{b+\varepsilon_{\textrm{\tiny AdS}} }\right)^{d/2} \,
		\!\!  _2F_1\!\left(\frac{d}{2},d\,;\frac{d}{2}+1 \,;\frac{b-\varepsilon_{\textrm{\tiny AdS}} }{b+\varepsilon_{\textrm{\tiny AdS}} }\right)
		\nn
		\\
		\rule{0pt}{1.1cm}
		&=&
		\frac{ L^d_{\textrm{\tiny AdS}}\Omega_{d-1}}{4  G_{\textrm{\tiny N}}}
		\;
		\sum_{n=0}^{\lfloor (d/2) -1 \rfloor } \!
		\frac{ (-1)^n }{ (2n)!! (d-2n -1) } 
		\frac{(d-2)!!}{(d-2-2n)!!}
		\left( 
		\frac{b}{\varepsilon_{\textrm{\tiny AdS}} }
		\right)^{d-1-2n}  
		\nonumber\\
		& &
		\rule{0pt}{1.2cm}
		+\, \frac{ L^d_{\textrm{\tiny AdS}} \, \Omega_{d-1}} {4  G_{\textrm{\tiny N}}}
		\left\{\begin{array}{ll}\displaystyle
			(-1)^{ \frac{d}{2} }    \,
			\frac{( d -2 )!! }{ (d-1)!!}
			\quad &
			\  d \textrm{ even}
			\\ 
			\rule{0pt}{.7cm}
			\displaystyle
			(-1)^{\frac{d-1}{2} } \;
			\frac{( d-2)!! }{ (d-1)!!} \;
			\log \!\big( b / \varepsilon_{\textrm{\tiny AdS}} \big)
			+O(1)
			\quad &
			\  d \textrm{ odd}
		\end{array}
		\right. 
		\eea
		where  $\Omega_{d-1} = \tfrac{2 \pi^{d/2} }{ \Gamma(d/2) }$ 
		is the area of the $(d-1)$-dimensional boundary of the unit ball in $\RR^d$. 
		It is noteworthy that 
		the final result for even values of $d$ is simply a polynomial in $b /\varepsilon_{\textrm{\tiny AdS}}$, 
		in contrast to the case of odd values of $d$, where an infinite series is encountered.

		The expression \eqref{HEE-SA-full-d} exactly reproduces the holographic entanglement entropy
		originally obtained in \cite{Ryu:2006bv, Ryu:2006ef}
		by evaluating the regularized area of $\gamma_A$, namely
		\bea
		\label{HEE-SA-RT}
		S_A
		&=&
		\frac{L^d_{\textrm{\tiny AdS}} \, \Omega_{d-1}}{4 G_{\textrm{\tiny N}} }
		\int_0^{\sqrt{b^2-\varepsilon_{\!\!\textrm{ \fontsize{1.5}{3}\selectfont AdS}}^2}} 
		\frac{\sqrt{1+w'_{m}(r_m)^2}}{w_{m}(r_m)^d} \; r_m^{d-1} \,\rd r_m 
		\nn
		\\
		\rule{0pt}{.9cm}
		&=&
		\frac{L^d_{\textrm{\tiny AdS}}\, \Omega_{d-1} }{4 G_{\textrm{\tiny N}} }
		\int_0^{\sqrt{b^2-\varepsilon_{\!\!\textrm{ \fontsize{1.5}{3}\selectfont AdS}}^2}}
		\frac{b \,r_m^{d-1}}{ \big(b^2-r_m^2\big)^{\frac{d}{2}+\frac{1}{2}}}  \, \rd r_m 
		\eea
		which becomes the radial integral in \eqref{HEE-SA-full-d} after the change of variable $r_m= 2 b^2 r /( b^2+r^2)$.
		
		By adapting this analysis to the complement $B = \RR^d \setminus A$, 
		for its holographic entanglement entropy we find 
		\be
		\label{HEE-SB-full-d}
		\begin{split}
		S_B
		\,&=\,
		\int_{B_\varepsilon }\!\!\! \mathcal{C}(\boldsymbol{r})\, 
		\rd^d \boldsymbol{r}
		\,=\, 
		\frac{L^d_{\textrm{\tiny AdS}} \, \Omega_{d-1}}{4 G_{\textrm{\tiny N}}} 
		\int_{b+\varepsilon_{\textrm{\tiny bdy}}^{\textrm{\tiny $B$}}}^{+\infty}
		\frac{(2b)^d }{\big| b^2 - r^2\big|^d}  \, r^{d-1} \rd r
		\\
		\,&=\,		\frac{L^d_{\textrm{\tiny AdS}} \,\Omega_{d-1}}{4 d G_{\textrm{\tiny N}}} \;
		2^d \left(\frac{b-\varepsilon_{\textrm{\tiny AdS}} }{b+\varepsilon_{\textrm{\tiny AdS}} }\right)^{d/2} \,
		_2F_1\! \left(\frac{d}{2},d\,;\frac{d}{2}+1 \,;\frac{b-\varepsilon_{\textrm{\tiny AdS}} }{b+\varepsilon_{\textrm{\tiny AdS}} }\right)
		\end{split}
		\ee
		where  
		$B_\varepsilon \equiv \big\{  (r, \boldsymbol{\Omega}) \,; \, r \geqslant b+\varepsilon_{\textrm{\tiny bdy}}^{\textrm{\tiny $B$}}  \big\} \subsetneq B$.
		Comparing (\ref{HEE-SA-full-d}) and (\ref{HEE-SB-full-d}) gives that $S_A=S_B$,
		as expected from the purity of the ground state of the dual CFT$_{d+1}$.
		The equality of the two radial integrals  follows directly from the fact that they are mapped into each other
		by the transformation $r\mapsto b^2/r$.  
		Notice also that the equality holds at finite values of the cutoff 
		because of  the  suitable choice of  $\varepsilon^{\textrm{\tiny $A$}}_{\textrm{\tiny bdy}}$ and $\varepsilon^{\textrm{\tiny $B$}}_{\textrm{\tiny bdy}}$.

		\subsection {Strip}
		\label{stripemptyads}

		In the following, we consider the bipartition of $\RR^d$ characterized by an infinite strip $A$ of width $2b$.
		A suitable choice of the Cartesian coordinates for $\RR^d$ allows us to describe $A$
		as the set of points on the boundary of the constant time slice of AdS$_{d+2}$
		such that
		$- b\leqslant y_1\leqslant b$ and $-b_\perp\leqslant y_j\leqslant b_\perp$ for $j>1$.
		Additionally, we assume that we are taking the limit as $b_\perp\to\infty$ to restore translation invariance in the directions of $y_j$ with $j>1$. 
		To enlighten the notation, hereafter, we will also write the first component of $\boldsymbol{y}$ simply as $y$ instead of $y_1$. 
		
		Th RT hypersurface $\gamma_A$ corresponding to this infinite strip $A$
		can be described either as $w=w(y)$ or $y=y(w)$
		and the result is \cite{Ryu:2006ef}
		\begin{equation}
			\pm \, y_{m}\left(w_{m}\right)
			=
			w_{*} \frac{\sqrt{\pi} \Gamma\left(\frac{d+1}{2 d}\right)}{\Gamma\left(\frac{1}{2 d}\right)}-\frac{w_{m}}{d+1}\left(\frac{w_{m}}{w_{*}}\right)^{d}
			{ }_{2} F_{1} \! \left(\frac{1}{2}, \frac{d+1}{2 d}, \frac{3}{2}+\frac{1}{2 d}, \frac{w_{m}^{2 d}}{w_{*}^{2 d}}\right) 
			\label{superficieanalitica}
		\end{equation}
		where  we have denoted with $P_m =(w_m,\boldsymbol{y}_m)$ a generic point of $\gamma_A$. 
		The plus (minus) sign describes the branch of $\gamma_A$ with $y_{m}>0~ (y_{m}<0)$. 
		However, 
		because of the obvious reflection symmetry w.r.t. the  hyperplane $y=0$,
		we can restrict to  $y_{m}>0$ without loss of generality. 
		In \eqref{superficieanalitica}, the integration constant $w_*$ represents the maximal height of $\gamma_A$ 
		in the holographic direction, and it reads
		\begin{equation}
			w_{*}=  \frac{d\;\Gamma\big(\frac{2d+1}{2 d}\big)}{\sqrt{\pi}\;\Gamma\big(\frac{1+d}{2 d}\big)} \; 2 b \,.
			\label{zstarstrisciavuoto}
		\end{equation}

		As observed in  \cite{Agon:2018lwq}, 
		the construction of the geodesics bit threads for the strip 
		does not work when $d>2$.
		Nevertheless, it is instructive to examine the reason of this failure. 
		The geodesics occurring in this analysis are still characterized 
		by the two-parameter family of circumferences  in \eqref{geodesicAdSd},
		but in this case the two constants  $b_0$ and $c_0$  are \cite{Agon:2018lwq}
		\be
		\label{b0strip}
		b_{0}=\frac{w_{*}^{d} \,w_m}{\sqrt{w_{*}^{2 d}-w_{m}^{2 d}}} 
		\;\;\;\qquad\;\;\;
		c_0=y_{m}(w_{m})+\frac{w_{m}^{d+1}}{\sqrt{w_{*}^{2 d}-w_{m}^{2 d}}} 
		\ee
		being $(w_m, y_{m})$ defined as the point in the plane $(w,y)$ 
		where the geodesic intersects $\gamma_A$  orthogonally. 
		Each of these geodesics intersects the boundary $w=0$  for  two different values of $y$ given by 
		\be
		\label{es1-2}
		y_{\textrm{\tiny $A$}}= y_m(w_m)
		+ \frac{w_m \left(w_m^d-w_*^d\right)}{\sqrt{w_*^{2 d}-w_m^{2 d}}}
		\;\;\;\qquad\;\;\;
		y_{\textrm{\tiny $B$}}=y_m(w_m)
		+\frac{w_m \left(w_m^d+w_*^d\right)}{\sqrt{w_*^{2 d}-w_m^{2 d}}}
		\ee
		where the subscripts $A$ and $B\equiv \RR^d \setminus A$  indicate the region where the intersection lies.

			\begin{wrapfigure}[11]{l}{0.45\textwidth}
		\vskip .35cm
		\hspace{-.3cm}
		\includegraphics[width=0.47\textwidth]{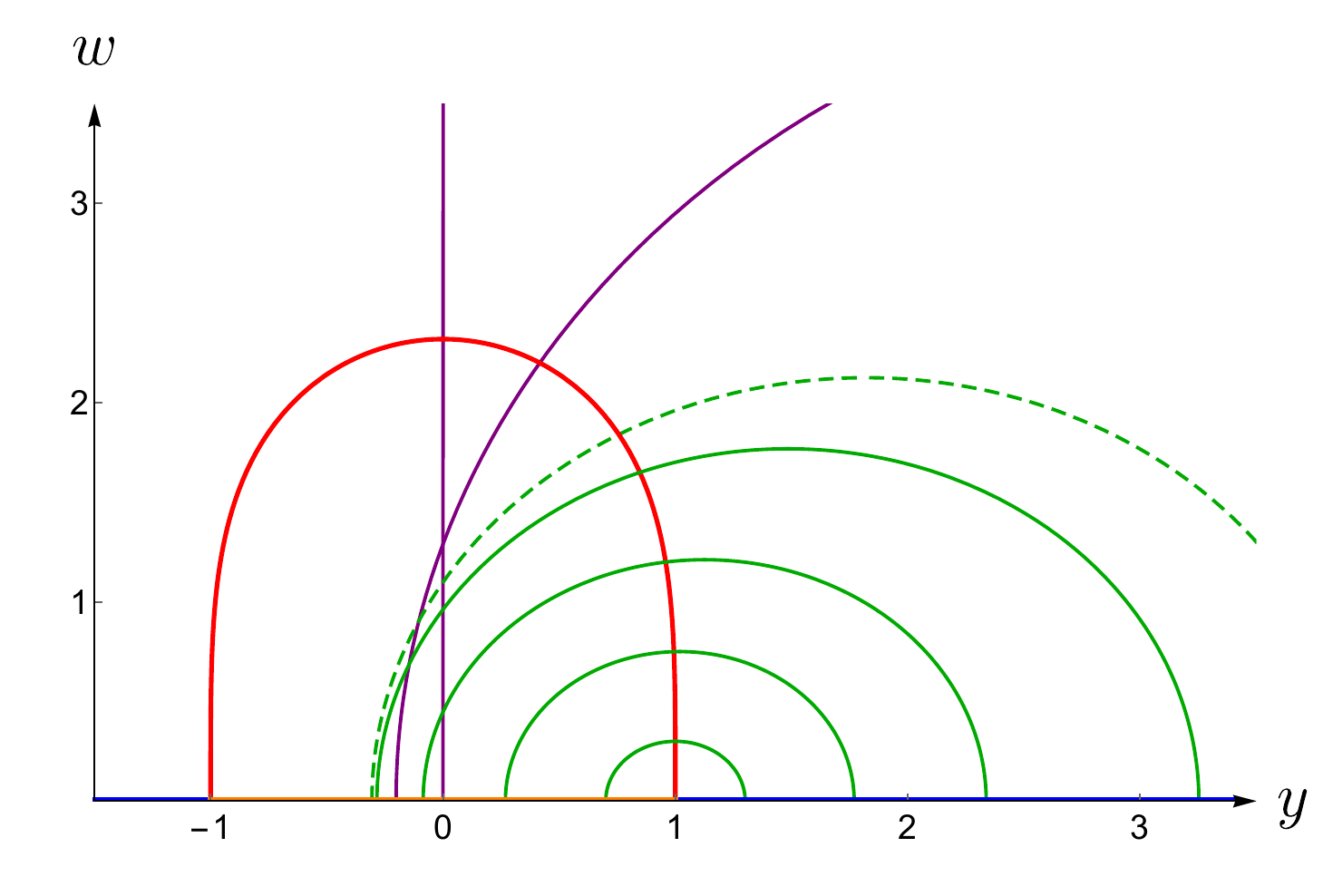}
		\vspace{-.9cm}
		\caption{\label{fig:Dequals4bitb}
			\small Geodesics (green and purple curves) in AdS$_5$ orthogonal to $\gamma_A$ (red curve) for the strip.
		}
	\end{wrapfigure}

		A necessary condition for geodesics to define consistent bit threads is that $y_{\textrm{\tiny $A$}}$ is a decreasing function of 
		$w_m$, while $y_{\textrm{\tiny $B$}}$ increases as $w_m$ increases, otherwise the geodesics will inevitably intersect. 
		This behavior has been called the nesting property in \cite{Agon:2018lwq}. 
		A simple way to check this property is to study the sign of the derivative of the two intersections with respect to $w_m$,
		which are given by  \cite{Agon:2018lwq} 
		\be
		\label{condizgeo}
		\begin{array}{l}
		\displaystyle
		\frac{\partial y_{\textrm{\tiny $A$}}}{\partial w_m} 
		\,=\,
		-\frac{w_*^d \left(w_*^d-(d-1) w^d_m \right)}{\sqrt{w_*^{2 d}-w_m^{2 d}} \left(w_m^d+w_*^d\right)} 
		\\
		\rule{0pt}{1.cm}
		\displaystyle
		\frac{\partial y_{\textrm{\tiny $B$}}}{\partial w_m} 
		\,=\,
		\frac{w_*^d \left(w_*^d+(d-1) w^d_m \right)}{\sqrt{w_*^{2 d}-w_m^{2 d}} \left(w_m^d+w_*^d\right)} \,.
		\end{array}
	\ee

	\noindent
	The derivative of $y_{\textrm{\tiny $B$}}$ is positive for any $d>1.$ 
	As for the derivative of $y_{\textrm{\tiny $A$}}$, it is always negative for $d=1,$ which explains why 
	the bit threads construction discussed in Sec.\,\ref{sec-AdS3} works. 
	When $d=2,$ this derivative is less than or equal to zero,
	and it vanishes when $w_m=w$;
	hence, the nesting condition is still satisfied. 
	However, for $d\geqslant 3$ this derivative changes sign at 
	$w_m= w_\ast / (d-1)^{1/d} $,
	and therefore $y_{\textrm{\tiny $A$}}$ is not a monotonically decreasing function of $w_m.$ 
	This means that the geodesics orthogonal to $\gamma_A$ 
	do intersect and therefore they do not define a family of bit threads. 
	This behavior is illustrated for the case of AdS$_5$ in Fig.\,\ref{fig:Dequals4bitb}.
	Moving from the smaller green semicircles to the larger ones, the $y$-coordinate of the endpoint in $A$ 
	(the orange interval in Fig.\,\ref{fig:Dequals4bitb}) 
	keeps decreasing up to the geodesic corresponding to the dashed green curve,
	where  the derivative of $y_{\textrm{\tiny $A$}}$ vanishes.
	If we keep moving along the minimal surface towards $w_\ast$ (see the purple geodesic), 
	 $y_{\textrm{\tiny $A$}}$ reverses its direction and begins to increase.
	
	These inherent limitations in constructing the geodesic bit threads for the strip naturally prompt us to explore alternative constructions. 
	However, we remark that in the specific case of AdS$_4$ 
	the geodesic bit threads for the infinite strip can be constructed
	and the details of this analysis are discussed in Appendix\;\ref{GeodAdS4}.

	\subsection {Alternative bit threads in AdS$_{d+2}$}
	\label{sec5}

	This section explores an alternative construction of bit threads for the infinite strip
	by introducing a set of integral curves called minimal hypersurface inspired bit threads.
	In pure AdS$_{d+2}$, because of the conformal symmetry of the background,
	these curves coincide with what we call translated and dilated bit threads in the following, 
	already proposed in \cite{Agon:2018lwq}.
	However, in more complicated backgrounds like the black brane geometries,
	these two constructions yield different results,
	as shown in Sec.\,\ref{Schwarzschild AdS black brane} and discussed further  in Appendix\;\ref{app-traslatipedraza}.

	For the bipartition of $\RR^d$ characterized by an infinite strip $A$ of width $2b$ (see Sec.\,\ref{stripemptyads}),
	also the construction of the minimal hypersurface inspired bit threads becomes a two-dimensional problem
	because of the symmetry of the configuration. 
	This construction provides a family of integral curves by employing the profile of the minimal hypersurface \eqref{superficieanalitica}.
	Each curve is characterized by two free parameters
	given by the depth $\tilde{w}_*$ and the center $c_0$, and
	consists of  the two following branches
	\begin{equation}
		y{\textrm{\tiny$\gtrless$}}(w) = c_0 \pm y \left(w;\tilde{w}_* \right)
		\label{msgeneralcurves}
	\end{equation}
	where the profile $y \left(w;\tilde{w}_\ast \right)$ is defined as \eqref{superficieanalitica}
	where the depth $w_\ast$ is replaced by the free parameter $\tilde{w}_\ast$. 
	The branch denoted by $y_{\mbox{\tiny$<$}}(w)$ corresponds to the minus sign and originates from the interval $A$;
	while the branch represented by $y_{\mbox{\tiny $>$}}(w)$ is associated with the plus sign and extends towards the complementary region $B$. 
	The two free parameters $c_0$ and $\tilde w_*$ are determined by requiring that the integral curves \eqref{msgeneralcurves} 
	intersect the minimal hypersurface \eqref{superficieanalitica} orthogonally at the point $(w_m,y_m(w_m))$; 
	hence we must impose
	\begin{equation}
		\left\{ \begin{array}{l}
			y_{\textrm{\tiny$<$}}(w_m) = y_{m}(w_m)\cr
			\rule{0pt}{.5cm}
			\big[ \,g_{ww} + g_{yy} \, y’_{\textrm{\tiny $<$}}(w) \,y’_{m}(w) \,\big] \big|_{(w,y)=(w_m,y_m(w_m))} = 0
		\end{array}\right.
		\label{ortomsads}
	\end{equation}
	where $g_{yy} = g_{ww} = L^2_{\textrm{\tiny AdS}} / w^2$ are the diagonal components of the metric \eqref{fAdS-metric}. 
	From \eqref{ortomsads}, we can determine the depth of each integral curve and its center,
	which read respectively
	\begin{equation}
		\tilde{w}_* = \frac{w_m w_*}{\left(w_*^{2d} - w_m^{2d}\right)^{\frac{1}{2d}}}
		\;\;\;\qquad\;\;\; 
		c_0 = y_m(w_m) + y(w_m; \tilde{w}_*)\,.
	\end{equation}
	As mentioned in Sec.\,\ref{stripemptyads}, 
	we can verify whether these putative bit threads obey the nesting property by analyzing the sign of the derivative w.r.t. $w_m$ of the two intersections with the boundary $w=0$,
	denoted by $y_{\textrm{\tiny $A$}}$ and $y_{\textrm{\tiny $B$}}$,
	where the subscript of $y$ indicates the subregion of the boundary where the intersection occurs. 
	This analysis yields 
	\bea
	\label{condizionems-A}
	\frac{\partial y_{\textrm{\tiny $A$}} }{\partial w_m} 
	&=&
	-\frac{1 \, }{(d+1) \big[1-(w_m /w_*)^{2 d} \big]^{1/2}} \;\,
	{}_2F_1\! \left(\frac{1}{2}\,,\frac{d+1}{2 d}\,;\frac{3d+1}{2d}\,;1-(w_m /w_*)^{2 d} \right)  
	\hspace{1cm}
	\\
	\label{condizionems-B}
	\rule{0pt}{.8cm}
	\frac{\partial y_{\textrm{\tiny $B$}}}{\partial w_m} 
	&=&
	\frac{1}{\big[ 1-(w_m /w_*)^{2 d}  \big]^{(2 d+1)/(2 d)}} 
	\\
		\rule{0pt}{.7cm}
	& & \times
	\left[\, \frac{1}{d}\left(\frac{w_m}{w_*}\right)^{d} {}
	{}_2F_1\!\left(\frac{1}{2}\, , \frac{1}{2} \left(1-\frac{1}{d}\right)\, ;\frac{3}{2}; (w_m /w_*)^{2 d} \right)
	+
	\frac{\sqrt{\pi } \;\Gamma \left(\frac{d+1}{2 d}\right)}{\Gamma \left(\frac{1}{2 d}\right)} \,\right] .
	\nonumber
	\eea
	Since the two hypergeometric functions in the r.h.s.'s
	are both positive for any $d$ when  $w_m / w_\ast \in (0,1)$, 
	we can conclude that  (\ref{condizionems-A}) remains always negative, while  (\ref{condizionems-B}) is always positive. 
	The positivity of these hypergeometric functions follows from their standard integral representation.
	This implies that this novel set of integral lines does not self-intersect, making them applicable even when $d > 2$. 

	\begin{figure}[t!]
		\vspace{-.5cm}
		\hspace{-1cm}
		\includegraphics[width=0.55\textwidth]{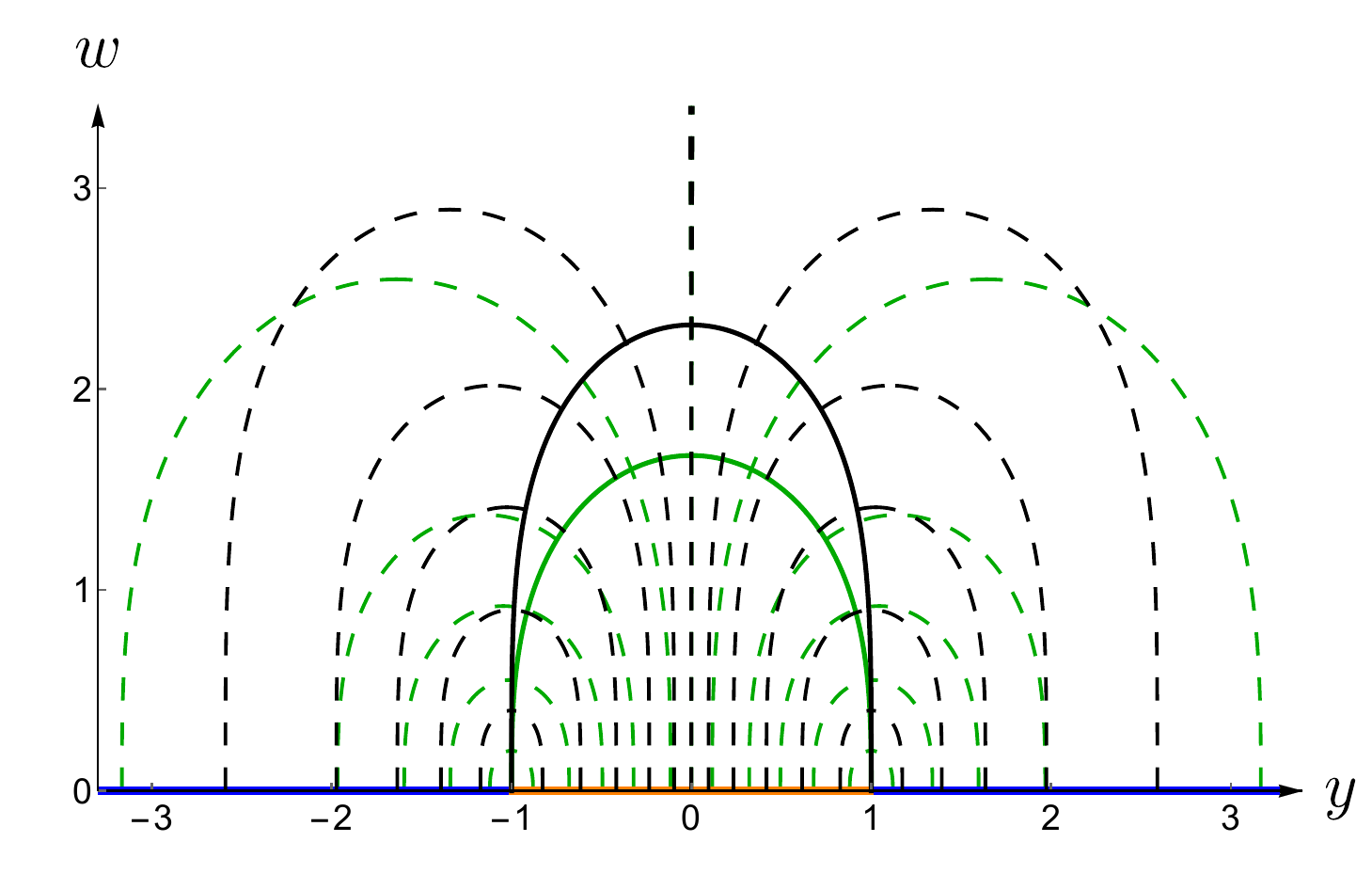}
		\hspace{0.2cm}
		\includegraphics[width=0.55\textwidth]{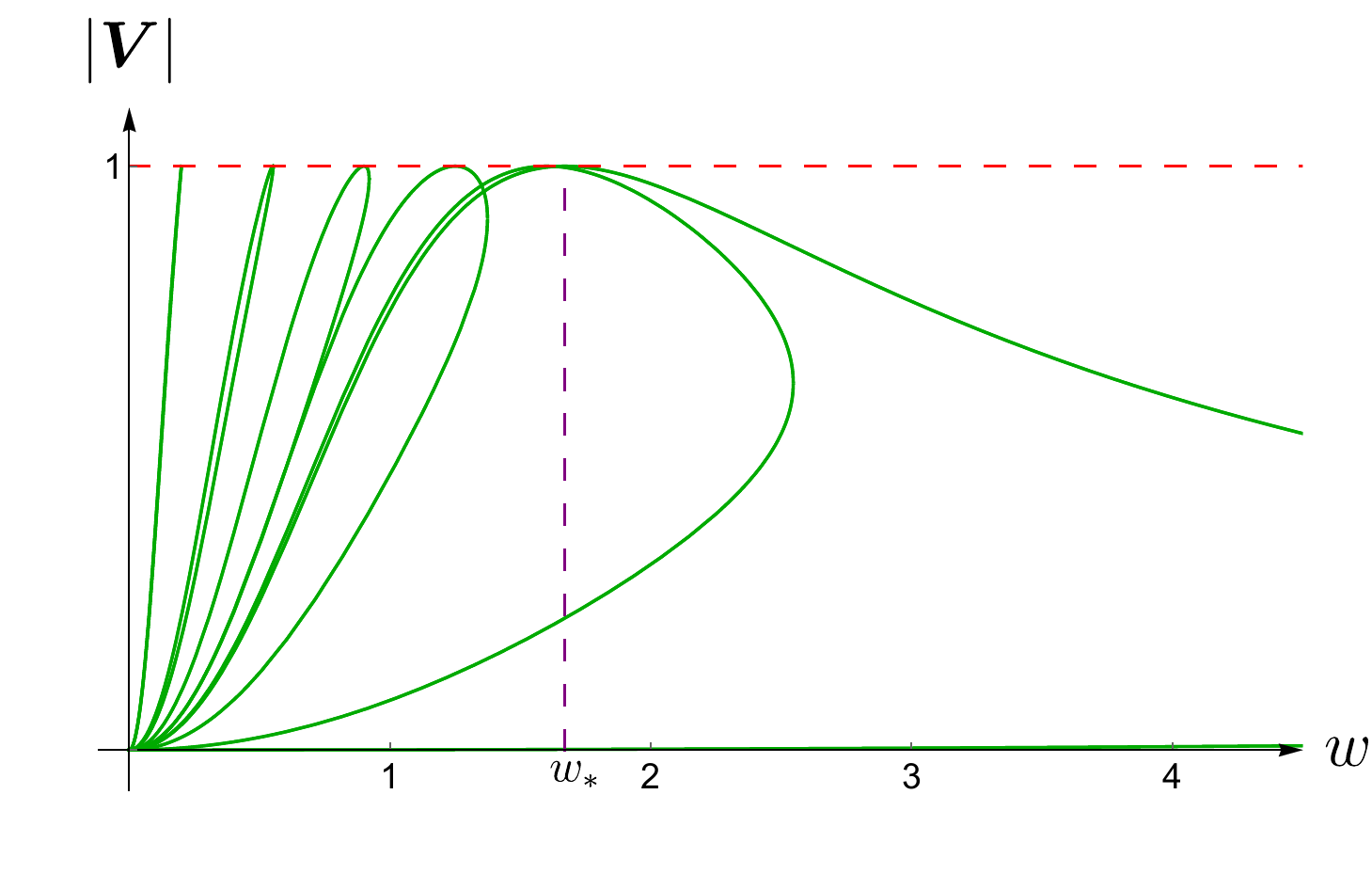}
		\vspace{-.5cm}
		\caption{\small 
			Left: Minimal hypersurface inspired bit threads for the infinite strip, 
			in the constant time slice of AdS$_4$ and AdS$_5$ 
			(dashed green and dashed black curves respectively),
			from \eqref{superficieanalitica}-\eqref{msgeneralcurves}.
			Right: $\big|\boldsymbol{V} \big|$ along these bit threads in AdS$_4$ 
			(see \eqref{moduloadsms}).
			}
		\label{fig:HigherDMSmoduli} 
	\end{figure}

	This class of bit threads is displayed in  the left panel of Fig.\,\ref{fig:HigherDMSmoduli} for AdS$_4$ and AdS$_5$ 
	(see the dashed green curves and dashed black curves respectively). 
	The corresponding RT hypersurface is denoted through the solid curve having the same color.  
	Notice that the height of the RT hypersurface increases with $d$,
	as the transverse width $2b$ of the strip remains constant.

	The vector field $\boldsymbol{V}$ associated with the minimal hypersurface inspired bit threads 
	can be found in terms of $w$ and $w_m$ 
	by following the procedure reviewed in Appendix\;\ref{app-modulus}.
	For its magnitude, we obtain
	\begin{equation}
		\big|\boldsymbol{V}_{\!\!\textrm{\tiny$\lessgtr$}}\big|
		=
		\left|\,
		\frac{w^d \, w_m^d}{\sqrt{w^{2 d} \left(w_m^{2 d}-w_*^{2 d}\right)+w_m^{2 d} \, w_*^{2 d}}} \; 
		\frac{ \big[ \partial_{w_m} y_{\textrm{\tiny$<$}}(w)\big]\!\big|_{w=w_m} }{\partial_{w_m} y_{\textrm{\tiny$\lessgtr$}}(w)} 
		\,\right|
		\label{moduloadsms}
	\end{equation}
	where $\boldsymbol{V}_{\!\!\!\textrm{\tiny$<$}}$ and $\boldsymbol{V}_{\!\!\!\textrm{\tiny$>$}}$
	refer to the minus and plus branches, respectively, in \eqref{msgeneralcurves}.
	The direction of the vector field is determined by the unit tangent vector $\boldsymbol{\tau}$ to the minimal hypersurface inspired bit threads,
	which, in the same notation, reads
	\begin{equation}
		\boldsymbol{\tau}_{\textrm{\tiny$\lessgtr$}}
		= \left(\, \tau^w_{\textrm{\tiny $\lessgtr$}}  , \tau^y_{\textrm{\tiny$\lessgtr$}}\,\right) 
		=
		\frac{w}{L_{\textrm{\tiny AdS}}\, \tilde{w}_*^{d}} 
		\Big( \pm \! \sqrt{\tilde{w}_*^{2 d}-w^{2 d}}  \, , \, w^d \, \Big)  \, .
	\end{equation} 
	The modulus $\big|\boldsymbol{V}_{\!\!\textrm{\tiny$\lessgtr$}}\big|$ in \eqref{moduloadsms} 
	cannot be expressed analytically only in terms of $y$ and $w$. 
	However, since \eqref{moduloadsms} and  \eqref{msgeneralcurves} provide a parametric representation of the modulus in terms of  $w$ and $w_m$,
	it is straightforward to plot $\big|\boldsymbol{V}_{\!\!\textrm{\tiny$\lessgtr$}}\big|$ along the different bit threads, verifying graphically that  $|\boldsymbol{V}|\leqslant 1$. 
	In the right panel of Fig.\,\ref{fig:HigherDMSmoduli}, we illustrate this property for AdS$_4$. 
	Each green curve corresponds to a fixed value of $w_m$ and 
	displays the behavior of $|\boldsymbol{V}|$ on a single bit thread as $w$ varies. 
	Similarly to the cases discussed above, 
	these curves reach their maximum value at 1 
	(see the dashed red horizontal line, denoting $|\boldsymbol{V}|=1$)
	only when $w$ corresponds to the RT hypersurface $\gamma_A$  and consistently remains below this value otherwise. 
	All of these curves are tangent to the red line for a value of $w$ that is always less than the maximum depth $w_\ast$ of the RT hypersurface 
	(see the purple dashed vertical line).
	These observations show that the RT hypersurface inspired integral curves are consistent bit threads.

	The contour $\mathcal{C}_{A} (\boldsymbol{y})$ for the infinite strip
	(we remind that the first component of $\boldsymbol{y}$ is $y$)
	is simply given by the density of flux of $\boldsymbol{V}_{\!\!\!\mbox{\tiny$<$}}$ through the region  $A$, namely
	\bea
	\label{contourms}
	\mathcal{C}_{A} (\boldsymbol{y}) 
	&=& 
	\lim_{w \to 0^+} \! \left(\frac{1}{4G_{\textrm{\tiny N}}} \big|\boldsymbol{V}_{\!\!\!\mbox{\tiny$<$}} \big| \  \tau_a \, n^a \,
	\frac{L_{\textrm{\tiny AdS}}^{d}}{w^{d}}   \right)  
	\\
	\rule{0pt}{.7cm}
	&=& 
	\frac{L_{\textrm{\tiny AdS}}^{d}}{4G_{\textrm{\tiny N}}}  \; 
	\frac{d+1 }{\ w_m(y_{\textrm{\tiny $A$}})^d\;  _2F_1 \big(\frac{1}{2},\frac{d+1}{2 d};\frac{3 d+1}{2d};1- (w_m(y_{\textrm{\tiny $A$}}) / w_* )^{2 d}\big)}
	\nn
	\eea
	where $|y_{\textrm{\tiny $A$}} | \leqslant b$,
	the factor $(L_{\textrm{\tiny AdS}}/w)^d$ comes from the square root of the determinant of the induced metric on the $w=\text{const}$ slice
	and $\boldsymbol{n}$ is the unit normal vector to the boundary $w=0$, 
	whose components are  $(n^w,n^y,n^{\boldsymbol{y}_\perp})=\frac{w}{L_{\textrm{\tiny AdS}}} (1,0,\boldsymbol{0})$.  
	
	The function $w_m(y_{\textrm{\tiny $A$}})$ is implicitly defined by reversing the relation that specifies the coordinate $y_{\textrm{\tiny $A$}}$ 
	of the point where the bit thread intersects the boundary.
	This relation reads
	\be
	\label{contourmsya}
	y_{\textrm{\tiny $A$}}
	=\, 
	c_0 - y (0 ; \tilde{w}_* )
	\,=\, 
	y_m(w_m) + y\left(w_m \,;  \frac{w_m \, w_*}{\big(w_*^{2d} - w_m^{2d}\big)^{1/(2d)}} \right)
	- 
	y \left( 0 \,;  \frac{w_m \, w_*}{\big(w_*^{2d} - w_m^{2d}\big)^{1/(2d)}}  \right)
	\ee
	which can be solved numerically. 
	Alternatively, we can consider \eqref{contourms} and \eqref{contourmsya} as a parametric representation of the contour function with respect to the parameter $w_m$. 
	It is worth noting that, for $d=2$, \eqref{contourms} and \eqref{contourmsya} yield a regular contour, unlike the geodesic bit threads discussed in Appendix\;\ref{GeodAdS4} 
	(see Fig.\,\ref{contourAdS4}).

	The holographic entanglement entropy for the strip can be computed from  \eqref{HEE-BT-intro}.
	The symmetry of the problem allows us to consider $y\in [0, b- \varepsilon_{\textrm{\tiny bdy}}^A ]$,
	where the UV cutoff $\varepsilon_{\textrm{\tiny bdy}}^A \ll 1$ is implicitly given by $w_m(\varepsilon_{\textrm{\tiny AdS}})$, finding 
	\bea
	\label{flussoms}
	S_A &=& 
	\int_A \ \mathcal{C}_A (\boldsymbol{y}) \, \rd^d \boldsymbol{y} 
	\\
	\rule{0pt}{.8cm}
	&=&
	\frac{2 L_{\textrm{\tiny AdS}}^{d} (2 b_\perp)^{d-1}}{4G_{\textrm{\tiny N}}} \int_{0}^{b-\varepsilon_{\textrm{\tiny bdy}}^A}\!\!\!\!\!  
	\frac{d+1 }{w_m^{d}(y_A)\; _2F_1\!\left(\frac{1}{2},\frac{d+1}{2 d};\frac{3 d+1}{2d}; 1 - ( w_m(y_A) / w_* )^{2 d}\right)} \; \rd y
	\nonumber\\
	\rule{0pt}{.8cm}
	&=&
	\frac{2 L_{\textrm{\tiny AdS}}^{d} (2 b_\perp)^{d-1}}{4G_{\textrm{\tiny N}}} \int_{w_*}^{\varepsilon_{\textrm{\tiny AdS}}} \frac{\rd y_{\textrm{\tiny $A$}}}{\rd w_m} \frac{(d+1) 
	}{
		w_m^{d}~  {}_2F_1\big(\frac{1}{2},\frac{d+1}{2 d};\frac{3 d+1}{2 d};1- ( w_m / w_*)^{2 d} \big)} \; \rd w_m
	\nonumber\\
	\rule{0pt}{.8cm}
	&=&
	\frac{2 L_{\textrm{\tiny AdS}}^{d} (2 b_\perp)^{d-1}}{4G_{\textrm{\tiny N}}}  
	\int_{w_*}^{\varepsilon_{\textrm{\tiny AdS}}}  
	\left( \! -\frac{w_*^d}{w_m^{d}\sqrt{w_*^{2 d}-w_m^{2 d}}} \right)  \rd w_m
	\nonumber
	\\
	\rule{0pt}{.8cm}
	&=&
	\frac{L_{\textrm{\tiny AdS}}^{d} (2 b_\perp)^{d-1}}{4 G_{\textrm{\tiny N}}} 
	\left[\, 
	\frac{2}{\varepsilon_{\textrm{\tiny AdS}}^{d-1} \ (d-1)} \;
	{}_2F_1\left(\tfrac{1}{2},\tfrac{1-d}{2d};\tfrac{d+1}{2 d}; \left(\varepsilon_{\textrm{\tiny AdS}} / w_* \right)^{2 d}\right)
	- \frac{2\,\sqrt{\pi } \;\Gamma \big(\frac{1+d}{2 d}\big) }{ (d-1) \ w_*^{d -1} \, \Gamma \big(\frac{1}{2 d}\big)}
	\,\right]
	\nonumber
	\eea
	which can be rewritten in terms of the width of the strip $2 b$ 
	by employing the relation between $w_*$ and $2 b$ in   \eqref{zstarstrisciavuoto}
	and this exactly reproduces the standard computation 
	for $S_A$ given in \cite{Ryu:2006ef}.

	The holographic entanglement entropy for the complement $B$ can be found similarly.
	The holographic contour function in $B$ can be written as 
	\begin{equation}
		\mathcal{C}_{B} (\boldsymbol{y}) 
		\equiv 
		\lim_{w \to 0^+} \left( -\frac{1}{4G_{\textrm{\tiny N}}}  \,\big|\boldsymbol{V}_{\!\!\mbox{\tiny$>$}} \big|  
		\,\tau_a \,n^a \ \frac{L_{\textrm{\tiny AdS}}^{d}}{w^{d}}\right)  
		\label{contourmspiu}
	\end{equation}
	where we remark that $\boldsymbol{V}_{\!\!\mbox{\tiny$>$}}$ occurs.
	Thus, the holographic entanglement entropy of $B$ is 
	\begin{equation}
		\label{StripSB}
		S_B=  
		\int_B  \ \mathcal{C}_B (y) \, \rd^d\boldsymbol{y}   
		\,=\,
		\frac{2 L_{\textrm{\tiny AdS}}^{d}(2 b_\perp)^{d-1}}{4G_{\textrm{\tiny N}}}  \,
		\int_{\varepsilon_{\textrm{\tiny AdS}}}^{w_*} 
		\frac{w_*^d}{w_m^{d} \,\sqrt{w_*^{2 d}-w_m^{2 d}}} \; \rd w_m  \,.
	\end{equation}
	Comparing this result with the second-last line of  \eqref{flussoms},
	one observes that $S_A=S_B$, 
	as expected from the purity of the ground state of the dual CFT$_{d+1}$.

	\section{Hyperbolic black hole}
	\label{sec-hyp-bh}

	In this section 
	we employ the map of \cite{Casini:2011kv} to obtain analytic results for 
	the geodesic bit threads of a sphere and the corresponding relevant fluxes
	when the gravitational background is a constant time slice of a 
	specific static hyperbolic black hole.

	We consider the following class of $(d+2)$-dimensional black holes  \cite{Emparan:1998he, Birmingham:1998nr, Emparan:1999gf}
	\be 
	\label{topological-bh-general-metric}
	ds^2 = 	\frac{ L_{\textrm{\tiny AdS}}^2 }{z^2} 
	\left( 
	- f_k(z) \, \rd t^2  + 
	\frac{ \rd z^2 }{ f_k(z) } + 
	d \boldsymbol{\Sigma}_{k,d}^2
	\right)
	\qquad 
	f_k(z) = 1 + \frac{k z^2}{\ell^2} - \frac{\mu z^{d+1} }{\ell^{2d}} 
	\ee 
	where
	\be 
	d \boldsymbol{\Sigma}_{k,d}^2 = 
	\begin{cases}
		\ell^2 \, \rd \boldsymbol{\Omega} ^2 _d  &  k =1 \\ 
		\sum_{i=1}^d \rd x_i^2  \hspace{1cm}& k =0 \\ 
		\rule{0pt}{.6cm}
		\ell^2 \, \rd \boldsymbol{H} ^2 _d   &  k =- 1  
	\end{cases}
	\ee 
	being $\rd \boldsymbol{\Omega} ^2 _d $ defined as the metric of the unit $d$-dimensional sphere
	(in Sec.\,\ref{HigherdimAdS} the notation $\rd \boldsymbol{\Omega}^2 = \rd \boldsymbol{\Omega} ^2 _{d-1} $ has been adopted)
	and $ \rd \boldsymbol{H} ^2 _d $ as the metric of the unit $d$-dimensional hyperbolic space $\mathbb{H}_d$.
	The boundary at $z \to 0 $ and the event horizon at $z=z_h$ (such that $f_k(z_h)=0$)
	have the topology of either a sphere or a plane or a hyperbolic plane, 
	for $k=+1$, $k=0$, and $k=-1$, respectively. 
	Thus, (\ref{topological-bh-general-metric}) is parameterized by $\mu \geqslant 0$,  $\ell^2 >0$ and $k \in \{-1,0,+1\}$.
	The inverse temperature of the dual CFT$_{d+1}$ is the standard Hawking temperature of the black hole,
	whose inverse reads
	(see e.g. Eq.\,(4) of \cite{Emparan:1999gf})
	\be
	\label{beta-topo}
	\beta = 
	\frac{4 \pi  \ell^2 z_h}{(d-1) \,k \, z_h^2 +(d+1) \,\ell^2} \,.	
	\ee

	The case $k = 0$ with $z_h =  (\ell^{2d} /\mu)^{1/(d+1)}  $
	corresponds to the Schwarzschild AdS$_{d+2}$ black brane
	discussed in Sec.\,\ref{Schwarzschild AdS black brane} (see \eqref{sch-ads-brane-metric}).

	In this section, we focus on the special case of the hyperbolic black hole 
	characterized by $k =-1$ and $ \mu =0$, whose horizon is located at $z_h=\ell$,
	because this spacetime can be mapped into a portion of AdS$_{d+2}$.
	We remark that the function $f_k(z)$ in (\ref{topological-bh-general-metric})
	becomes independent of $d$
	for these specific values of the parameters.
	The metric induced on a constant time slice of this specific hyperbolic black hole reads
	\be 
	\label{hyp-metric}
	ds^2 = 
	\frac{ L_{\textrm{\tiny AdS}}^2 }{z^2}
	\left( 
	\frac{\rd z^2}{1 - (z/z_h)^2}
	+  
	\rd \boldsymbol{u}^2
	\right) 
	\ee 
	where $z>0$, the $d$-dimensional vector $\boldsymbol{u} \equiv (u, \boldsymbol{\Omega} )$ has been introduced
	and 
	\be 
	\label{hyp-metric-dHd}
	\rd \boldsymbol{u}^2  \equiv 
	z_h^2 \, \rd \boldsymbol{H}^2_d
	\equiv 
	\rd u^2 + 
	z_h^2 \, [\sinh ( u/z_h )]^2 \, \rd \boldsymbol{\Omega} ^2
	\ee 
	being $u \geqslant 0$ defined as the radial coordinate.
	The metric (\ref{hyp-metric-dHd})  is equivalent to \eqref{Hd-metric} 
	(see e.g. (2.22)-(2.23) and (2.26)-(2-27) in \cite{Aharony:1999ti}).
	The boundary at $z\to 0^+$ of the constant time slice 
	of the hyperbolic black hole described  by (\ref{hyp-metric})
	is equipped with the hyperbolic metric (\ref{hyp-metric-dHd}).
	When $d=1$, we have that  $\rd \boldsymbol{u}^2  $ becomes $\rd x^2$ where $x\in \RR$;
	hence (\ref{hyp-metric})  reduces to the metric of the constant time slice of the BTZ black brane given in (\ref{btz-brane-metric}).

	\subsection{Geodesic bit threads}
	\label{sec-hyp-bh-gbt}

	In the following, we study the geodesic bit threads of  a sphere
	when the gravitational background is the hyperbolic black hole
	whose constant time slice is described by (\ref{hyp-metric}).
	

	A straightforward extension of \eqref{chm-btz-inside}-\eqref{chm-btz-inside-inverse} allows us to map
	two copies of the spatial region outside the horizon in the constant time slice of 
	the hyperbolic black hole above described,
	parameterized by the two sets of coordinates $(z_\pm, \boldsymbol{u}_{\pm})$ 
	(see (\ref{hyp-metric})-(\ref{hyp-metric-dHd})),
	and that we denote by HYP$^{\pm}_{d+1}$ hereafter,
	into two complementary regions of $\mathbb{H}_{d+1}$ \cite{Casini:2011kv, Espindola:2018ozt},
	parameterized by the two sets of coordinates $(w_\pm, \boldsymbol{r}_{\pm})$ (see (\ref{Hd-metric})).
	These mappings read
	\be 
	\label{chm-hyp-inside}
	r_\pm  =   \frac{z_h\, \sinh(u_\pm /z_h)}{ \cosh(u_\pm /z_h) \pm \sqrt{1-(z_\pm/z_h)^2} }
	\;\;\; \qquad \;\;\;
	w_\pm =  \frac{z_\pm}{ \cosh(u_\pm /z_h) \pm \sqrt{1-(z_\pm /z_h)^2} }
	\ee
	where $u_\pm \geqslant 0$ and whose inverse are
	\be
	\label{chm-hyp-inside-inverse}
	u_\pm = \frac{z_h}{2}\, \log\! \left(  \frac{(z_h + r_\pm)^2 + w_\pm^2}{(z_h - r_\pm)^2 + w_\pm^2}  \right)
	\;\;\qquad\;\;
	z_\pm = \frac{2z_h^2\, w_\pm}{\sqrt{ z_h^4+2 z_h^2\big(w_\pm^2 - r_\pm^2\big) +\big(w_\pm^2 + r_\pm^2\big)^2 }} \,.
	\ee
	These maps
	send the background (\ref{hyp-metric}) for the coordinates $(z_\pm,  \boldsymbol{u}_{\pm}) = (z_\pm,  u_{\pm}, \boldsymbol{\Omega}_{\pm})$
	to the metric (\ref{Hd-metric}) in the coordinates $(w_\pm, \boldsymbol{r}_{\pm}) = (w_\pm, r_{\pm}, \boldsymbol{\Omega}_{\pm}) $
	constrained by $r_+^2 + w_+^2 < z_h^2$ and $r_-^2 + w_-^2 > z_h^2$,
	which define two $(d+1)$-dimensional  regions 
	that provide a specific bipartition of $\mathbb{H}_{d+1}$.
	We remark that the maps in (\ref{chm-hyp-inside}) send the horizons at $z_\pm = z_h$  onto the hemisphere $r_\pm^2 + w_\pm^2 = z_h^2$,
	which corresponds to the hypersurface separating the two subregions in the above bipartition of $\mathbb{H}_{d+1}$.
	When $d=1$, (\ref{chm-hyp-inside})-(\ref{chm-hyp-inside-inverse}) 
	reduce to \eqref{chm-btz-inside}-\eqref{chm-btz-inside-inverse}, as expected.

	Let us consider the bipartition of the boundary of HYP$^{+}_{d+1}$ 
	defined by the $d$-dimensional sphere $A \equiv \big\{ u_+ \leqslant b\big\} $ and its complement $B$.
	The map in \eqref{chm-hyp-inside} sends $A$ into 
	the sphere $\tilde{A} \equiv \big\{ r_+ \leqslant \tilde{b}\, \big\} $,
	whose radius is given by \eqref{interval-before-chm},
	in the boundary of the part of $\mathbb{H}_{d+1}$ where $r_+^2 + w_+^2 < z_h^2$.
	Since the RT hypersurface corresponding to $\tilde{A} $ is the hemisphere with radius $\tilde{b}$,
	we can employ (\ref{chm-hyp-inside-inverse}) to obtain the RT hypersurface associated with $A$:
	it is described by \eqref{RT-curve-btz-brane}
	with $x_m \in A$ replaced by $u_{+,m} \in A$.

	The vector field characterizing the geodesic bit threads for $A$ can be found
	by applying the maps (\ref{chm-hyp-inside-inverse})
	to the vector field of the geodesic bit threads for $\tilde{A}$,
	which is obtained  by replacing $b$ with $\tilde{b}$ in \eqref{Vd}.
	This provides both the vector field $\boldsymbol{V}_{\!\! +} $ for the geodesic bit threads in HYP$^{+}_{d+1}$
	and the vector field $\boldsymbol{V}_{\!\! -} $ for the auxiliary geodesics in HYP$^{-}_{d+1}$.
	They read
	\be
	\label{vector-field-hyp}
	\boldsymbol{V}_{\!\! \pm} 
	=  
	\big( \, V_\pm^{z_\pm} , \, V_\pm^{u_\pm} \,\big) 
	=  
	|\boldsymbol{V}_{\!\! \pm}| \,\boldsymbol{\tau}^\mu_\pm   
	=
	|\boldsymbol{V}_{\!\! \pm}|
	\,\big( \,\tau^{z_\pm}_\pm  , \, \tau^{u_\pm}_\pm \,\big) 
	\ee
	where the amplitudes  $|\boldsymbol{V}_{\!\! \pm}|  $ and the  unit tangent vectors $\boldsymbol{\tau}_\pm   $ are given respectively by
	\be 
	\label{mod-gbt-hyp}
	|\boldsymbol{V}_{\!\! \pm}|  
	=  
	\frac{ \big[ (z_\pm /z_h) \sinh (b/z_h )\big]^d
	}{ 
		\left\{
		\left[\, \cosh \!\big( (b-u_\pm )/z_h \big)  \mp \sqrt{1-(z_\pm /z_h)^2} \; \right] 
		\left[\, \cosh \! \big((b+u_\pm )/z_h \big) \mp \sqrt{1-(z_\pm /z_h)^2} \;\right]
		\right\}^{d/2}}
	\ee 
	and 
	\bea
	\label{tau-gbt-hyp-components-2}
	\tau_\pm^{z_\pm}
	&=&  
	\frac{ |\boldsymbol{V}_{\!\! \pm} |^{1/d} }{L_{\textrm{\tiny AdS}} }\;
	\frac{ \sqrt{z_h^2-z_\pm^2} }{\sinh (b/z_h) }
	\left[\,
	\pm \cosh (b/z_h )  - \sqrt{1-(z_\pm/z_h)^2} \, \cosh (u_\pm/z_h) 
	\,  \right]
	\\
	\rule{0pt}{1cm}
	\label{tau-gbt-hyp-components-1}
	\tau_\pm^{u_\pm}
	&=& 
	\frac{
		|\boldsymbol{V}_{\!\! \pm} |^{1/d}  
	}{L_{\textrm{\tiny AdS}} }\;
	\frac{ z_\pm \sinh (u_\pm/z_h) }{ \sinh (b/z_h)} \;.
	\eea
	Setting $d=1$ in these expressions, the vector fields described by (\ref{mod-gbt-btz})-(\ref{tau-gbt-btz-components-1}) are recovered, as expected. 
	By adapting the analysis reported in Sec.\,\ref{subsec-BTZ-plane-GBT} to these vector fields,
	it is straightforward to observe that the expressions in \eqref{mod-gbt-hyp} satisfy $|\boldsymbol{V}_{\!\! \pm}| \leqslant 1$,
	where the equality holds only on the RT hypersurface, as required for consistent bit threads.
	By restricting the bottom left panel and bottom right panel of  Fig.\,\ref{fig:BTZ-brane-main}
	to their halves characterized by $x_+ \geqslant 0$ and $x_{-} \geqslant 0$ respectively,
	one obtains the integral lines of (\ref{vector-field-hyp}) in HYP$^{+}_{d+1}$ and HYP$^{-}_{d+1}$ 
	respectively.
	They include the critical line arriving at the horizon at infinity,
	obtained from the rightmost magenta curve 
	in the bottom left panel of  Fig.\,\ref{fig:BTZ-brane-main}.
	Furthermore, 
	for HYP$^{+}_{d+1}$ and only at qualitative level, 
	the pattern of these geodesic bit threads is similar to the one displayed 
	in Fig.\,\ref{fig:HigherDPBBspherebitthreds} with $r$ replaced by $u$.
	This critical thread intersects the RT hypersurface at $(z_{m,\beta}\,, u_{m,\beta}  )$ and the boundary at $u= b_{\beta}$, 
	which are given respectively 
	by \eqref{x-z-beta-coord-from-b} with $x_{m,\beta}$ replaced by $u_{m,\beta}$
	and by \eqref{bbeta}.
	%
	Notice that the dimensionality parameter $d$ does not occur in the profile of the integral lines of the geodesics that we are considering.

	
	In the limit $b \to + \infty$, the spherical domain $A$ becomes the entire boundary of HYP$^{+}_{d+1}$ 
	and the corresponding RT hypersurface displays a plateau that approximates the entire horizon of HYP$^{+}_{d+1}$.
	In $\mathbb{H}_{d+1}$, from \eqref{interval-before-chm} for $\tilde{A}$ we have that $\tilde{b} \to z_h$ in this limit;
	hence the corresponding RT hypersurface becomes $r_+^2 + w_+^2 = z_h^2$,
	namely the hypersurface characterizing the above partition of $\mathbb{H}_{d+1}$.
	In this limiting regime, the integral lines of the geodesic bit threads for $A$ in HYP$^{+}_{d+1}$ 
	and of the auxiliary geodesics in HYP$^{-}_{d+1}$ become vertical straight lines 
	whose moduli are given by 
	\bea
	\lim_{b \to +\infty}
	|\boldsymbol{V}_{\!\! \pm}|
	= 
	\big(z_\pm /z_h \big)^d
	\label{d1limit-modulus-hyp}
	\eea 
	which is obtained from (\ref{mod-gbt-hyp}). 
	This provides a higher dimensional generalization of the 
	$d=1$ results shown in Fig.\,\ref{fig:ads3-main}; 
	indeed \eqref{d1limit-modulus-hyp} reduces to \eqref{eq:CHMmodulusADS} for $d=1$,
	with $z_h$ playing the role of $b$, as expected.

	\subsection{Fluxes through the boundary and the horizon}

	In the following, we study the fluxes of $\boldsymbol{V}_{\!\! \pm} $ 
	in (\ref{vector-field-hyp})-(\ref{tau-gbt-hyp-components-1})
	through either the boundary or the horizon of HYP$_{d+1}^\pm$.

	%
	
	The flux densities $\Phi_\pm (\boldsymbol{u}_\pm ; z_{\pm,0})  $ of $\boldsymbol{V}_{\!\! \pm} $ through a generic slice at constant $z_\pm = z_{\pm,0}$
	can be obtained from (\ref{hyp-metric}) and (\ref{vector-field-hyp})-(\ref{tau-gbt-hyp-components-1}),
	finding 
	\bea
	\label{hyp-flow-zfixed}
	& &
	\hspace{-.4cm}
	\frac{4 G_{\textrm{\tiny N}}}{L_{\textrm{\tiny AdS}}^d }\;
	\Phi_\pm (\boldsymbol{u}_\pm ; z_{\pm,0})  
	\, \equiv \,
	\frac{
		\; g_{z_\pm z_\pm}  V^{z_\pm}_\pm  \, n^{z_\pm} }{ L_{\textrm{\tiny AdS}}^d\, z_\pm^{d}} \, \bigg|_{z_\pm = z_{\pm,0}}
	\\
	\rule{0pt}{.9cm}
	& & 
	=
	\frac{
		\big(  \sinh(b/z_h) / z_h \big)^d \,
		\big[ 	
		\pm \cosh (b/z_h )  - \sqrt{1-(z_{\pm,0}/z_h)^2}\, \cosh (u_\pm /z_h) 
		\big] 
	}{ 
		\left(
		\left[\, \cosh ( (b-u_\pm)/z_h )  \mp \sqrt{1-(z_{\pm,0}/z_h)^2} \; \right] 
		\left[ \, \cosh ((b+u_\pm)/z_h) \mp \sqrt{1-(z_{\pm,0}/z_h)^2} \; \right]
		\right)^{\frac{d+1}{2}}
	}
	\nn
	\eea 
	where
	the non-vanishing elements $g_{zz}$ and $g_{uu}$ of the metric \eqref{hyp-metric} 
	and 
	the unit vector $ \boldsymbol{n}  = (n^{z_\pm}, n^{u_\pm} ) $  normal to the constant $z_\pm$ slice
	have been used.
	When $d=1$, the expression \eqref{hyp-flow-zfixed} reduces to \eqref{btz-flow-zfixed}, as expected.
	The holographic contour function for the entanglement entropy of the sphere $A$ in the boundary of HYP$^{+}_{d+1}$ 
	can be found by taking the limit $z_{+,0} \,\to\, 0$ of $\Phi_+ (\boldsymbol{u}_+ ; z_{+,0}) $ in \eqref{hyp-flow-zfixed}, for $\boldsymbol{u}_+\in A$.
	Instead, 	the limit $z_{-,0} \,\to\, 0$ of $\Phi_{-} (\boldsymbol{u}_- ; z_{-,0}) $ in \eqref{hyp-flow-zfixed}
	provides the flux density on the boundary of HYP$^{-}_{d+1}$.
	By using $\beta = 2 \pi z_h $, we arrive to 
	\be
	\label{contour-hyp-bdy}
	\mathcal{C}_{A}^\pm (\boldsymbol{u}_\pm  )\,
	\equiv
	\lim_{z_{\pm,0} \,\to\, 0} 
	\big| \Phi_\pm (\boldsymbol{u}_\pm ; z_{\pm,0})   \big| \,
	=\,
	\frac{L_{\textrm{\tiny AdS}}^d }{4 G_{\textrm{\tiny N}}}  
	\left( 
	\frac{ 2 \pi \,\sinh (2 \pi  b / \beta )}{\beta\, \big| \cosh (2 \pi  b / \beta ) \mp \cosh (2 \pi  u_\pm  /  \beta  ) \, \big| 
	}
	\right)^d .
	\ee 
	The zero temperature limit $\beta \to +\infty$ of (\ref{contour-hyp-bdy}) gives
	\be
	\label{contour-hyp-betainfinite}
	\mathcal{C}_{A}^+ (\boldsymbol{u}_+  ) \to 
	\frac{L_{\textrm{\tiny AdS}}^d }{4 G_{\textrm{\tiny N}}}  \;
	\frac{(2b)^d}{\big| b^2 - u_+^2 \big|^d}
	\;\;\;\;\;\qquad\;\;\;\;\;
	\mathcal{C}_{A}^- (\boldsymbol{u}_-  ) \to \,0
	\ee
	hence, the zero temperature limit of $\mathcal{C}_A^+ (\boldsymbol{u}_+  )$ 
	gives (\ref{contour-function-vacuum}) for $d=1$, as expected,
	and also (\ref{contour-function-vacuum-ddim}) if $u_+$ is replaced by $r$.

	Another limiting regime that is worth exploring is defined by the limit $z_{\pm ,0} \,\to\, z_h$,
	where the flux densities \eqref{hyp-flow-zfixed} become
	\be 
	\label{contour-hyp-hor}
	\mathcal{C}_h (\boldsymbol{u}_\pm; \beta) 
	\equiv
	\lim_{z_{\pm,0} \,\to\,z_h} \!
	\big| \Phi_\pm (\boldsymbol{u}_\pm ; z_{\pm,0})   \big|
	= \,
	\frac{  L_{\textrm{\tiny AdS}}^d  }{4  G_{\textrm{\tiny N}} } \;
	\frac{ 
		\big[( 2 \pi / \beta)  \sinh (2 \pi b / \beta ) \big]^d
		\cosh (2 \pi b / \beta ) 
	}{
		\big[\, \tfrac{1}{2}
		\big(
		\cosh (4 \pi b / \beta )
		+    
		\cosh (4 \pi u_\pm  / \beta )
		\big)
		\big]^{ \frac{d+1}{2}}
	}\,.
	\ee

	Taking the limit $b/\beta \to +\infty$ in the expressions in \eqref{contour-hyp-bdy} and \eqref{contour-hyp-hor},
	one finds 
	\be
	\label{contour-hyp-binfinite}
	\mathcal{C}_{A}^+  (\boldsymbol{u}_+  ) \to 
	s_{\textrm{\tiny th}}
	\;\;\qquad\;\;
	\mathcal{C}_{h} (\boldsymbol{u}_\pm  ) \to 
	s_{\textrm{\tiny th}}
	\; \frac{1+\chi_A(\boldsymbol{u}_\pm)}{2}
	\;\;\qquad\;\;
	\mathcal{C}_{A}^-  (\boldsymbol{u}_-  ) \to 
	s_{\textrm{\tiny th}}
	\; \frac{1+\chi_A(\boldsymbol{u}_-)}{2}
	\ee 
	where $\boldsymbol{u}_+  \in A$ for $\mathcal{C}_{A}^+  (\boldsymbol{u}_+  )$,
	the step function $\chi_A(\boldsymbol{u} )$ 
	is equal to $+1$ for $\boldsymbol{u} \in A$ and to $-1$ for $\boldsymbol{u} \in B$,
	and 
	the holographic density of thermal entropy is defined as follows \cite{Emparan:1999gf}
	\be 
	\label{SBthermalentropy-ddim}
	s_{\textrm{\tiny th}} 
	\equiv 
	\frac{1}{4 G_{\textrm{\tiny N}}} \; \frac{L_{\textrm{\tiny AdS}}^{d}}{z_h^d}
	=
	\frac{L_{\textrm{\tiny AdS}}^d }{4 G_{\textrm{\tiny N}}} 
	\left( 
	\frac{2 \pi}{\beta}
	\right)^d\,.
	\ee
	This thermal entropy density can also be found from the factor multiplying the integral in Eq.\,(3.4) of \cite{Casini:2011kv}, 
	by employing the expression for $a_d^*$ that can be extracted from Eq.\,(7.10) of \cite{Ryu:2006ef}.
	For $d=1$, the thermal density (\ref{SBthermalentropy-ddim}) becomes \eqref{SBthermalentropy} with $c$ given by the Brown-Henneaux central charge (\ref{BH-central-charge}), as expected.
	Thus, the holographic entanglement entropy of a sphere in AdS$_{d+2}$ 
	can be interpreted as the holographic thermal entropy in the hyperbolic black hole,
	as first highlighted in \cite{Casini:2011kv}.

	
	To evaluate the holographic entanglement entropy of $A$ through its holographic contour function in (\ref{contour-hyp-bdy}),
	a suitable UV regularisation is needed. 
	Also, in this case, it is convenient to adopt the  entanglement wedge cross-section regularization 
	\cite{Dutta:2019gen,Han:2019scu,Headrick:2022nbe, Nguyen:2017yqw,Takayanagi:2017knl},
	following the straightforward generalization to higher dimensions
	of the procedure described in Sec.\,\ref{subsec-flows-btz-planar}. 
	By introducing the cutoff at $z= \varepsilon_{\textrm{\tiny H}}$ in the holographic direction,
	one can define $A_{\varepsilon} \equiv \big\{u_+ < b - \varepsilon_{\textrm{\tiny bdy}}^{\textrm{\tiny $A$}} \big\} \subsetneq A $
	and $B_{\varepsilon} \equiv \big\{ u_+ > b + \varepsilon_{\textrm{\tiny bdy}}^{\textrm{\tiny $B$}} \big\} \subsetneq B$, 	
	where $ \varepsilon_{\textrm{\tiny bdy}}^{\textrm{\tiny $A$}}$ and $ \varepsilon_{\textrm{\tiny bdy}}^{\textrm{\tiny $B$}}$ 
	are obtained through \eqref{pBTZ1-eps-AB} with $\varepsilon_{\textrm{\tiny BTZ}}$ replaced by $ \varepsilon_{\textrm{\tiny H}}$.

			\begin{figure}[t!]
			\vspace{-.2cm}
			\hspace{-1.5cm}
			\begin{minipage}{0.32\textwidth}
				\centering
				\includegraphics[width=1.4\textwidth]{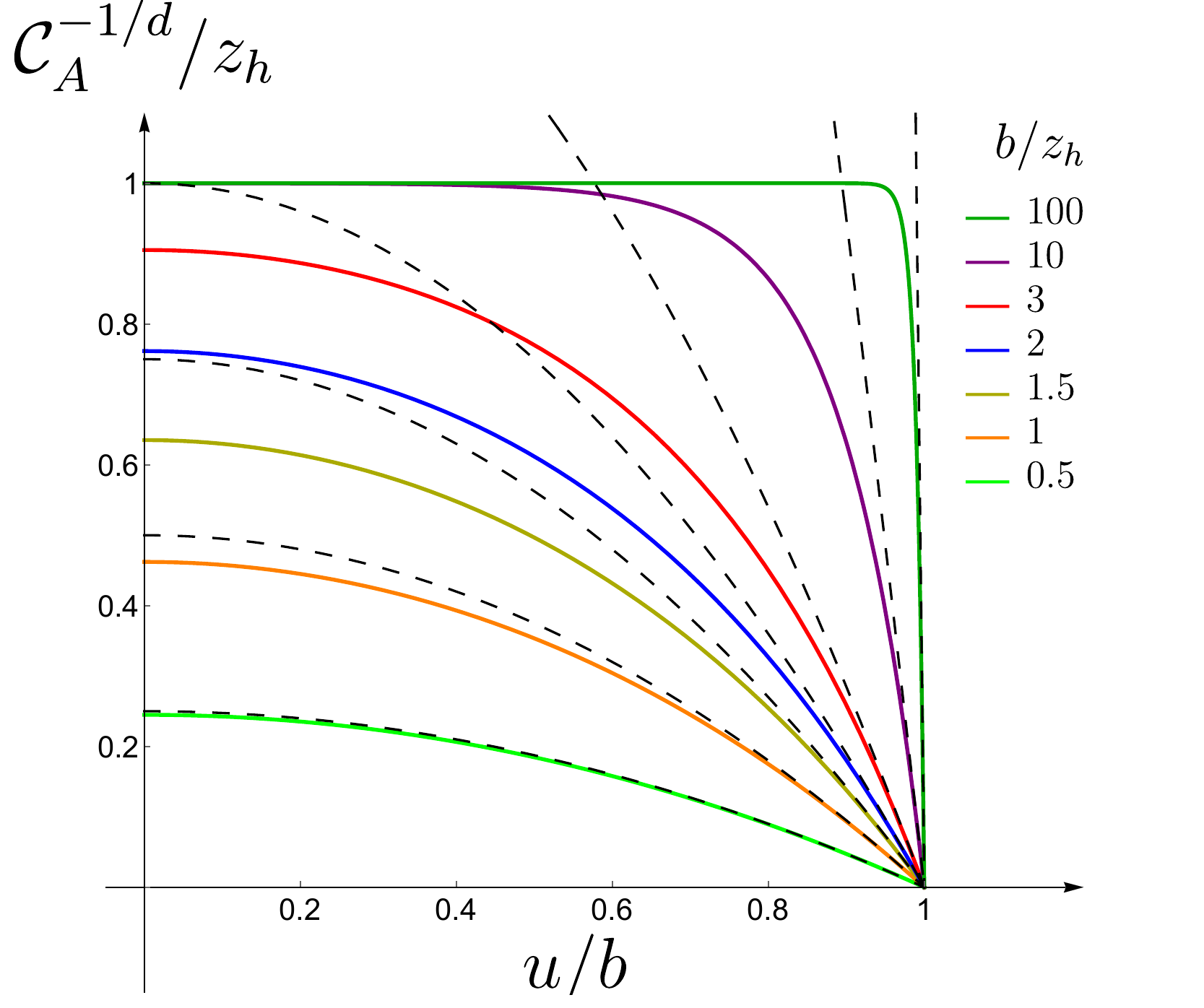}
			\end{minipage}
			\hspace{.8cm}
			\begin{minipage}{0.32\textwidth}
				\centering
				\includegraphics[width=1.4\linewidth]{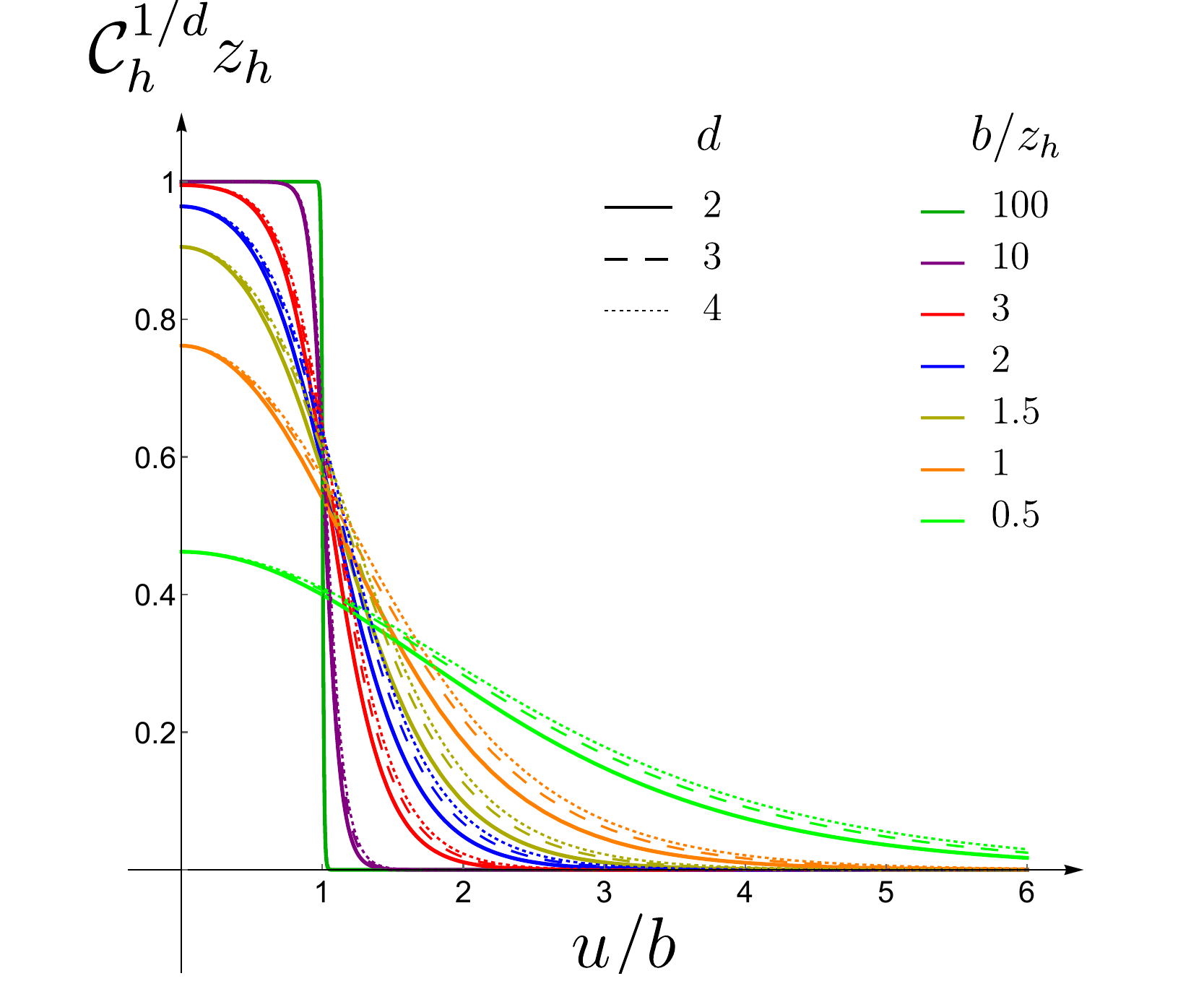}
			\end{minipage}
			\hspace{.8cm}
			\begin{minipage}{0.32\textwidth}
				\centering
				\includegraphics[width=1.4\linewidth]{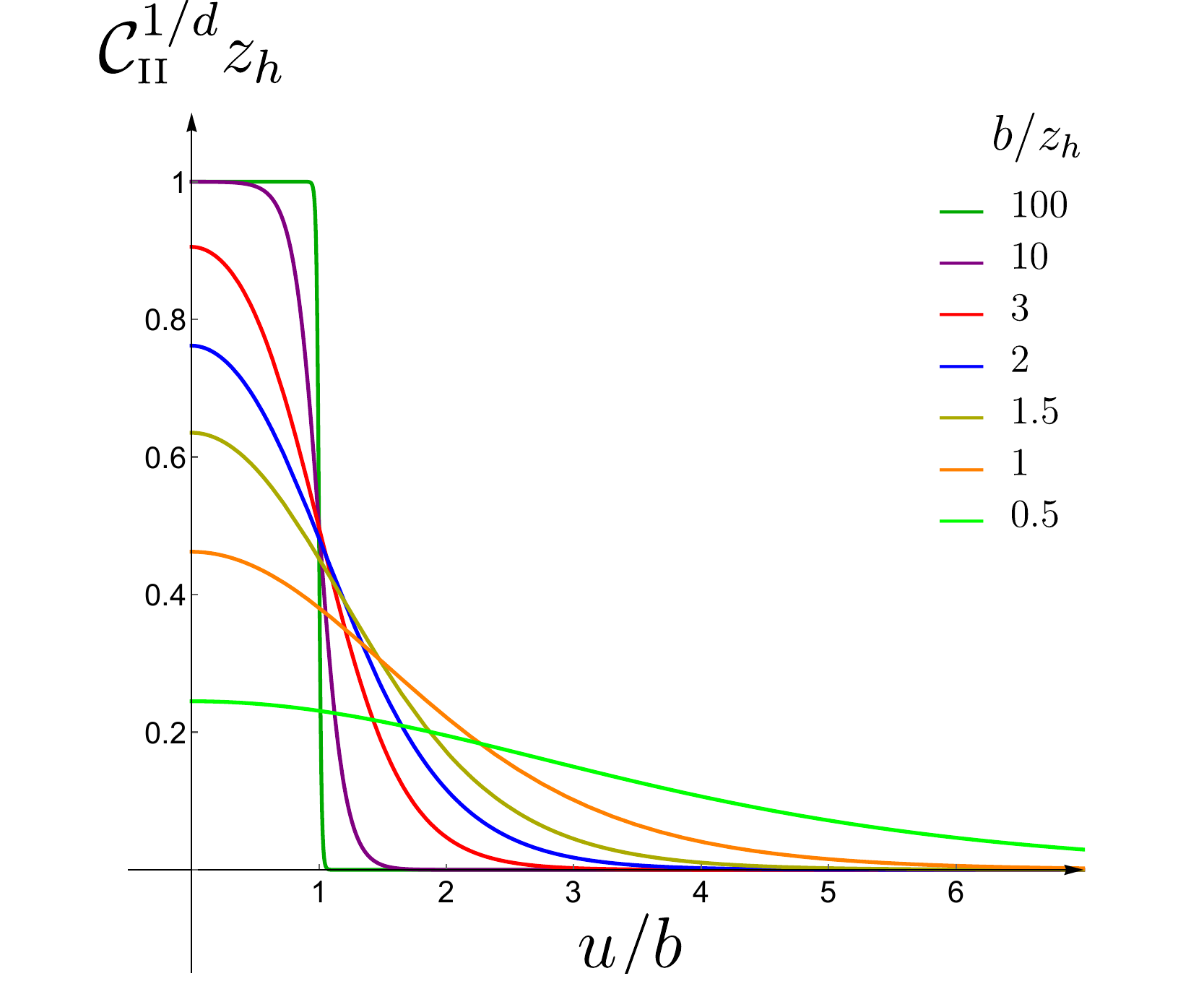}
			\end{minipage}
			\vspace{.1cm}
			\caption{\small 
				Holographic contour functions for a sphere $A$  induced by the geodesic bit threads 
				in the constant time slice of the $(d+2)$-dimensional hyperbolic black hole (\ref{hyp-metric}),
				from \eqref{contour-hyp-bdy} and (\ref{contour-hyp-hor}):
				in $A$ (left), on the horizon (middle)
				and for the auxiliary geodesics on the whole boundary (right).
			}
			\label{fig:HYPcontourbordoorizsfera}
		\end{figure}

		In Fig.\,\ref{fig:HYPcontourbordoorizsfera},
		by setting $ L_{\textrm{\tiny AdS}}^d /(4 G_{\textrm{\tiny N}}) =1$  for the sake of simplicity,
		we show $z_h \big[ \mathcal{C}_{A}^\pm (\boldsymbol{u}_\pm  ) / \tfrac{L_{\textrm{\tiny AdS}}^d }{4 G_{\textrm{\tiny N}}}   \big]^{1/d}$ 
		from \eqref{contour-hyp-bdy} 
		(left panel and right panel respectively)
		and $z_h \big[ \mathcal{C}_h (\boldsymbol{u}_\pm)  / \tfrac{L_{\textrm{\tiny AdS}}^d }{4 G_{\textrm{\tiny N}}}   \big]^{1/d}$ 
		from (\ref{contour-hyp-hor})
		(middle panel)
		as functions of $u_\pm/b$ 
		and for various choices of $b/z_h$.
		In particular, in the left panel, we have considered 
		$\tfrac{1}{z_h} \big[ \mathcal{C}_{A}^\pm (\boldsymbol{u}_\pm  ) / \tfrac{L_{\textrm{\tiny AdS}}^d }{4 G_{\textrm{\tiny N}}}   \big]^{-1/d}$ 
		in order to highlight properly the regime $u_\pm/b \to 1$,
		which is responsible of the area law in the holographic entanglement entropy. 
		 From \eqref{contour-hyp-bdy} and (\ref{contour-hyp-hor}),
		 it is straightforward to realize that 
		 $z_h \big[ \mathcal{C}_{A}^\pm (\boldsymbol{u}_\pm  ) / \tfrac{L_{\textrm{\tiny AdS}}^d }{4 G_{\textrm{\tiny N}}}   \big]^{1/d}$ 
		 are independent of $d$,
		 while $z_h \big[ \mathcal{C}_h (\boldsymbol{u}_\pm)  / \tfrac{L_{\textrm{\tiny AdS}}^d }{4 G_{\textrm{\tiny N}}}   \big]^{1/d}$ 
		 displays a mild dependence on $d$,
		 as one observes by comparing 
		 the solid, dashed, and dotted lines having the same color in the middle panel of Fig.\,\ref{fig:HYPcontourbordoorizsfera}.
		 In the left panel of Fig.\,\ref{fig:HYPcontourbordoorizsfera}, 
		 the dashed curves correspond to $b\big[ 1 -(u/b)^2\big]/(2z_h)$,
		 which are obtained from the zero temperature regime (\ref{contour-hyp-betainfinite})
		 and capture the expected behavior as $u/b \to 1^-$. 
In the right panel of Fig.\,\ref{fig:HYPcontourbordoorizsfera}, 
in the label of the vertical axis $\mathcal{C}_{A}^- (\boldsymbol{u}_-)$ 
has been denoted by $\mathcal{C}_{\text{\tiny II}} (\boldsymbol{u})$ 
in order to facilitate comparison with the results presented in Sec.\,\ref{Schwarzschild AdS black brane}.  
Indeed, in that case the analog of the mapping \eqref{chm-hyp-inside} does not exist 
and $\mathcal{C}_{\text{\tiny II}} (\boldsymbol{u})$ denotes the holographic contour function associated with the auxiliary branch of the geodesics reaching the horizon;
hence it is natural to associate  $\mathcal{C}_{\text{\tiny II}} (\boldsymbol{u})$ with $\mathcal{C}_{A}^- (\boldsymbol{u}_-)$.
Notice that the curves corresponding to large $b/z_h$ in Fig.\,\ref{fig:HYPcontourbordoorizsfera}
are compatible with (\ref{contour-hyp-binfinite}).


	By employing the fact that on the slice at fixed $z_\pm =  z_{\pm,0}$, from \eqref{hyp-metric-dHd},
	the volume element is 
	$\rd^d \boldsymbol{u} \equiv \big(z_h  \sinh{(u/z_h)} \big)^{d-1} \rd u \wedge \rd \boldsymbol{\Omega} $,
	the holographic entanglement entropy of $A$ can be found as 
	the flux of $\mathcal{C}_A^+ (\boldsymbol{u} _+  )$ in \eqref{contour-hyp-bdy} through  $A_{\varepsilon}$, namely
	\bea
	\label{HEE-hyp-bdy}
	S_{A }
	&=&
	\int_{A_{\varepsilon}} 
	\mathcal{C}_{A}^+ (\boldsymbol{u} _+  )  \, 
	\rd ^d \boldsymbol{u}_+
	\,=\,
	\frac{ L^d_{\textrm{\tiny AdS}}}{4  G_{\textrm{\tiny N}}} \;
	\frac{2^d}{d} 
	\big( \widetilde{\mathcal{P}}_{-} / \widetilde{\mathcal{P}}_{+} \big)^{d/2}
	\;  _2F_1 \! \left(d/2\, , d \,; (d+2)/2 \,;	\widetilde{\mathcal{P}}_{-} / \widetilde{\mathcal{P}}_{+}	\right) 
	\\ 
	\label{HEE-hyp-bdy-step1}
	\rule{0pt}{1.cm}
	&=&	
	\frac{ L^d_{\textrm{\tiny AdS}}}{4  G_{\textrm{\tiny N}}} \;
	\frac{ 2 \pi^{d/2}}{ \Gamma(d/2) }\,
	\Large\Bigg[ \,
	\sum_{n=0}^{\lfloor (d/2) -1 \rfloor  }
	\sum_{j=0}^{\lfloor (d/2) -1 -n \rfloor  }
	Q_{d; n,j}	\big(\sinh(b/z_h)\big)^{d-1-2n}
	\left(
	\frac{z_h}{\varepsilon_{\textrm{\tiny H}}}
	\right)^{d- 2j -2n -1} 
	\nn
	\\ 
	\rule{0pt}{1.3cm}
	&&  \hspace{3cm}
	+ \,
	\begin{cases}
		\displaystyle
		(-1)^{d/2}    \ 
		\frac{( d -2 )!! }{ (d-1)!!}
		&
		\textrm{even $d$}
		\\
		\rule{0pt}{.8cm}
		\displaystyle
		(-1)^{\frac{d-1}{2} }
		\frac{( d-2)!! }{ (d-1)!!}\,
		\log ( z_h  /\varepsilon_{\textrm{\tiny H}} )
		+O(1)
		&
		\textrm{odd $d$} 
	\end{cases}  
	\;+ o(1)
	\, \Large\Bigg] 
	\hspace{1.2cm}
	\eea
	in terms of 
	\be
	\widetilde{\mathcal{P}}_\pm \equiv \sqrt{z_h^2-\varepsilon_{\textrm{\tiny H}}^2} \, \sinh (b/z_h) \pm \varepsilon_{\textrm{\tiny H}}
	\ee
	(see also (\ref{calP-pm-def}))  and
	\be
	Q_{d; n,j} \equiv
	(-1)^{n+j}\,
	\frac{(d-2)!!}{(2n)!! \, (2j)!! } \;
	\frac{(d-3 -2n)!! }{(d-2-2n)!! \, (d-2-2n-2j)!!}\,.
	\ee

	Let us briefly discuss the derivation of the formula \eqref{HEE-hyp-bdy}.
	Given the domains $A$ and $\tilde{A}$ introduced in Sec.\,\ref{sec-hyp-bh-gbt},
	the RT hypersurface of $\tilde{A}$
	intersects 
	the UV cutoff hyperplane in $\mathbb{H}_{d+1}$ at $w = \varepsilon_{\textrm{\tiny AdS}}$
	along the hypersphere whose points have coordinates
	$w_m =  \varepsilon_{\textrm{\tiny AdS}}$ and $r_m = \big( \tilde{b}^2- \varepsilon^2_{\textrm{\tiny AdS}} \big)^{1/2}$. 
	By applying \eqref{chm-hyp-inside-inverse} to this hypersphere, we obtain the hypersphere $(z_m^+, u_m^+)$  in HYP$_{d+1}^+$. 
	Then, identifying the UV cutoff hyperplane at $z =   \varepsilon_{\textrm{\tiny H}} $ in the hyperbolic geometry 
	and the hyperplane crossing $\gamma_{A}$
	along the hypersphere $(z_m^+, u_m^+)$, we find
	\be 
	\label{HYP-ADS-cutoff-relation}
	\varepsilon_{\textrm{\tiny AdS}} 
	= 
	\frac{\varepsilon_{\textrm{\tiny H}} }{\sqrt{1- ( \varepsilon_{\textrm{\tiny H}} / z_h)^2} \; \big[ 1 +  \cosh  (b/ z_h) \big]} \,.
	\ee 
	In order to check this result, we observe that,
	plugging \eqref{interval-before-chm} and \eqref{HYP-ADS-cutoff-relation} into \eqref{ads-EWCS-regEndp}, 
	one obtains \eqref{pBTZ1-eps-AB}-(\ref{calP-pm-def}) 
	with $\varepsilon_{\textrm{\tiny BTZ}} $ and $b$ 
	replaced by $ \varepsilon_{\textrm{\tiny H}}$ and $\tilde{b}$ respectively. 
	The holographic contour function for $\tilde{A}$ is given by 
	\eqref{contour-function-vacuum-ddim} with $b$ replaced by $\tilde{b}$ introduced in \eqref{interval-before-chm}.
	Then, the integral in the first line of \eqref{HEE-hyp-bdy} is obtained by
	applying \eqref{interval-before-chm} and \eqref{HYP-ADS-cutoff-relation} 
	to such holographic contour function and its associated volume element.
	The analytic expression containing the hypergeometric function in \eqref{HEE-hyp-bdy} is found 
	by plugging \eqref{interval-before-chm} and \eqref{HYP-ADS-cutoff-relation} 
	into the definite integral occurring in the first line of \eqref{HEE-SA-full-d}, 
	once $b$ has been replaced with $\tilde{b}$ in \eqref{interval-before-chm}.

	Finally, the expansion of $S_{A}$ as $\varepsilon_{\textrm{\tiny H}} \to 0$ in (\ref{HEE-hyp-bdy-step1}) 
	is obtained by employing \eqref{interval-before-chm} and (\ref{HYP-ADS-cutoff-relation}), 
	observing that 
	\bea
	\label{HYP-ADS-btilde-over-epsilon}
	\big( \,\tilde{b} / \varepsilon_{\textrm{\tiny AdS}}  \big)^k
	&=&
	\left[ \sqrt{ (  z_h / \varepsilon_{\textrm{\tiny H}}  )^2 -1 } \sinh  (b/ z_h) \right]^k 
	\\
	&=&
	\big[ \sinh  (b/ z_h) \big]^k  
	\sum_{j=0}^{\lfloor k/2 \rfloor }
	\frac{(-1)^j}{(2j)!!}\;
	\frac{k!!}{(k- 2j)!!} 
	\bigg( 
	\frac{\varepsilon_{\textrm{\tiny H}} }{z_h}
	\bigg)^{2j -k}
	\! + O(\varepsilon_{\textrm{\tiny H}}) 
	\nn
	\eea
	which can be plugged into the last line of \eqref{HEE-SA-full-d}, 
	once $b$ has been replaced with $\tilde{b}$ in \eqref{interval-before-chm}.

	For the spherical region $A$, we find it worth considering also
	the integral of the flux density $\mathcal{C}_A^+ (\boldsymbol{u}_+  )$ in \eqref{contour-hyp-bdy} 
	through the spherical region $A_{\beta } \equiv \{ u \leqslant b_\beta \} \subsetneq A$ with $b_\beta$ given by \eqref{bbeta},
	extending to this higher dimensional case, the analysis performed in Sec.\,\ref{subsec-flows-btz-planar} for the interval.
	The result is 
	\be 
	\label{Gibbs-ent-definition-hyp}
	S_{A, \textrm{\tiny th}}
	\equiv 
	\int_{A_{\beta} } 
	\mathcal{C}^+_A (\boldsymbol{u}_+  )\, \rd ^d \boldsymbol{u}_+
	= 
	\,s_{\textrm{\tiny th}} 
	\textrm{Vol} (A)
	\ee 
	where $s_{\textrm{\tiny th}} $ has been defined in (\ref{SBthermalentropy-ddim}) and
	\be 
	\textrm{Vol} ( A) 
	\,= 	
	\int_{A } \rd ^d \boldsymbol{u}_+
	\,=\, 
	\frac{2^d}{d}\;
	\frac{2 \pi ^{d/2}}{\Gamma(d/2)}
	\left[ \frac{\beta }{ 2 \pi} \tanh ( \pi b / \beta ) \right]^d
	{}_2F_1 \bigg(\frac{1}{2} \,,\frac{d}{2} \,;\frac{d+2}{2} \,; [\tanh ( \pi b/ \beta )]^2 \bigg) \, .
	\ee 
	Notice that $S_{A, \textrm{\tiny th}}$  in \eqref{Gibbs-ent-definition-hyp} is the holographic thermal entropy corresponding to the spherical region $A$,
	which becomes \eqref{Gibbs-ent-definition} in the special case of $d=1$, as expected. 
	
	The holographic thermal entropy $S_{A, \textrm{\tiny th}}$ in \eqref{Gibbs-ent-definition-hyp} can be found  also 
	by integrating  either $\mathcal{C}_h (\boldsymbol{u}_\pm; \beta) $ in \eqref{contour-hyp-hor} over the horizon 
	or $\mathcal{C}_A^- (\boldsymbol{u}_-  )$ in \eqref{contour-hyp-bdy}  over the entire boundary of HYP$_{d+1}^-$, 
	namely
	\be
	\label{eq:PhiHorizonIntegral-hyp}
	S_{A, \textrm{\tiny th}}
	= \int_{\mathbb{H}_d  } \mathcal{C}_h (\boldsymbol{u}_\pm  ) \, \rd ^d \boldsymbol{u}_\pm
	\, = \int_{\mathbb{H}_d  } \mathcal{C}^-_A (\boldsymbol{u}_-  )\,  \rd ^d \boldsymbol{u}_-
	\, = 
	\,s_{\textrm{\tiny th}} 
	\textrm{Vol} ( A)
	\ee 
	where the horizon and the boundary are $d$-dimensional hyperbolic spaces $\mathbb{H}_d$ (see \eqref{hyp-metric-dHd}).
	Notice that \eqref{eq:PhiHorizonIntegral-hyp} becomes \eqref{eq:PhiHorizonIntegral} for $d=1$, as expected.

	The integral of the flux density $\mathcal{C}_A^+ (\boldsymbol{u}_+  )$ in \eqref{contour-hyp-bdy} over $B_{\varepsilon}$ reads
	\bea
	\label{SB-tilde-hyp}
	\widetilde{S}_{B}
	&\equiv &
	\int_{B_{\varepsilon} } \!
	\mathcal{C}^+_{A} (\boldsymbol{u}_+  )\,   \rd ^d \boldsymbol{u}_+
	= 
	S_{A} - S_{A, \textrm{\tiny th}}
	\eea  
	where  $S_A$ is the holographic entanglement entropy of the spherical region $A$ given in (\ref{HEE-hyp-bdy})
	and $S_{A, \textrm{\tiny th}}$ is the corresponding holographic thermal entropy (\ref{Gibbs-ent-definition-hyp}).	
	We remark that (\ref{SB-tilde-hyp}) is not  the entanglement entropy $S_B$ of the complementary domain $B =  \mathbb{H}_d \setminus A$, 
	as discussed  in Appendix\;\ref{app-entropy-complementary-btz} for the BTZ black brane.

	\section{Schwarzschild AdS$_{d+2}$ black brane}
	\label{Schwarzschild AdS black brane}

	In this section, we study the geodesics bit threads 
	and the minimal hypersurface inspired bit threads
	for the sphere (Sec.\,\ref{sec-Sch-AdS-sphere}) and the strip (Sec.\,\ref{sec-Sch-AdS-strip})
	when the gravitational background is the 
	constant time slice of the Schwarzschild AdS$_{d+2}$ black brane,
	whose metric is
	\begin{equation}
		\label{sch-ads-brane-metric}
		ds^2 = \frac{L_{\textrm{\tiny AdS}}^2}{z^2} 
		\left(\,
		\frac{\rd z^2}{f(z)} 
		+
		\rd \boldsymbol{x}^2 
		\right)
	\end{equation}
	where $f(z) \equiv 1 - (z/z_h)^{d+1}$, the holographic direction corresponds to $z>0$
	and $\boldsymbol{x} \in \RR^d$ parameterizes any translation invariant  $z=\textrm{const}$ slice
	(e.g. the boundary and the horizon, at $z=0$ and $z=z_h$ respectively).
	When $d=1$, the metric (\ref{sch-ads-brane-metric}) becomes the one equipping 
	the constant time slice of the BTZ black brane in (\ref{btz-brane-metric}).
	The dual CFT$_{d+1}$ living in the $(d+1)$-dimensional Minkowski space 
	on the boundary of the Schwarzschild AdS$_{d+2}$ black brane
	has finite inverse temperature $\beta= 4\pi z_h /(d+1)$ \cite{Emparan:1999gf}.

	\subsection {Sphere}
	\label{sec-Sch-AdS-sphere}

	When the constant time slice of the CFT$_{d+1}$ living on the boundary
	is bipartite by a sphere $A$  centered at the origin with radius $b>0$ 
	and its complementary region  $B \equiv \RR^d \setminus A$,
	the spherical symmetry suggests adopting the polar coordinates $ \boldsymbol{r} =(r, \boldsymbol{\Omega})$ for $\RR^d$ in the boundary
	and consequently $(z,\boldsymbol{r})$ parameterize the points on the constant time slice
	of the Schwarzschild AdS$_{d+2}$ black brane
	(see (\ref{sch-ads-brane-metric})).

\begin{figure}[t!]
		\vspace{-.5cm}
		\hspace{.3cm}
		\includegraphics[width=0.9\textwidth]{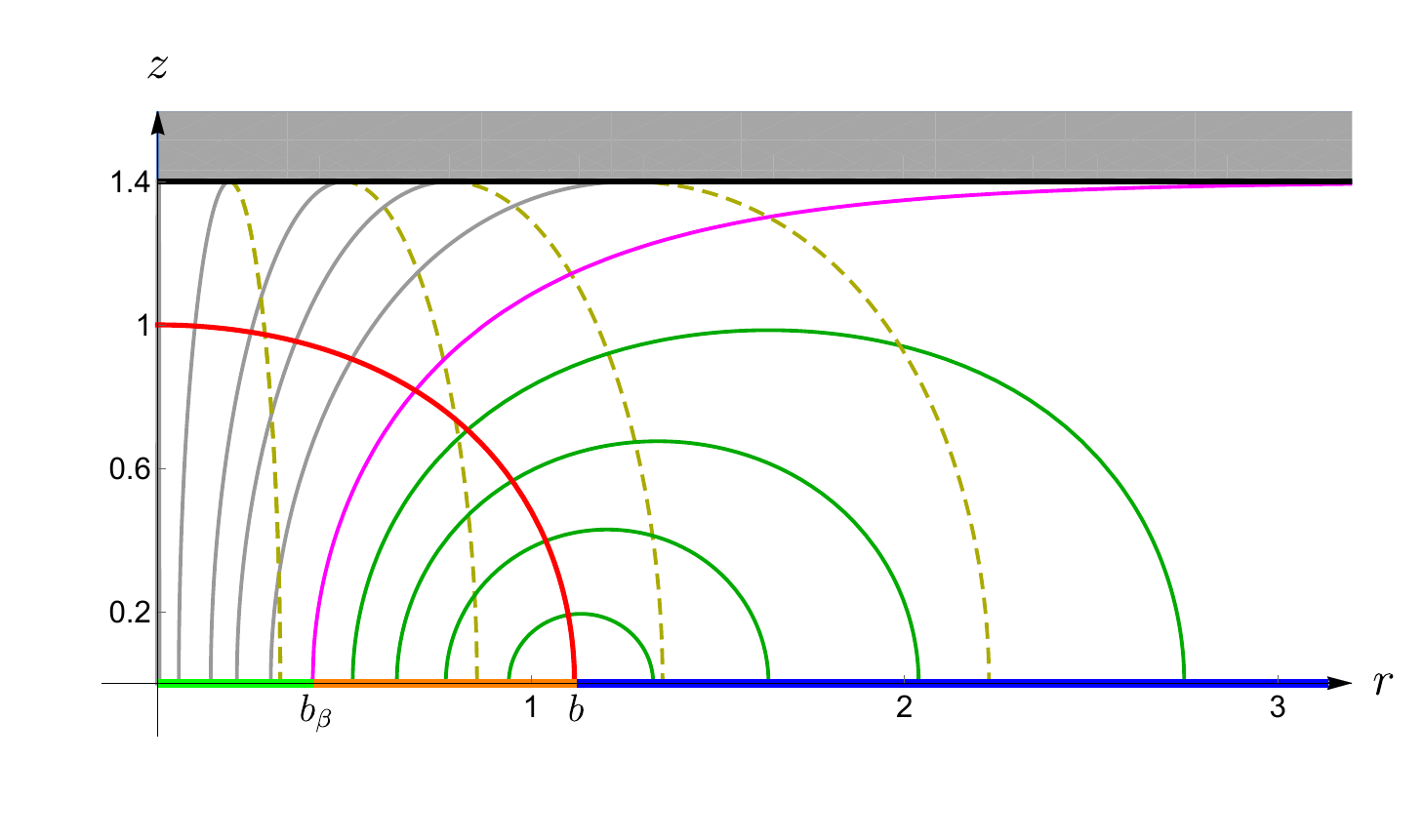}
		\vspace{-.8cm}
		\caption{\small 
			Geodesic bit threads (solid green, magenta and grey curves) for a disk in $\RR^2$,
			in the constant time slice of  the Schwarzschild AdS$_4$ black brane 
			(see (\ref{sch-ads-brane-metric}) with $z_h =1.4$).
			The RT surface $\gamma_A$ 
			corresponds to the red curve.
			The critical geodesic bit thread (magenta curve) 
			intersects $\gamma_A$ at $\widetilde{P}_\beta$ (whose radial coordinate is $r_{m,\beta}$),
			identifying the spherical dome $\tilde{\gamma}_{A, \beta}\subsetneq \gamma_A$
			and the spherical region $A_\beta$ on the boundary with radius $b_\beta$ (see the green segment),
			which provide $\widetilde{S}_{A,\textrm{\tiny th}}$ in (\ref{SA-th-gbt-final}).
			The data points for $r_{m,\beta}$ and $b_\beta$ are shown in Fig.\,\ref{fig:bbetasphere}.
			The black dot $P_\beta$ identifies the spherical dome $\gamma_{A, \beta}\subsetneq \gamma_A$
			whose area is equal to the thermal entropy of $A$ in (\ref{HigherDStefanBoltzmann}).
		} 
		\label{fig:HigherDPBBspherebitthreds}
	\end{figure}

	The RT hypersurface $\gamma_A$
	displays a rotational invariance around the axis at $r=0$;
	hence its profile is characterized by $z=z(r)$,
	which can be found by minimizing the area functional
	\begin{equation}
		\label{AreaSphereinBHfunctional}
		\operatorname{Area}[\,\gamma\,]
		\,=\,
		\Omega_{d-1} \int_{0}^{\sqrt{b^2-\varepsilon_{\!\!\textrm{ \fontsize{1.5}{3}\selectfont AdS}}^2}} \ \frac{r^{d-1}}{z(r)^d} \ \sqrt{1+\frac{{z}^\prime(r)^{2}}{f(z)}} \;\rd r 
	\end{equation}
		where $\gamma$ is a $d$-dimensional hypersurface 
		which is rotationally invariant around the axis at $r=0$
		and anchored to $\partial A$.
	Unlike the $d=1$ case and the RT hypersurface for the infinite strip (see Sec.\,\ref{sec-Sch-AdS-strip}),
	neither $r$ nor $z$ serves as a cyclic coordinate in the functional \eqref{AreaSphereinBHfunctional};
	hence, we cannot exploit any conservation law 
	and therefore $\gamma_A$ can be found by solving 
	the Euler-Lagrange equation, which reads
	\begin{equation}
		\label{ode-RT-sphere-ddim}
		\frac{z''(r)}{f(z)}
		+(d-1) \; \frac{z'(r)}{r \,f(z)} \left(\frac{z'(r)^2}{f(z)}+1\right)
		+ \frac{d}{z(r)} \left(\frac{z'(r)^2}{f(z)}+1\right)
		-\frac{z'(r)^2 }{2 f(z)^2} \,f'(z)
		\,= \,0\,.
	\end{equation}
	To the best of our knowledge, an analytic solution for this equation is not available in the literature;
	therefore, we must rely on numerical techniques to determine $\gamma_A$.
	%
	Due to spherical symmetry, $\gamma_A$ reaches its maximum at $r=0$,
	where $z(0)=z_\ast$ and $z'(0)=0$ are imposed, 
	being $z_\ast$ defined as the maximum height of $\gamma_A$.
	In the numerical determination of $\gamma_A$, 
	we cannot enforce the initial conditions precisely at $r=0$,
	but we set  $z(\delta_0)=z_*-\delta_1$ and $z^\prime(\delta_0)=\delta_1$, 
	where $\delta_{0}$ and $\delta_{1}$ are very small 
	(around $10^{-15}$ in our analysis). 
	Then, the radius $b$ of the sphere $A$ on the boundary is determined by solving $z(b)=0$. 
	Following the same convention of the previous cases, 
	we denote by $z_m(r_m)$ the solution obtained from this numerical procedure, 
	which corresponds to the solid red curve in Fig.\,\ref{fig:HigherDPBBspherebitthreds}.

	To construct the geodesic bit threads, which are orthogonal to $\gamma_A$,
	one observes that both the coordinate $r$ and the angular variables $\boldsymbol{\Omega}$ are cyclic in the length functional  (i.e. the functional  \eqref{AreaSphereinBHfunctional} for $d=1$).  
	We restrict our analysis to radial geodesics, where the angular variables are held constant. 
	Consequently, the only non-trivial conserved momentum is the radial one, which allows us to simplify the problem into the following first-order differential equation
	\begin{equation}
		z'(r)=\pm\, \frac{1}{z(r)} \; \sqrt{f(z(r))\big[C^{2}-z(r)^2\big]} 
		\label{geodesicdifferentialequation}
	\end{equation} 
	where 
	$C$ is the integration constant associated with the conservation of the momentum along the radial direction. 
	The equation (\ref{geodesicdifferentialequation}) can be solved through the separation of variables. 
	However, since the resulting integral cannot be expressed in terms of known special functions for a generic value of $d$, 
	we must rely on numerical integration again.
	The boundary conditions for the geodesic we are looking for
	are enforced at the intersection point with $\gamma_A$, corresponding to $r=r_m$, 
	where $z(r_m)=z_m(r_m)$ holds. 
	Instead, the condition that the integral line of the geodesic $z(r)$ remains orthogonal to $\gamma_A$ at $r=r_m$ provides $z'(r_m)$.
	Specifically, considering the unnormalized tangent vectors $\tilde{\boldsymbol{\tau}}_m=\big(1,z'_m(r_m)\big)$ 
	and $\tilde{\boldsymbol{\tau}}=\big(1,z'(r_m)\big)$ 
	associated to $\gamma_A$ and to the geodesic at the point $(z_m,r_m)$, 
	the orthogonality condition $\tilde{\boldsymbol{\tau}}_m\cdot \tilde{\boldsymbol{\tau}}=1+\tfrac{z'_m(r_m) z'(r_m)}{f(z_m(r_m))}= 0$ gives
	\be
	\label{ortogonalityconditionBHhigher}
	z'(r_m)= -\frac{ f(z_m(r_m))}{z'_m(r_m)}\,.
	\ee
	The integration constant $C$ in \eqref{geodesicdifferentialequation} is determined by comparing the value of $z'(r_m)$ from \eqref{geodesicdifferentialequation} 
	with the value obtained from \eqref{ortogonalityconditionBHhigher}, finding
	\be
	C=-\frac{z_m (r_m)\,
		\sqrt{ f(z_m(r_m)) + z_m^{\prime}(r_m)^2}}{
		z_m' (r_m)} \,.
	\label{z*bhsphere}
	\ee

	The geodesics intersecting $\gamma_A$ (i.e. such that $z(r_m)=z_m(r_m)$)
	orthogonally (i.e. satisfying \eqref{ortogonalityconditionBHhigher})
	naturally fall into two distinct categories
	(see Fig.\,\ref{fig:HigherDPBBspherebitthreds}).
	The first class is made by the geodesics whose maximum height, denoted by $\tilde z_\ast$, is strictly below $z_h$
	(see the solid green curves in Fig.\,\ref{fig:HigherDPBBspherebitthreds})
	and their value of $\tilde z_\ast$ coincides with $C$, given by \eqref{z*bhsphere}. 
	The second class includes all geodesics with a maximum height $\tilde z_\ast$ exactly equal to $z_h$. 
	These geodesics exhibit two branches
	corresponding to the solid  grey curves in Fig.\,\ref{fig:HigherDPBBspherebitthreds}, 
	which extend from the green region in the boundary to the horizon,
	and to the dashed dark yellow curves in Fig.\,\ref{fig:HigherDPBBspherebitthreds}, 
	going from the horizon back to the boundary.
	The solid part of the trajectories belonging to the second class represents the actual geodesic bit thread, 
	while the dashed part corresponds to the auxiliary geodesics
	and provides the higher dimensional analog (for the sphere)
	of the dashed dark yellow curves in the bottom left panel of Fig.\,\ref{fig:BTZ-brane-main}. 
	Following \cite{Mintchev:2022fcp}, here it is straightforward to observe that
	the integral lines of the geodesic bit threads combined with the curves associated with their auxiliary geodesics
	(see the solid and the dashed curves in Fig.\,\ref{fig:HigherDPBBspherebitthreds}  respectively)
	naturally define a bijective map between $A$ and $B \cup \RR^d$, 
	where $\RR^d$ corresponds 
	either to the horizon if only the geodesic bit threads are considered
	or to a second copy of the space where the dual CFT$_{d+1}$ lives
	if the auxiliary geodesics are also taken into account. 
	This map could provide the gravitational counterpart 
	of a possible map in the  dual CFT$_{d+1}$
	that implements the geometric action of the modular conjugation. 
	We are unaware of the existence in the literature of such a map in this setup, 
	i.e. for a CFT$_{d+1}$ with $d>1$ at finite temperature 
	and the bipartition of $\RR^d$ given by a sphere. 

	The critical geodesic bit thread corresponding to the magenta curve in Fig.\,\ref{fig:HigherDPBBspherebitthreds}
	provides the natural geodesic bit thread separating the two classes of geodesic bit threads mentioned above;
	indeed, it reaches its maximal height $z_h$ as $r\to +\infty$.
	This behavior occurs when the two real zeros of the square root in \eqref{geodesicdifferentialequation} coincide, implying that $C=z_h$. 
	Such property allows us to determine the value  of $r_m$ (denoted by $r_{m,\beta}$)
	where the magenta geodesic intersects $\gamma_A$ (the red curve),
	which is obtained by first replacing $C$ with $z_h$ in the l.h.s. of \eqref{z*bhsphere} 
	and then solving the resulting equation for $r_m$ numerically. 
	The intersection of this critical geodesic bit thread with $\gamma_A$
	characterizes the spherical dome that we denote by $\tilde{\gamma}_{A,\beta}$.
	The endpoint of this critical geodesic bit thread  in $A$, at $r=b_\beta$,
	identifies the spherical region $A_\beta \equiv \big\{ r \leqslant b_\beta \big\} \subsetneq A $
	corresponding to the green interval in Fig.\,\ref{fig:HigherDPBBspherebitthreds}.
	
		\begin{figure}[t!]
		\vspace{-.5cm}
		\hspace{-.5cm}
		\includegraphics[width=1.\textwidth]{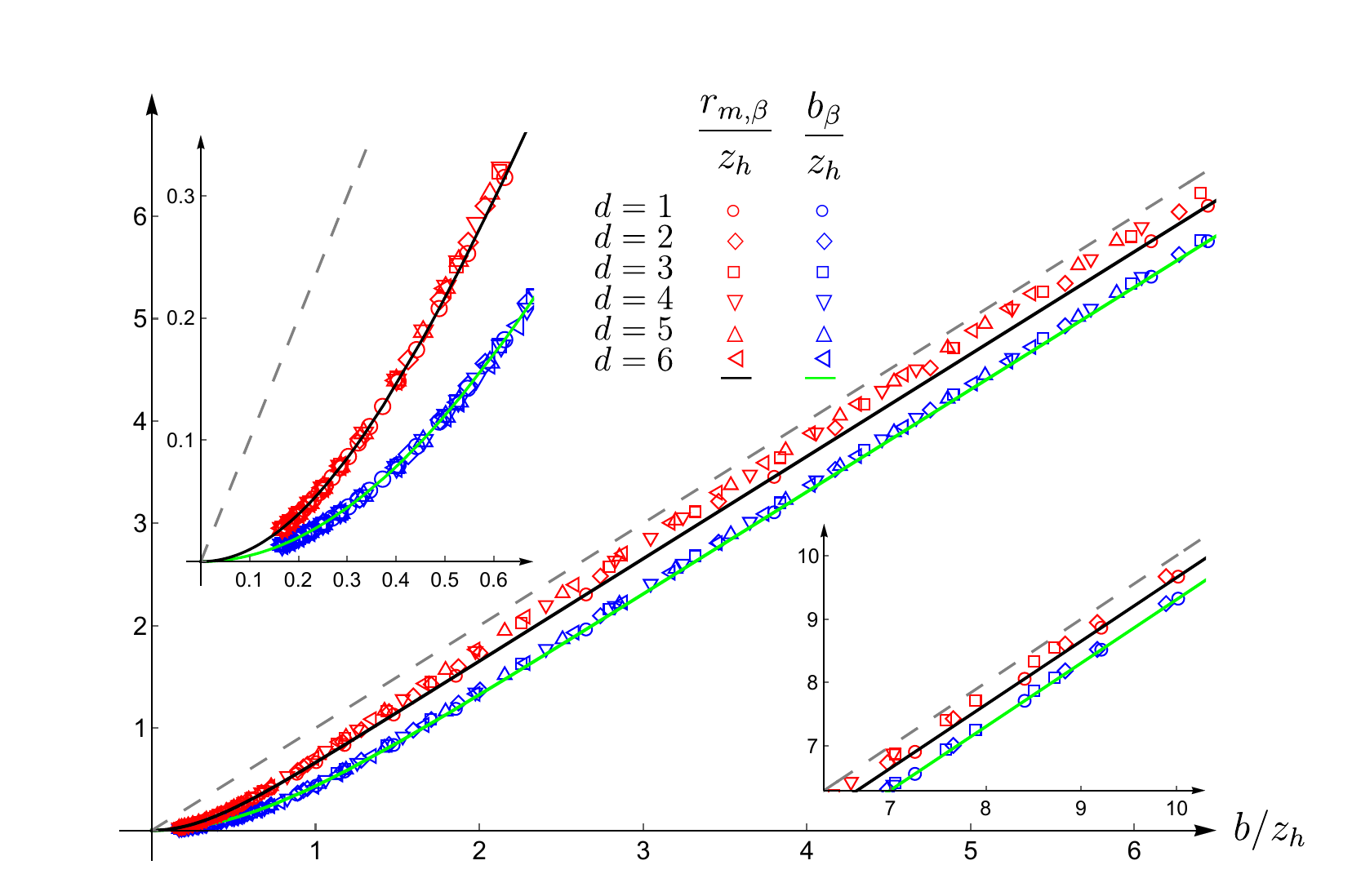}
		\vspace{-.1cm}
		\caption{ \small 
		Ratios $b_\beta/z_h$ (blue markers and green curve) 
		and $r_{m,\beta}/z_h$ (red markers and black curve) 
		as functions of $b/z_h$ for different values of the dimensionality parameter $d$.
		The solid curves correspond to the analytic expressions in (\ref{x-z-beta-coord-from-b}), 
		obtained for $d=1$.
		The dashed grey curve indicates the straight line whose slope equals one. 
		The left inset zooms in on small values of $b/z_h$,
		while the right inset displays the data corresponding to large values of $b/z_h$ 
		that are not reported in the main plot. 
		}
		\label{fig:bbetasphere}
	\end{figure}
	\noindent

	In Fig.\,\ref{fig:bbetasphere}, we report our numerical results for the dimensionless ratios 
	 $r_{m,\beta}/z_h$ (red points) and $b_\beta/z_h$ (blue points) 
	 in terms of $b/z_h$, for different values of $d$.
	 The black and green solid curves correspond respectively  to $r_{m,\beta}/z_h$ and $b_\beta/z_h$
	 for $d=1$, whose analytic expressions are given in (\ref{x-z-beta-coord-from-b}).
	 Comparing the data points obtained numerically against the corresponding solid curves,
	 we conclude that $b_\beta / z_h$  might be  independent of $d$,
	 while $r_{m,\beta}/z_h$ is either  independent of $d$ 
	or could display a mild dependence on the dimensionality parameter.
	It is worth performing a more precise numerical analysis 
	to establish these results.


	To determine the modulus of the divergenceless vector field $\boldsymbol{V}$ characterizing the geodesic bit threads,
	it is convenient to use \eqref{geodesicdifferentialequation} and write the geodesic bit thread intersecting $\gamma_A$ in $(z_m ,r_m)$ as follows
	\begin{equation}
		\label{bitpiumenosfera}
		r_{\textrm{\tiny$\gtrless$}}(z)
		= 
		r_m(z_m)
		+\int_{z_m}^{\tilde{z}_{*}} \! \frac{v}{\sqrt{f(v)\big(C^2-v^2\big)}} \, \rd v
		\pm
		\int_{z}^{\tilde{z}_{*}} \! \frac{v}{\sqrt{f(v)\big(C^2-v^2\big)}} \, \rd v
	\end{equation}
	where we recall that $\tilde z_\ast$ is the maximal height of the geodesic bit thread.
	The plus sign corresponds to the right branch of the curve ($r_{\mbox{\tiny$>$}}$), ending in $B$, 
	while the minus sign corresponds to the left branch ($r_{\mbox{\tiny$<$}}$), starting in $A$. 

	Considering the expression of $r_{\mbox{\tiny$<$}}$ in (\ref{bitpiumenosfera}) 
	specialized to the critical geodesic bit thread,
	which has $C=z_h$, $r_m(z_m) =r_{m,\beta}$ and $\tilde{z}_\ast =z_h$
	(see the magenta curve in Fig.\,\ref{fig:HigherDPBBspherebitthreds}),
	and for $z=0$,
	by using that $r_{\mbox{\tiny$<$}}(z=0) = b_\beta$ we get 
	\be
	\label{diff-rmbeta-bbeta-check}
	\frac{r_{m,\beta}}{z_h} - \frac{b_\beta}{z_h} 
	\,=
	\int_0^{z_{m,\beta}/z_h} 
	\!\! \frac{q}{\sqrt{\big( 1- q^{d+1} \big) \big( 1 - q^2\big)}}\; \rd q
	\ee
	which is a positive quantity. 
	For $d=1$, this relation becomes 
	\be
	\label{diff-rmbeta-bbeta-check-d1}
	\frac{x_{m,\beta}}{z_h} - \frac{b_\beta}{z_h} 
	=
	\frac{1}{2}\log\!\bigg(1-\frac{z^2_{m,\beta}}{z_h^2}\bigg)
	=\,
	\log\!\bigg(\frac{\cosh(b/z_h)}{\sqrt{\cosh(2b/z_h)}}
	\bigg)
	\ee
	where the second relation in (\ref{x-z-beta-coord-from-b}) has been used in the last step.
	We can easily check that (\ref{diff-rmbeta-bbeta-check-d1}) is satisfied by the expressions reported in 
	(\ref{bbeta}) and (\ref{x-z-beta-coord-from-b}).
	This relation also occurs for the setup considered in Sec.\,\ref{sec-hyp-bh}, for a specific hyperbolic black hole. 
	For $d=3$ the relation (\ref{diff-rmbeta-bbeta-check}) is also simple
	\be
	\frac{r_{m,\beta}}{z_h} - \frac{b_\beta}{z_h} 
	=
\frac{1}{\sqrt{2}}\;
\text{arctanh}\!
\left(\sqrt{ \frac{1+(z_{m,\beta   }/z_h)^2}{2}} \,\right) .
	\ee
	For $d=5$, the integral in (\ref{diff-rmbeta-bbeta-check}) can be evaluated in terms of elliptic functions, 
	but its explicit form is quite cumbersome. 
	Instead, for higher odd  $d$ and  for even $d$
	this integral cannot be computed in closed form. 
	%

	By exploiting the integral representation \eqref{bitpiumenosfera}
	and following  the strategy outlined in Appendix\;\ref{app-modulus} 
	to calculate the magnitude of $\boldsymbol {V}$ on the two branches, we obtain
	\begin{equation}
		\label{modulopiumeno}
		\big|\boldsymbol{V}_{\textrm{\tiny$\gtrless$}} \big|
		=
		\left(\frac{z}{z_m}\right)^d 
		\frac{\sqrt{C^2-z_m^2}}{\sqrt{C^2-z^2}}  \left(\frac{r_m}{r_{\textrm{\tiny$\gtrless$}}}\right)^{d-1} \,
		\frac{(\partial_{z_m} r_{\mbox{\tiny$<$}} )\big|_{z=z_m}}{\partial_{z_m} r_{\textrm{\tiny$\gtrless$}}} \,.
	\end{equation}
		The  unit vector field $\boldsymbol{\tau}$, specifying the direction of the vector field $\boldsymbol{V}$
		along the bit thread passing through $(z_m(r_m), r_m) \in \gamma_A$ reads
	\be
	\boldsymbol{\tau}
	=
	\big( \, \tau^z, \tau^r ,\boldsymbol{0} \, \big) 
	=\,
	\frac{1}{L_{\textrm{\tiny AdS}} \, C }
	\left( z \,\sqrt{f(z) \big(C^2-z^2\big)} \, , {z^2},\boldsymbol{0}
	\right)
	\label{tausfera}
	\ee
	where $C$ is given by \eqref{z*bhsphere}. 
	%
	To remove the dependence on $r_m$   from   \eqref{modulopiumeno} and \eqref{tausfera}
	in order to obtain a vector field depending only on the point $(z,r)$, 
	one must first solve \eqref{bitpiumenosfera} to express $r_m$ as a function of $r$ and $z$ and then substitute the result back into  \eqref{modulopiumeno} and \eqref{tausfera}. 
	
		\begin{figure}[t!]
		\vspace{-.5cm}
		\hspace{-1.7cm}
		\includegraphics[width=0.57\textwidth]{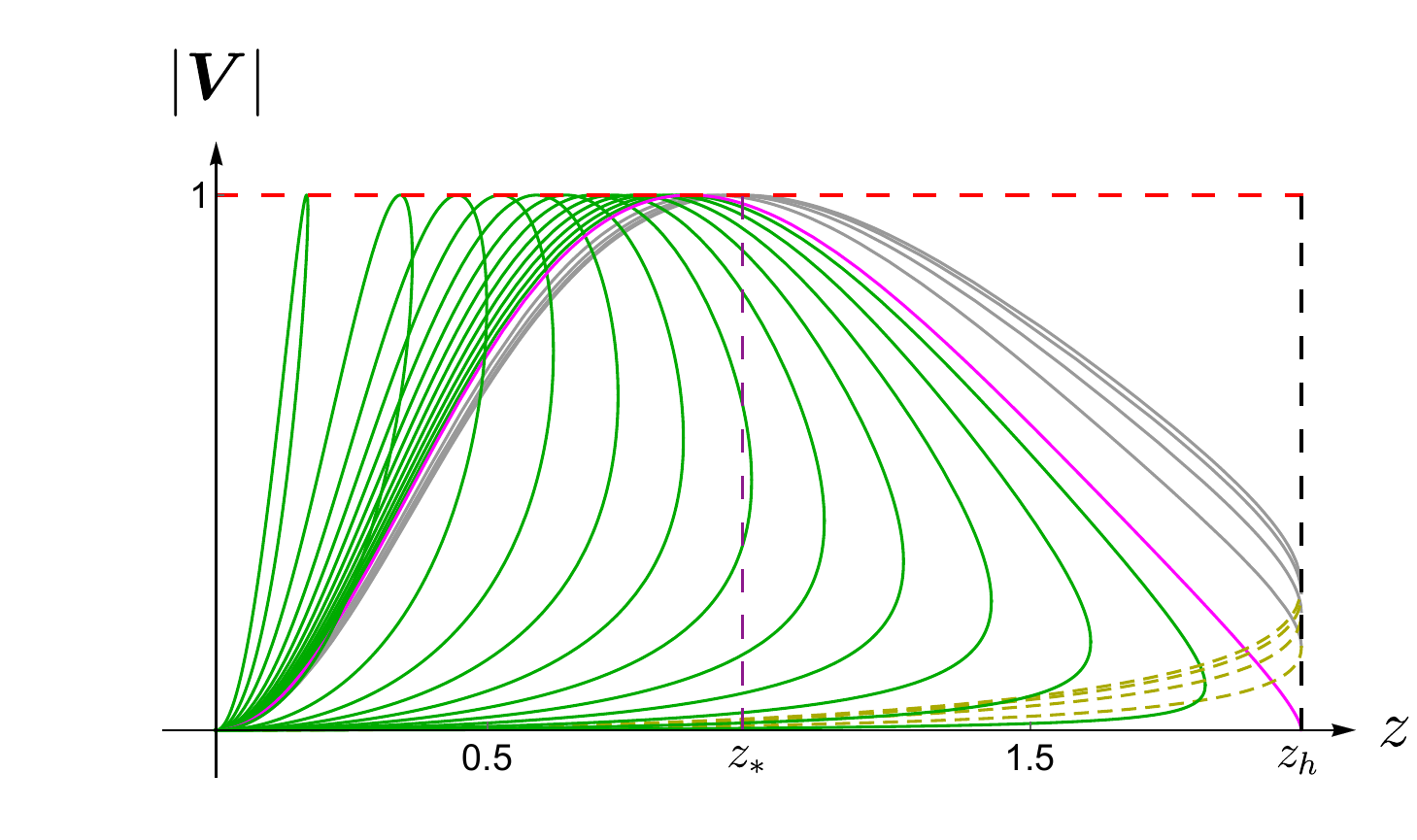} 
		\hspace{0.3cm}
		\includegraphics[width=0.57\textwidth]{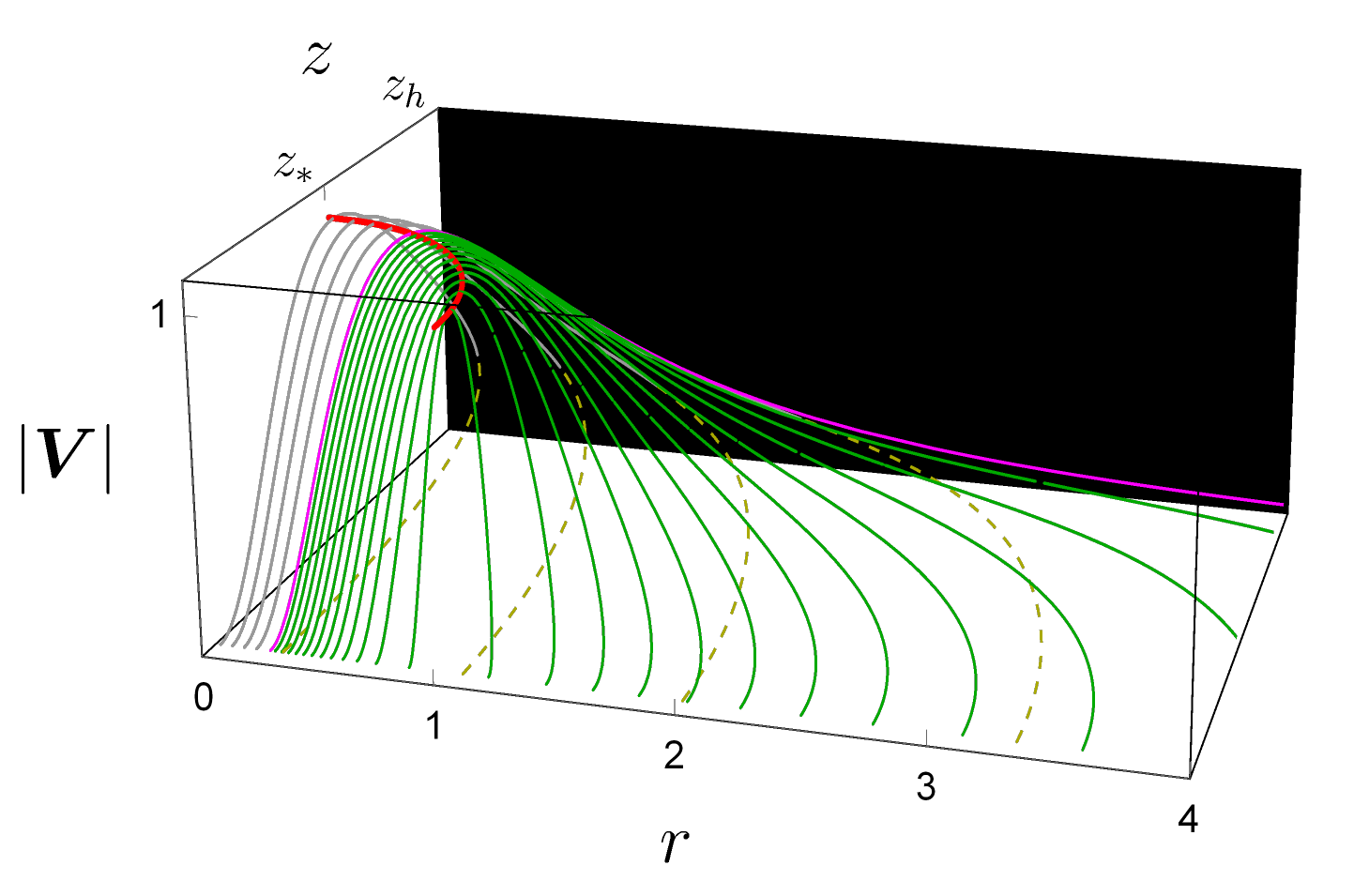}
		\vspace{-.6cm}
		\caption{\small
			Two-dimensional (left) and three-dimensional (right) representation of 
			$|\boldsymbol{V}|$ for the geodesic bit threads of a sphere 
			in the constant time slice of the Schwarzschild AdS$_4$ black brane.
		}
		\label{fig:HigherDPBBmodulussphere}
	\end{figure}

	In the left panel of Fig.\,\ref{fig:HigherDPBBmodulussphere},
	where each distinct curve corresponds to a different geodesic bit thread,
	we have depicted  the curves representing $|\boldsymbol {V}|$ as the coordinate $z$ varies along a single geodesic bit thread. 
	The green closed curves are associated with geodesic bit threads
	that originate from the boundary and return to the boundary without intersecting the horizon;
	indeed, they do not reach the black dashed vertical line at $z=z_h$.
	The magenta curve provides $|\boldsymbol {V}|$ for the first geodesic bit thread that reaches the horizon. 
	Finally, the grey solid curves give $|\boldsymbol {V}|$ for the geodesic bit threads 
	that reach the horizon at finite values of the radial coordinate. 
	For these curves, we have shown through dark yellow dashed curves
	also the putative value of $|\boldsymbol {V}|$ along the corresponding auxiliary branch, which extends from the horizon back to the boundary. 
	It is also evident from the figure that these curves consistently lie below the horizontal dashed red line at $|\boldsymbol {V}|=1$.  
	Importantly, the values of $z$ for which the solid curves in the left panel of Fig.\,\ref{fig:HigherDPBBmodulussphere}
	touch the horizontal line at $|\boldsymbol {V}|=1$ correspond to $z_m$ on $\gamma_A$ and, in fact, 
	all of them lie before the dashed purple vertical line 
	representing the maximum height $z_\ast$ of  $\gamma_A$  in the holographic direction. 
	An equivalent three-dimensional picture of this situation is shown  in the right panel of Fig.\,\ref{fig:HigherDPBBmodulussphere}.

	Given the vector field $\boldsymbol{V}$, it is worth considering   its flux through constant $z=z_0$ hyperplanes, with $0\leqslant z_0\leqslant z_h$,
	whose density is denoted by ${\Phi} (\boldsymbol{r} ;z_0) $ in the following. 
	The normal vector to these  hyperplanes  is $\boldsymbol{n}=\big(n^z,n^r,n^{\boldsymbol{\Omega}} \big) =\frac{1}{L_{\textrm{\tiny AdS}}} \big(z \sqrt{f(z)}\,,0,\bf{0} \big)$.   
	To define a positive flux density through a constant $z=z_0$ hyperplane,
	first, we must identify the geodesic bit thread whose maximum height $\tilde z_*$ is precisely equal to $z_0$. 
	Then, denoting by $c_0$, with $(c_0>b)$, the value of $r$ where this geodesic reaches its maximum,
	by  ${c}_0\pm b_0$  the radial coordinates of its endpoints on the boundary
	and by  $\chi_{ c_0}(r)$ the step function that is 
	$\chi_{ c_0}(r)=1$ for $0\leqslant r\leqslant  c_0$ and $\chi_{ c_0}(r)=-1$ otherwise,
	the flux density can be written as
	\begin{equation}
	\label{Phi-z0-slice}
		\Phi (\boldsymbol{r}; z_0) =
		\lim_{z \to z_0} \left(\frac{1}{4G_{\textrm{\tiny N}} }  \, \big|\boldsymbol{V} \big| \, \tau_a \, n^a \, \frac{L_{\textrm{\tiny AdS}}^{d}}{z^{d}} \; \chi_{c_0}(r) \right)
	\end{equation}
	which provides information only for the regions on the boundary 
	where either $0\leqslant r\leqslant c_0-{b}_0 $ or $r\geqslant {c}_0+{b}_0 $;
	indeed, all the contributions of the threads with maximum height $\tilde z_* <z_0$ are missed. 
	In the following, we focus on relevant choices for $z_0$.

\begin{figure}[t!]
			\vspace{-1.cm}
			\hspace{-1.5cm}
			\vspace{.5cm}
			\begin{minipage}{0.32\textwidth}
				\centering
				\includegraphics[width=1.4\textwidth]{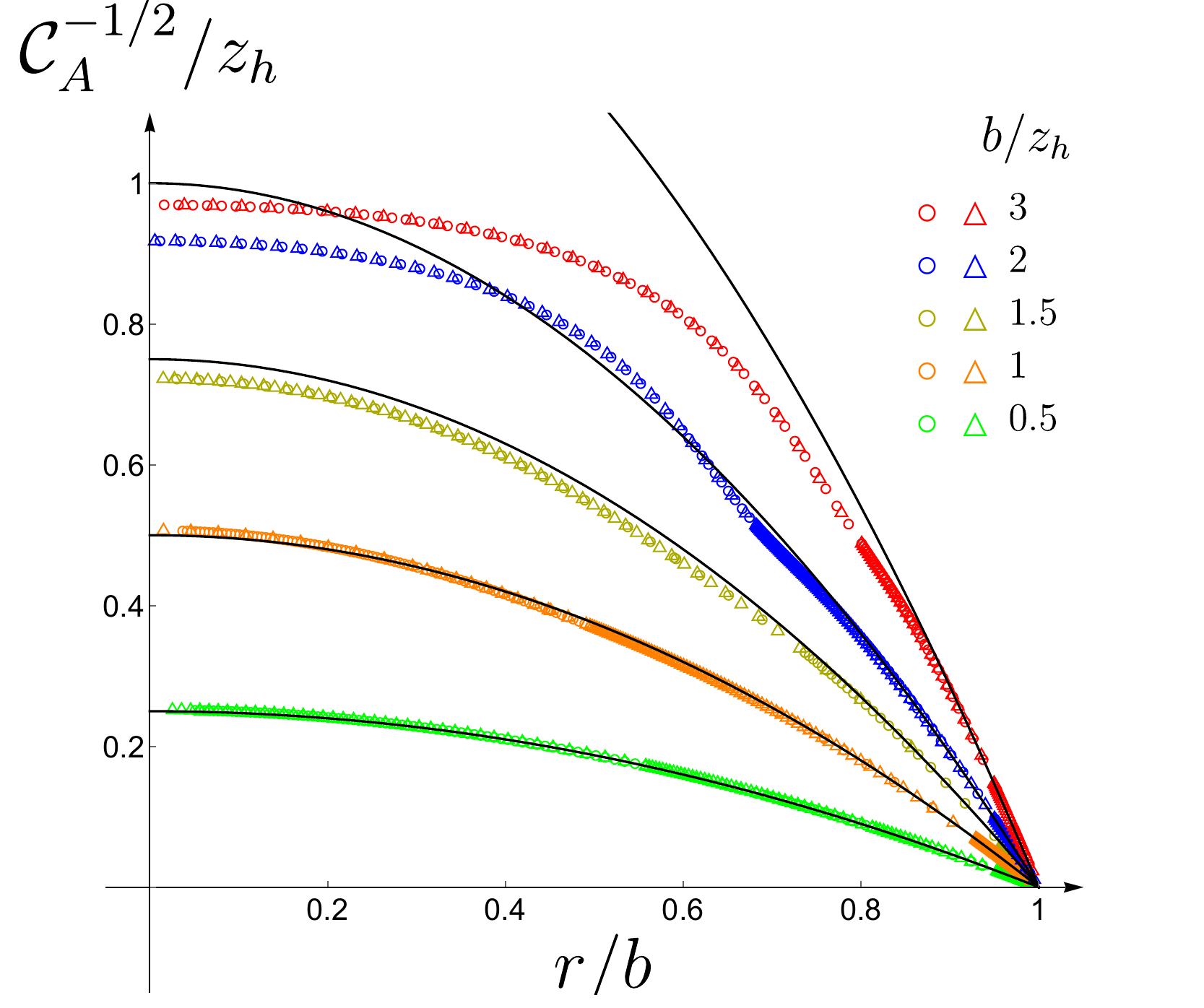}
			\end{minipage}
			\hspace{.8cm}
			\begin{minipage}{0.32\textwidth}
				\centering
				\includegraphics[width=1.4\linewidth]{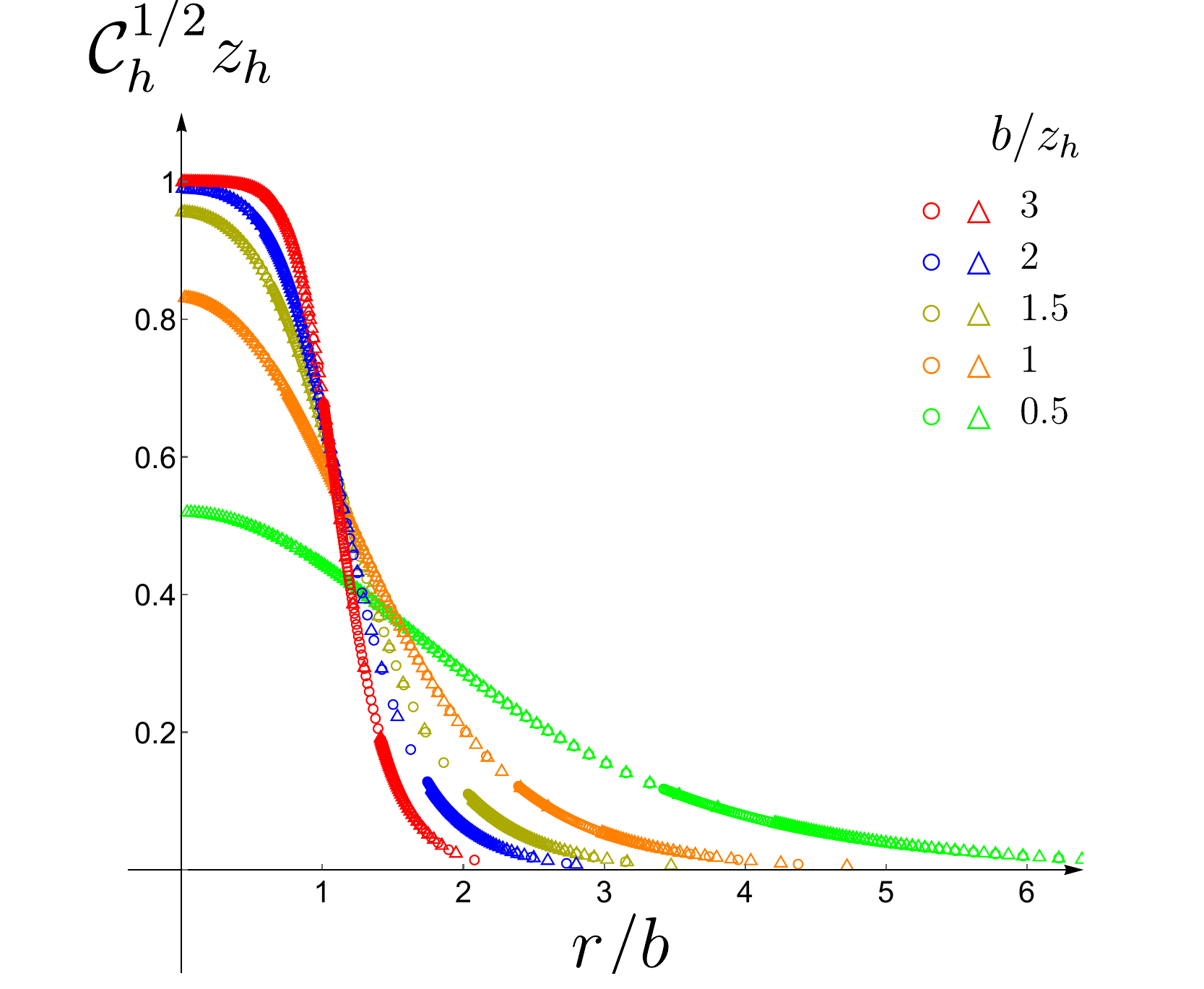}
			\end{minipage}
			\hspace{.8cm}
			\begin{minipage}{0.32\textwidth}
				\centering
				\includegraphics[width=1.4\linewidth]{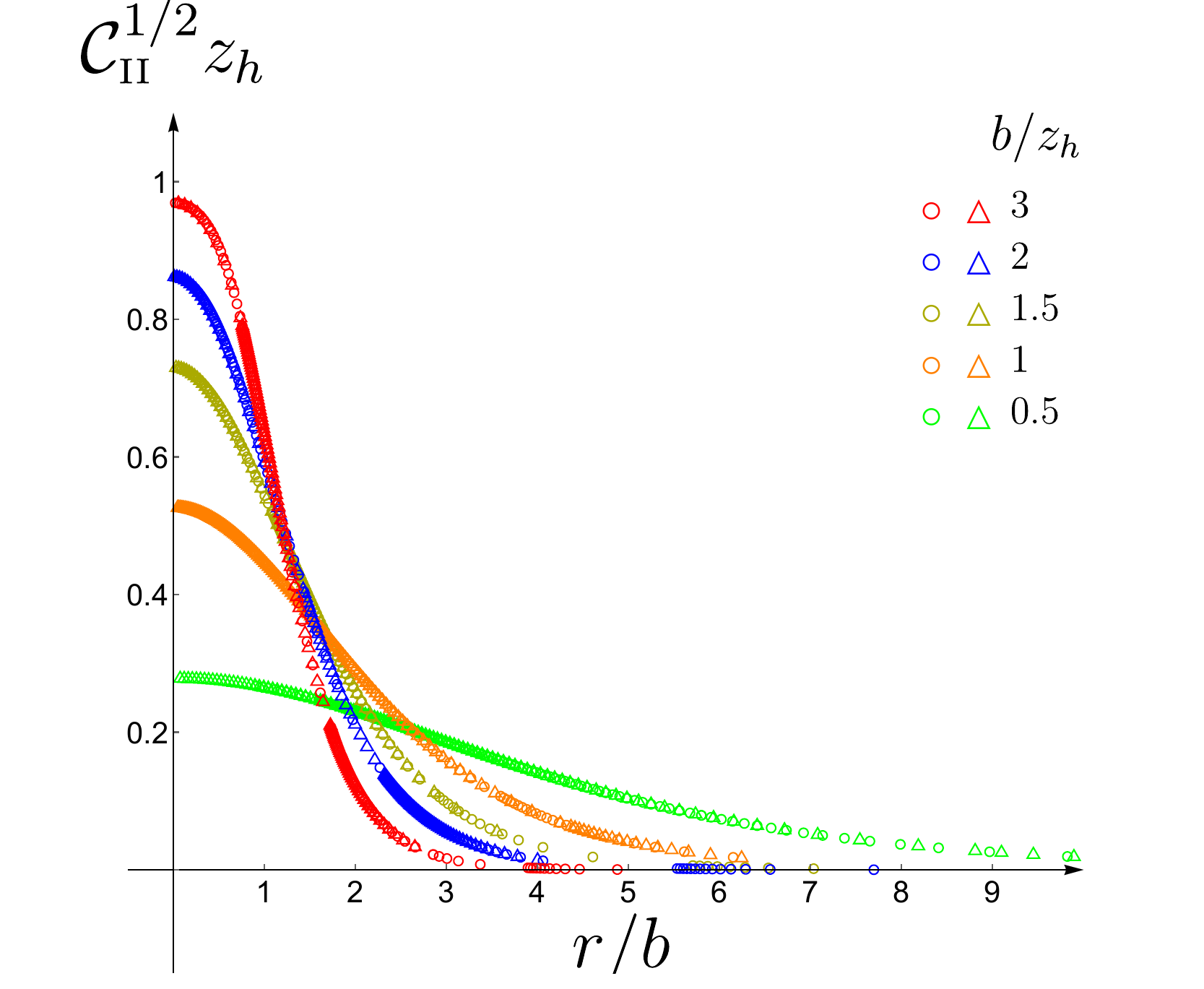}
			\end{minipage}
			\\
			\vspace{.5cm}
			\hspace{-1.6cm}
			\begin{minipage}{0.32\textwidth}
				\centering
				\includegraphics[width=1.4\textwidth]{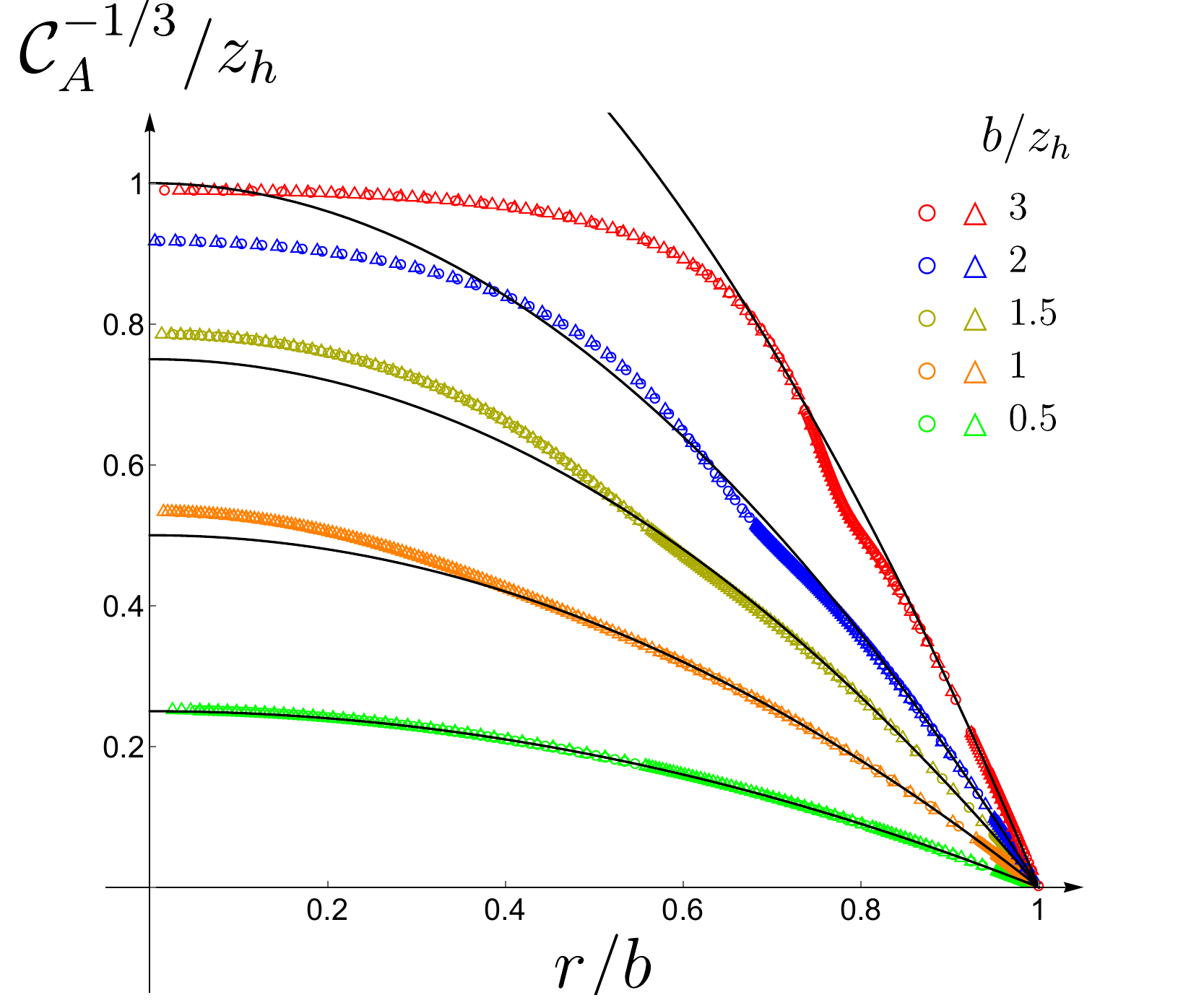}
			\end{minipage}
			\hspace{.8cm}
			\begin{minipage}{0.32\textwidth}
				\centering
				\includegraphics[width=1.4\linewidth]{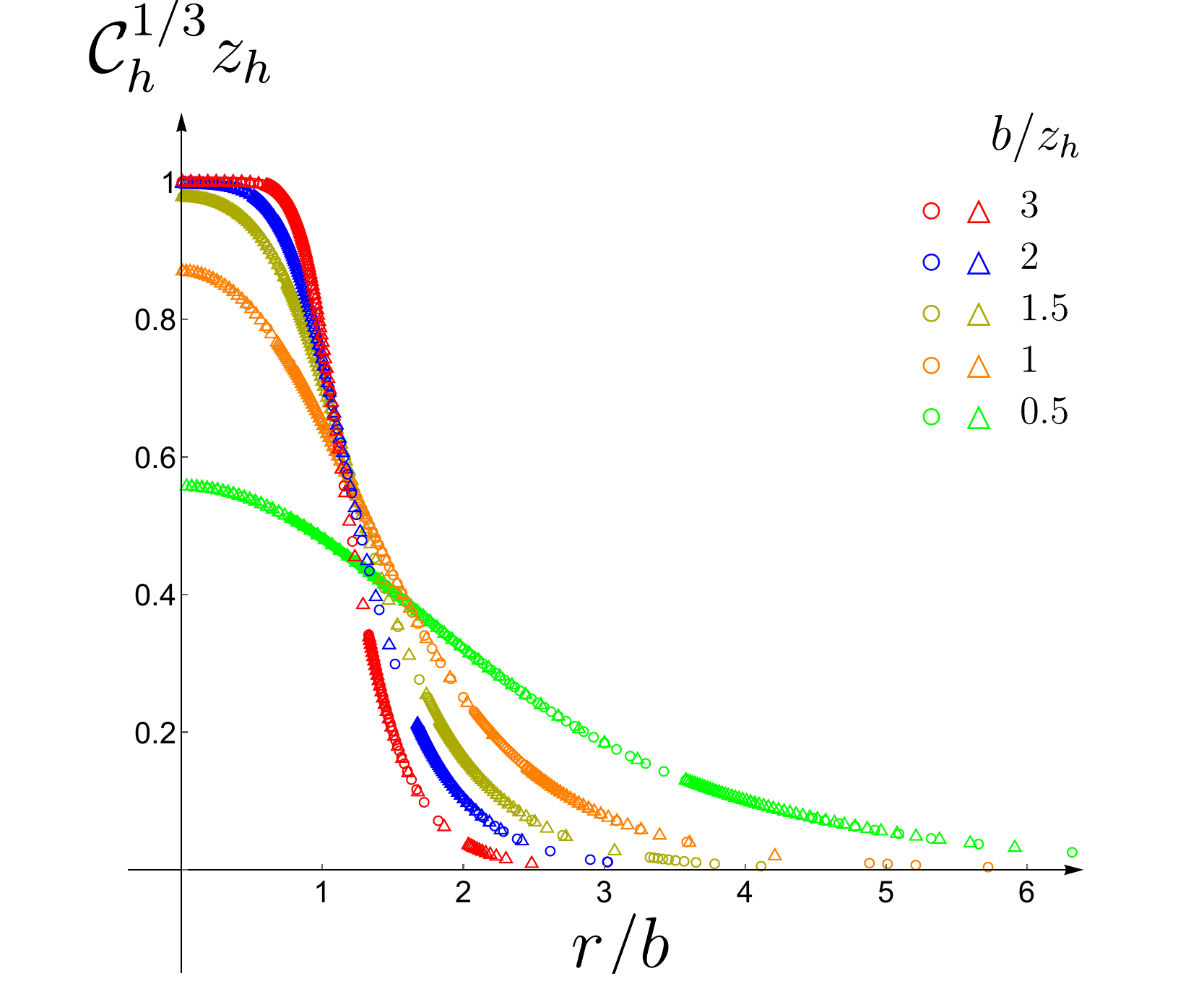}
			\end{minipage}
			\hspace{.8cm}
			\begin{minipage}{0.32\textwidth}
				\centering
				\includegraphics[width=1.4\linewidth]{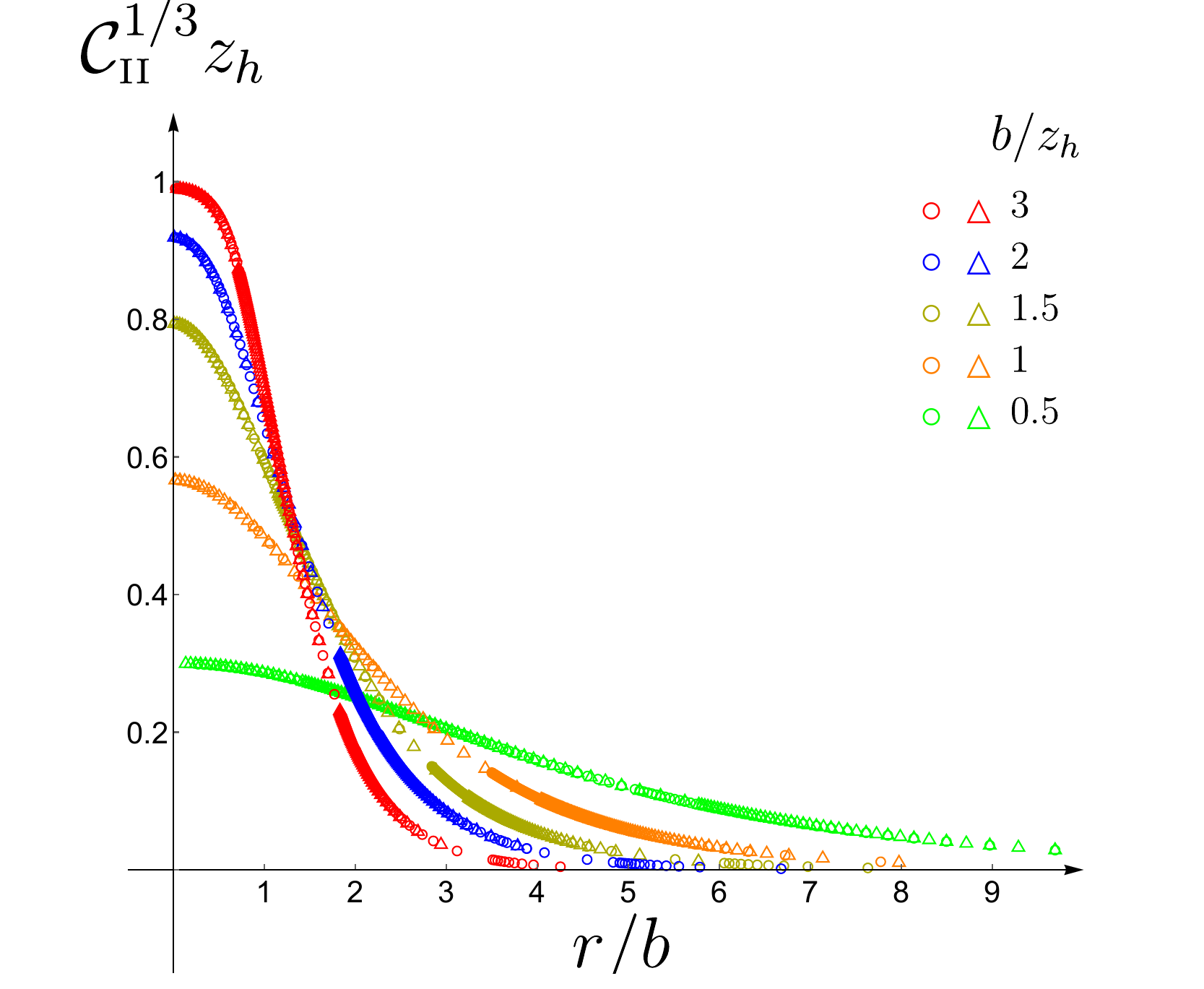}
			\end{minipage}
			\\
			\vspace{.5cm}
			\hspace{-1.6cm}
			\begin{minipage}{0.32\textwidth}
				\centering
				\includegraphics[width=1.4\textwidth]{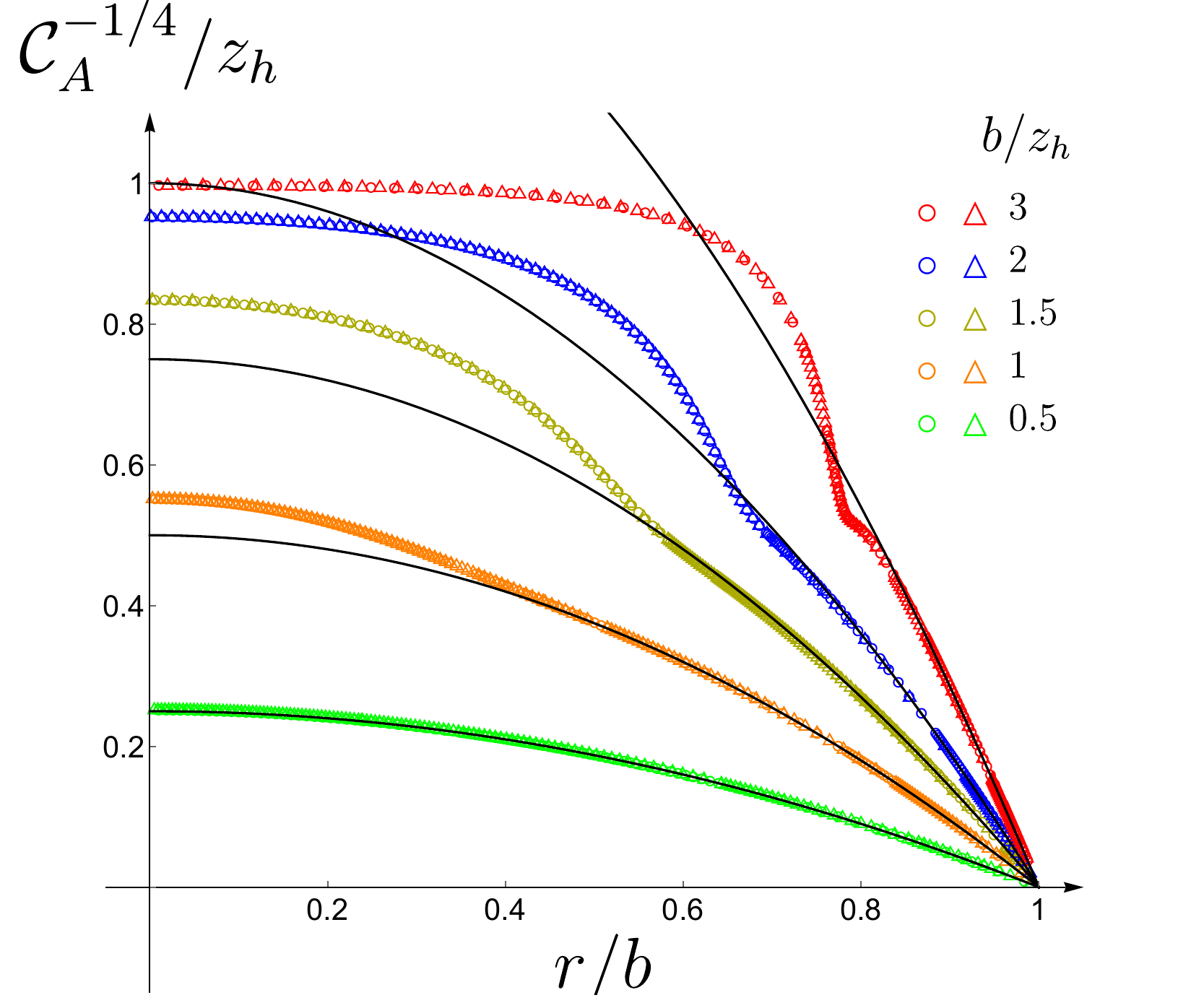}
			\end{minipage}
			\hspace{.8cm}
			\begin{minipage}{0.32\textwidth}
				\centering
				\includegraphics[width=1.4\linewidth]{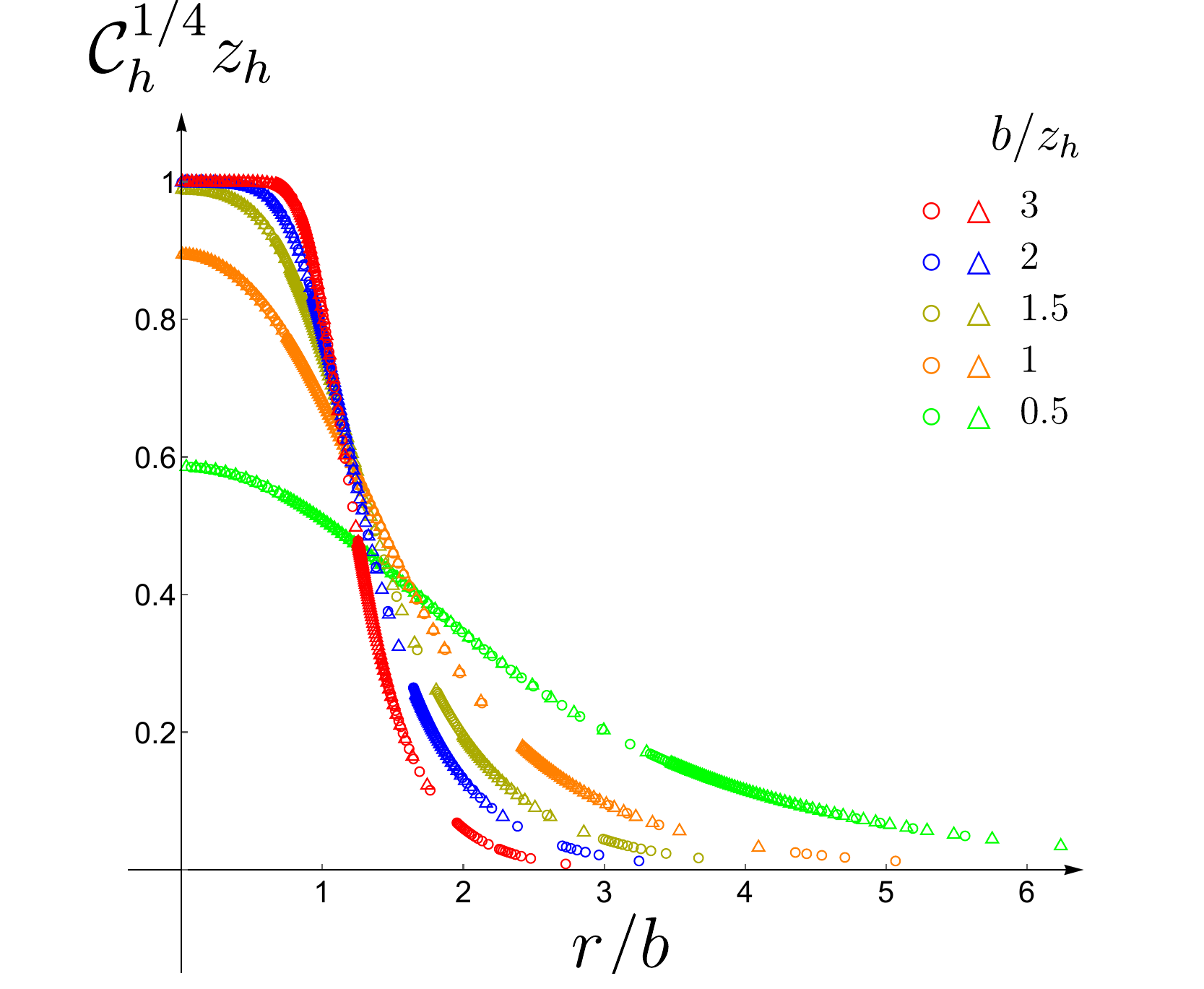}
			\end{minipage}
			\hspace{.8cm}
			\begin{minipage}{0.32\textwidth}
				\centering
				\includegraphics[width=1.4\linewidth]{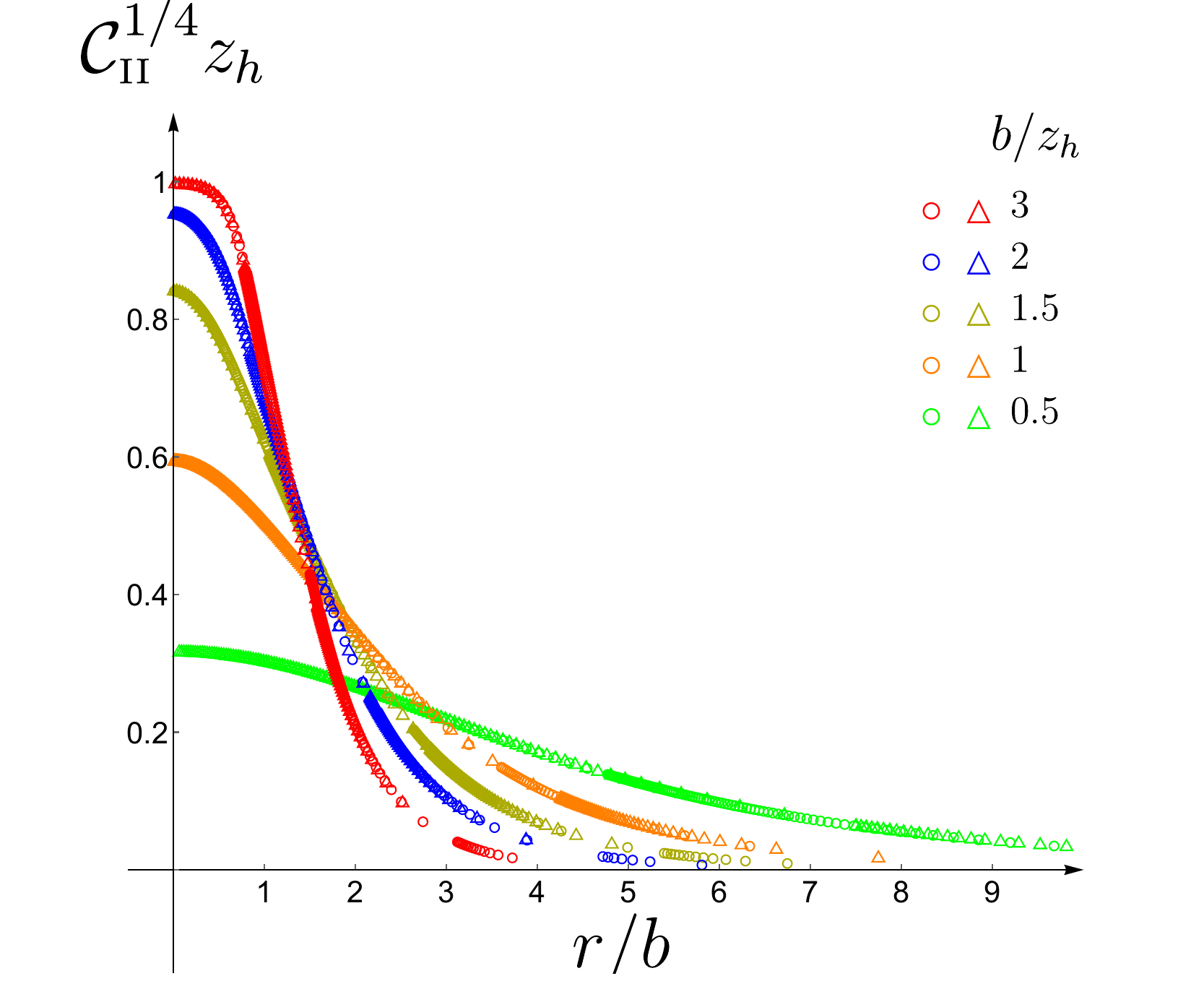}
			\end{minipage}
			\vspace{-.3cm}
			\caption{\small 
				Holographic contour functions for a sphere $A$  with radius $b$ induced by the geodesic bit threads 
				in the constant time slice of  the Schwarzschild AdS$_{d+2}$ black brane
				for either $d=2$ (top panels)
				or $d=3$ (middle panels)
				or $d=4$ (bottom panels)		
				and
				either in $A$ (left panels)
				or on the horizon (middle panels)
				or for the auxiliary geodesics on the whole boundary (right panels).
			}
			\label{fig:contourbordoorizsfera}
		\end{figure}

	When $z_0=0$, the flux density ${\Phi} (\boldsymbol{r}; 0)$ becomes the holographic contour function 
	$\mathcal{C}_A(\boldsymbol{r} ) \equiv \mathcal{C}(\boldsymbol{r} ) |_{\boldsymbol{r} \in A}$ 	
	induced by the geodesic bit threads of the sphere $A$ with radius $b$.
	This function can be determined only through a numerical analysis. 
	Inspired by the analytic results reported in Fig.\,\ref{fig:HYPcontourbordoorizsfera}
	and setting $ L_{\textrm{\tiny AdS}}^d /(4 G_{\textrm{\tiny N}}) =1$ for the sake of simplicity,
	in the left panels of Fig.\,\ref{fig:contourbordoorizsfera} 
	we show our numerical results for 
	$\tfrac{1}{z_h} \big[ \mathcal{C}_A(\boldsymbol{r} )  / \tfrac{L_{\textrm{\tiny AdS}}^d }{4 G_{\textrm{\tiny N}}}   \big]^{-1/d}$ 
	for $d=2$ (top left panel), $d=3$ (middle left panel) and $d=4$ (bottom left panel).
	The solid curves 
	are the same curves represented by the dashed lines in Fig.\,\ref{fig:HYPcontourbordoorizsfera}
	and correspond to $b\big[ 1 -(u/b)^2\big]/(2z_h)$,
	which is compatible with the area law of the holographic entanglement entropy.
	They nicely capture the behavior of the numerical data for $r/b \to 1^-$, 
	as expected from the fact that the horizon does not influence this regime. 
	Numerical data corresponding to the same value of $b/z_h$ nicely collapse, 
	as expected from conformal invariance. 
	The largest value of $b/z_h$ that we have been able to study is $b/z_h =3$.
	However, the plateau expected by analogy with Fig.\,\ref{fig:HYPcontourbordoorizsfera}
	is already visible at this value. 
	We remark that a dependence on $d$ is observed in the left panels of Fig.\,\ref{fig:contourbordoorizsfera} 
	and this is an important difference w.r.t. the case considered in Fig.\,\ref{fig:HYPcontourbordoorizsfera}.
	In particular, we highlight the interesting behavior
	of the crossover regime of $r/b$ between $r/b\simeq 0$ and $r/b\simeq 1$,
	for large values of $b/z_h$ as $d$ increases.
	It would be interesting to understand such a crossover 
	through analytic expressions
	and to perform more precise numerical analyses.

		Another relevant hyperplane to consider is the horizon $z_0=z_h$,
	whose flux density is denoted as $\mathcal{C}_h(\boldsymbol{r})$.
	The central panels of Fig.\,\ref{fig:contourbordoorizsfera} 
	report our numerical results for 
	$z_h \big[ \mathcal{C}_h (\boldsymbol{r})  / \tfrac{L_{\textrm{\tiny AdS}}^d }{4 G_{\textrm{\tiny N}}}   \big]^{1/d}$
	when either $d=2$ (top central panel) or $d=3$ (middle central panel) or $d=4$ (bottom central panel),
	taking inspiration from the central panel of Fig.\,\ref{fig:HYPcontourbordoorizsfera}.
	The results in these two figures are qualitatively very similar,
	displaying a mild dependence on $d$.
	
		It is also instructive to examine the vector field associated with the auxiliary geodesics going from the whole horizon to the entire boundary
	(see $r_{\mbox{\tiny$>$}}$ in (\ref{bitpiumenosfera}) with $\tilde{z}_{*} = z_h$
	and the dashed dark yellow curves in Fig.\,\ref{fig:HigherDPBBspherebitthreds}),
	denoted by $\boldsymbol{V}_{\! \textrm{\tiny II}}$ in the following. 
	For this vector field, the same analysis carried out for $\boldsymbol{V}$ remains valid 
	once $\tilde z_*$ is identified with $z_h$ in \eqref{bitpiumenosfera} and \eqref{modulopiumeno}. 
	This naturally leads to the introduction of the following holographic contour function
	\be
	\mathcal{C}_{\textrm{\tiny II}}(\boldsymbol{r} )
	\equiv 
	\lim_{z\to 0^+}\left( \frac{ 1 }{4 G_{\textrm{\tiny N}}} \, \boldsymbol{V}_{\! \textrm{\tiny II}} \, \tau_a \,n^a \, \frac{L_{\textrm{\tiny AdS}}^{d}}{z^d} \right)
	\ee
	which provides the flux density of $\boldsymbol{V}_{\! \textrm{\tiny II}}$ through the whole boundary.
	
	\clearpage

	Inspired by the right panel of Fig.\,\ref{fig:HYPcontourbordoorizsfera},
	in the right panels of Fig.\,\ref{fig:contourbordoorizsfera}
	we report our numerical results for 
	 $z_h \big[ \mathcal{C}_{\textrm{\tiny II}}(\boldsymbol{r} ) / \tfrac{L_{\textrm{\tiny AdS}}^d }{4 G_{\textrm{\tiny N}}}   \big]^{1/d}$ 
	 as function of $r/b \in [0, +\infty)$ for various $b/z_h$,
	 when either $d=2$ (top right panel) or $d=3$ (middle right panel) or $d=4$ (bottom right panel).
	 A mild dependence on $d$ is observed
	 and this is a crucial difference w.r.t. the right panel of Fig.\,\ref{fig:HYPcontourbordoorizsfera}.
	 For a given $d$ and $b/z_h$, the corresponding curves in the central panels and the right panels
	 of Fig.\,\ref{fig:contourbordoorizsfera} are qualitatively very similar, 
	 like in Fig.\,\ref{fig:HYPcontourbordoorizsfera}.
	 The  function $\mathcal{C}_{\textrm{\tiny II}}(\boldsymbol{r} )$ keeps track of the flux through the whole horizon;
	hence, it represents a sort of holographic thermal entropy density associated with $A$ spread all over the boundary.

	In the following, we discuss the main result of this manuscript, 
	namely the relation between the geodesic bit threads 
	of the spherical region $A$ in the Schwarzschild AdS$_{d+2}$ black brane discussed so far
	and the thermal entropy of $A$.
	This extends to higher dimensions the $d=1$ analysis performed in \cite{Mintchev:2022fcp}
	and further discussed  in Sec.\,\ref{subsec-flows-btz-planar}.

	The holographic thermal entropy density for a holographic  CFT$_{d+1}$ is \cite{Emparan:1999gf}
	\be
	\label{fig:HigherDStefanBoltzmann}
	s_{\textrm{\tiny th}}
	=
	\frac{1}{4 G_{\textrm{\tiny N}}} \; \frac{L_{\textrm{\tiny AdS}}^{d}}{z_h^d}
	\,=\,
	\frac{L_{\textrm{\tiny AdS}}^{d}}{4 G_{\textrm{\tiny N}} }  \left(\frac{4\pi }{(d+1)\, \beta} \right)^d 
	\ee
	hence the holographic  thermal entropy of the $d$-dimensional ball $A$ of radius $b$ reads
	\be
	\label{HigherDStefanBoltzmann}
	S_{A,\textrm{\tiny th}}
	=
	\frac{L_{\textrm{\tiny AdS}}^{d}}{4 G_{\textrm{\tiny N}} } \; V_d \left( \frac{b}{z_h} \right)^d 
	\;\;\;\qquad\;\;\;
	V_d \equiv \frac{\pi^{d/2}}{\Gamma(d/2+1)}  
	\ee
	where $V_d\, b^d$ is the volume of $A$.
	The difference between the last expressions in (\ref{fig:HigherDStefanBoltzmann})
	and in (\ref{SBthermalentropy-ddim}) 
	can be understood by using (\ref{beta-topo}).
	We find it worth introducing the spherical dome $\gamma_{A,\beta} \subsetneq \gamma_A$ 
	whose area is equal to the holographic thermal entropy of $A$ given in (\ref{HigherDStefanBoltzmann}).
	Because of the axial symmetry, 
	$\gamma_{A,\beta}$ is the portion of $\gamma_A$ 
	whose profile is characterized by the portion of the profile of $\gamma_A$
	enclosed by $(z_{\ast},0)$ and the point $P_\beta$,
	denoted by a black dot in Fig.\,\ref{fig:HigherDPBBspherebitthreds}.

	Given the vector field $\boldsymbol{V}$ characterizing the 
	geodesic bit threads of the sphere $A$ in the 
	Schwarzschild AdS$_{d+2}$ black brane (see Fig.\,\ref{fig:HigherDPBBspherebitthreds}),
	let us consider the critical geodesic bit thread 
	corresponding to the magenta curve in Fig.\,\ref{fig:HigherDPBBspherebitthreds}
	and evaluate
	either the flux through the spherical dome 
	$\tilde{\gamma}_{A,\beta} \subsetneq \gamma_A$, which is equal to its area 
	because $|\boldsymbol{V}|=1$ on $\gamma_A$,
	or the flux of $\boldsymbol{V}$ through the horizon,
	or  the flux of $\boldsymbol{V}$ through $A_\beta$
	(see the green segment in Fig.\,\ref{fig:HigherDPBBspherebitthreds}),
	or the flux of $\boldsymbol{V}$ extended along the auxiliary geodesics
	through the whole boundary.
	All these fluxes are equal and  can be written respectively as follows
	\be
		\label{SA-th-gbt-final}
		\widetilde{S}_{A,\textrm{\tiny th}}
		\, =
		\frac{ 1 }{4 G_{\textrm{\tiny N}}}
		\int_{\tilde{\gamma}_{A, \beta}} \!\! \sqrt{h} \; \rd ^d\boldsymbol{r} 
		\, =
		\int_{\RR^d} 
		\! \mathcal{C}_h(\boldsymbol{r} )\; \rd ^d\boldsymbol{r} 
		\, =
		\int_{A_\beta} 
		\! \mathcal{C}_A(\boldsymbol{r} )\; \rd ^d\boldsymbol{r} 
		\, =
		\int_{\RR^d} 
		\! \mathcal{C}_{\textrm{\tiny II}}(\boldsymbol{r}) \; \rd ^d\boldsymbol{r} 
		\ee
	in terms of the corresponding holographic contour functions introduced above
	(see Fig.\,\ref{fig:contourbordoorizsfera}).
	We denote by $\widetilde{P}_\beta$
	the intersection between the profile of $\gamma_A$ (see the red curve in Fig.\,\ref{fig:HigherDPBBspherebitthreds})
	and the critical geodesic bit threads (see the magenta curve in Fig.\,\ref{fig:HigherDPBBspherebitthreds}).
	
		\begin{figure}[t!]
		\vspace{-.5cm}
	\hspace{-1.5cm}
		\includegraphics[width=0.57\textwidth]{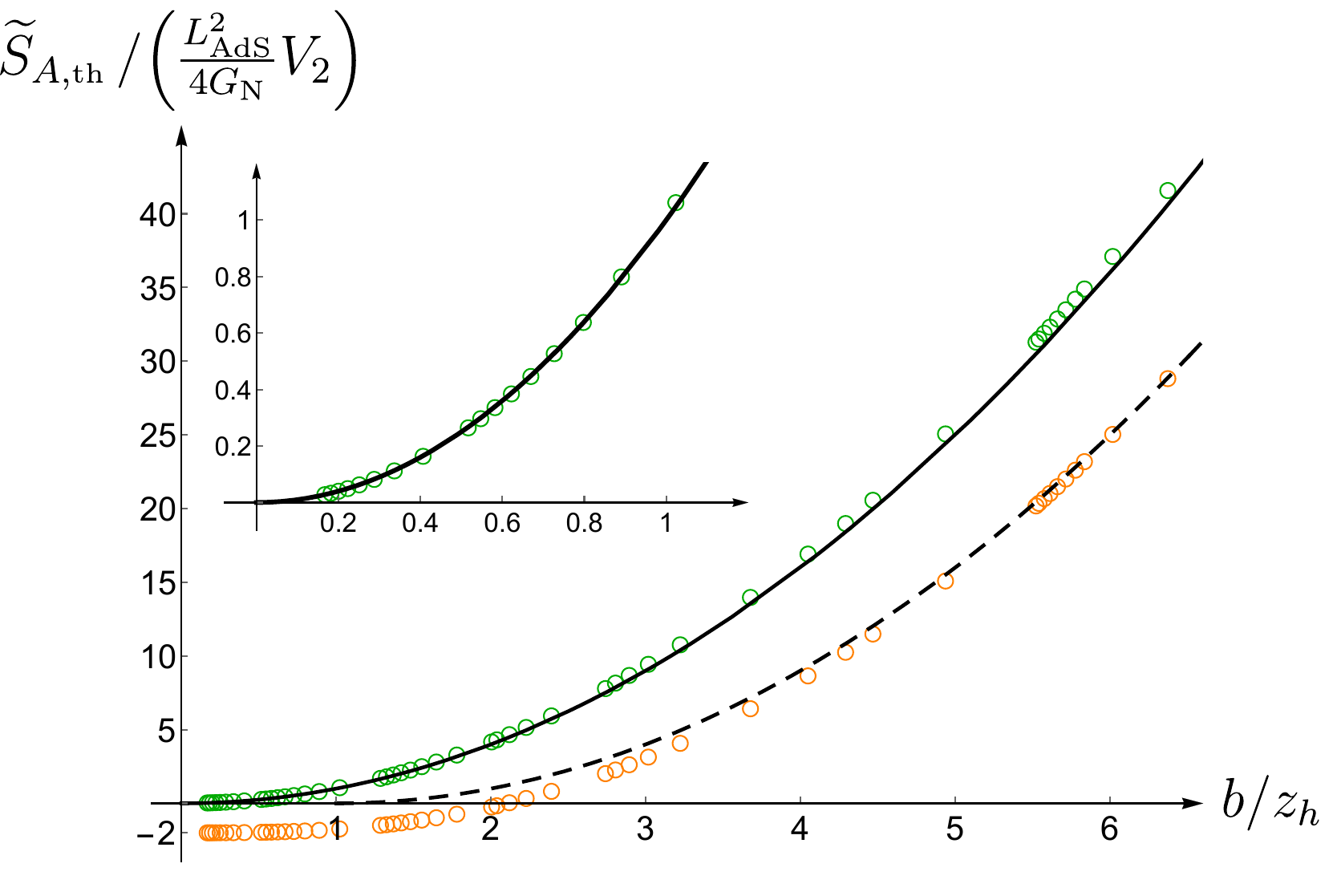} 
		\hspace{0.1cm}
		\includegraphics[width=0.57\textwidth]{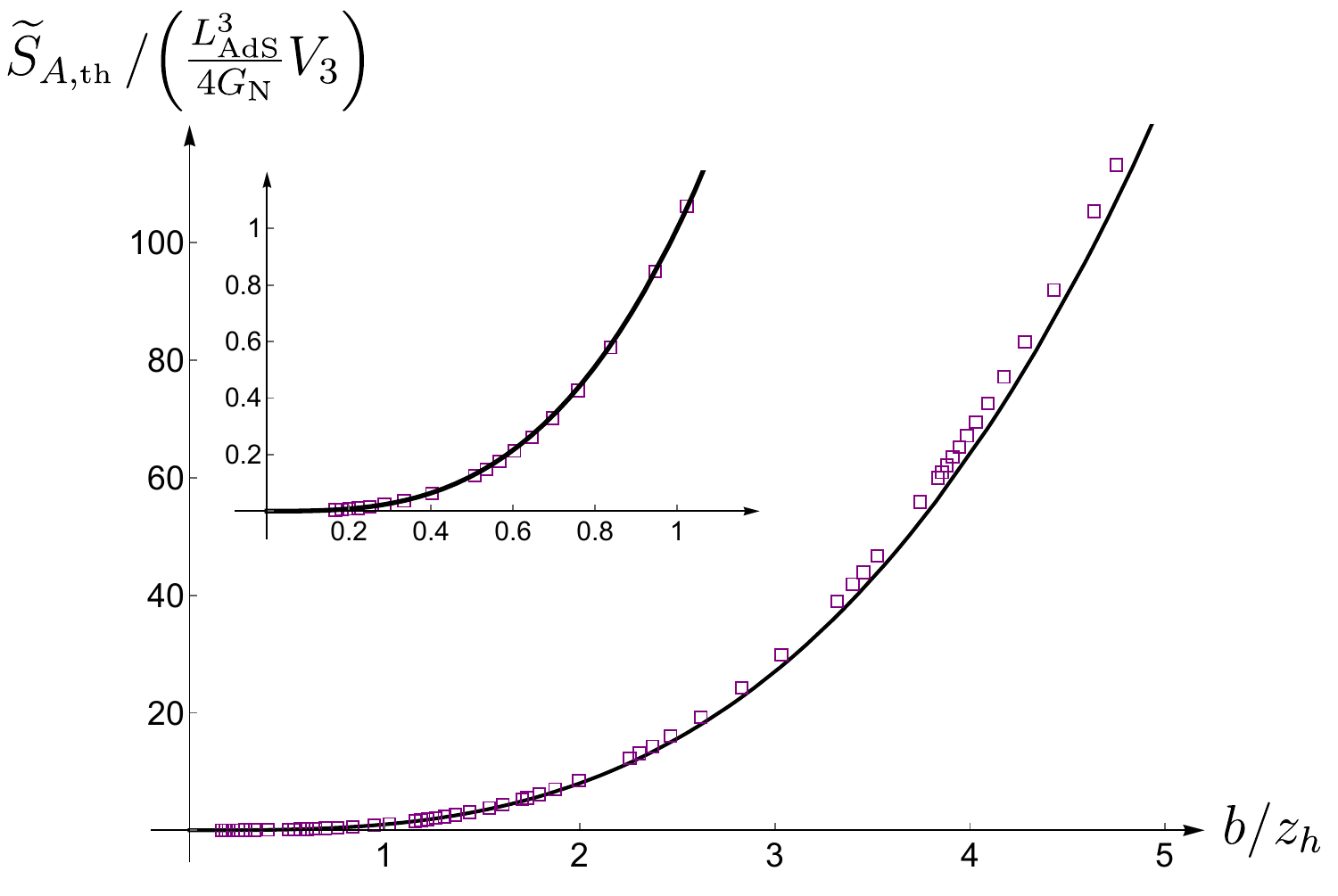}
		\\
		\\
		\vspace{1cm}
		\hspace{-1.5cm}
		\includegraphics[width=0.57\textwidth]{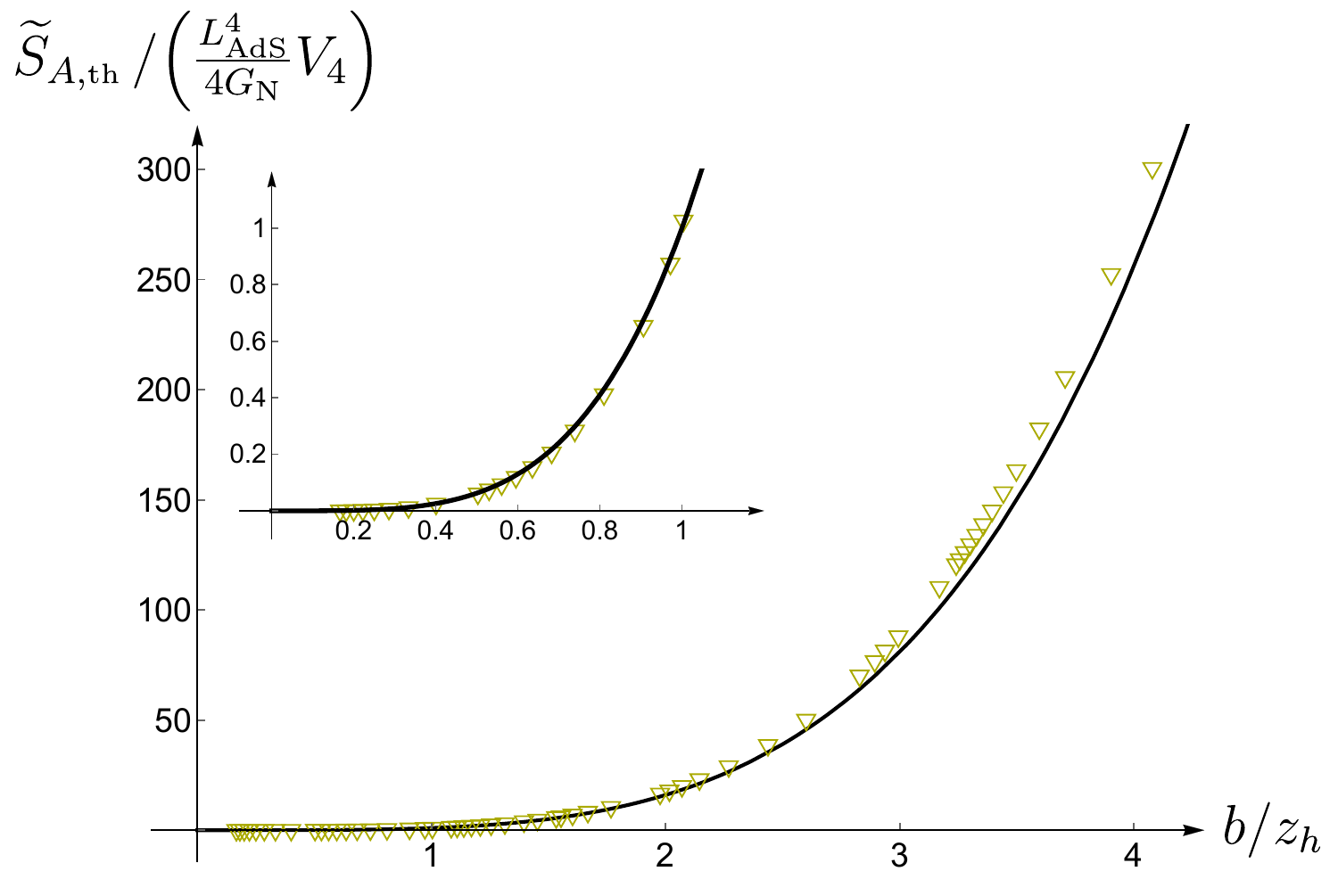} 
		\hspace{0.1cm}
		\includegraphics[width=0.57\textwidth]{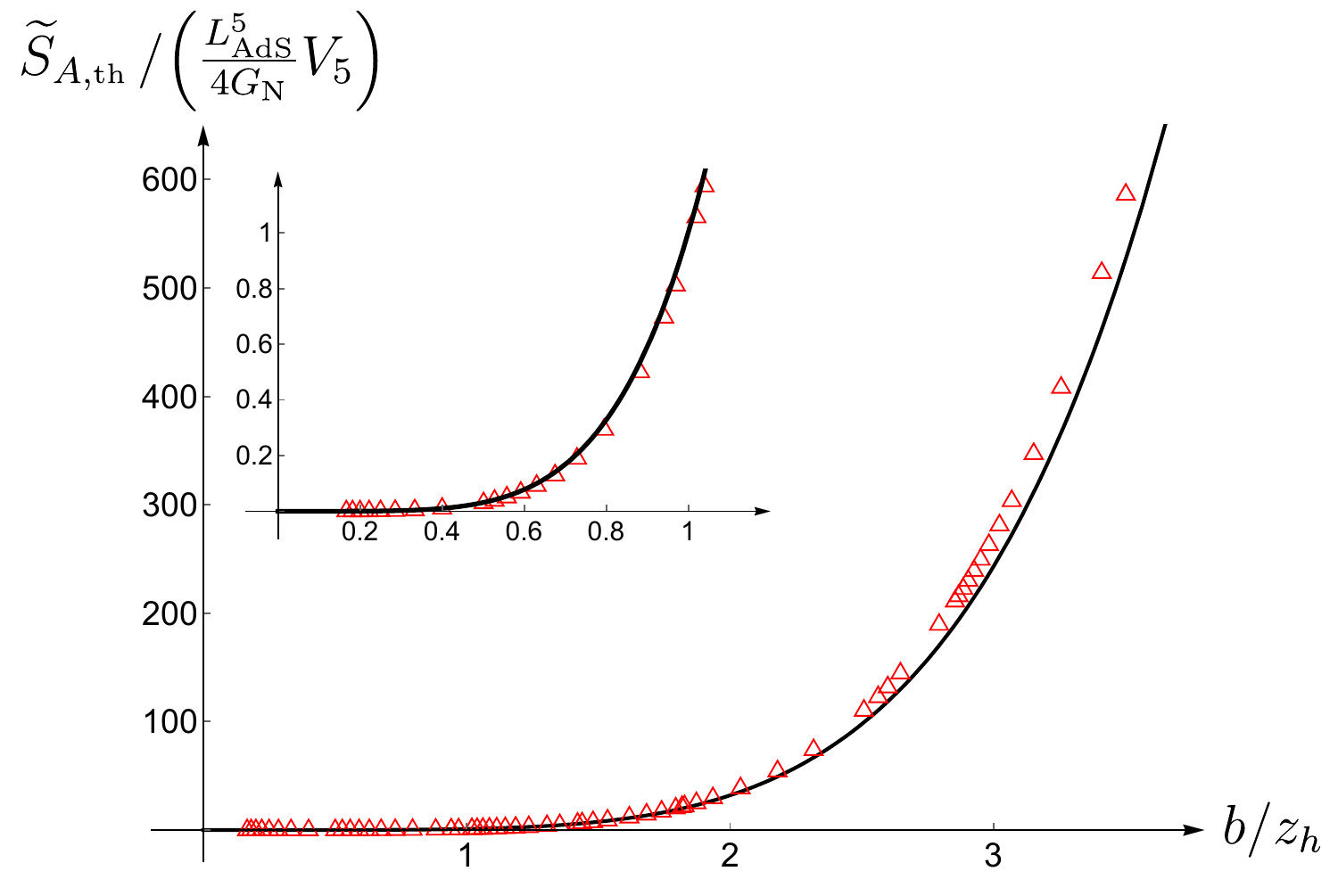}
		\vspace{-1.2cm}
		\caption{\small
		$\widetilde{S}_{A,\textrm{\tiny th}}$ providing the flux of the geodesic bit threads through the horizon,
		obtained numerically from (\ref{SA-th-gbt-final}),
		compared with $S_{A,\textrm{\tiny th}}$,
		i.e. the holographic thermal entropy of the sphere $A$ (solid black lines),
		given by (\ref{HigherDStefanBoltzmann}),
		for $d=2$ (top left), $d=3$ (top right), $d=4$ (bottom left) and $d=5$ (bottom right). 
				The insets zoom in on small values of $b/z_h$. 
				In the case of $d=2$ (top left panel), the orange data points correspond to $-F_A/V_2$,
				obtained from (\ref{FA-disk-bh}).
		}
		\label{fig:SA-th-various-d}
	\end{figure}

\begin{figure}[t!]
		\vspace{-.5cm}
		\hspace{-.2cm}
		\includegraphics[width=1.\textwidth]{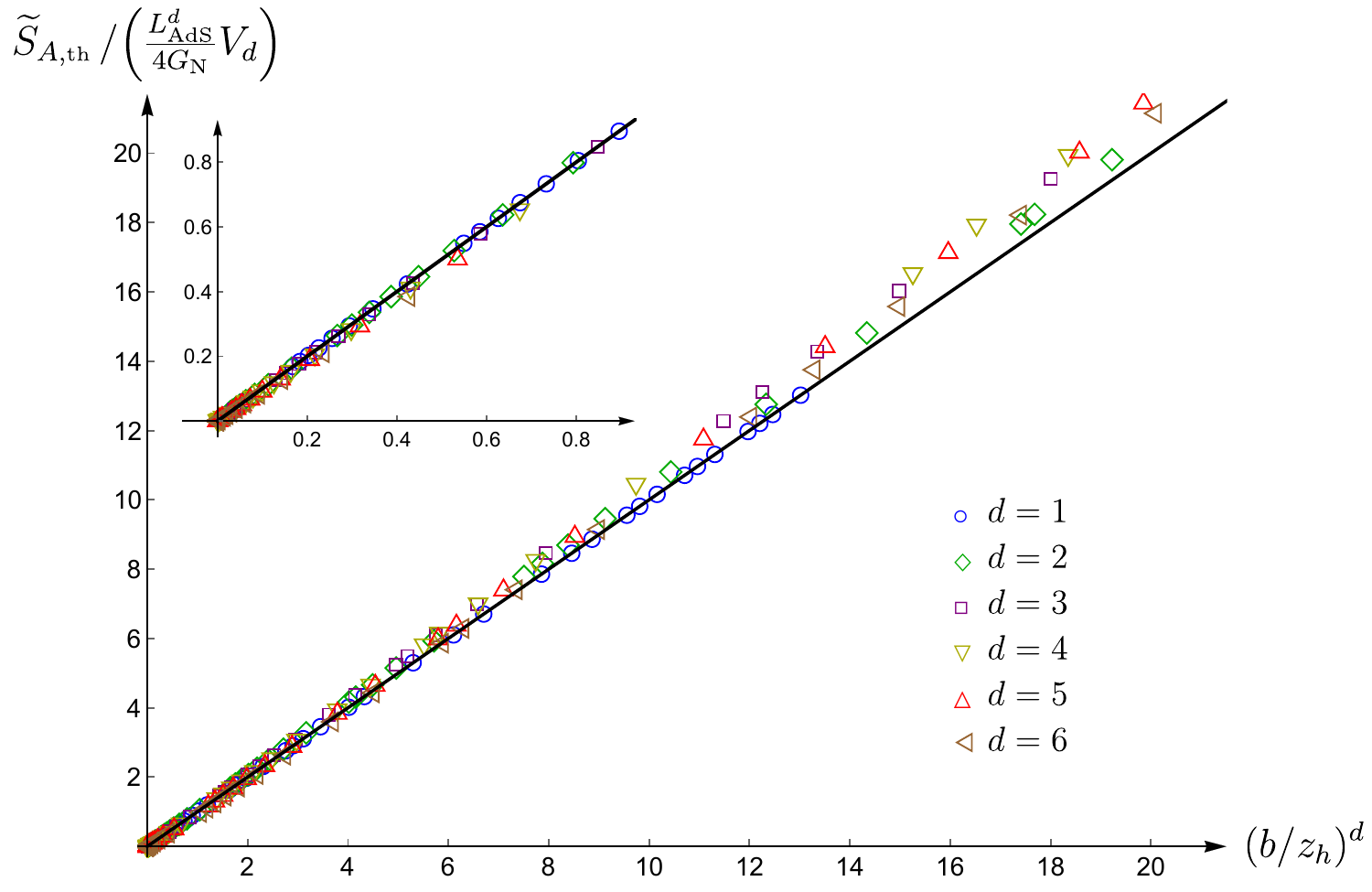}
		\vspace{.1cm}
		\caption{
			{\small  
			The quantity $\widetilde{S}_{A,\textrm{\tiny th}}$ for $1\leqslant d \leqslant 6$, 
			obtained numerically from (\ref{SA-th-gbt-final}),
				compared with the holographic thermal entropy of the sphere $A$ (solid black line),
				from (\ref{HigherDStefanBoltzmann}).
				The data corresponding to $2\leqslant d \leqslant 5$ 
				are also reported in Fig.\,\ref{fig:SA-th-various-d}.
				The inset zooms in on small values of $(b/z_h)^d$.}
		}
		\label{fig:FTHHigherD}
	\end{figure}

	In Fig.\,\ref{fig:SA-th-various-d} and Fig.\,\ref{fig:FTHHigherD}
	we compare 
	our numerical results for (\ref{SA-th-gbt-final}), given by the data points, 
	with the analytic expression (\ref{HigherDStefanBoltzmann}) for the holographic thermal entropy of $A$,
	(solid black curves). 
	In Fig.\,\ref{fig:SA-th-various-d} each panel corresponds to a fixed $d$, with $2\leqslant d \leqslant 5$,
	and the results are shown as functions of $b/z_h$,
	while in Fig.\,\ref{fig:FTHHigherD} we show the data for $1\leqslant d \leqslant 6$ all together in the same plot,
	including the data already displayed in Fig.\,\ref{fig:SA-th-various-d}, 
	in terms of $(b/z_h)^d$.
	The numerical results reported in Fig.\,\ref{fig:SA-th-various-d} and Fig.\,\ref{fig:FTHHigherD}
	have been obtained through the first integral in (\ref{SA-th-gbt-final}),
	i.e. the area of $\tilde{\gamma}_{A,\beta}$,
	but their compatibility with the other integrals has been checked. 
	From Fig.\,\ref{fig:SA-th-various-d}, 
	we observe that, 
	while the agreement between $S_{A,\textrm{\tiny th}}$
	and the data points for $\widetilde{S}_{A,\textrm{\tiny th}}$
	is very good for small values of $b/z_h$,
	it becomes worse for large values of $b/z_h$.
	This is probably due to the numerical difficulties
	occurring in the regime where $\gamma_A$
	is very close to the horizon. 
	Let us remind that,
	in the regime where the size of $A$ is large w.r.t. the position of the horizon,
	it is expected that 
	the finite term in the expansion of the holographic entanglement entropy as $\varepsilon_{\textrm{\tiny AdS}} \to 0$
	grows like the holographic thermal entropy of $A$.

	We find it instructive to focus on the $d=2$ case, where
	\be
	\label{hee-d=2}
	S_A = \frac{L_{\textrm{\tiny AdS}}^{2}}{ 4 G_{\textrm{\tiny N}} }
	\left( \, \frac{2\pi \, b}{\varepsilon_{\textrm{\tiny AdS}}} - F_A + o(1) \right)
	\ee
	as $\varepsilon_{\textrm{\tiny AdS}} \to 0$,
	being $F_A$ the $O(1)$ term,
	which can be written as follows \cite{Fonda:2015nma}
	\be
	\label{FA-disk-bh}
	F_A =
	 2\pi 
	\int_0^b
	\frac{1}{z^2}
	\left[\,
	\left( f(z) +\frac{z\, f'(z)}{2} \right) \frac{f(z)}{f(z)+(z')^2}
	+
	f(z) -\frac{z\, f'(z)}{2} -1
	\,\right]
	\sqrt{ \frac{ f(z)+(z')^2 }{f(z)} }\;
	r\, \rd r
	\ee
	where $f(z) = 1-(z/z_h)^2$ and $z' \equiv \partial_r z$, 
	being $z(r)$ defined as the profile of the RT surface. 
	The UV finite expression (\ref{FA-disk-bh})
	can be evaluated through a numerical integration, 
	once the numerical solution of (\ref{ode-RT-sphere-ddim}) has been obtained. 
	In the top left panel of Fig.\,\ref{fig:SA-th-various-d},
	the orange data points correspond to our numerical results for $-F_A/V_2$, 
	where $V_2 =\pi$ from (\ref{HigherDStefanBoltzmann}). 
	The dashed curve in the same panel is $(b/z_h - 1)^2$,
	which is obtained by shifting the independent variable in 
	(\ref{HigherDStefanBoltzmann}) for $d=2$.
	It is worth performing a more accurate numerical analysis of (\ref{FA-disk-bh})
	to capture the expected asymptotic behavior for large values of $b/z_h$.

	The numerical results in Fig.\,\ref{fig:SA-th-various-d} and Fig.\,\ref{fig:FTHHigherD}
	strongly suggest 
	that the fluxes in (\ref{SA-th-gbt-final})
	provide the holographic thermal entropy of the sphere $A$,
	or, equivalently, 
	that $P_\beta = \widetilde{P}_\beta$.
	We remark that this relation holds at any value of $b/ z_h$
	and not only for large values of $b/ z_h$,
	as expected from the above mentioned relation in this regime (see also Sec.\,\ref{sec-intro})
	between the finite term of the holographic entanglement entropy of $A$
	and the holographic thermal entropy of $A$.

	The geodesic bit threads of the sphere $A$ of radius $b$ naturally provide
	the ball $A_\beta \subsetneq A$ with radius $b_\beta < b$
	which encapsulates the information about the holographic thermal entropy of $A$
	because any geodesic bit thread of $\boldsymbol{V}$ arriving on the horizon originates from $A_\beta$,
	and vice versa. 
	This extends to higher dimensions the result obtained in $d=1$ for the interval \cite{Mintchev:2022fcp},
	which has been further discussed also in Sec.\,\ref{subsec-flows-btz-planar}.

	We stress that all the expressions in (\ref{SA-th-gbt-final}) except for the first integral, 
	which gives the area of $\tilde{\gamma}_{A,\beta}$,
	are intimately related to the geodesic bit threads characterized by the vector field $\boldsymbol{V}$.
	Indeed, considering a bit thread configuration $\widetilde{\boldsymbol{V}} \neq \boldsymbol{V}$
	displaying the axial symmetry dictated by $A$ for simplicity,
	the specific bit thread of $\widetilde{\boldsymbol{V}}$ intersecting $\gamma_A$ at $\widetilde{P}_\beta$
	identifies a spherical domain $\mathcal{A}_\beta \subsetneq A$ different from $A_\beta$
	such that the flux of $\widetilde{\boldsymbol{V}}$ through $\mathcal{A}_\beta $
	gives the holographic thermal entropy of $A$
	because this flux is equal to the area of $\gamma_{A,\beta}$, 
	by construction. 
	It would be interesting to find a way to construct $\widetilde{\boldsymbol{V}}$
	analytically, as done for $\boldsymbol{V}$.

	We conclude this subsection with a general remark about the possibility of constructing 
	minimal hypersurface inspired bit threads for the sphere $A$.
	Since the differential equation of the RT hypersurface is not invariant under translations in $r$,
	the translated profile does not qualify as a solution of the original differential equation
	and this makes the analysis more difficult.
	Instead, if we define the translated profile as the solution of the differential equation, 
	with the boundary condition determined by the requirement that it intersects the RT hypersurface orthogonally, 
	we immediately encounter numerical issues
	because these solutions develop a point where all the derivatives diverge. 
	For these reasons, we do not explore further 
	the construction of minimal hypersurface inspired bit threads for the sphere.

		\subsection {Strip}
		\label{sec-Sch-AdS-strip}

		Considering a holographic CFT$_{d+1}$ in a thermal state
		whose gravitational dual is given by the planar black hole geometry \eqref{sch-ads-brane-metric},
		let us investigate the case where 
		the bipartition of the boundary is given by an infinite strip $A$ with transverse width $2b$,
		adopting conventions similar to those outlined in Sec.\,\ref{stripemptyads}.

		A generic point on the constant time slice is described by the coordinates $(z, x, \boldsymbol{x}_\perp)$,
		where $z>0$ represents the holographic dimension, 
		$x$ spans the direction along which the strip has finite width $2b$ 
		and $\boldsymbol{x}_\perp$ parameterize the remaining transverse directions along which the strip extends
		(in the following we use also $\boldsymbol{x} \equiv (x, \boldsymbol{x}_\perp)$).  
		Leveraging the symmetry of this configuration, we can choose the ansatz $z = z(x)$ for the RT hypersurface $\gamma_A$. 
		The symmetry under reflection w.r.t. the hyperplane at $x=0$
		allows us to restrict our analysis to $x\geqslant 0$.
		Then, the area functional to be extremized reads
		\cite{Ryu:2006ef, Tonni:2010pv}
		\be
		\label{area-functional-strip}
		\text{Area}[\,\gamma\,]
		\,=\, 
		2 \, L_{\text{\tiny  AdS}}^d \, (2b_\perp)^{d-1} \int_{0}^{b_\varepsilon} \!\! \mathcal{L}_{\text{\tiny strip}}(x) \, \mathrm{d}x 
		\,= \,
		2 \, L_{\text{\tiny  AdS}}^d \, (2 b_\perp)^{d-1} \int_{0}^{b_\varepsilon} \! \frac{1}{z^d(x)} \, \sqrt{1+\frac{z'(x)^2}{f(z(x))}}\; \mathrm{d}x
		\ee
		where $\gamma$ is a $d$-dimensional hypersurface 
		which is translationally invariant along  $\boldsymbol{x}_\perp$
		and anchored to $\partial A$
		and, as usual,  the holographic UV cutoff $\varepsilon_{\text{\tiny AdS}}$ bounds the integration region 
		through the condition $z\left(b_\varepsilon\right) = \varepsilon_{\text{\tiny AdS}}$. 
		Since the integrand $\mathcal{L}_{\text{\tiny strip}}(x)$ in (\ref{area-functional-strip})
		does not explicitly depend on $y$, we can readily derive the following first integral of motion
		\be
		\mathcal{H}_{\text{\tiny strip}}(x) 
		= 
		z'(x)\, \frac{\partial \mathcal{L}_{\text{\tiny strip}}(x)}{\partial z'(x)} - \mathcal{L}_{\text{\tiny strip}}(x) 
		= 
		-\frac{1}{z(x)^d \, \sqrt{1+ z^{\prime}(x)^2 / f(z(x)) } } \;.
		\ee
		By setting $\mathcal{H}_{\text{\tiny strip}}(x) = 1/z_{\ast}^d$, the problem of finding $\gamma_A$ reduces to the following ordinary differential equation (ODE)
		\be
		\label{ode-RT-strip-ddim}
		z'(x) = - \frac{\sqrt{f(z(x))\big[z_\ast^{2d} - z(x)^{2d} \big]}}{z(x)^d}
		\ee
		where $z_{\ast}$ represents an integration constant defining the maximum height of $\gamma_A$. 
		For generic values of $d$, the ODE (\ref{ode-RT-strip-ddim}) cannot be integrated in terms of known special functions. 
		In order to establish the relationship between the width $2b$ of the strip $A$ 
		and the maximum height $z_{\ast}$ of $\gamma_A$,
		we separate the variables in \eqref{ode-RT-strip-ddim} and perform the integration over $x$ from $0$ to $b$ 
		and over $z$ from $z_{\ast}$ to $0$, finding 
		\be
		\label{www}
		b \,= \int_0^{z_{\ast}}  \frac{z^d}{\sqrt{f(z) \big(z_\ast^{2d} - z^{2d} \big)}} \; \rd z.
		\ee
		To avoid any confusion with the threads considered in the following,
		which share a structure similar to the RT hypersurface $\gamma_A$, 
		hereafter we refer to $\gamma_A$
		(that satisfies the boundary condition $z(0) = z_\ast$ and the integral condition \eqref{www})
		as $z_m = z_m(x_m)$.

		\subsubsection{Geodesic bit threads}
		\label{sec-gbt-strip-ddim}

		In the following, we investigate the family of geodesics originating from $A$ at $z=0$ 
		and intersecting orthogonally with the RT hypersurface. 
		The symmetry of this configuration allows us to focus only on geodesics that lie in the plane where the vector $\boldsymbol{x}_\perp$ remains a constant. 
		This property simplifies our analysis to an effective two-dimensional problem, where only the coordinates $(z,x)$ are relevant;
		hence we can consider geodesics parameterized by $z=z(x)$.
		The independence of $x$ in the line element naturally leads to the following ODE
		\be
		\label{ode-geodesic}
		{z}(x) \sqrt{  1+ \frac{z'(x)^2}{f(z(x))} } = {C}
		\ee
		where the integration constant $C$ can be determined by imposing the orthogonality condition 
		between the bit thread $z(x)$ and the RT hypersurface $z_m(x_m)$. 
		For the geodesics intersecting $\gamma_A$ at $(z(x_m)=z_m(x_m) , x_m)$, the orthogonality condition becomes
		\be
		\label{BHstriporthogonalityinHigherD}
		1+ \frac{z'_m(x_m) \,z' (x_m)  }{f(z_m)} =0\,.
		\ee
		By employing (\ref{ode-RT-strip-ddim}) and (\ref{ode-geodesic}) to eliminate the dependence on $z'_m(x_m)$ and $z' (x_m) $ in \eqref{BHstriporthogonalityinHigherD}, 
		this condition simplifies to
		\be
		\label{eq-alg-zast}
		\frac{\big( z_\ast^{2d} - z_m^{2d} \big)\, \big( C^2 - z_m^2 \big)}{z_m^{2d+2}} =1
		\ee
		which leads to
		\be
		\label{CCC}
		C =\frac{z_m}{ \sqrt{ 1 - (z_m/z_\ast)^{2d} }} \, .
		\ee

\begin{figure}[t!]
			\vspace{-.5cm}
			\hspace{-.9cm}
			\includegraphics[width=0.55\textwidth]{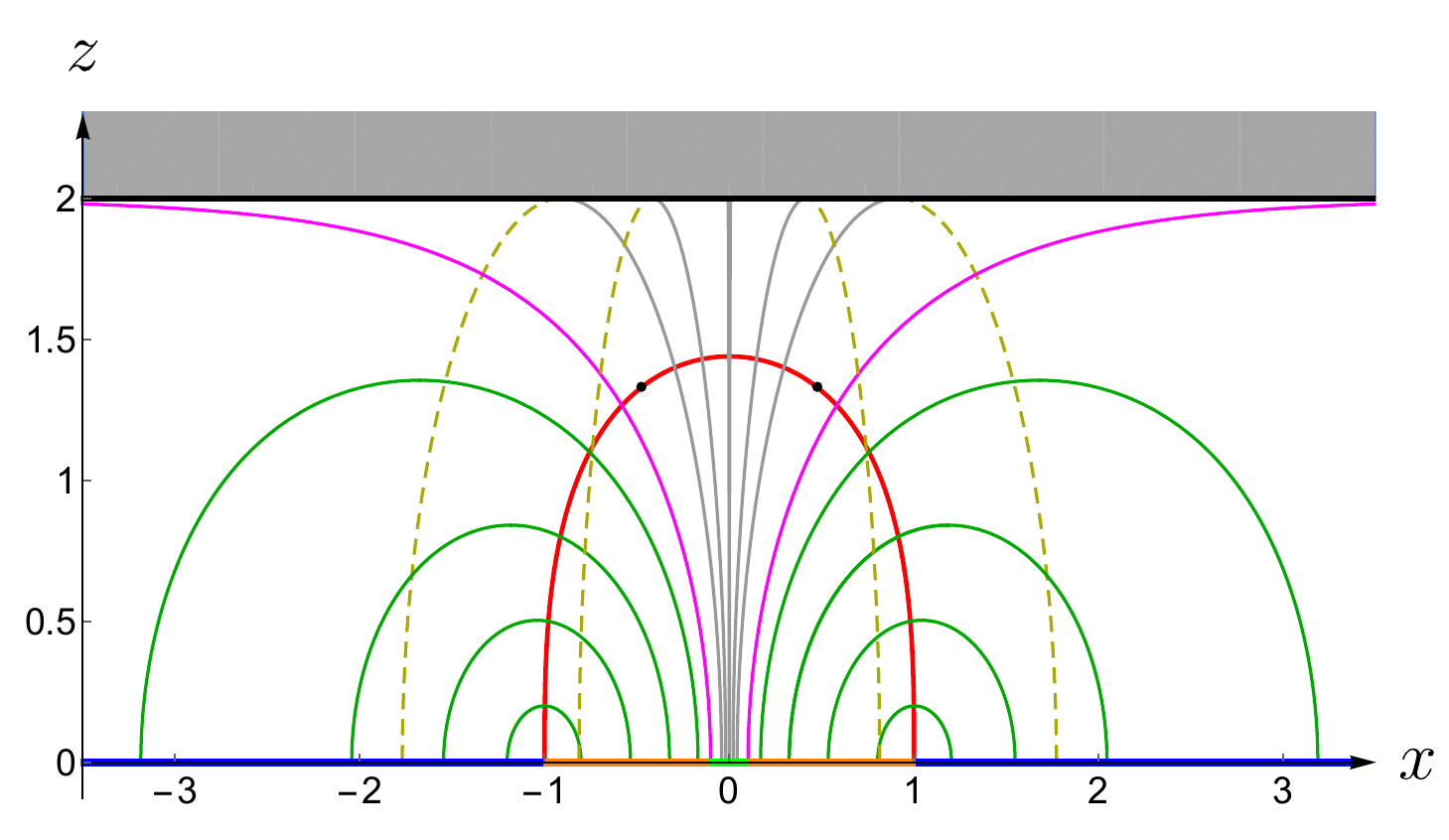}
			\hspace{.2cm}
			\includegraphics[width=0.55\textwidth]{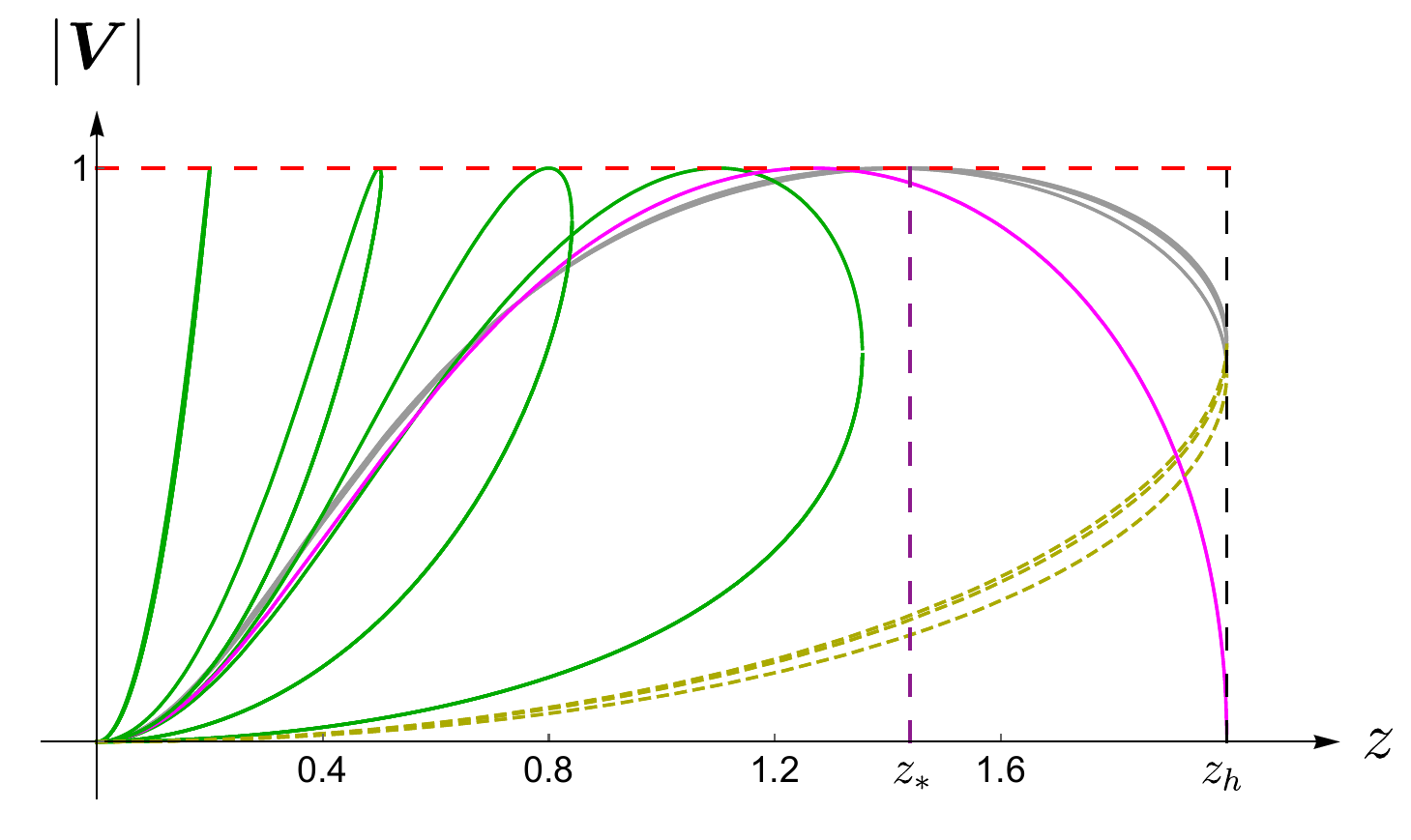}
			\vspace{-.5cm}
			\caption{\small 
				Geodesic bit threads for the strip $A$, in the constant time slice of the Schwarzschild AdS$_4$ black brane:
				trajectories (left) and modulus of the vector field (right).
				The black dots in the left panel identify the portion $\gamma_{A,\beta} \subsetneq \gamma_A$ 
				whose area is equal to $S_{A,\textrm{\tiny th}}$, i.e. the holographic thermal entropy of $A$.
			}
			\label{GeoandModulusBHD}
			\label{fig:integralcurvesgeodbh}
		\end{figure}

		Similarly to the case of the sphere,  
		we can split this class of geodesics in the $(z,x)$ plane  into two groups,
		displayed with different colors in the left panel of Fig.\,\ref{GeoandModulusBHD} for the case of AdS$_4$. 
		A group consists of the geodesics with a maximum height $\tilde z_* < z_h$,
		which can be identified with the integration constant $C$  given by \eqref{CCC}
		(see the solid green curves).    
		The other group includes all the geodesics whose maximum height is exactly equal to $z_h$ 
		(see the solid grey curves).  
		These geodesics possess a second branch (see the dashed dark yellow curves)
		which does not play a direct role in constructing the geodesic bit threads.
		In the left panel of Fig.\,\ref{GeoandModulusBHD},
		these two groups are geometrically separated by two (critical) geodesics
		that reach their maximum height $z_h$  only when $x\to \pm \infty$
		(see the magenta curves).
		These two geodesics intersect $\gamma_A$ at the points $(z_{m,\beta},\pm \,x_{m,\beta})$,
		where $z_{m,\beta}$ satisfies
		\be
		\label{CCC1}
		z_h = \frac{z_{m,\beta} }{ \sqrt{1 - (z_{m,\beta}/z_\ast)^{2d} }} \;.
		\ee
		  The infinite strip $\tilde{\gamma}_{A,\beta} \subsetneq \gamma_A$ 
		  is identified as
		  the portion of $\gamma_A$ enclosed by the critical geodesic bit threads
		  corresponding to the magenta curves in Fig.\,\ref{fig:integralcurvesgeodbh}.
		  The black dots in Fig.\,\ref{fig:integralcurvesgeodbh} provide the infinite strip $\gamma_{A,\beta} \subsetneq \gamma_A$ 
		  whose area is equal to the holographic thermal entropy of $A$, denoted by $S_{A, \textrm{\tiny th}}$.
		  For the setup shown in Fig.\,\ref{fig:integralcurvesgeodbh},
		  the area of  $\tilde{\gamma}_{A,\beta}$ is strictly larger than the area of  $\gamma_{A,\beta}$;
		  hence the flux of the geodesic bit threads through $\tilde{\gamma}_{A,\beta}$
		   is strictly larger than $S_{A, \textrm{\tiny th}}$.
		  The flux through $\tilde{\gamma}_{A,\beta}$ 
		  is equal to the flux through $A_\beta$, 
		  corresponding to the green interval in Fig.\,\ref{fig:integralcurvesgeodbh}.

		It is necessary to verify the nesting property to determine whether these geodesics constitute a proper family of bit threads. 
		To do so, the first step is to integrate  \eqref{ode-geodesic} and find $x$ as a function of $z$. 
		Since  \eqref{ode-geodesic} is quadratic in $z'(x)$, we have two branches. 
		The first one runs from a point of coordinate $x_{\textrm{\tiny $A$}} \in A$ to the maximum height $\tilde z_\ast$ of this geodesic, namely
		\begin{equation}
			x_{\mbox{\tiny$<$}}(z)
			\,=\,
			x_m(z_m) 
			- 
			\int_{z}^{z_m}   \frac{v \, z_h^{(d+1)/2} \, \sqrt{z_{\ast}^{2d}-z_m^{2d}}}{ \sqrt{ \big(z_h^{d+1}-v^{ d+1} \big) \big[ z_\ast^{2d} \, z^2_m- \big(z_{\ast}^{2d}-z_m^{2d} \big)v^{2}\big]}} \; \rd v
			\label{xmenobuconerogeodetico} 
		\end{equation}
		where we have imposed that the solution intersects the RT hypersurface at $(z_m, x_m)$. 
		This provides the intersection of this geodesic with the boundary as
		\be
		\label{yAstripBH}
		x_{\textrm{\tiny $A$}}
		\,=\,
		x_{\mbox{\tiny$<$}}(0)
		\,=\,
		x_m(z_m) 
		- 
		\int_{0}^{z_m}   \frac{v \, z_h^{(d+1)/2} \, \sqrt{z_{\ast}^{2d}-z_m^{2d}}}{ \sqrt{ \big(z_h^{d+1}-v^{ d+1} \big) \big[ z_\ast^{2d} \, z^2_m- \big(z_{\ast}^{2d}-z_m^{2d} \big)v^{2}\big]}} \; \rd v
		\ee
		which determines $x_{\textrm{\tiny $A$}}$ as a function of $z_m$. 
		In order to have the nesting property fulfilled, 
		$\partial_{z_m} x_A $ must be negative
		and in Appendix\;\ref{Nestingtotal} we show analytically that this derivative remains everywhere negative only for $d\leqslant 2$.

		To complete  our analysis of the nesting property, 
		we also need to consider the second branch for the geodesics with maximum height  $ \tilde z_\ast < z_h$. 
		For the geodesic intersecting $\gamma_A$ at $(z_m, x_m)$, this branch is given by 
		\begin{equation}
			x_{\mbox{\tiny$>$}}(z)
			=
			x_m(z_m) + 
			\int_{z_m}^{\tilde z_*}  \!
			\frac{v }{ \sqrt{\big[ 1- (v/z_h)^{d+1}\big] \big(\tilde z_\ast^{2} -v^{ 2} \big)}} \, \rd v
			+
			\int_{z}^{\tilde z_*}  \!
			\frac{v }{ \sqrt{\big[ 1- (v/z_h)^{d+1}\big] \big(\tilde z_\ast^{2} -v^{ 2} \big)}} \, \rd v \,.
			\label{xpiubuconerogeodetico}
		\end{equation}
		Since $\tilde z_*=C$ for this class of geodesics, 
		it is straightforward to calculate the coordinate $x_{\textrm{\tiny $B$}}\in B$ of the endpoint, 
		which satisfies $x_{\textrm{\tiny{$B$}}}=x_{\mbox{\tiny$>$}}(0)$. 
		To verify the nesting property, we must also ensure that $ \partial_{z_m} x_{\textrm{\tiny{$B$}}}$ is always positive. 
		In Appendix\;\ref{Nestingtotal} we analytically show that this is the case for any $d\geqslant 1$. 
		Hence, combining this result with the corresponding one for $\partial_{z_m} x_A $ discussed above, 
		we conclude that
		the construction of the geodesic bit threads for the strip 
		in the constant time slice of the Schwarzschild AdS$_{d+2}$ black brane
		fails for $d\geqslant 3$, 
		like for the case of AdS$_{d+2}$ (see Sec.\,\ref{stripemptyads}).
		Since the $d=1$ case has been already discussed in Sec.\,\ref{sec-BTZ},
		in the following, we consider the case of the infinite strip when $d=2$.

		For $d=2$, we evaluate $| \boldsymbol{V} |$ for the geodesic bit threads
		by employing the approach outlined in Appendix\;\ref{app-modulus}. 
		The computation is carried out separately for the two branches of the geodesics, resulting in the following expression
		\begin{equation}
			\big| \boldsymbol{V}_{\!\!\textrm{\tiny$\lessgtr$}} \big| 
			= 
			\frac{z_m \, z^2}{\sqrt{z^2 \left(z_m^{4} - z_*^{4}\right) + z_m^2 z_*^{4}}} \;
			\frac{\big(\partial_{z_m} x_{\textrm{\tiny$<$}}\big) \big|_{z=z_m}}{\partial_{z_m} x_{\textrm{\tiny$\gtrless$}}} \,.
			\label{modulo}
		\end{equation}
		Then, the dependence on $z_m$ is eliminated by inverting the relation \eqref{yAstripBH}, 
		thereby expressing $z_m$ as a function of $x$. 
		However, this analysis can be carried out only numerically.

		In the  right panel of Fig.\,\ref{GeoandModulusBHD},  
		where each distinct curve corresponds to a different geodesic bit thread,
		we show the curves describing $|\boldsymbol {V}|$ as $z$ varies along a single bit thread. 
		The modulus $|\boldsymbol {V}|$ is described 
		by the green closed orbits for the geodesics that do not reach  the horizon,
		by the magenta curve for the geodesics that reach the horizon at infinity
		and, finally, by the grey solid curves for the geodesics that reach the horizon
		(indeed, all of them are tangent to the vertical dashed black line $z=z_h$). 
		For completeness, we have also included the  dashed dark yellow curves associated to the grey solid curves,
		which provide the putative value of $|\boldsymbol {V}|$ along the auxiliary branch of the geodesic bit threads reaching the horizon, 
		which extend from the horizon back to the boundary. 
		From the right panel of Fig.\,\ref{GeoandModulusBHD}, 
		we can easily infer that the green curves and the grey curves 
		consistently lie below the horizontal dashed red line $|\boldsymbol {V}|=1$, thereby obeying the constraint at $|\boldsymbol {V}|\leqslant 1$.  
		The values of $z$ for which these curves touch the horizontal line $|\boldsymbol {V}|=1$  correspond to points lying on $\gamma_A$.  
		In fact, all of these points lie before the dashed purple vertical line representing $z= z_\ast$.

			\begin{figure}[t!]
		\vspace{-.5cm}
		\hspace{1.5cm}
			\includegraphics[width=0.82\textwidth]{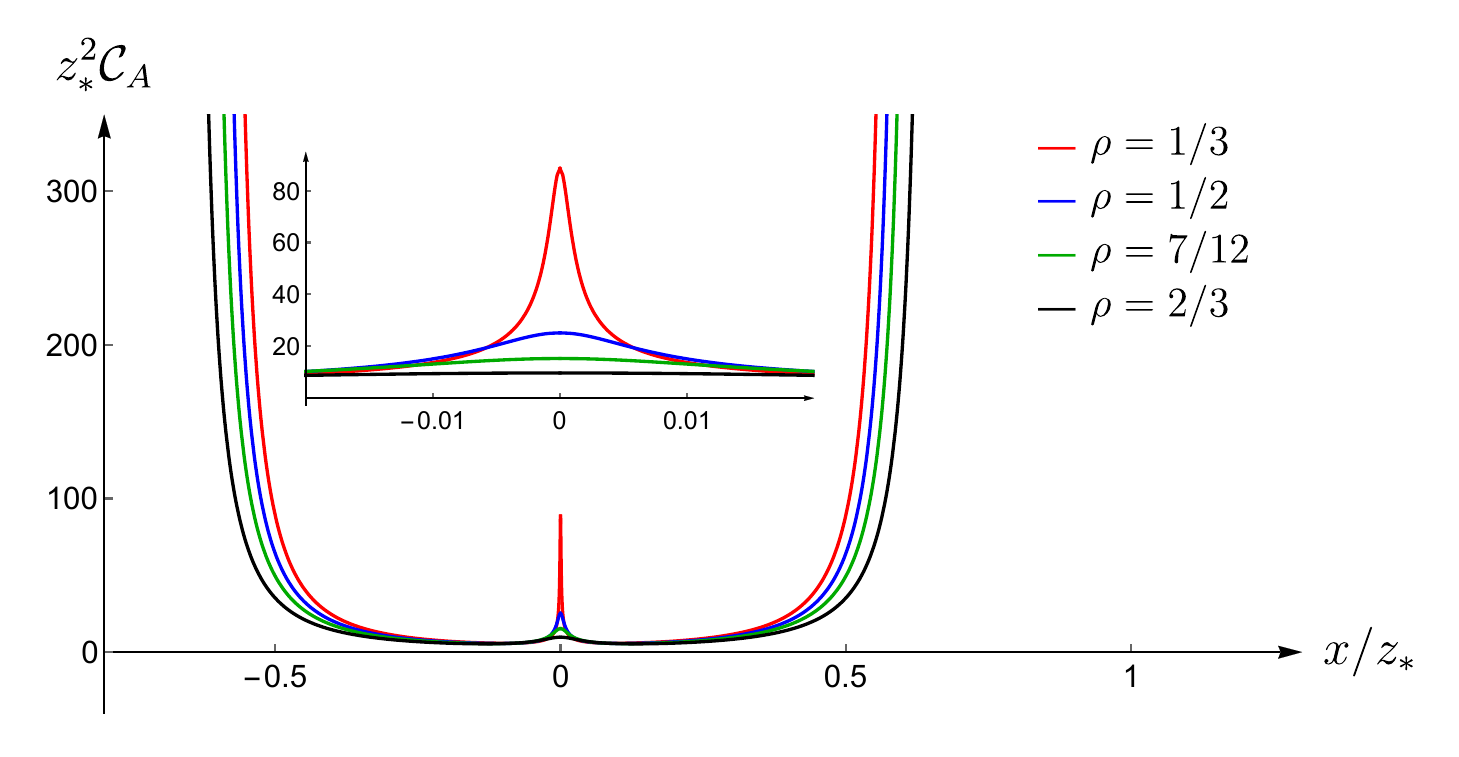}
		\vspace{-.2cm}
			\caption{\small 
				Holographic contour function $\mathcal{C}_A(\boldsymbol{x})$ for the strip
				 from (\ref{contour-strip-d-gen})-(\ref{x-strip-from-xi}), 
				 for the case of AdS$_4$ and for different values of $\rho$.
				The inset zooms in on the center of the strip. 
				}
				\label{fig:HigherDstripcontour}
		\end{figure}

		The holographic contour function $\mathcal{C}_A(\boldsymbol{x} )$ for  $\boldsymbol{x}\in A$
		is obtained as usual  as the density of flux 
		through this region on the boundary, namely
		\begin{equation}
		\label{contour-strip-d-gen}
		\mathcal{C}_A(\boldsymbol{x} )=
		\lim_{z \to 0}\left. \left(\frac{1}{4G_{\textrm{\tiny N}} }  \, \big|\boldsymbol{V} \big| \, \tau_a \, n^a \, \frac{L_{\textrm{\tiny AdS}}^{d}}{z^{d}}\right)\right|_{\boldsymbol{x}\in A}
	\end{equation}
where $ |\boldsymbol{V} |$ is given  in	\eqref{modulo}. 
The unit vector $\boldsymbol{\tau}$ providing the direction of $\boldsymbol{V}$ is once again determined by \eqref{tausfera}, 
where $C$ is now defined by \eqref{CCC}. Finally, the unit vector orthogonal to the surface of constant $z$ is 
$\boldsymbol{n}=\frac{1}{L^2_{\textrm{\tiny AdS}}}\big(z \sqrt{f(z)},0,\boldsymbol{0} \big)$.	
		The final result can be written in the parametric form w.r.t. $\xi \equiv z_m/z_\ast$ given by 
		\be
		\label{CA-strip-xi}
		\mathcal{C}_A(\boldsymbol{x} )
		\,=\, 
		\frac{1}{z_*^2}  \left[\, 1-\sqrt{1-( \xi \, \rho)^{d+1}} 
		\int_0^1 
		\frac{ \xi^d \, \big[(d-1)  \xi ^{2 d}+1\big] \,t  }{
			\big[   (\xi ^{2 d}-1 )\, t^2  +1\big]^{3/2} \sqrt{1-(\xi  \,\rho \, t)^{d+1}}} \;\rd t
		\,\right]^{-1}
		\ee
		where $t \equiv v/z_m$ (see (\ref{xmenobuconerogeodetico}) and (\ref{xpiubuconerogeodetico}))
		and $\rho\equiv  z_\ast / z_h$.
		Then, 
		$\xi$ can be expressed in terms of $\boldsymbol{x} \in A$ (hence $|x| \leqslant 2b$)
		and $\rho$ by inverting the following expression
		\be
		\label{x-strip-from-xi}
		x \,= \,
		z_\ast 
		\left[\, \int_\xi^1
		\frac{t^d}{\sqrt{1-t^{2 d}} \; \sqrt{1-(\rho \, t)^{d+1}}} \;\rd t 
		-
		\int_0^1 
		\frac{\xi  \,\sqrt{1-\xi ^{2 d}} \; t}{ \sqrt{(\xi ^{2 d}-1 ) \, t^2 +1}  \;\sqrt{1-(\xi \rho  t)^{d+1}}} \; \rd t 
		\,\right] 
		\ee
		where $t=v/z_\ast$ and  $t=v/z_m$ in the first and second integral respectively,
		and whose r.h.s. simply represents $x_{\mbox{\tiny$<$}}(0)$ expressed in terms of the dimensionless variables $\xi$ and $\rho$. 
		Here, \eqref{ode-RT-strip-ddim}  has been used to write an integral representation of $x_m(z_m)$.

		In Fig.\,\ref{fig:HigherDstripcontour} we show $ z_*^2 \, \mathcal{C}_A(\boldsymbol{x})$
		for $d=2$ and different values of $\rho$. 
		Notice that the presence of the horizon serves to smoothen the divergence at $x=0$ occurring in AdS$_4$ 
		(see Appendix\;\ref{GeodAdS4} and Fig.\,\ref{contourAdS4}). 
		As $z_h$ approaches infinity, meaning that $\rho$ becomes smaller and smaller, 
		the maximum at $x=0$ becomes progressively more pronounced and higher, 
		ultimately reproducing the singularity observed in AdS$_4$.

		When the subregion $A$ is a sphere, in Sec.\,\ref{sec-Sch-AdS-sphere}, we found that the flux of its geodesic bit threads 
		through the horizon
		in the constant time slice of Schwarzschild AdS$_{d+2}$ black brane provides the holographic thermal entropy of $A$.
		It is worth investigating whether this feature
		remains valid for the strip in $d=2$. 
		In order to calculate the flux $\Phi (\boldsymbol{x}; z_h)$ through the horizon
		of the geodesic bit threads for the strip  
		in the constant time slice of Schwarzschild AdS$_{4}$ black brane,
		the most straightforward approach consists in 
		determining the area of the RT hypersurface corresponding to 
		the infinite strip contained in $A$
		and identified by the interval $(-x_{m,\beta}, x_{m,\beta})$ 
		in the $x$-direction,
		associated with
		\be
		z_{m,\beta}= \frac{z_*}{\sqrt{2}\, z_h} \; 
		\sqrt{ \sqrt{4 z_h^4+z_*^4} - z_*^2}
		\ee
		which has been obtained by inverting \eqref{CCC1}.

		In the discussion of Fig.\,\ref{fig:integralcurvesgeodbh} (see the text below (\ref{CCC1})) it has been observed that  
		the holographic thermal entropy $S_{A,\textrm{\tiny th}}$ is not reproduced by the flux through the horizon, denoted by $\widetilde{S}_{A,\textrm{\tiny th}}$.
		Nevertheless, we find it interesting to compare these two quantities anyway;
		hence, we consider 
  \be
  \label{RSB0}
  R_{\textrm{\tiny th}}
  \equiv
  \frac{ \widetilde{S}_{A,\textrm{\tiny th}} }{S_{A,\textrm{\tiny th}}}
  \;\;\;\qquad\;\;\;
  S_{A,\textrm{\tiny th}}
  =\frac{1}{4G_{\textrm{\tiny N}} } \; \frac{L^2_{\textrm{\tiny AdS}}}{z_h^2} (2b) (2b_\perp)
  \ee
  where $(2b) (2b_\perp)$ is the volume of $A$.
  To simplify the evaluation of $R_{\textrm{\tiny th}}$, we alternatively assess $4G_{\textrm{\tiny N}} \,\widetilde{S}_{A,\textrm{\tiny th}}$ as the area of $\tilde{\gamma}_{\textrm{\tiny$A,\beta$}}$. 
Then, by exploiting the expression for $b$ given in \eqref{www} 
and redefining the integration variable from $v$ 
(see (\ref{xmenobuconerogeodetico}) and (\ref{xpiubuconerogeodetico}))
to $t=v/z_\ast$, 
  we  find the following result
		\be
		\label{RSB}
		R_{\textrm{\tiny th}} (\rho)
		=
		\left[\, \int_0^1 \frac{\rho\, t^2}{\sqrt{1-t^{4}} \;\sqrt{1-\rho ^{3} t^{3}}} \;\rd t \, \right]^{-1}
		\! \int_{\tilde z_{m\beta}}^1 \frac{1}{\rho \,t^2\; \sqrt{1-t^{4}} \; \sqrt{1-\rho ^{3} t^{3}}} \; \rd t
		\ee
  where $\tilde z_{m,\beta}=\tfrac{1}{\sqrt{2}}\, \sqrt{ \sqrt{4+\rho^4} -\rho^2 }$.
Consequently, $R_{\textrm{\tiny th}} $ depends only on the dimensionless variable $\rho$ (see the text below \eqref{CA-strip-xi}).
 The flux through the horizon would reproduce the holographic thermal entropy of $A$ 
 if  $R_{\textrm{\tiny th}}(\rho)=1$  for all values of $\rho$.

			\begin{wrapfigure}[12]{l}{0.5\textwidth}
			\vskip -.5cm
			\hspace{-.5cm}
			\includegraphics[width=0.52\textwidth]{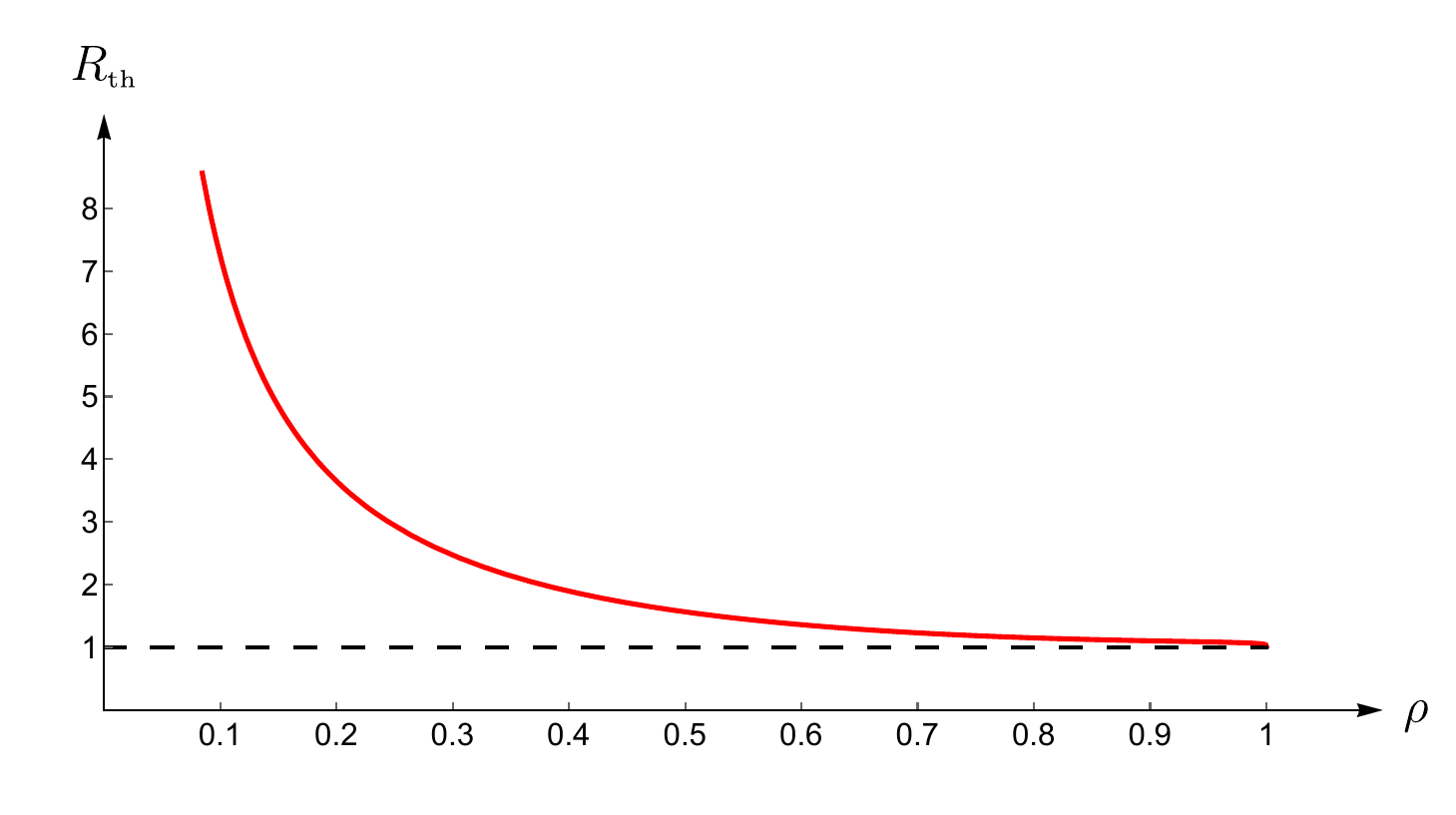}
			\vspace{-.2cm}
			\caption{\small Ratio (\ref{RSB0}) 
			for the geodesic bit threads of a strip,
			in the constant time slice of the Schwarzschild AdS$_4$ black brane,
			from (\ref{RSB}).}
				\label{fig:SBstrip}
		\end{wrapfigure}

		In Fig.\,\ref{fig:SBstrip}, the red curve has been obtained by evaluating (\ref{RSB}) numerically
		and it differs significantly from the constant 1.
The unexpected divergence for $\rho \sim 0$ stems from 
the peculiar behavior of the numerator in \eqref{RSB} in this regime. 
The integral remains finite despite the integration region shrinking to zero as $\rho^2$ for $\rho \to 0$. 
One might initially think that the integrand scales only as $1/\rho$, leading to a numerator that vanishes as $\rho$ 
and consequently to a finite ratio in this limit (the denominator decreases linearly with $\rho$). 
However, as $\rho\to 0$, the fact that $t\simeq 1+O(\rho^2)$ introduces an additional $1/\rho$ factor arising from the term $\sqrt{1-t^4}$  in the denominator. 
Hence, the numerator remains finite as $\rho$ approaches zero, resulting in a divergent ratio because of the vanishing of the denominator.
On the other hand, $R_{\textrm{\tiny th}}(\rho) \to 1$ as $\rho \to 1$. 
This implies that $\widetilde{S}_{A,\textrm{\tiny th}}$ captures the holographic thermal entropy of $A$
when $\gamma_A$ gets closer and closer to the horizon,
which is an expected feature, as mentioned in Sec.\,\ref{sec-intro} and Sec.\,\ref{sec-Sch-AdS-sphere}.

		\subsubsection{Minimal hypersurface inspired bit threads} 
		\label{sec-strip-ddim-mhbt}

		The unsuccessful attempt with the geodesic bit threads for generic $d$ 
		prompts us to explore alternative constructions, as done e.g. in Sec.\,\ref{sec5} in pure AdS$_{d+2}$.
		In the following, we expand upon the minimal hypersurface inspired bit threads initially applied in the constant time slice of AdS$_{d+2}$, 
		showing its applicability in the constant time slice of Schwarzschild AdS$_{d+2}$ black brane for any value of $d$.
		
		This analysis employs the profile of the RT hypersurface $\gamma_A$ solving \eqref{ode-RT-strip-ddim} to generate a candidate family of curves.
		More precisely, these curves are defined by deforming the profile of  $\gamma_A$ through two parameters:
		an arbitrary constant $c_0$ parameterizing a uniform translation
		and the maximum height  $\tilde z_\ast$ of the curve describing the deformation along the holographic direction. 
		Thus, these curves are
		\begin{equation}
			x_{\textrm{\tiny $\lessgtr$}}(z)=c_0 \pm x (z; \tilde{z}_* ) 
			\label{msbitbuconero}
		\end{equation}
		where $x (z ; \tilde{z}_* )$  solves  \eqref{ode-RT-strip-ddim} with $z_*$ replaced by $\tilde{z}_*$
		and the constant  $c_0$ represents the center of each curve. 
		Thus, the curve (\ref{msbitbuconero}) is symmetric under reflection w.r.t. the hyperplane corresponding to $x=c_0$.  
		Notice that understanding $\tilde{z}_\ast$ as the depth or the maximal height is slightly inaccurate
		because this is the case only when $\tilde{z}_\ast < z_h$,
		while for $\tilde{z}_\ast \geqslant z_h$ the depth of (\ref{msbitbuconero}) consistently remains equal to $z_h$.

		The two independent parameters $c_0$ and $\tilde{z}_*$ can be found by following the standard procedure,
		which requires  that the candidate bit thread in \eqref{msbitbuconero} intersects $\gamma_A$ orthogonally at the point $\big( z_m, x_m(z_m) \big)$. 
		This leads to the following set of conditions
		\begin{equation}
			\left\{ \begin{array}{l}
				x_{\textrm{\tiny $<$}}(z_m) = x_{m}(z_m)
				\\ 
				\rule{0pt}{.5cm}
				\big[ \,g_{zz} +  g_{xx} \, x_{\textrm{\tiny $<$}}’(z) \,x’_{m}(z) \,\big] \big|_{(z,x)=(z_m,x_m(z_m))} = 0
			\end{array}\right.
			\label{ortostrisciamsbh}
		\end{equation}
		being $g_{xx} = L^2_{\textrm{\tiny AdS}} / z^2$ and $g_{zz}= L^2_{\textrm{\tiny AdS}} /( f(z) z^2)$ defined as the diagonal components of \eqref{sch-ads-brane-metric}. 
		Solving \eqref{ortostrisciamsbh} allows us to determine $\tilde{z}_*$ and $c_0$ of each integral curve, finding 
		\begin{equation}
			\label{zsandc0}
			\tilde{z}_*= \frac{z_m z_*}{\big(z_*^{2d}-z_m^{2d}\big)^{1/(2d)}} 
			\;\;\;\qquad\;\;\;
			c_0= x_m(z_m)+x(z_m,\tilde{z}_*) 
		\end{equation}
		where $x_m(z_m)$ represents  $\gamma_A$ satisfying \eqref{ode-RT-strip-ddim}
		and $x(z_m,\tilde{z}_*)$ shares the same functional form of $x_m(z_m)$ 
		but with $z_\ast$ replaced by $\tilde{z}_\ast$. 
		The putative bit threads in \eqref{msbitbuconero} consist of two branches sharing an endpoint at $\tilde{z}_\ast$. 
		Let us denote by $x_{\mbox{\tiny$<$}}(z)$ the branch corresponding to the minus sign, originating from $A$
		and by $x_{\mbox{\tiny$>$}}(z)$ the branch associated with the plus sign, originating from $\tilde z_\ast$ and extending to the complementary region $B$.

								\begin{figure}[t!]
			\vspace{-.5cm}
			\hspace{-.9cm}
			\includegraphics[width=0.55\textwidth]{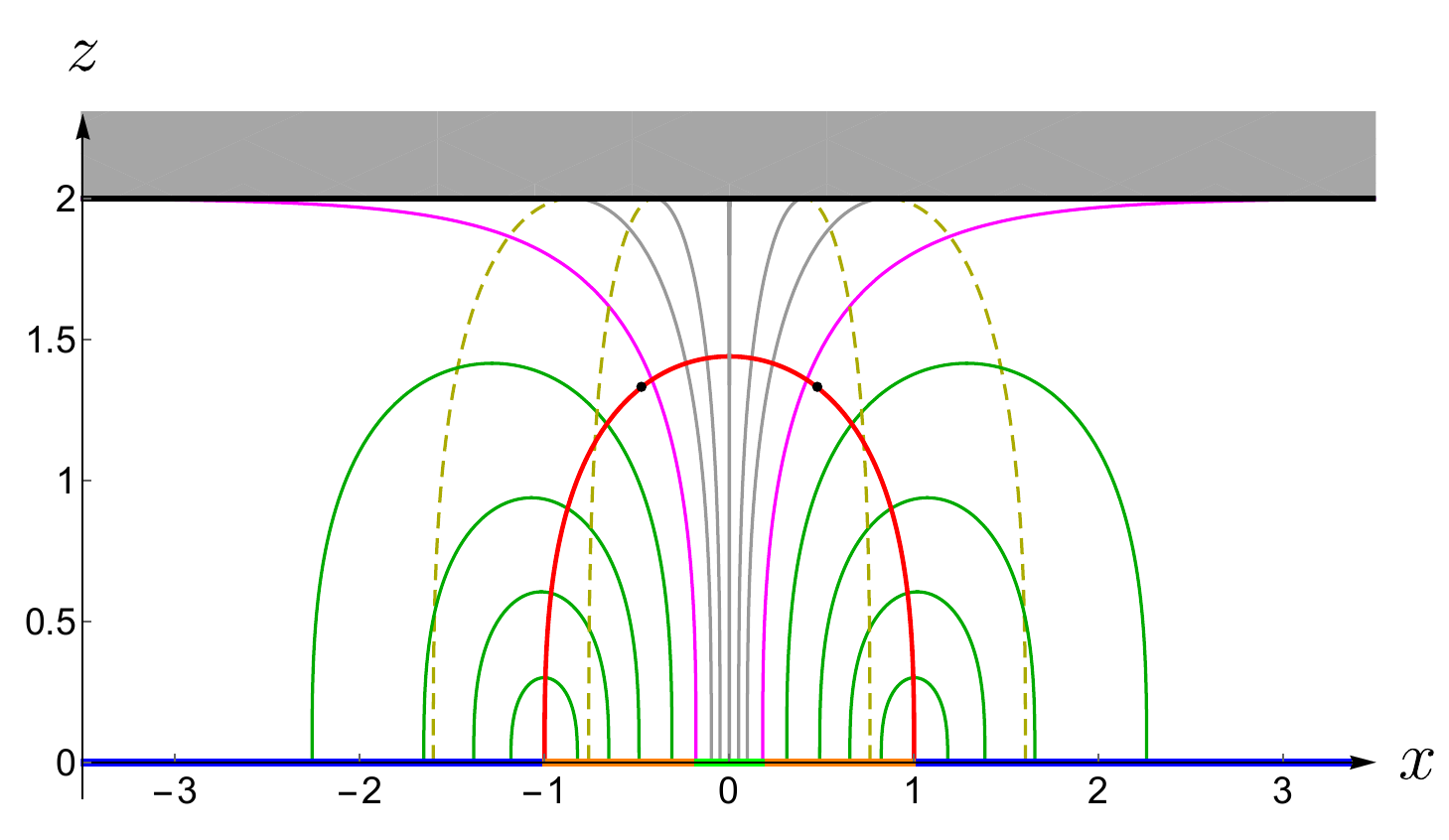}
			\hspace{.1cm}
			\includegraphics[width=0.55\textwidth]{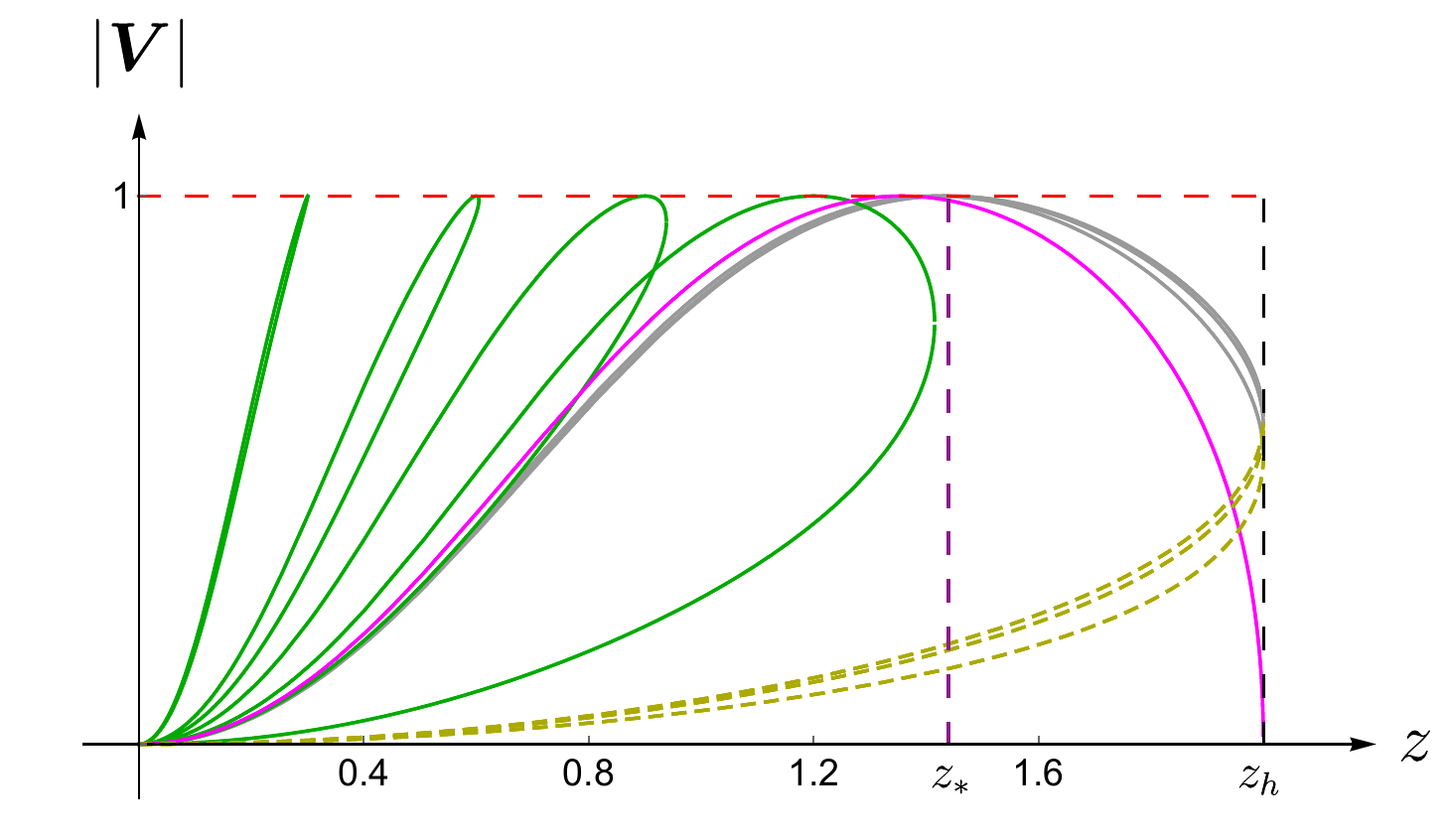}
			\vspace{-.5cm}
			\caption{\small
				Minimal hypersurface inspired bit threads for the strip $A$, in the constant time slice of the Schwarzschild AdS$_4$ black brane:
				trajectories (left) and modulus of the vector field (right).
				The black dots in the left panel identify the portion $\gamma_{A,\beta} \subsetneq \gamma_A$
				whose area is equal to $S_{A,\textrm{\tiny th}}$.
			}
			\label{fig:msbitintegrallinesbh} 
		\end{figure}

		This family of curves orthogonal to $\gamma_A$ must satisfy the nesting property in order to provide a consistent set of bit threads. 
		To perform this crucial check, 
		we examine the sign of the partial derivatives w.r.t. $z_m$ of the two intersections with the boundary at $z=0$, 
		denoted by $x_{\textrm{\tiny $A$}}\equiv x_{\mbox{\tiny$<$}}(0)$ and $x_{\textrm{\tiny $B$}}=x_{\mbox{\tiny$>$}}(0)$.
		Evaluating these two derivatives is intricate but relatively straightforward
		and in Appendix\;\ref{Nestingtotal} we discuss their signs,
		showing that the nesting property is satisfied for any $d\geqslant 1$.

		Also, in this setup, we can classify the trajectories in the $(z, x)$ plane into two distinct classes:
		the curves having a maximum height strictly below the horizon
		(see the green lines in the left panel of Fig.\,\ref{fig:msbitintegrallinesbh})
		and the curves touching the horizon, i.e. whose maximum height is precisely equal to $z_h$
		(see the grey lines in the left panel of Fig.\,\ref{fig:msbitintegrallinesbh}).
		Additionally, the latter curves possess an auxiliary branch that extends from the horizon back to the boundary
		(see the dashed dark yellow lines in the left panel of Fig.\,\ref{fig:msbitintegrallinesbh}).
		These two groups are geometrically separated by the two curves 
		that reach their maximum height $z_h$ only when $x$ approaches infinity
		(see the magenta lines in the left panel of Fig.\,\ref{fig:msbitintegrallinesbh}).
		These two critical lines intersect $\gamma_A$ at $(z_{m,\beta}, \pm \,x_{m,\beta})$, 
		where 
		\be
		\label{extremMS}
		z_{m,\beta}=\frac{z_*  z_h}{ \left(z_h^{2 d}+z_*^{2 d}\right){}^{1/(2d)}}
		\ee
		which is obtained by first setting $\tilde z_* = z_h$ in the first relation of \eqref{zsandc0} and then solving for $z_m$.
		The intersections between $\gamma_A$ and the magenta curves fix
		the extrema $\pm b_\beta$ of the green interval in the left panel of Fig.\,\ref{fig:msbitintegrallinesbh},
		which characterizes the infinite strip $A_\beta \subsetneq A$.
		Furthermore, these intersections identify the infinite strip $\tilde{\gamma}_{A, \beta} \subsetneq \gamma_A$.
		The flux through $\tilde{\gamma}_{A,\beta}$ equals the flux through $A_\beta$.
		 
		 Comparing Fig.\,\ref{fig:integralcurvesgeodbh} with Fig.\,\ref{fig:msbitintegrallinesbh},
		 where the same $z_h$ and $b$ have been chosen, 
		 we observe that $\gamma_{A, \beta} \subsetneq \tilde{\gamma}_{A, \beta}$ in the former one, 
		 while $\tilde{\gamma}_{A, \beta} \subsetneq \gamma_{A, \beta}$ in the latter one,
		 where we remind that $\gamma_{A, \beta} $ is determined by $S_{A,\textrm{\tiny th}}$ 
		 (see the black dots on $\gamma_A$ in the figures).
		 Moreover, $A_\beta$ (see the green segment on the boundary in the figures) 
		 is larger in Fig.\,\ref{fig:msbitintegrallinesbh}.
		 For both these bit thread configurations, 
		 the holographic thermal entropy of $A$ is not captured 
		 by their fluxes either through the horizon or through $A_\beta$.

				\begin{figure}[t!]
				\hspace{-1.2cm}
			\includegraphics[width=0.55\textwidth]{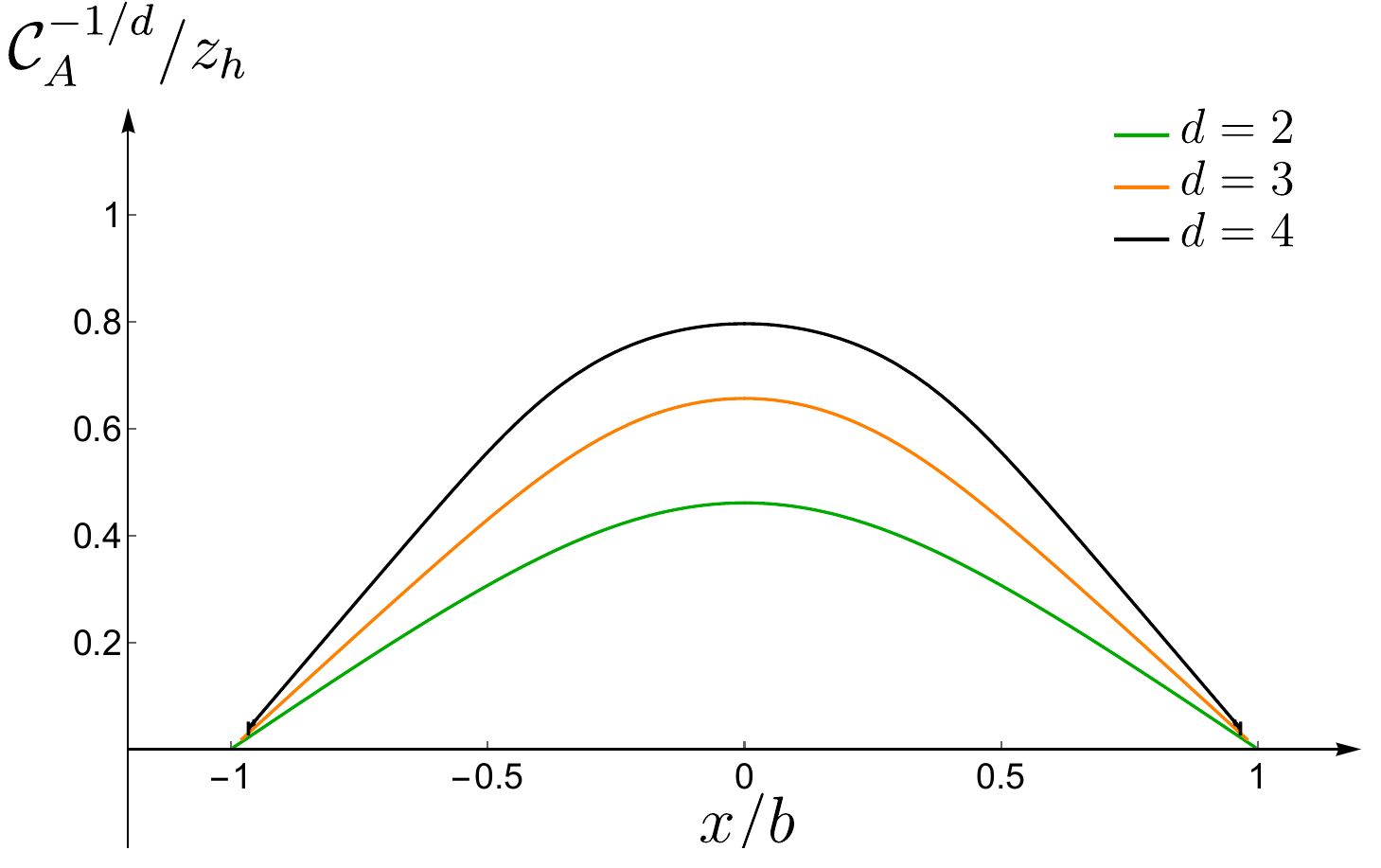}
			\includegraphics[width=0.55\textwidth]{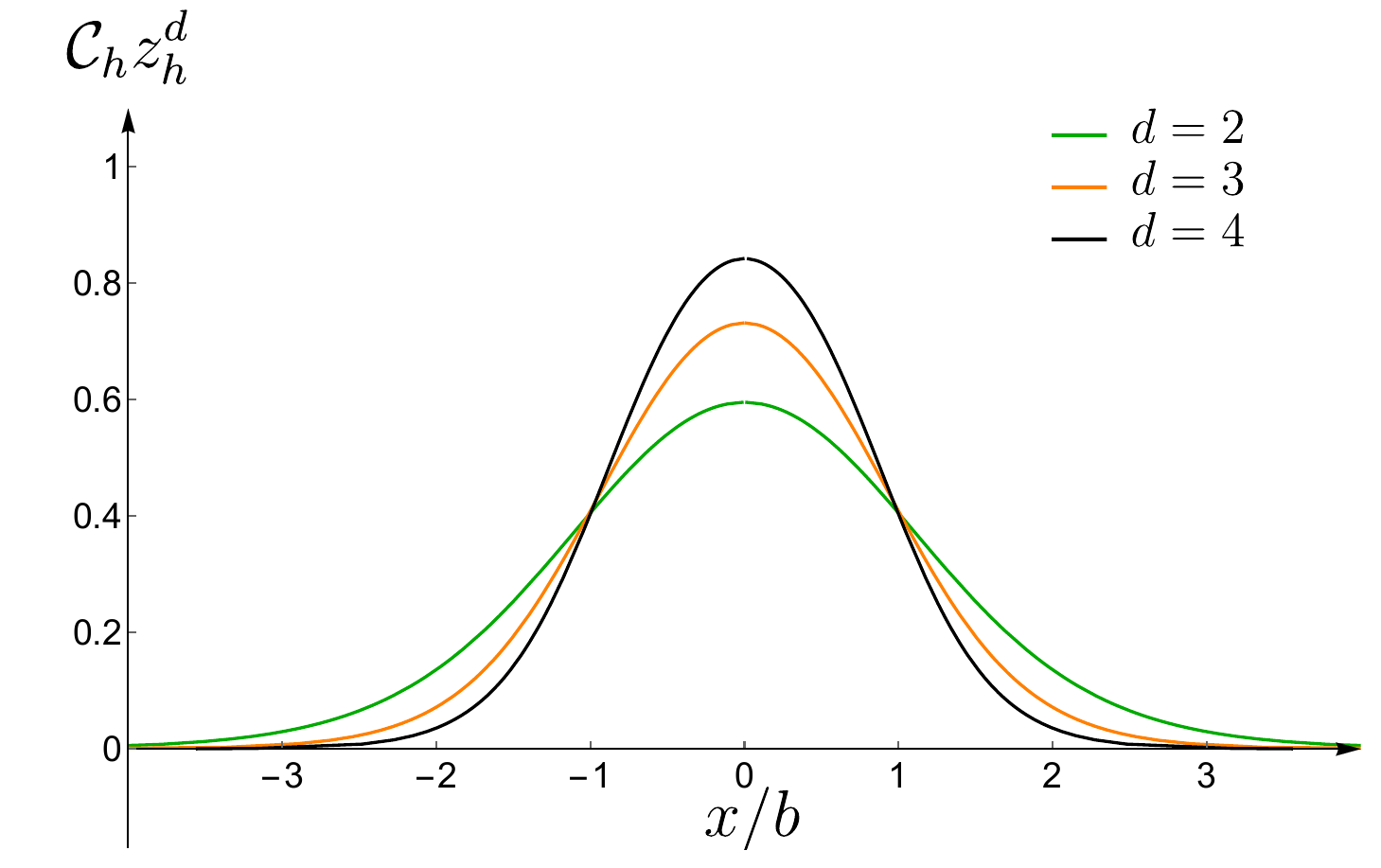}
			\caption{
				\small  
				Holographic contour function 
				$\mathcal{C}_{A} (\boldsymbol{x})$ (left)
				and $\mathcal{C}_{h} (\boldsymbol{x})$ (right)
				induced by the minimal hypersurface inspired bit threads 
				for the strip, in the constant time slice of the Schwarzschild AdS$_{d+2}$ black brane
				(see (\ref{contourbordomsbh}) and (\ref{contourorizzmsbh}) respectively).
				}
					\label{fig:contouromsbh}
		\end{figure}

		The modulus of the corresponding vector field  $\boldsymbol{V}$ can be obtained in terms of both $z$  and $z_m$ by following the standard
		procedure outlined in Appendix\;\ref{app-modulus}. We find
		\begin{equation}
			\big|\boldsymbol{V}_{\!\!\textrm{\tiny$\lessgtr$}}\big|
			= 
			\left| \,
			\frac{z^d \, z_m^d}{\sqrt{z^{2 d} \big(z_m^{2 d}-z_*^{2 d}\big)+z_m^{2 d} \,z_*^{2 d}}} \;
			\frac{\big(\partial_{z_m} x_{\textrm{\tiny$<$}}\big)\big|_{z=z_m} }{\partial_{z_m} x_{\textrm{\tiny$\lessgtr$}}} \,\right|
			\label{modulomsbh}
		\end{equation}
		where $|\boldsymbol{V}_{\!\!\!\textrm{\tiny$<$}}|$ and $|\boldsymbol{V}_{\!\!\!\textrm{\tiny$>$}}|$
		refer to the magnitude associated with the minus and plus branch in \eqref{msbitbuconero} respectively.  
		The direction of  $\boldsymbol{V}$ is provided by the unit vector $\boldsymbol{\tau}$ 
		tangent to the minimal hypersurface inspired bit threads
		and it reads
		\begin{equation}
			\boldsymbol{\tau}_{\textrm{\tiny$\lessgtr$}}
			=
			\left(\tau^z_{\textrm{\tiny$\lessgtr$}},\tau^x_{\textrm{\tiny$\lessgtr$}} \right)
			= 
			\frac{z}{L_{\textrm{\tiny AdS}} \, \tilde{z}^d_*} 
			\left(  \frac{\sqrt{z_h^{d+1}-z^{d+1}} \, \sqrt{\tilde{z}_*^{2 d}-z^{2 d}}}{z_h^{(d+1)/2}} \,, \, z^d  \right) .
		\end{equation}
		The magnitude in \eqref{modulomsbh} cannot be expressed analytically in terms of $x$ and $z$
		because \eqref{msbitbuconero}  cannot be inverted in closed form. 
		However, \eqref{modulomsbh} and \eqref{msbitbuconero} provide a parametric representation of the magnitude in terms of $z$ and $z_m$. 
		In the right panel of Fig.\,\ref{fig:msbitintegrallinesbh} we show the trajectories representing $|\boldsymbol {V}|$ as $z$ varies along a single bit thread
		(each distinct curve corresponds to a different bit thread).  
		The pattern of these curves  follows closely the one obtained for the geodesics bit threads in $d=2$ 
		(see the right panel of Fig.\,\ref{GeoandModulusBHD}).
		This type of representation also allows us to verify graphically that $|\boldsymbol{V}|\leqslant 1$ 
		and that the saturation $|\boldsymbol{V}| = 1$ is reached only on the RT hypersurface.
		This feature and the fact that the nesting property is respected
		tell us  that the integral curves in \eqref{msbitbuconero},
		with the proper parameters obtained as described above,
		provide proper bit threads.

		These results can be employed to compute (at least parametrically) the holographic contour function $\mathcal{C}_A(\boldsymbol{x})$ on the boundary for $\boldsymbol{x}\in A$
		by using  the density of flux of $|\boldsymbol{V}_{\!\!\!\textrm{\tiny$<$}}|$ through $A$.
		It reads
		\bea
		\label{contourbordomsbh}
		\mathcal{C}_{A} (\boldsymbol{x}) 
		&=& 
		\lim_{z \to 0^+} \left(\frac{1}{4G_{\textrm{\tiny N}} } | \, \boldsymbol{V}_{\!\!\!\mbox{\tiny$<$}}| \  \tau_a \, n^a \,\frac{L_{\textrm{\tiny AdS}}^{d}}{z^{d}}   \right)
		\\
		\rule{0pt}{.85cm}
		&=&
		\frac{L^d_{\textrm{\tiny AdS}}}{4 G_{\textrm{\tiny N}} \, z_*^d} 
		\left[\,
		1- d \int_0^1  \frac{ \xi^d \sqrt{1-(\xi\, \rho) ^{d+1}} \;  t^d}{ \big[ 1-\big(1-\xi ^{2 d}\big) t^{2 d}\big]^{3/2}
			\sqrt{1-(\xi \,\rho \,t)^{d+1}}} \; \rd t
		\,\right]^{-1}
		\nonumber
		\eea
		where $(L_{\textrm{\tiny AdS}}/z)^d$ comes from the square root of the determinant of the induced metric on the $z=\text{const}$ slice
		and $\boldsymbol{n}$ is the unit vector normal to the boundary,
		whose components are  $(n^z,n^x)=\frac{z}{L_{\textrm{\tiny AdS}}} \,\big(1,0\big)$.  
		The coordinate $x$ as a function of $\xi$ and $\rho$ is obtained by setting $z=0$ in \eqref{msbitbuconero}.
		In the left panel of Fig.\,\ref{fig:contouromsbh}
		we  show $\mathcal{C}_{A} (\boldsymbol{x})^{-1/d}/z_h$,
		to facilitate the comparison with the same quantity
		for the cases considered in Fig.\,\ref{fig:HYPcontourbordoorizsfera} 
		and Fig.\,\ref{fig:contourbordoorizsfera}.

		Another flux that we find it worth evaluating is the one through the horizon,
		whose holographic contour function  is 
		\begin{equation}
			\mathcal{C}_{h} (\boldsymbol{x})
			= 
			\lim_{z \to z_h^-} \left(\frac{1}{4G_{\textrm{\tiny N}}} \, \big|\boldsymbol{V}_{\!\!\!\mbox{\tiny$<$}} \big| \, \tau_a \, n^a \, \frac{L_{\textrm{\tiny AdS}}^{d}}{z^{d}} \,\right)
			\label{contourorizzmsbh}
		\end{equation}
		where $\boldsymbol{n}$ represents the unit vector normal to the horizon.
		Again, we can study (\ref{contourorizzmsbh}) numerically
		and some results of this analysis
		are reported in the right panel of  Fig.\,\ref{fig:contouromsbh}.

		\begin{figure}[t!]
		\vspace{-.5cm}
		\hspace{1.5cm}
			\includegraphics[width=0.8\textwidth]{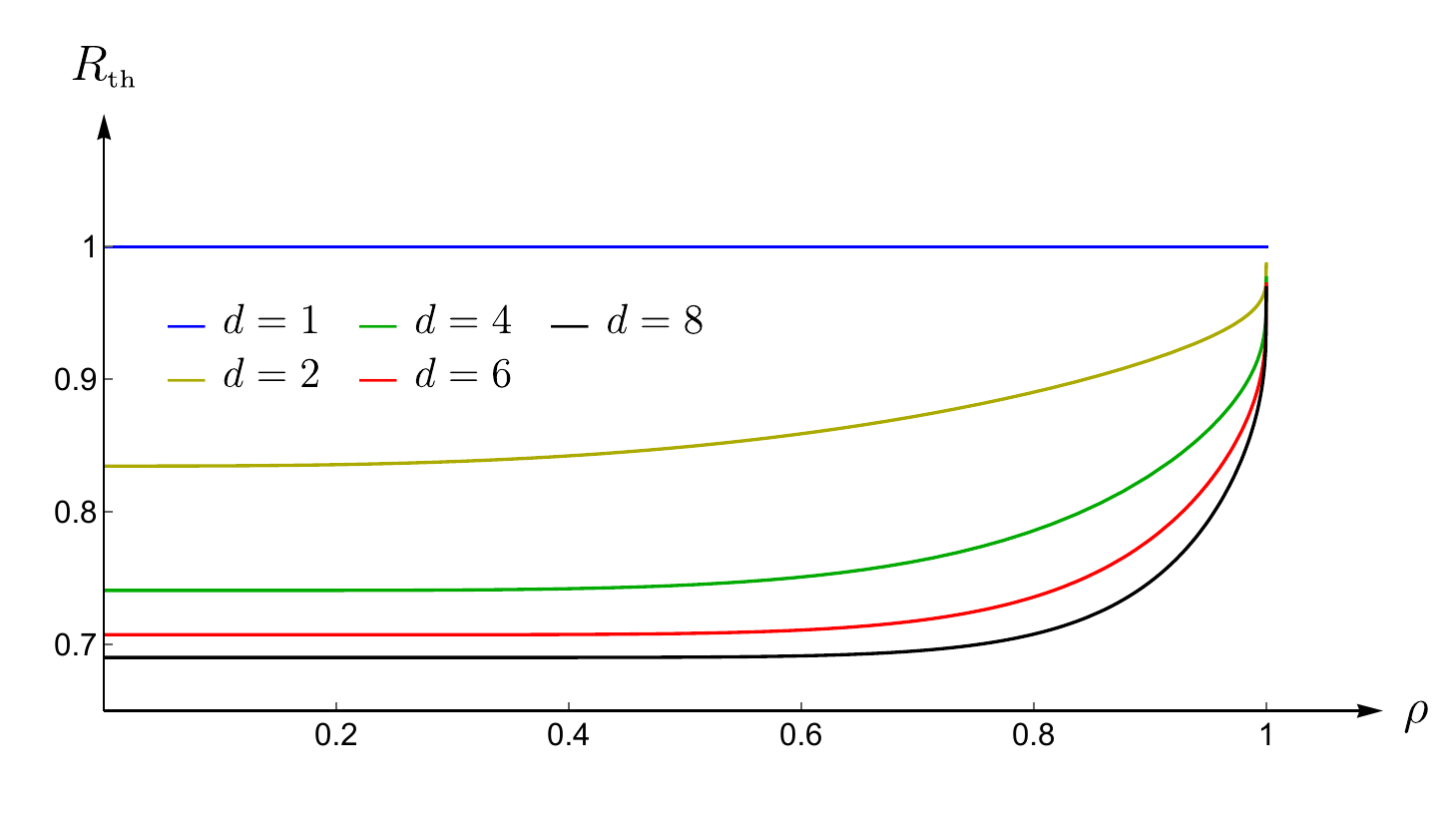}
		\vspace{-.2cm}
			\caption{
				\small 
				Ratio $R_{\textrm{\tiny th}}$ in (\ref{RSB0}) 
				for the minimal hypersurface inspired bit threads of the strip,
				as a function of $\rho$, 
				for different values of the dimension $d$ (see the final paragraph of Sec.\,\ref{sec-strip-ddim-mhbt}).
			}
			\label{fig:SBMSstrip} 
		\end{figure}

		In the discussion of Fig.\,\ref{fig:msbitintegrallinesbh}, we have highlighted that $S_{A, \textrm{\tiny th}}$ is not captured by the flux through the horizon. 
		However, it is interesting to quantify this failure. 
		By considering the ratio $R_{\textrm{\tiny th}}$ defined in \eqref{RSB0} for these minimal hypersurface inspired bit threads
		and adapting the steps used in Sec.\,\ref{sec-gbt-strip-ddim} to this case in a straightforward way, 
		we arrive at the obvious higher dimensional generalization of \eqref{RSB},
		where the only difference is that the extremum of integration called $\tilde{z}_{m,\beta}$ is now given by $(1+\rho^{2d})^{-1/(2d)}$.
		In Fig.\,\ref{fig:SBMSstrip}, we have plotted  $R_{\textrm{\tiny th}} $ in this setup for some values of $d$.
		The resulting curves are regular and finite but differ from the constant value $1$ when $d>1$.  
		As $d$ increases, the dependence of $\rho$  progressively mild when $\rho$ is far from $1$
		and $R_{\textrm{\tiny th}}$ almost takes a constant value,
		which decreases with $d$.  
		The dependence $\rho$ becomes significant when $\rho\sim 1$, 
		where all the curves converge to the asymptotic value $1$, as expected. 
		Indeed, when $A$ is large w.r.t.  $z_h$,
		the RT hypersurface gets closer to the horizon
		and the $O(1)$ term in the $\varepsilon_{\textrm{\tiny AdS}} \to 0$ expansion of  $S_A$
		grows like $S_{A, \textrm{\tiny th}}$,
		as already discussed 
		in Sec.\,\ref{sec-intro} and in Sec.\,\ref{sec-Sch-AdS-sphere}.

		\section{BTZ black hole}
		\label{sec-BTZ-black-hole}

		Considering the BTZ black hole 
		and the bipartition of the circle on the boundary of its constant time slice
		given by an arc $A$,
		in Sec.\,\ref{subsec-BTZ-bh-geodesics} 
		we show that the construction of the geodesic bit threads fails. 
		However, 
		by using fake geodesic bit threads that are allowed to intersect, 
		the holographic thermal entropy of $A$ is recovered,
		as discussed in Sec.\,\ref{subsec-flows-btz-bh}.


		A constant time slice of the non-rotating BTZ black hole 
		is equipped with the following metric
		\cite{Banados:1992wn,Banados:1992gq}
		(see also e.g. \cite{Carlip:1994gc, Carlip:1995qv})
		\be 
		\label{btz-global-metric}
		ds^2 =   \frac{L_{\textrm{\tiny AdS}}^2} {  r^2 - r_h^2 } \,\rd r^2 + r^2 \,\rd \phi^2
		\ee 
		where $r \geqslant 0$ is the radial coordinate, 
		$\phi \in [-\pi , \pi ]$ is the angular coordinate with period of $2\pi$
		and the horizon corresponds to $r=r_h$.
		In these coordinates, the boundary of the non-rotating BTZ black hole corresponds to $r \to +\infty$,
		whose constant time slice is the circle parameterized by $\phi$.
		According to the AdS/CFT correspondence, 
		on this boundary, we can find a dual CFT$_2$ on a circle 
		and at finite inverse temperature $\beta =  2 \pi  L_{\textrm{\tiny AdS}}^2 / r_h$.
		%
		

		The gravitational background \eqref{btz-global-metric} 
		can be derived from the metric \eqref{btz-brane-metric}
		corresponding to a constant time slice of the BTZ black brane
		by performing first
		the quotient  $ x \sim x +  2  \pi k  L_{\textrm{\tiny AdS}} $
		with  $k \in \mathbb{Z}$ 
		and then the change of coordinates given by 
		\be
		\label{btz-brane-to-btz-global-coord-change}
		z=  L_{\textrm{\tiny AdS}}^2 / r
		\;\;\;\qquad\;\;\;\;
		x= L_{\textrm{\tiny AdS}}\,\phi \, .
		\ee
		Since this procedure involves only a coordinate transformation at least locally, 
		we can quickly establish a relation between the geodesics and the RT curves 
		in both geometries depicted in
	 	Fig.\,\ref{fig-global-btz-threads-before} and Fig.\,\ref{fig-global-to-planar-btz-before} 
		as well as  
		in Fig.\,\ref{fig-global-btz-threads-after} and Fig.\,\ref{fig-global-to-planar-btz-after}.
		To ensure a finite range for the radial coordinate 
		in Fig.\,\ref{fig-global-btz-threads-before} and Fig.\,\ref{fig-global-btz-threads-after}
		we have employed $\rho \equiv \arctan r$. 
		More specifically, 
		the mapping (\ref{btz-brane-to-btz-global-coord-change}) provides 
		a one-to-one correspondence between 
		a domain of finite width $2\pi L_{\textrm{\tiny AdS}}$
		 in the constant time slice of the BTZ black brane  
		and the whole constant time slice of the BTZ black hole.
		This domain is selected by  fixing
		a point on the boundary, a point on the horizon, and a curve connecting them
		on the constant time slice of the BTZ black hole
		(see the dashed grey curves in Fig. \ref{fig-global-btz-threads-before} and Fig.\,\ref{fig-global-btz-threads-after}),
		whose preimages partition the domain outside the horizon
		of the planar BTZ  black brane into equivalent domains of width $2  \pi L_{\textrm{\tiny AdS}} $.

		We remark that the BTZ black hole background \eqref{btz-global-metric} 
		can also be obtained from a certain domain of the constant time slice of AdS$_3$,
		whose metric is \eqref{H2-metric}, as discussed in Appendix\;\ref{app-BTZ-global-CTmap}.

		\subsection{Geodesics}
		\label{subsec-BTZ-bh-geodesics}

		%

		The geodesics in the BTZ black hole
		have been largely discussed in the AdS/CFT literature for various applications
		(see e.g. \cite{Cruz:1994ir, Headrick:2007km, Hubeny:2013gta, Arefeva:2017pho,Balasubramanian:2012tu}),
		including the ones in its time slice equipped with the metric (\ref{btz-global-metric}),
		that are employed throughout this section. 

		Following the analysis performed in Sec.\,\ref{subsec-BTZ-plane-GBT} on
		the time slice of the BTZ black brane,
		also in the BTZ black hole geometry (\ref{btz-global-metric})
		two classes of geodesics must be considered:
		the geodesics having both the endpoints on the boundary (type I)
		and the geodesics with one endpoint on the boundary and the other one on the horizon (type II).
		The geodesics belonging to these classes
		can be obtained by applying the map (\ref{btz-brane-to-btz-global-coord-change})
		to the corresponding geodesics discussed in Sec.\,\ref{subsec-BTZ-plane-GBT}.
		From \eqref{geod-btz-gen-2bdy} and (\ref{btz-brane-to-btz-global-coord-change})
		with $z_h = L^2_{\textrm{\tiny AdS}} / r_h$, $c_0 = L_{\textrm{\tiny AdS}}\, \phi_0$ and $b_0 = L_{\textrm{\tiny AdS}} \, \theta_0$,
		we find that the geodesic having both the endpoints on the boundary,
		at angular coordinates $\phi_0 + \theta_0$ and $\phi_0 - \theta_0$ with $0 < \theta_0 < \pi / 2$,
		which is given by  
		\be
		\label{geo-btz-global-bh-2bdy}
		r_{\textrm{\tiny I}}(\phi) 
		\,=\,
		r_h \left[ \, 1 - \left( \frac{\cosh(r_h (\phi - \phi_0)/L_{\textrm{\tiny AdS}})}{\cosh( (r_h \theta_0 )/L_{\textrm{\tiny AdS}} )} \right)^2 \, \right]^{-1/2}.
		\ee
		Similarly, from \eqref{geod-btz-gen-bdy-hor} and (\ref{btz-brane-to-btz-global-coord-change})
		we get a geodesic with one endpoint on the boundary and the other one on the horizon, which reads
		\be
		\label{geo-btz-global-bh-hor}
		r_{\textrm{\tiny II}}(\phi) 
		\,=\,
		r_h \left[ \, 1 - \left( \frac{\sinh (r_h (\phi - \phi_0)/L_{\textrm{\tiny AdS}}) }{\sinh((r_h \theta_0 )/L_{\textrm{\tiny AdS}} )} \right)^2 \, \right]^{-1/2}
		\ee
		where $\phi_0$ is the angular coordinate of the endpoint on the horizon, 
		while the angular coordinate of the endpoint on the boundary is either $\phi_0 + \theta_0$ or $\phi_0 - \theta_0$.

						\begin{figure}[t!]
			\vspace{-.2cm}
			\begin{minipage}{0.5\textwidth}
				\centering
				\includegraphics[width=.9\textwidth]{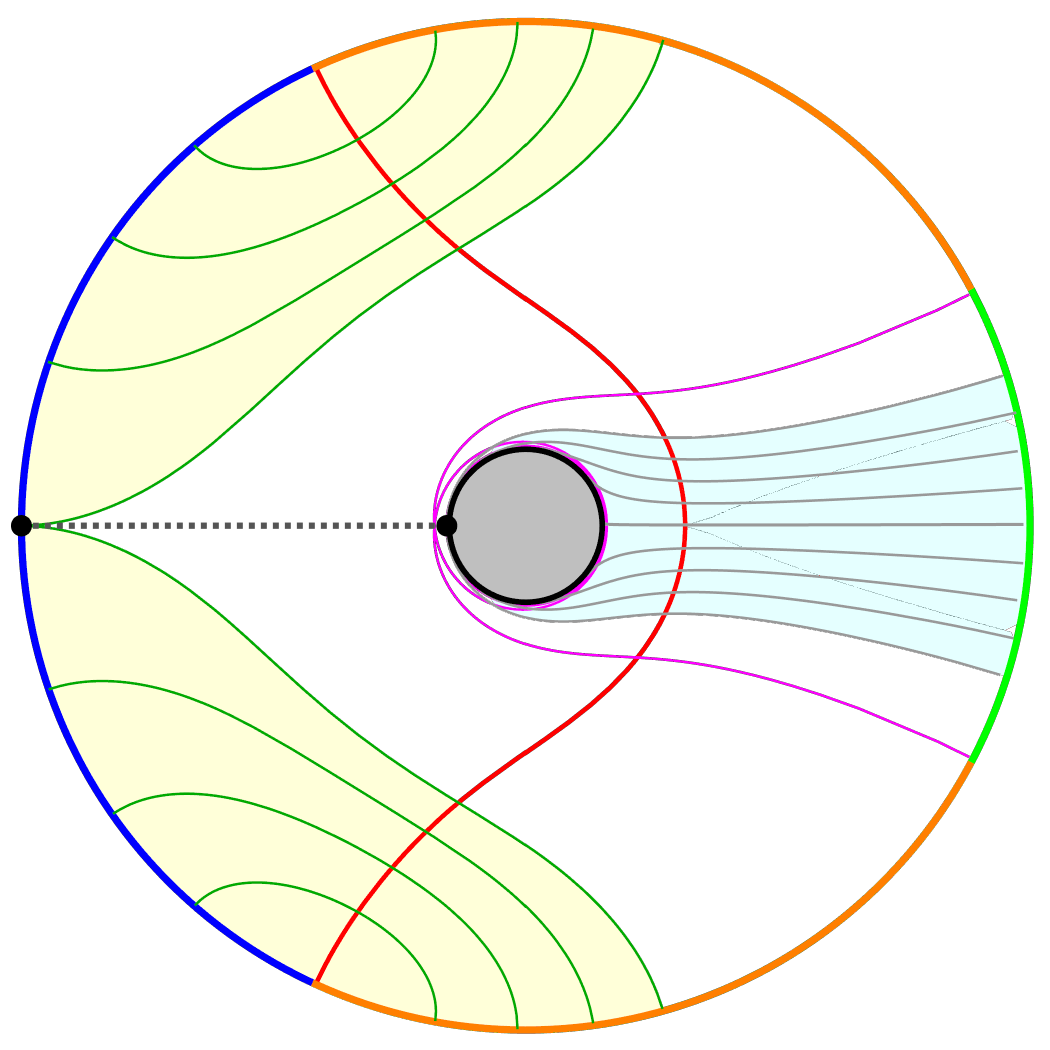}
			\end{minipage}\hfill
			\begin{minipage}{0.5\textwidth}
				\centering
				\includegraphics[width=.9\linewidth]{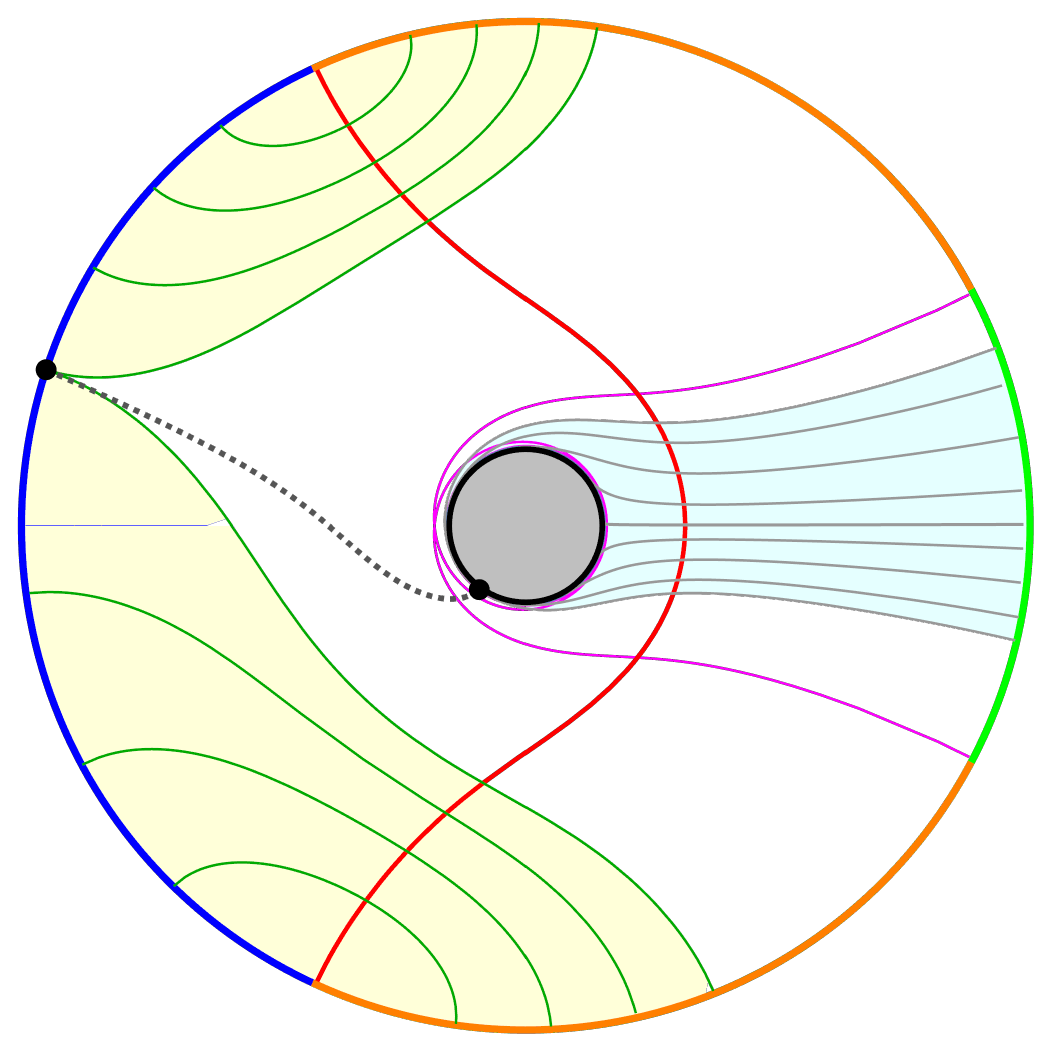}
			\end{minipage}
			\vspace{.3cm}
			\caption{\small 
				BTZ black hole and  $\phi_{b} < \phi_{b}^*$\,:
				Maximal sets of non-intersecting geodesics (green and grey curves) 
				that intersect orthogonally the RT curve $\gamma_A =\gamma_{A,1}$ (red curve).
				The two panels correspond to
				two different choices of $P_{\textrm{\tiny bdy}}$ and $P_{\textrm{\tiny hor}}$
				(see the black dot on the boundary and the horizon respectively).
			}
			\label{fig-global-btz-threads-before}
		\end{figure}
		
		\begin{figure}[t!]
			\vspace{.8cm}
			\hspace{-10mm}
			\begin{minipage}{0.55\textwidth}
				\centering
				\includegraphics[width=.95\textwidth]{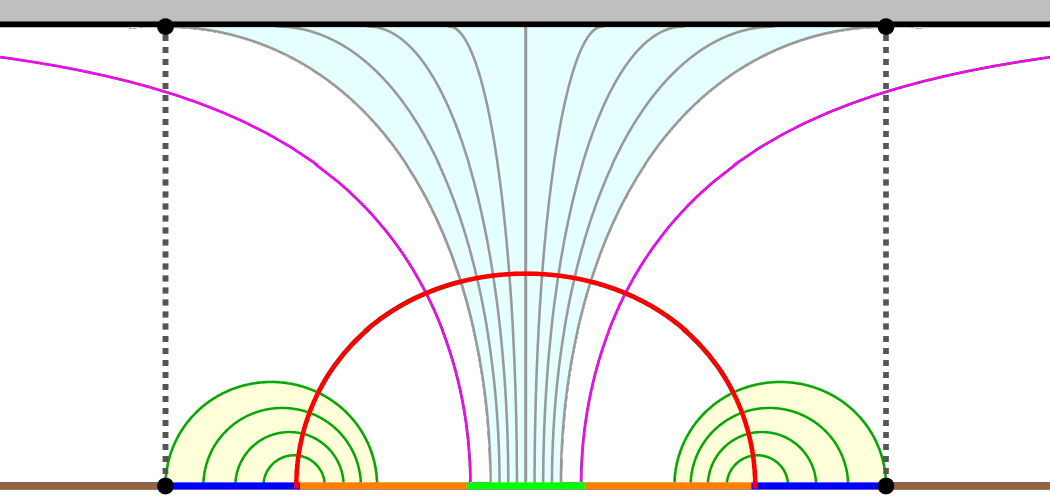}
			\end{minipage}
			\hfill
			\begin{minipage}{0.55\textwidth}
				\centering
				\includegraphics[width=0.95\linewidth]{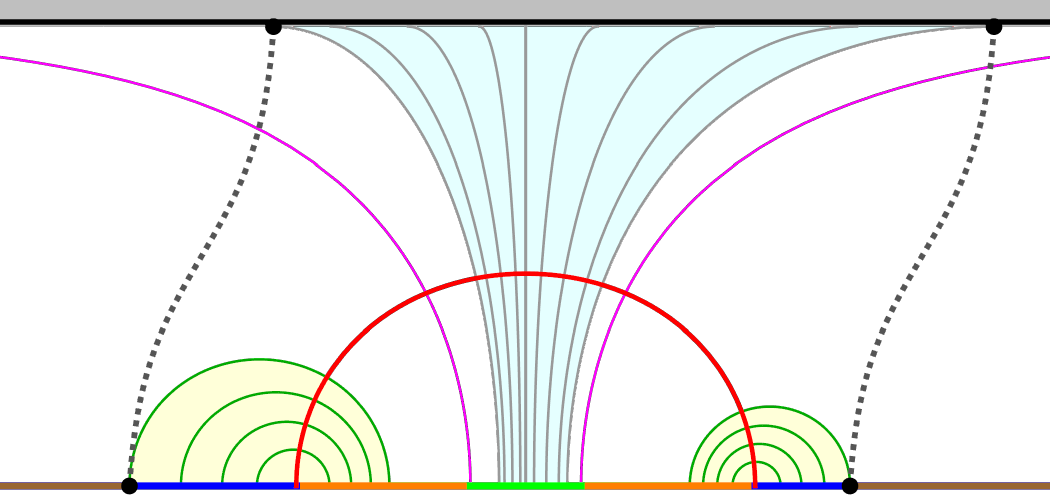}
			\end{minipage}
			\vspace{.3cm}
			\caption{\small
				Images through (\ref{btz-brane-to-btz-global-coord-change})
				of the  geodesics shown in Fig.\,\ref{fig-global-btz-threads-before} 
				in a portion of the BTZ black brane determined by the dashed grey curves.
			}
			\label{fig-global-to-planar-btz-before}
		\end{figure}

							The expressions (\ref{geo-btz-global-bh-2bdy}) and (\ref{geo-btz-global-bh-hor}) 
		also provide geodesics whose length is non-minimal and have non-vanishing winding numbers, 
		as discussed in  Appendix\;\ref{app-LongGeodesics-GlobalBTZ}
		(see Fig.\,\ref{fig-global-btz-long-geod-bdy-to-bdy} and Fig.\,\ref{fig-global-btz-long-geod-bdy-to-hor1}).
		For these geodesics having nontrivial winding numbers,
		$\phi_0$ and $\theta_0$ are promoted to parameters varying on the whole real line. 
			Then, taking $\phi_0 \to \pm \infty $ and $\theta_0 \to + \infty $ in \eqref{geo-btz-global-bh-2bdy} and \eqref{geo-btz-global-bh-hor},
			we obtain the geodesic given by
			\be 
			r_{\textrm{\tiny I/II}}^\pm (\phi)  
			\,=\,
			\frac{r_h}{ \sqrt{1 - e^{\pm 2 r_h ( \phi_s- \phi)}} }
			\label{geo-btz-global-bh-bdy-infty} 
			\ee 
			that start from the boundary at $\phi_s = \phi_0 \mp \theta_0 $ 
			and wind infinitely many times around the horizon,
			which is  reached as $\phi \to \pm \infty$.
			The geodesics \eqref{geo-btz-global-bh-bdy-infty} in the BTZ black hole
			can be found also from the geodesics (\ref{geod-btz-gen-bdy-inf}) in the BTZ black brane,
			by applying \eqref{btz-brane-to-btz-global-coord-change},
			that implies $ s = L_{\textrm{\tiny AdS}}\,\phi_s $ and $ z_h =  L_{\textrm{\tiny AdS}}^2 / r_h$.


							\begin{figure}[t!]
				\vspace{-.2cm}
				\begin{minipage}{0.5\textwidth}
					\centering
					\includegraphics[width=.9\textwidth]{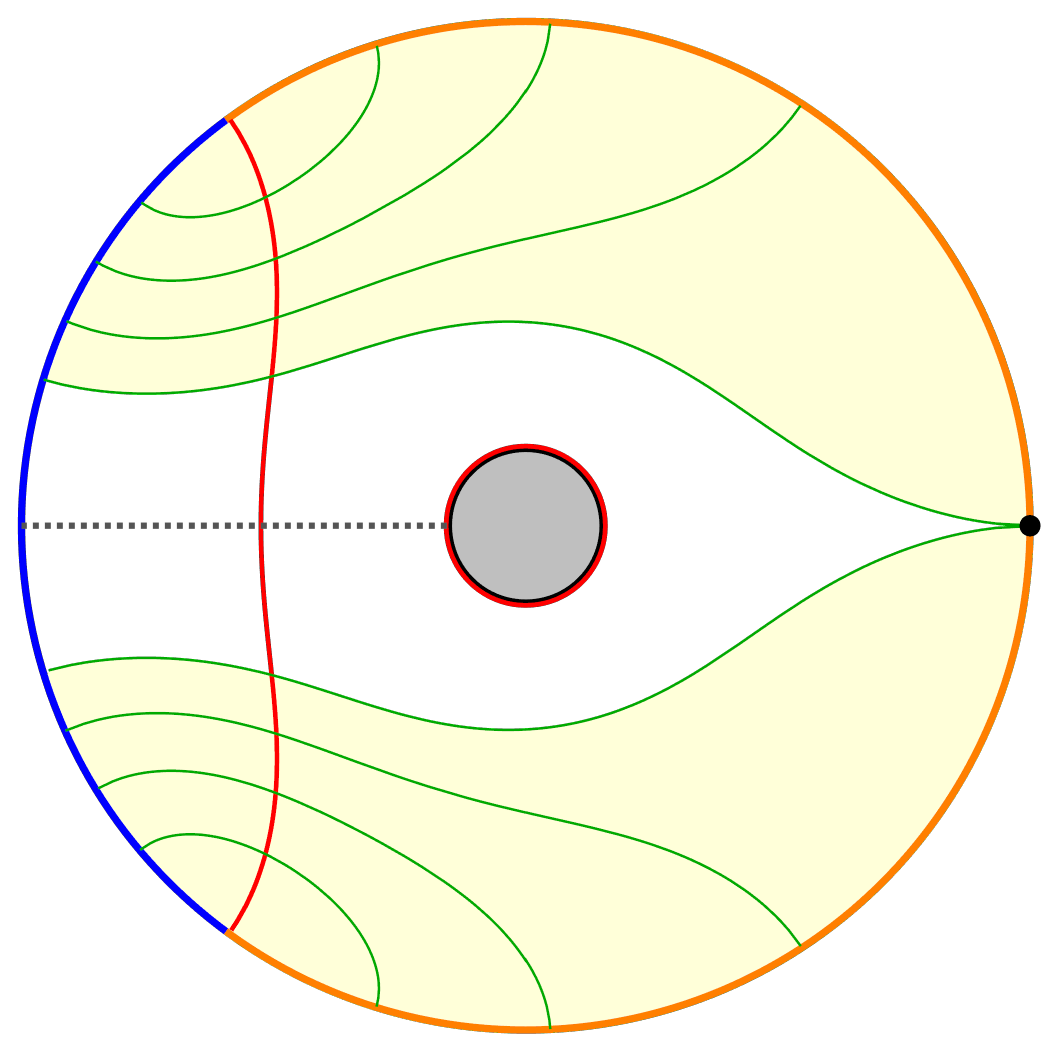}
				\end{minipage}\hfill
				\begin{minipage}{0.5\textwidth}
					\centering
					\includegraphics[width=.9\linewidth]{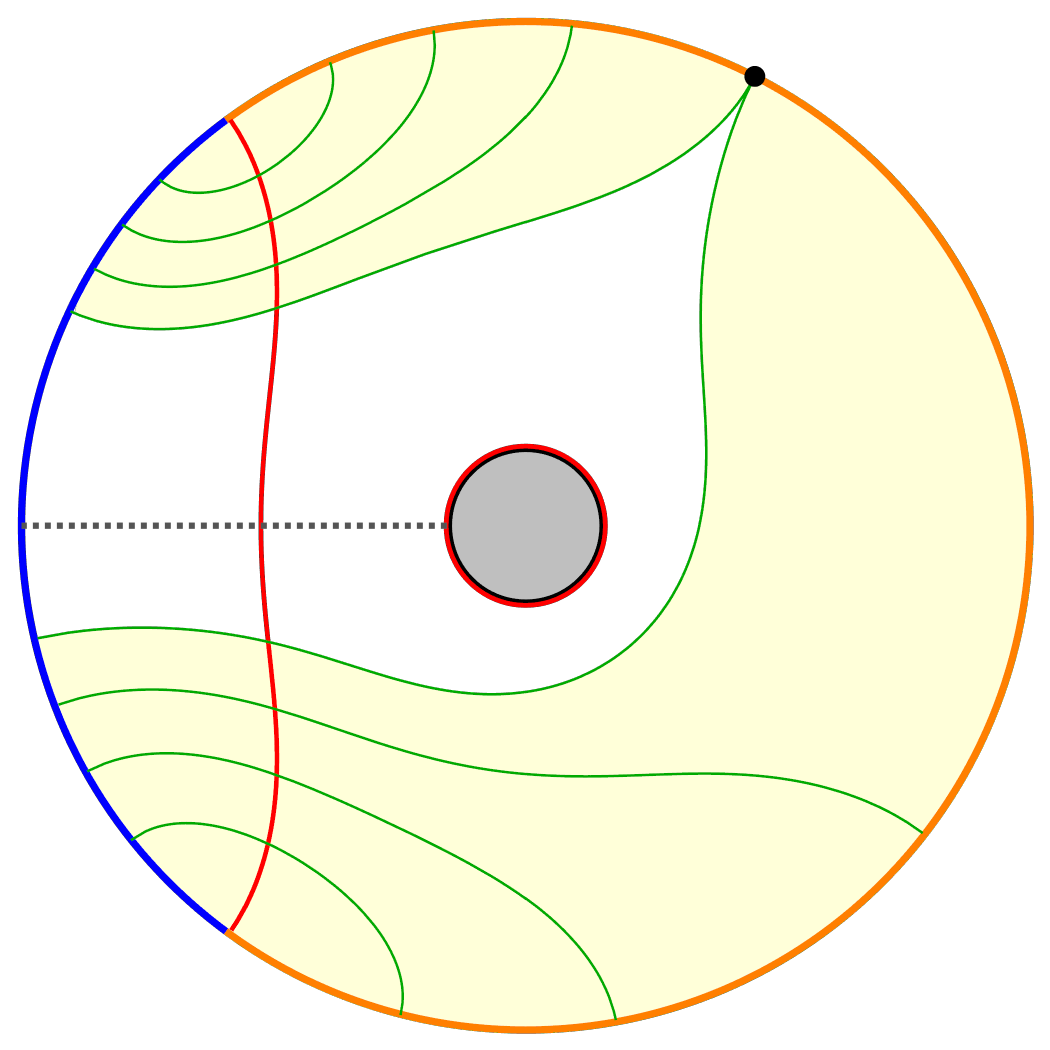}
				\end{minipage}
				\vspace{.3cm}
				\caption{\small
					BTZ black hole and  $\phi_{b} > \phi_{b}^*$\,:
					The RT curve $\gamma_A$ is the union of $\gamma_{A,2}$ and of the horizon (red curves).
					Maximal sets of non-intersecting geodesics with both the endpoints on the boundary 
					that intersect orthogonally $\gamma_{A,2}$ (green curves).
					The two panels correspond to
					two different choices of $P_{\textrm{\tiny bdy}}$ (black dot on the boundary).
					}
					\label{fig-global-btz-threads-after}
			\end{figure}
			\begin{figure}[t!]
				\vspace{.8cm}
				\hspace{-10mm}
				\begin{minipage}{0.55\textwidth}
					\centering
					\includegraphics[width=.95\textwidth]{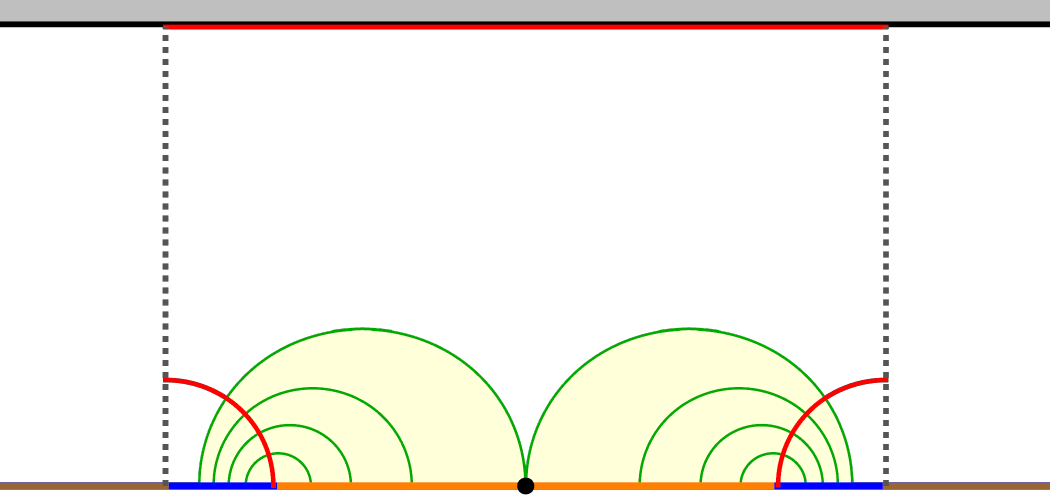}
				\end{minipage}
				\hfill
				\begin{minipage}{0.55\textwidth}
					\centering
					\includegraphics[width=.95\linewidth]{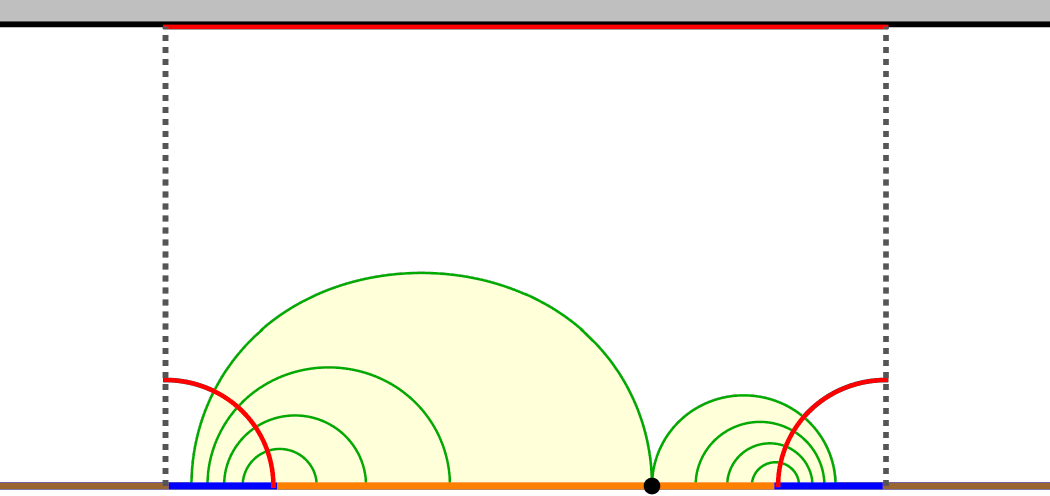}
				\end{minipage}
				\vspace{.3cm}
				\caption{\small
					Images through (\ref{btz-brane-to-btz-global-coord-change})
					of the geodesics shown in Fig.\,\ref{fig-global-btz-threads-after} 
					in a portion of the BTZ black brane determined by the dotted grey curves.
				}
				\label{fig-global-to-planar-btz-after}
			\end{figure}

			In the time slice of the BTZ black hole (see (\ref{btz-global-metric})),
			let us consider the bipartition of the boundary given by the circular arc
			$A = (-\phi_b , \phi_b)$ with $0< \phi_b < \pi $ 
			(see the union of the orange and of the green circular arcs in Fig.\,\ref{fig-global-btz-threads-before}
			and the orange circular arcs in Fig.\,\ref{fig-global-btz-threads-after})
			and its complement $B$
			(see the blue circular arcs in Fig.\,\ref{fig-global-btz-threads-before} and Fig.\,\ref{fig-global-btz-threads-after}).
			The homology constraint plays a crucial role 
			to determine the configuration of geodesics providing the holographic entanglement entropy 
			\cite{Headrick:2007km,Hubeny:2013gta}.
			Indeed, we have two geodesics $\gamma_{A,1}$ and $\gamma_{A,2}$ of the form (\ref{geo-btz-global-bh-2bdy}),
			anchored to the endpoints of $A$
			and with vanishing winding number 
			which are homologous to $A$ and $B$ respectively.
			The homology constraint requires that only the curves homologous to $A$
			must be considered in the extremization procedure 
			providing the holographic entanglement entropy of $A$;
			hence also, the horizon must be taken into account together with $\gamma_{A,2}$.
			The minimal length prescription and  the homology constraint 
			define the configuration $\gamma_A$ of curves 
			providing the holographic entanglement entropy.
			For our choice of $A$, we have that $\gamma_A$ 
			is given by $\gamma_A=\gamma_{A,1}$ when $\phi_b < \phi_b^*$
			(see the red curves in Fig.\,\ref{fig-global-btz-threads-before},
			described by (\ref{geo-btz-global-bh-2bdy}) with $\phi_0 = 0$ and $\theta_0 = \phi_b$)
			and by the union of $\gamma_{A,2}$ and of the horizon when $\phi_b > \phi_b^*$
			(see the set of red curves in Fig.\,\ref{fig-global-btz-threads-after}, 
			which includes the horizon and
			where $\gamma_{A,2}$ is described by (\ref{geo-btz-global-bh-2bdy}) with $\phi_0 = \pi$ and $\theta_0 = \pi- \phi_b$),
			where the critical angle $\phi_b^* $ is
			\be 
			\label{crit-angle-RT-global-btz}
			\phi_b^* 
			\equiv 
			\frac{L_{\textrm{\tiny AdS}}}{r_h} \;
			\textrm{arcoth} 
			\big[ \, 2 \coth ( \pi r_h / L_{\textrm{\tiny AdS} } ) - 1 \, \big] \,.
			\ee 
			Hence, the holographic entanglement entropy of  the circular arc $A$ reads 
			\be 
			\label{hee-global-BTZ}
			S_A = 
			\left\{
			\begin{array}{ll}
				\displaystyle
				\frac{L_{\textrm{\tiny AdS}}}{2 G_{\textrm{\tiny N}}} \, 
				\log \! \left[
				\frac{2 r_\infty }{r_h} \, \sinh\! \left( \frac{ r_h \phi_b}{L_{\textrm{\tiny AdS}}}
				\right) \right]
				&
				\phi_b \in (0, \phi_b^* )
				\\ 
				\rule{0pt}{.95cm}
				\displaystyle
				\frac{ \pi  r_h }{2 G_{\textrm{\tiny N}}} 
				+ 
				\frac{L_{\textrm{\tiny AdS}}}{2 G_{\textrm{\tiny N}}}  \, \log \! \left[ \frac{2 r_\infty }{r_h} \, \sinh \! \left( \frac{ r_h ( \pi - \phi_b}{L_{\textrm{\tiny AdS}}} \right) \right]
				\hspace{1cm}
				&
				\phi_b \in ( \phi_b^*,\pi )
			\end{array}
			\right.
			\ee 
			where $r_\infty  \gg   r_h$ is the UV cutoff.

			The existence of two competing configurations for the holographic entanglement entropy can be understood also
			by examining the fundamental domains $x \in [- L_{\textrm{\tiny AdS}} \pi , L_{\textrm{\tiny AdS}} \pi ]$
			considered e.g. in the left panels of Fig.\,\ref{fig-global-to-planar-btz-before} and Fig.\,\ref{fig-global-to-planar-btz-after} 
			(i.e. the regions delimited by the grey dotted vertical lines), 
			which are equipped with the BTZ black brane metric \eqref{btz-brane-metric}. 
			The homology constraint, jointly with the choice of the fundamental domain, prompts us to consider two possible candidates. 
			The first one is the geodesic connecting the two endpoints of the image of the arc $A$
			obtained as the image of $\gamma_{A,1}$ through (\ref{btz-brane-to-btz-global-coord-change}) 
			(see the red curves in Fig.\,\ref{fig-global-to-planar-btz-before}). 
			For the choice of the cut shown in Fig.\,\ref{fig-global-btz-threads-after} and in Fig.\,\ref{fig-global-to-planar-btz-after},
		  	the second one is formed by two half geodesics, each starting  from one endpoint of the image of the arc $A$ 
			and intersecting orthogonally the dotted vertical lines 
		    	at $x= \pm L_{\textrm{\tiny AdS}} \pi$, along with a horizontal segment wrapping the piece of horizon between
	    		 $x= - L_{\textrm{\tiny AdS}} \pi$ and $x=  L_{\textrm{\tiny AdS}} \pi$ 
			 (see red curves in Fig.\,\ref{fig-global-to-planar-btz-after}).
			The latter configuration is the image of $\gamma_{A,2}$ and of the circle surrounding the horizon in Fig.\,\ref{fig-global-btz-threads-after} 
		   	 through  (\ref{btz-brane-to-btz-global-coord-change}).
			 The RT curve is determined by the first configuration when the width of the image of the arc $A$  
			 is less than $2L_{\textrm{\tiny AdS}} \phi_b^*$,  
			 where $\phi_b^*$ is defined in \eqref{crit-angle-RT-global-btz} 
			 with $r_h$ replaced by  $L_{\textrm{\tiny AdS}}^2 / z_h$; 
			 otherwise, it is determined by the second configuration
			 (see Fig.\,\ref{fig-global-to-planar-btz-before} and Fig.\,\ref{fig-global-to-planar-btz-after} respectively).
			
			%

			
			Denoting by $r_m(\phi_m)$ the geodesic of the form (\ref{geo-btz-global-bh-2bdy})
			corresponding to either $\gamma_{A,1}$ or $\gamma_{A,2}$,
			the integral lines of all the geodesics that intersect such geodesic orthogonally
			can be obtained from the integral lines of the geodesic bit threads constructed in Sec.\,\ref{subsec-BTZ-plane-GBT}
			by employing (\ref{btz-brane-to-btz-global-coord-change}).
			They have one endpoint in $A$ and the other endpoint either in $B$ or on the horizon;
			hence, they are described 
			by either (\ref{geo-btz-global-bh-2bdy}) or (\ref{geo-btz-global-bh-hor}) respectively. 
			For these geodesics
			the parameters $\phi_0 = c_0/ L_{\textrm{\tiny AdS}} $ and $ \theta_0 = b_0 / L_{\textrm{\tiny AdS}} $
			occurring in (\ref{geo-btz-global-bh-2bdy}) and (\ref{geo-btz-global-bh-hor}) 
			are obtained from (\ref{c0-btz-brane}) and (\ref{b0-btz-brane})
			with $x_m = L_{\textrm{\tiny AdS}}\, \phi_m$, $z_h = L_{\textrm{\tiny AdS}}^2/r_h$,
			$b = L_{\textrm{\tiny AdS}} \, \phi_b$
			and $z_m(x_m) = L_{\textrm{\tiny AdS}}^2 / r_m(\phi_m)$.

			We remark that, in this BTZ black hole setup, 
			the set made by all the geodesics intersecting orthogonally $\gamma_A$
			does not provide a proper configuration of geodesic bit threads.
			Indeed,  it contains intersecting curves,
			while the integral curves of proper bit thread configuration must be non-intersecting. 

			Consider the maximal set of non-intersecting geodesics that intersect orthogonally $\gamma_A$.
			When $\phi_{b} < \phi_{b}^*$
			(see Fig.\,\ref{fig-global-btz-threads-before} and Fig.\,\ref{fig-global-to-planar-btz-before}),
			we have $\gamma_A = \gamma_{A,1}$ and 
			the maximal sets of non-intersecting geodesics orthogonal to $\gamma_{A}$
			contain
			both geodesics of type I
			(see the green curves in Fig.\,\ref{fig-global-btz-threads-before}, that foliate the yellow region)
			and geodesics of type II
			(see the grey curves in Fig.\,\ref{fig-global-btz-threads-before}, that foliate the light blue region). 
			The geodesics of type I can be constructed
			by first selecting a point $P_{\textrm{\tiny bdy}}$ 
			(see the black dot on the boundary in Fig.\,\ref{fig-global-btz-threads-before})
			and then drawing  all the geodesics orthogonal to $\gamma_{A}$ 
			starting from both endpoints of $\gamma_{A}$ 
			until two geodesics originating from the two different endpoints of $\gamma_A$
			share an endpoint in $P_{\textrm{\tiny bdy}}$ 
		        (see the two green geodesics defining the boundary of the yellow regions in Fig.\,\ref{fig-global-btz-threads-before}).
%
			%
			The geodesics of type II  are instead  obtained by first choosing a point $P_{\textrm{\tiny hor}}$ on the horizon
			(see the black dot on the horizon in Fig.\,\ref{fig-global-btz-threads-before})
			and then determining
			 the two geodesics of the form (\ref{geo-btz-global-bh-hor}) 
			 that intersect $\gamma_A$ orthogonally
			and connect $P_{\textrm{\tiny hor}}$ to $A$,
			within the green arc in Fig.\,\ref{fig-global-btz-threads-before}.
			These two geodesics delimit the light blue region in Fig.\,\ref{fig-global-btz-threads-before}.
			These two geodesics, 
			completed through the analog of the auxiliary geodesics discussed in Sec.\,\ref{sec-BTZ}
			for the planar BTZ brane, 
			could have nontrivial winding around the horizon.
			This is the case, e.g. for the  limiting geodesic 
			defining the highest boundary of the light blue region  
			in the right panel of  Fig.\,\ref{fig-global-btz-threads-before}.	
			The geodesics with nontrivial winding around the horizon are discussed in Appendix\;\ref{app-LongGeodesics-GlobalBTZ}.

			In the two panels of Fig.\,\ref{fig-global-btz-threads-before}
			we show two different maximal sets of non-intersecting geodesics
			corresponding to two different choices of the pair made by $P_{\textrm{\tiny bdy}}$ and $P_{\textrm{\tiny hor}}\,$.
			In the left (right) panel of Fig.\,\ref{fig-global-to-planar-btz-before} 
			we show the image of the left (right) panel of Fig.\,\ref{fig-global-btz-threads-before}
			through (\ref{btz-brane-to-btz-global-coord-change}),
			where the choice of the portion of the constant time slice of BTZ black brane 
			is determined by the dashed grey curves.
			In Fig.\,\ref{fig-global-btz-threads-before} we have chosen a dashed grey curve
			that connects $P_{\textrm{\tiny bdy}}$ to $P_{\textrm{\tiny hor}}\,$
			and does not intersect the green geodesics.
			It is always possible to make such a choice. 
			In each panel of Fig.\,\ref{fig-global-to-planar-btz-before},
			the union of the green and grey geodesics gives a maximal set of non-intersecting geodesics orthogonal to $\gamma_{A}$
			spanning the union of the yellow  and the light blue regions, 
			which is properly contained into the spatial domain outside the horizon.
			We remark that,
			since a maximal set of non-intersecting geodesics orthogonal to $\gamma_{A}$ constructed as explained above
			depends on the choice of $P_{\textrm{\tiny bdy}} \in B$ and $P_{\textrm{\tiny hor}}\,$,
			infinitely many different maximal sets can be constructed. 
			
			In Fig.\,\ref{fig-global-btz-threads-before} 
			it is straightforward to identify a portion $\gamma_{A,0} \subsetneq \gamma_A$ made by two disjoint arcs
			(see the parts of $\gamma_A$ in the white regions)
			that are not crossed by any curve belonging to the
			maximal set of non-intersecting geodesics orthogonal to $\gamma_{A}$.
			This tells us that the geodesic bit threads cannot be constructed for this setup;
			indeed, the vector field of proper bit threads does not vanish on $\gamma_A$.

			We highlight that two particular geodesics orthogonal to $\gamma_{A}$ occur
			that intersect $\gamma_{A}$ in $\gamma_{A,0}$
			and are characterized by the fact that they arrive at the horizon
			after wrapping around the horizon infinitely many times.
			These two critical geodesics do not depend on the choice of 
			$P_{\textrm{\tiny bdy}} \in B$ and $P_{\textrm{\tiny hor}}\,$
			and correspond to the magenta curves in Fig.\,\ref{fig-global-btz-threads-before}.
			They are described by \eqref{geo-btz-global-bh-bdy-infty} 
			with  $\phi_s = \pm b_\beta  / L_{\textrm{\tiny AdS}} $ 
			where $b_\beta$ is obtained from \eqref{bbeta} with 
			$b$ and $z_h$ replaced by $L_{\textrm{\tiny AdS}}\,\phi_b$ and $L_{\textrm{\tiny AdS}}^2 / r_h$ respectively. 
			These two critical geodesics 
			are mapped through \eqref{btz-brane-to-btz-global-coord-change} 
			into the geodesic hitting the horizon at infinity in the planar BTZ geometry
			(see the magenta curves in Fig.\,\ref{fig-global-to-planar-btz-before}).
			Their endpoints on the boundary define the green arc contained in $A$,
			which is the image through \eqref{btz-brane-to-btz-global-coord-change} of
			the green interval in Fig.\,\ref{fig-global-to-planar-btz-before},
			denoted by $A_\beta$ in Sec.\,\ref{sec-BTZ}.

			Any maximal set of geodesics constructed as explained above for $\phi_{b} < \phi_{b}^*$
			does not provide the integral lines of geodesic bit threads for $\gamma_A = \gamma_{A,1}$.
			Indeed, while any point in $B$ is connected to a point in $A$ through a geodesic of this maximal set, the opposite is not true. 
			Circular arcs in $A$ occur that are not connected to $B$ through the geodesics belonging to this maximal set
			and any geodesic having one endpoint in these arcs and intersecting $\gamma_{A,1}$ orthogonally
			necessarily intersects the geodesics of the maximal set. 
			We remind you that a necessary condition for a consistent bit thread configuration 
			is that the corresponding divergenceless vector field is nonvanishing on the RT hypersurface. 
			This straightforwardly tells us that the maximal set of geodesics in Fig.\,\ref{fig-global-btz-threads-before}
			cannot be the integral lines of geodesic bit threads.

			When $\phi_{b} > \phi_{b}^*$ (see Fig.\,\ref{fig-global-btz-threads-after} and Fig.\,\ref{fig-global-to-planar-btz-after}),
			$\gamma_A$ is the union of $\gamma_{A,2}$ and the horizon
			(see the red curves in Fig.\,\ref{fig-global-btz-threads-after}).
			This implies that the maximal sets of non-intersecting geodesics that intersect orthogonally $\gamma_A$
			include only geodesics having both their endpoints on the boundary
			(indeed, we cannot include geodesics that intersect $\gamma_A$ twice). 
			These geodesics can be obtained by considering their intersection point with $\gamma_{A,2}$,
			starting from both the endpoints of $A$ until two of them share a common endpoint $P_{\textrm{\tiny bdy}} \in A$
			(in Fig.\,\ref{fig-global-btz-threads-after}, see the black dot on the boundary).
			In the two panels of Fig.\,\ref{fig-global-btz-threads-after},
			we show two different maximal sets 
			which correspond to two different choices of $P_{\textrm{\tiny bdy}}\,$.
				Notice that $P_{\textrm{\tiny bdy}} \in B$ when $\phi_{b} < \phi_{b}^*$,
				while $P_{\textrm{\tiny bdy}} \in A$ when $\phi_{b} > \phi_{b}^*$.
			We remark that,
			for $\phi_{b} > \phi_{b}^*$,
			the point $P_{\textrm{\tiny hor}}$ 
			and the geodesics corresponding to the magenta curves in Fig.\,\ref{fig-global-btz-threads-before} 
			do not occur
			because the horizon is part of the RT curve $\gamma_A$.
			In fact, any geodesic running through 			
			the white region in Fig.\,\ref{fig-global-btz-threads-after} and  reaching the horizon 
			would intersect both $\gamma_{A,2}$ and the horizon,
			i.e. it would intersect $\gamma_A$ twice,
			which is forbidden for proper holographic bit threads.

			The left (right) panel of Fig.\,\ref{fig-global-to-planar-btz-after} displays
			the image of the left (right) panel of Fig.\,\ref{fig-global-btz-threads-after} 
			through (\ref{btz-brane-to-btz-global-coord-change}),
			where the choice of the portion of a constant time slice of BTZ black brane is determined by the dashed grey curves.
			In each panel of Fig.\,\ref{fig-global-btz-threads-after},
			the green geodesics provide a maximal set of non-intersecting geodesics orthogonal to $\gamma_{A,2}$
			spanning the yellow region, which is properly contained into the spatial domain outside the horizon
			and depends on the choice of $P_{\textrm{\tiny bdy}} \in A$;
			hence, infinitely many maximal sets of this kind can be found.
			Any maximal set of geodesics constructed in this way does not provide a configuration of geodesic bit threads. 
			Indeed, a circular arc properly contained in $B$ occurs that is not connected to $A$ through the geodesics belonging to this maximal set
			(in Fig.\,\ref{fig-global-btz-threads-after}, see the portion of the blue arc belonging to the boundary of the white region 
			of the domain outside the horizon).
			Consequently, a finite portion of $\gamma_{A,2}$ can be identified 
			along which the divergenceless vector field defined by this maximal set 
			vanishes
			(in Fig.\,\ref{fig-global-btz-threads-after}, see the portion of $\gamma_{A,2}$ intersecting the white region 
			of the domain outside the horizon),
			and this is not allowed for well defined bit thread configurations.

			%

		\subsection{Holographic thermal entropy}
		\label{subsec-flows-btz-bh}
In Sec.\,\ref{subsec-BTZ-bh-geodesics} 
			we have investigated the geodesics that intersect $\gamma_A$ orthogonally
			corresponding to a circular arc $A$ with angular width $2\phi_b$
			(see Fig.\,\ref{fig-global-btz-threads-before} and Fig.\,\ref{fig-global-btz-threads-after}),
			finding that they can also be obtained from the trajectories of the geodesic bit threads 
			of an interval of length $2b$ in the BTZ black brane (discussed in Sec.\,\ref{subsec-BTZ-plane-GBT})
			through \eqref{btz-brane-to-btz-global-coord-change}.
			In the following, this relation between these two BTZ geometries 
			is employed for the modulus  \eqref{mod-gbt-btz} 
			and the corresponding holographic contour functions \eqref{contour-bdy-btz} and \eqref{contour-horizon-btz}
			to discuss the holographic entanglement entropy $S_A$
			and the holographic thermal entropy $S_{A,\textrm{\tiny th}}$ 
			of $A$ in the BTZ black hole background (\ref{btz-global-metric}).

			When $\phi_{b} < \phi_{b}^*$, where $\phi_{b}^*$ is given in \eqref{crit-angle-RT-global-btz}, 
			the RT curve $\gamma_A$ is the red curve in Fig.\,\ref{fig-global-btz-threads-before},
			that has been denoted by $\gamma_{A,1}$ in Sec.\,\ref{subsec-BTZ-bh-geodesics}.
			Given the holographic contour function  $\mathcal{C}_A^+ (x_+)$
			for the interval $A$  on the boundary of the planar  BTZ black brane 
			(see \eqref{contour-bdy-btz} and the left panel of Fig.\,\ref{fig-global-to-planar-btz-before}), 
			we can introduce a holographic contour function $\widetilde{\mathcal{C}}_A^+ (\phi   )$  
			for the arc $A$ in Fig.\,\ref{fig-global-btz-threads-before}
			as the image of $\mathcal{C}_A^+ (x_+)$ through the coordinate transformation \eqref{btz-brane-to-btz-global-coord-change}. 			
			We obtain
			\be
			\label{CApBTZ}
			\widetilde{\mathcal{C}}_A^+ (\phi   ) 
			\equiv 
			L_{\textrm{\tiny AdS}} \, \mathcal{C}_A^+ (L_{\textrm{\tiny AdS}} \phi   )
			=
			\frac{c_\text{\tiny BH} \, r_h}{6 L_{\text{\tiny AdS}}} \;
			\frac{ \sinh ( r_h\phi_b / L_{\text{\tiny AdS}})
			}{ 
			\cosh ( r_h\phi_b / L_{\text{\tiny AdS}} )
			-
			\cosh
  			 ( r_h\phi  / L_{\text{\tiny AdS}} )
			 }			
			 \ee
			where $\beta =  2 \pi  L_{\textrm{\tiny AdS}}^2 / r_h$ and $b=  L_{\textrm{\tiny AdS}} \, \phi_b$ have been used.  
			Under \eqref{btz-brane-to-btz-global-coord-change}, the holographic cutoff  $\varepsilon_{\textrm{\tiny BTZ}}$ 
			is mapped to $r_\infty = L_{\textrm{\tiny AdS}}^2 / \varepsilon_{\textrm{\tiny BTZ}}$, 
			while the boundary cutoff  $\varepsilon_{\textrm{\tiny bdy}}^{\textrm{\tiny $A$}}$ for the case of the planar BTZ black brane
			 is replaced by $\varepsilon_{\textrm{\tiny bdy}}^{\textrm{\tiny $A$}}  / L_{\textrm{\tiny AdS}} $.
			Then, as $r_\infty  \to +\infty$ 
			the integral of $\widetilde{\mathcal{C}}_A^+ (\phi   )$ over the regularized arc 
			$A_\varepsilon \equiv \big[ \!- \phi_b + \varepsilon_{\textrm{\tiny bdy}}^{\textrm{\tiny $A$}} / L_{\textrm{\tiny AdS}} \,  , \phi_b - \varepsilon_{\textrm{\tiny bdy}}^{\textrm{\tiny $A$}}  / L_{\textrm{\tiny AdS}}\big] $ 
			reproduces the expected result (\ref{hee-global-BTZ})  for the entanglement entropy of $A$ when $\phi_{b} < \phi_{b}^*$. 
			Indeed, by employing the analog of  (\ref{pBTZ1-eps-AB})  and (\ref{hee-thermal-from-C}), it is straightforward to find
			\be 
			\label{SA-globalBTZ}
			S_A = 
			\int_{- \phi_b + \varepsilon_{\textrm{\tiny bdy}}^{\textrm{\tiny $A$}}  / L_{\textrm{\tiny AdS}}  } ^{\phi_b -  \varepsilon_{\textrm{\tiny bdy}}^{\textrm{\tiny $A$}}  / L_{\textrm{\tiny AdS}}   }
			\! \widetilde{\mathcal{C}}_A^+ (\phi   )\,   \rd \phi
			\,=\,
			\frac{c_{\textrm{\tiny BH}}}{3} \,
			\log \! \left[
			\frac{2 r_\infty }{r_h} \, \sinh\! \left( \frac{ r_h \phi_b}{L_{\textrm{\tiny AdS}}}
			\right) \right]
			+o(1) \, .
			\ee 
			
			By exploiting again  the coordinate transformation \eqref{btz-brane-to-btz-global-coord-change},
			we can define the following holographic contour  functions 
			\be
			\label{ChorCIIBTZ}
			\widehat{\mathcal{C}}_h(\phi   ) \equiv L_{\textrm{\tiny AdS}} \; \mathcal{C}_h (L_{\textrm{\tiny AdS}} \phi   ) 
			\;\;\;\;\qquad\;\;\;\;
			\widehat{\mathcal{C}}_A^- (\phi   ) \equiv L_{\textrm{\tiny AdS}} \; \mathcal{C}_A^- (L_{\textrm{\tiny AdS}} \phi   )
			\ee
			 on the horizon and on the boundary of the constant time slice of the BTZ black hole respectively,
			as the images of  \eqref{contour-horizon-btz}  and  \eqref{contour-bdy-btz}  respectively
			under the transformation \eqref{btz-brane-to-btz-global-coord-change}.
			However, the natural range for the coordinate $\phi$, spanning either the horizon in $\widehat{\mathcal{C}}_h(\phi)$ or the boundary in $\widehat{\mathcal{C}}_A^- (\phi)$, is not $[-\pi,\pi]$ but rather the entire real line. This is because we have not implemented the quotient $x \sim x + 2 \pi k L{\textrm{\tiny AdS}}$ with $k \in \mathbb{Z}$. 
			This issue can be taken into account by performing the quotient and replacing the two holographic contour functions
			in (\ref{ChorCIIBTZ}) respectively by 
\begin{equation}
\label{ChorCIIBTZperiodic}
\widetilde{\mathcal{C}}_h(\phi) = \sum_{n\in\mathbb{Z}}\widehat{\mathcal{C}}_h(\phi +2\pi n) 
\;\;\;\qquad\;\;\;
\widetilde{\mathcal{C}}_A^- (\phi) = \sum_{n\in\mathbb{Z}} \widehat{\mathcal{C}}_A^- (\phi +2\pi n) \, .
\end{equation}
From a geometrical point of view, 
we can interpret each term in the sum as the contribution to the density of flux due to the geodesics wrapping the horizon with winding number $n$ 
(see Appendix\;\ref{app-LongGeodesics-GlobalBTZ} for a detailed discussion on geodesics wrapping multiple times around the horizon). 
Then,  integrating the 
contour $\widetilde{\mathcal{C}}_A^+ (\phi   )$ over $[-b_\beta/L_{\text{\tiny AdS}},b_\beta/L_{\text{\tiny AdS}}]$ (see the green arc in Fig.\,\ref{fig-global-btz-threads-before}), 
$\widetilde{\mathcal{C}}_h(\phi)$ over the horizon
and $\widetilde{\mathcal{C}}_A^- (\phi)$ over the boundary 
yields the  holographic thermal entropy of the arc $A$, i.e. 
\bea
			\label{Sth-BHglobal-p2}
			S_{A,\textrm{\tiny th}}
			&=&
			\int_{-b_\beta/ L_{\textrm{\tiny AdS}} }^{b_\beta / L_{\textrm{\tiny AdS}} }  
			\! \widetilde{\mathcal{C}}_A^+ (\phi   )\,   \rd \phi
			\,=
			\int_{-\pi}^{+ \pi} 
			\! \widetilde{\mathcal{C}}_h (\phi   )\,   \rd \phi
			\, =
			\int_{-\pi}^{+ \pi }  
			\! \widetilde{\mathcal{C}}_A^- (\phi   )\,   \rd \phi
			\\
			\label{Sth-BHglobal-p3}
			\rule{0pt}{.6cm}
			&=&
			s_{\textrm{\tiny th}} \,  2 L_{\textrm{\tiny AdS} } \phi_b 
			=\,
			\frac{\pi c_{\textrm{\tiny BH}} }{3 \beta }   \big( 2 L_{\textrm{\tiny AdS} } \phi_b   \big)\,.
			\eea
			This set of equalities follows directly from the integrals in  (\ref{Gibbs-ent-definition}) and (\ref{eq:PhiHorizonIntegral}) 
			and from  \eqref{SBthermalentropy}.

			When $\phi_{b} > \phi_{b}^*$,
			where the critical angle is given in \eqref{crit-angle-RT-global-btz}, 
			the RT curve  is the union of $\gamma_{A, 2}$ and the horizon
			(see the red curve and the red circle wrapping the horizon in Fig.\,\ref{fig-global-btz-threads-after}). 
			Considering $\gamma_{A,2}$, 
			let us introduce the holographic contour function $\widetilde{\mathcal{C}}_B^+(\phi)$, 
			where $B$ is the region corresponding to the blue arc in Fig.\,\ref{fig-global-btz-threads-after},
			given by $B = [ - \pi , - \phi_b ] \cup [\phi_b, \pi )$.
			In order to write the holographic contour function $\widetilde{\mathcal{C}}_B^+(\phi)$,
			first  we consider the holographic contour function \eqref{CApBTZ}  
			associated with the  arc $\bar A=(-\pi+\phi_b,\pi-\phi_b)$, 
			which is the projection of $B$ around  $\phi=0$. 
			Then, we perform the change of variables
			given by $\phi\to-\phi-\pi$ when $\phi\in (-\pi+\phi_b,0)$
			and by $\phi\to\pi-\phi$ when $\phi\in (0,\pi-\phi_b)$. 
			Summarizing, we get
			\be
			\label{CBpBTZ}
			\widetilde{\mathcal{C}}_B^+(\phi)=\left\{
			\begin{array}{lll}
			\widetilde{\mathcal{C}}_{\bar A}^+(-\pi-\phi) & \hspace{1cm} &\phi\in (-\pi,-\phi_b)\cr
				\rule{0pt}{.6cm}
			\widetilde{\mathcal{C}}_{\bar A}^+(\pi-\phi)& & \phi\in (\phi_b,\pi)  \; .\end{array}\right. 
			\ee		
	To evaluate the holographic entanglement entropy $S_B$ of the arc $B$ by means  of the holographic contour function  \eqref{CBpBTZ}, 
	the boundary cutoff  $\tilde{\varepsilon}_{\textrm{\tiny bdy}}^{\textrm{\tiny $B$}}$ must be specified.  
	Combining the  map \eqref{btz-brane-to-btz-global-coord-change}
	with the above sequence of transformations, we find that $\tilde{\varepsilon}_{\textrm{\tiny bdy}}^{\textrm{\tiny $B$}}$ is obtained  from $\varepsilon_{\textrm{\tiny bdy}}^{\textrm{\tiny $A$}}$ in \eqref{pBTZ1-eps-AB} 
			by replacing  $\varepsilon_{\textrm{\tiny BTZ}} $, $b$ and $z_h$ 
			with $L_{\textrm{\tiny AdS}}^2 / r_\infty$, $L_{\textrm{\tiny AdS}} \, ( \pi - \phi_b ) $ 
			and $ L_{\textrm{\tiny AdS}}^2 / r_h$ respectively. 
			Thus, $S_B$ is evaluated as follows
					\bea
			\label{global-horizon-StildeB}
			{S}_B  
			&=&
			\frac{L_{\textrm{\tiny AdS}}}{4 G_{\textrm{\tiny N}}}
			\left(  \,
			\int^{- \phi_b -\tilde{\varepsilon}_{\textrm{\tiny bdy}}^{\textrm{\tiny $B$}}  / L_{\textrm{\tiny AdS}} }_{-\pi} 
			\! \widetilde{\mathcal{C}}_B^+ (\phi   )\,   \rd \phi
			\; + 
			\int^{ \pi }_{  \phi_b + \tilde{\varepsilon}_{\textrm{\tiny bdy}}^{\textrm{\tiny $B$}} / L_{\textrm{\tiny AdS}} }
			\! \widetilde{\mathcal{C}}_B^+ (\phi   )\,   \rd \phi
			\right)\nn
			\\
			\rule{0pt}{.8cm}
			&=&
			\frac{L_{\textrm{\tiny AdS}}}{2 G_{\textrm{\tiny N}}}  \, \log \! \left[ \,\frac{2 r_\infty }{r_h} \, \sinh \! \left( \frac{ r_h ( \pi - \phi_b)}{L_{\textrm{\tiny AdS}}} \right)  \right]
			+ o(1) \,.
			\eea
			The holographic entanglement entropy of the arc  $A$ in this regime is recovered by exploiting the relation 
			$S_A    = S_{\textrm{\tiny th}}  + {S}_B  $,
			where $S_{\textrm{\tiny th}} = 2 \pi  r_h / (4 G_{\textrm{\tiny N}})   $
			is the holographic thermal entropy of the whole CFT$_2$.

		       Given the blue arc $B$ in Fig.\,\ref{fig-global-btz-threads-before}, let us define the analog of the green arc for this subsystem and denote it by $B_\beta$.
		       This arc is spanned by $\phi \in [-\pi,-\pi+\tilde b_\beta/L_{\textrm{\tiny AdS}}]\cup[\pi-\tilde b_\beta/L_{\textrm{\tiny AdS}},\pi)$. 
		       From \eqref{btz-brane-to-btz-global-coord-change}, $\tilde b_\beta$ can be evaluated   by  replacing  
		       $b$ with $ L_{\textrm{\tiny AdS}} \, ( \pi - \phi_b ) $ and $z_h$  with  $L_{\textrm{\tiny AdS}}^2 / r_h$ in \eqref{bbeta}. 
		       Then, the holographic thermal entropy ${S}_{B,\textrm{\tiny th}}$ of $B$ is recovered  as follows
		       \be 
			S_{B,\textrm{\tiny th}} = 
			\frac{L_{\textrm{\tiny AdS}}}{4 G_{\textrm{\tiny N}}}
			\int_{B_\beta}
			\, 
			\! \widetilde{\mathcal{C}}_B^+ (\phi   )\,   \rd \phi =
			\frac{\pi c_{\textrm{\tiny BH}} }{3 \beta }   \; 2 L_{\textrm{\tiny AdS} } ( \pi -\phi_b)   \,.
			\ee   
			Finally, the holographic thermal entropy $S_{A,\textrm{\tiny th}}$ of  the arc $A$ is  obtained as $ S_{A,\textrm{\tiny th}} = S_{\textrm{\tiny th}}  - {S}_{B,\textrm{\tiny th}}$.

			\newpage
			\section{Conclusions}
			\label{sec-conclusions}
			
			
			In this manuscript we have explored a connection between the geodesic bit threads 
			\cite{Freedman:2016zud, Agon:2018lwq}
			associated with the spatial domain $A$ 
			providing the bipartition of the space where a holographic CFT$_{d+1}$ is defined 
			and the holographic thermal entropy $S_{A, \textrm{\tiny th}}$ of the subsystem $A$.
			Our results extend to higher dimensions the analysis for $d=1$ reported in \cite{Mintchev:2022fcp}.
			The gravitational backgrounds that we have considered are
			Poincar\'e AdS$_3$, BTZ black brane, Poincar\'e AdS$_{d+2}$,
			a specific $(d+2)$-dimensional hyperbolic black hole,
			Schwarzschild AdS$_{d+2}$ black brane
			and BTZ black hole.
			Since all these spacetimes are static, our calculations have been performed 
			in the geometries defined by their constant time slices
			(see (\ref{H2-metric}), (\ref{btz-brane-metric}), (\ref{fAdS-metric}), 
			(\ref{hyp-metric}), (\ref{sch-ads-brane-metric}) and (\ref{btz-global-metric})
			respectively).
			As for the bipartition induced by $A$, 
			we focus on the simplest choices
			that make the calculations accessible;
			namely an interval of length $2b$ for $d=1$
			and either a sphere of radius $b$ 
			or an infinite strip of finite width $2b$ for $d>1$.

			
			In Poincar\'e AdS$_3$ (see Sec.\,\ref{sec-AdS3}), 
			our main improvement with respect to the corresponding analysis of \cite{Agon:2018lwq}
			consists in the observation that the map introduced in \cite{Casini:2011kv}
			(see (\ref{chm-ads-in-out}) and (\ref{chm-ads-inside-inverse}))
			can be employed to interpret the whole configuration of geodesic bit threads
			through simple geodesics in two identical suitable BTZ black branes
			(see Fig.\,\ref{fig:ads3-main}).  
			In this holographic CFT$_2$ setup,
			the endpoints of the geodesic bit threads serve as the map 
			that implements the geometric action of modular conjugation,
			thereby identifying a potential gravitational dual for this map in CFT$_2$ \cite{Mintchev:2022fcp}.

			
			In pure AdS$_{d+2}$ in Poincar\'e coordinates (see Sec.\,\ref{HigherdimAdS}), 
			when $A$ is a sphere (see Sec.\,\ref{sec-ads-gbt-sphere})
			the geodesic bit threads can be constructed for any $d$ 
			(see Fig.\,\ref{fig:HigherDspherebitthreads}) \cite{Agon:2018lwq}.
			We have extended to higher dimensions 
			the observation made in \cite{Mintchev:2022fcp} for $d=1$,
			showing that, in this setup, 
			the geodesic bit threads provide a gravitational dual
			of the map implementing the geometric action of the modular conjugation 
			in  the dual CFT$_{d+1}$ in its ground state
			(see (\ref{modconjsphere})). 
			Instead,
			when $A$ is an infinite strip (see Sec.\,\ref{stripemptyads}),
			the construction of the vector field characterising the geodesic bit threads
			fails when $d \geqslant 3$
			because the nesting property is not satisfied (see Fig.\,\ref{fig:Dequals4bitb}),
			as also noted in \cite{Agon:2018lwq}.
			The vector field for the geodesic bit threads of the infinite strip $A$ when $d=2$
			has been discussed in Appendix\;\ref{GeodAdS4} (see Fig.\,\ref{AdS4vuoto})
			and the corresponding holographic contour function in $A$  
			develops an integrable singularity in the center of the strip (see Fig.\,\ref{contourAdS4})
			that is rather unexpected in comparison with the other known cases in CFT.

			In Poincar\'e AdS$_{d+2}$ 
			and when the subregion $A$ is an infinite strip,	
			we have introduced
			a bit thread construction that we called minimal hypersurface inspired bit threads (see Sec.\,\ref{sec5}) 
			through a slight modification of the ansatz providing the translated and dilated bit threads 
			constructed in \cite{Agon:2018lwq} (see Fig.\,\ref{fig:HigherDMSmoduli}). 
			These two kinds of bit threads coincide in pure AdS$_{d+2}$
			but they are different in black hole backgrounds.  
			When $A$ is a sphere and in pure AdS$_{d+2}$, 
			both
			the minimal hypersurface inspired bit threads and 
			the translated and dilated bit threads 
			become the geodesic bit threads.

			In pure AdS$_{d+2}$,
			we have  also studied the holographic contour functions associated 
			with the geodesic bit threads for the sphere,
			recovering the results of \cite{Kudler-Flam:2019oru,Han:2019scu},
			and with the minimal hypersurface inspired bit threads for the strip. 
			In the latter case, the holographic contour function (\ref{contourms})
			can be constructed parametrically and remains regular for any value of $d$.
			 By integrating these holographic contour functions over the subsystem 
			 regularised in a suitable way,
			 a match is obtained with 
			 the conventional computation of the regularized area of the RT hypersurface \cite{Ryu:2006bv, Ryu:2006ef}.
			 We remark that this agreement occurs
			 only if an appropriate UV regularisation procedure 
			 (called entanglement wedge cross-section regularisation \cite{Dutta:2019gen,Han:2019scu,Headrick:2022nbe, Nguyen:2017yqw,Takayanagi:2017knl})
			 is chosen on the boundary.
			 A different regularisation procedure on the boundary would lead to results
			 that do not even reproduce the pattern of divergences found in  \cite{Ryu:2006bv, Ryu:2006ef} for these setups.
			 This tells us that the actual map between the cutoffs in the bulk and in the boundary 
			 can be highly nontrivial (see e.g. \eqref{ads-EWCS-regEndp}).

			In the BTZ black brane (see Sec.\,\ref{sec-BTZ}),
			we revisited the analyses reported in \cite{Agon:2018lwq, Mintchev:2022fcp} for the geodesic bit threads
			by employing the map of \cite{Casini:2011kv} (see (\ref{chm-btz-inside})-(\ref{chm-btz-inside-inverse})).
			The occurrence of the auxiliary geodesics naturally leads us to introduce a second copy of the BTZ black brane,
			as shown in Fig.\,\ref{fig:BTZ-brane-main}.
			In this setup, the portions of the RT curve $\gamma_A$ 
			given by $\tilde{\gamma}_{A,\textrm{\tiny th}}$, 
			determined by the critical geodesic bit threads (see the magenta curves in Fig.\,\ref{fig:BTZ-brane-main}),
			and by $\gamma_{A,\textrm{\tiny th}}$ (determined by $S_{A,\textrm{\tiny th}}$) 
			coincide
			and are characterized by (\ref{bbeta}) and (\ref{x-z-beta-coord-from-b}),
			as already observed in \cite{Mintchev:2022fcp}.
			Here, we found analytically that 
			the fluxes of the geodesic bit threads through 
			either $A_\beta \subsetneq A$, 
			or $\tilde{\gamma}_{A, \textrm{\tiny th}}$ 
			or the entire planar horizon 
			provide the holographic thermal entropy of the interval $A$ (see Sec.\,\ref{subsec-flows-btz-planar}). 
			Moreover, we have observed that the flux density in $A$
			of the vector field characterizing the geodesic bit threads 
			provides the CFT$_2$ results for the contour functions discussed in \cite{Cardy:2016fqc, Coser:2017dtb}
			specialised to this holographic setup, where the Brown-Henneaux central charge occurs. 
			In this analysis, we have employed the relation \eqref{pBTZ1-eps-AB} 
			between the cutoffs in the bulk and in the boundary,
			obtained through the entanglement wedge cross section regularisation.

			For $d >1$,
			considering the specific hyperbolic black hole defined by (\ref{hyp-metric})-(\ref{hyp-metric-dHd}),
			in Sec.\,\ref{sec-hyp-bh}
			we have employed the map of \cite{Casini:2011kv} (see (\ref{chm-hyp-inside})-(\ref{chm-hyp-inside-inverse}))
			to write analytic expressions for the geodesic bit threads of a sphere 
			and the corresponding fluxes.
			The analytic expressions for the relevant holographic contour functions 
			(namely in $A$, on the whole horizon and for the auxiliary geodesics on the whole boundary)
			are \eqref{contour-hyp-bdy} and (\ref{contour-hyp-hor}),
			which provide the curves in Fig.\,\ref{fig:HYPcontourbordoorizsfera}.
			%

			The main findings of this manuscript concern 
			the geodesic bit threads for the sphere 
			and the minimal hypersurface inspired bit threads for the strip
			in Schwarzschild AdS$_{d+2}$ black brane (see Sec.\,\ref{Schwarzschild AdS black brane}).
			They are numerical results
			about the comparison between 
			$\widetilde{S}_{A, \textrm{\tiny th}}$ 
			(i.e. the flux of the bit threads through the entire horizon)
			and $S_{A, \textrm{\tiny th}}$
			or, equivalently, 
			between the area of $\tilde{\gamma}_{A, \textrm{\tiny th}} \subsetneq \gamma_A$ 
			and of $\gamma_{A, \textrm{\tiny th}} \subsetneq \gamma_A$ respectively,
			for different types of bit threads,
			when the subregion $A$ is either a sphere (see Sec.\,\ref{sec-Sch-AdS-sphere}) 
			or an infinite strip (see Sec.\,\ref{sec-Sch-AdS-strip}).
			Notice that the RT hypersurface $\gamma_A$ in these setups 
			is not known analytically for generic $d>1$. 
			Our results 
			about the geodesic bit threads of a sphere
			are shown in Fig.\,\ref{fig:SA-th-various-d} and Fig.\,\ref{fig:FTHHigherD}
			and suggest that $\widetilde{S}_{A, \textrm{\tiny th}} = S_{A, \textrm{\tiny th}}$
			for the values of $d$ that we have explored, 
			which are $2 \leqslant d \leqslant 6$.
			In the case of $d=2$, 
			we have compared our numerical results for $\widetilde{S}_{A, \textrm{\tiny th}}$
			with the ones obtained numerically
			from the integral representation 
			of the finite term in the expansion of the holographic entanglement entropy 
			as the UV cutoff vanishes \cite{Fonda:2015nma}
			(see the top left panel of Fig.\,\ref{fig:SA-th-various-d}). 
			It would be insightful to interpret the difference between these two UV finite quantities. 
			%
			An interesting related numerical analysis about the radius of $\tilde{\gamma}_{A, \textrm{\tiny th}}$
			has been reported in Fig.\,\ref{fig:bbetasphere}
			and suggests that the ratio $b_\beta / z_h$  might be  independent of $d$, while
			the ratio  $r_{m,\beta}/z_h$ is  
			either  independent of $d$ 
			or mildly dependent on the dimensionality.
			However, a more precise numerical investigation 
			or, even better, an analytic study of these ratios 
			is an interesting future development.

			We remark that the property   
			$\widetilde{S}_{A, \textrm{\tiny th}} = S_{A, \textrm{\tiny th}}$ holds for an infinite class of bit threads
			that includes the geodesic bit threads.  
			Indeed, considering the case of the sphere for simplicity,
			it is not difficult to draw deformations of the integral lines of the geodesic bit threads (see Fig.\,\ref{fig:HigherDPBBspherebitthreds})
			corresponding to other bit thread configurations 
			whose flux through the horizon is equal to the holographic thermal entropy of $A$.
			More generically, considering a subregion $A$ having a finite volume
			and independently of its shape, 
			for any assigned bit thread configuration 
			we can compare its flux $\widetilde{S}_{A, \textrm{\tiny th}} $ through the entire horizon
			with  $S_{A, \textrm{\tiny th}}$, i.e. the holographic thermal entropy of $A$.
			When $\widetilde{S}_{A, \textrm{\tiny th}} = S_{A, \textrm{\tiny th}}$, 
			it is natural to consider the projections 
			$\tilde{\gamma}_{A,\beta} \subsetneq \gamma_A$ on the RT hypersurface $\gamma_A$
			and 
			$A_\beta \subsetneq  A$ on the boundary,
			obtained by following all the bit threads connecting the horizon to the boundary. 
			It would be insightful to find a way to characterize analytically 
			the bit threads satisfying $\widetilde{S}_{A, \textrm{\tiny th}} = S_{A, \textrm{\tiny th}}$.

			In the Schwarzschild AdS$_{d+2}$ black brane
			and when the subsystem $A$ is a sphere,
			we find it worth highlighting the numerical results in Fig.\,\ref{fig:contourbordoorizsfera}
			about the holographic contour functions 
			in $A$, on the whole horizon and for the auxiliary geodesics on the whole boundary.
			Despite their qualitative similarity with the corresponding quantities 
			for  the hyperbolic black hole explored in Sec.\,\ref{sec-hyp-bh}
			(see Fig.\,\ref{fig:HYPcontourbordoorizsfera}), 
			quantitative differences occur 
			and it would be very insightful to find analytic results for these
			holographic contour functions.

			When $A$ is an infinite strip,
			the construction of the corresponding geodesic bit threads in the Schwarzschild AdS$_{d+2}$ black brane fails for $d>2$.
			The $d=2$ case has been discussed in Appendix\;\ref{GeodAdS4},
			where, interestingly, we find that
			the singularity occurring in the holographic contour function in $A$ for pure AdS$_4$
			is smoothened out by the occurrence of the horizon 
			(see Fig.\,\ref{fig:HigherDstripcontour} and Fig.\,\ref{fig:HigherDstripcontoura})
			and that $\widetilde{S}_{A, \textrm{\tiny th}} \neq S_{A, \textrm{\tiny th}}$
			(see Fig.\,\ref{GeoandModulusBHD} and Fig.\,\ref{fig:SBstrip}).
			As for the minimal hypersurface inspired bit threads
			for the strip in the Schwarzschild AdS$_{d+2}$ black brane
			(see Sec.\,\ref{sec-strip-ddim-mhbt}),
			we find that they are well defined bit threads in any dimension,
			but $\widetilde{S}_{A, \textrm{\tiny th}} \neq S_{A, \textrm{\tiny th}}$
			also in this case
			(see Fig.\,\ref{fig:msbitintegrallinesbh} and Fig.\,\ref{fig:SBMSstrip}).

			
			In the BTZ black hole (see Sec.\,\ref{sec-BTZ-black-hole}),
			the global properties of the background are crucial in the bit thread construction.
			Indeed,
			the geodesic bit threads of an arc $A$ cannot be constructed
			(see Fig.\,\ref{fig-global-btz-threads-before} and Fig.\,\ref{fig-global-btz-threads-after}).
			However, 
			by considering fake geodesic bit threads that are allowed to intersect,
			the holographic thermal entropy of the arc $A$ can be recovered
			by taking into account 
			the integral curves winding around the horizon multiple times.

			
			Our analyses can be extended in various directions.
			The geodesic bit threads or alternative bit thread constructions
			could be explored in more complicated gravitational backgrounds.
			It would be interesting to understand 
			the stability of our results for the sphere
			under shape deformations
			by considering small perturbations \cite{Klebanov:2012yf, Allais:2014ata} 
			or even finite regions with arbitrary shape 
			\cite{Fonda:2014cca, Fonda:2015nma, Cavini:2019wyb}.
			This includes spatial domains with singularities
			like e.g. corners for $d=2$ \cite{Bueno:2015xda}.
			In the investigation of the shape dependence, 
			the analogy with magnetism discussed in  \cite{Gursoy:2023tdx}
			could be insightful.
			A relevant direction to explore consists of considering gravitational backgrounds
			with boundaries in the bulk,
			e.g. in the setup of  \cite{Takayanagi:2011zk, Fujita:2011fp}
			(see \cite{ Seminara:2017hhh, Seminara:2018pmr} for the shape dependence 
			of the holographic entanglement entropy in this context).
			As for the relation between bit threads and CFT quantities, 
			it would be interesting to develop further
			the connection with the modular conjugation \cite{Mintchev:2022fcp},
			with the entanglement of purification \cite{Harper:2019lff} 
			and with the contour functions for the entanglement related quantities,
			e.g. by considering 
			integrations over regions strictly contained in $A$ \cite{DiGiulio:2019lpb},
			contour functions in inhomogeneous spaces \cite{Tonni:2017jom} 
			or the contour functions explored in \cite{SinghaRoy:2019urc, Roy:2021qxf, Santalla:2022ygq}.
			Finally, we find it worth investigating all these directions 
			also in time-dependent gravitational backgrounds 
			by employing the covariant bit thread constructions
			discussed in \cite{Headrick:2022nbe}.


			\vskip 20pt 
			\centerline{\bf Acknowledgments} 
			\vskip 5pt 
			
			It is our pleasure to acknowledge 
			Chris Herzog,
			Veronika Hubeny, 
			Manuela Kulaxizi,
			Hong Liu, 
			Stam Nicolis, 
			Andrei Parnachev,
			Juan Pedraza, 
			Giuseppe Policastro, 
			Jacopo Sisti, 
			Sergey Solodukhin
			and especially Matthew Headrick, 
			Mihail Mintchev 
			and Diego Pontello 
			for very useful discussions. 
			ET acknowledges
			Center for Theoretical Physics at MIT (Boston) and University of Florence 
			for warm hospitality and financial support during part of this work.

\appendix

\section{Magnitude of the vector fields} 
\label{app-modulus}

In this appendix, we review the general procedure to compute the magnitude of the 
divergenceless vector field $\boldsymbol{V}$ occurring in the prescription (\ref{HEE-BT-intro})
proposed in \cite{Agon:2018lwq}.

In  the constant time slice of an asymptotically AdS$_{d+2}$ static spacetime
equipped with the metric $g_{ab}$,
let us consider its foliation in $\lambda=\textrm{const}$ hypersurfaces  generated by a family of integral lines
$Y(\boldsymbol{q}_m, \lambda)$.
Here the $d$-dimensional vector $\boldsymbol{q}_m$ identifies a point on the RT hypersurface $\gamma_A$
and consequently, also the integral line passing through it,
while $\lambda$ is the parameter running along such an integral line.

It is straightforward to obtain the unit tangent vector $\boldsymbol{\tau} $ to a given integral line 
passing through a certain point of the spacetime identified by $(\boldsymbol{q}_m,\lambda)$. 
The transverse metric $h_{ab}$, 
i.e. the induced metric on a hyperplane orthogonal to the integral line at the point $(\boldsymbol{q}_m,\lambda)$,
and the transverse area element $\delta A (\boldsymbol{q}_m,\lambda)$ are given respectively by 
\begin{equation}
	h_{ab}= g_{ab} -  \tau_a  \tau_b  
	\;\;\;\qquad \;\;\;
	\delta A (\boldsymbol{q}_m, \lambda)  = \sqrt{ h(\boldsymbol{q}_m,\lambda)} \,.
	\label{metricaortogonale}
\end{equation}
For example, for the Poincar\'e AdS$_3$ explored in Sec.\,\ref{sec-AdS3}, 
the configuration of integral lines is the set of geodesics shown in Fig.\,\ref{fig:ads3-main} 
and the parametrization is given by \eqref{gen-geo-ads3}-\eqref{c0-b-gbt-ads3},
with $y$ and $y_m$ 
playing the role of $\lambda$ and $\boldsymbol{q}_m$ respectively.

The evaluation of $|\boldsymbol{V}|$ relies on the fact that $\boldsymbol{V}$ is divergenceless 
and satisfies
\begin{equation}
	\big|\boldsymbol{V}(\boldsymbol{q}_m, \lambda_m)\big|=1 
	\label{saturazionemodulo}
\end{equation}
where $\lambda=\lambda_m$ identifies the position of $\gamma_A$. 
The vector field $\boldsymbol{V}$ is divergenceless if and only if 
the flux through the transverse area element is conserved along the integral line, namely
\be
\big|\boldsymbol{V}(\boldsymbol{q}_m, \lambda)\big| \,\delta A (\boldsymbol{q}_m,\lambda) 
= 
\big|\boldsymbol{V}(\boldsymbol{q}_m, \lambda_m)\big| \, \delta A (\boldsymbol{q}_m,\lambda_m) 
\;\;\;\qquad \;\;\;
\forall \lambda\,.
\ee 
Combining this condition with \eqref{saturazionemodulo}, we obtain 
\begin{equation}
\big|\boldsymbol{V}(\boldsymbol{q}_m, \lambda)\big|
	= 
	\frac{\sqrt{h(\boldsymbol{q}_m,\lambda_m)}}{\sqrt{h(\boldsymbol{q}_m,\lambda)}} \ .
	\label{formulageneralemodulo}
\end{equation}
This construction guarantees that the vector field is divergenceless and that its magnitude is equal to one on $\gamma_A$,
but it does not guarantee that $|\boldsymbol{V}| < 1$ everywhere away from $\gamma_A$.
The latter property  must be checked case by case,
ruling out the possibility that $\boldsymbol{V}$ describes well defined bit threads
whenever it is not satisfied.

In the main text, we have applied \eqref{formulageneralemodulo} to various cases where the spatial domain $A$ on the boundary 
is either an interval (see Sec.\,\ref{subsec-gbt-ads3}) 
or an infinite strip (see Sec.\,\ref{sec5} and Sec.\,\ref{sec-Sch-AdS-strip})
or a sphere (see Sec.\,\ref{sec-ads-gbt-sphere} and Sec.\,\ref{sec-Sch-AdS-sphere}).

When $A$ is an infinite strip (see Sec.\,\ref{stripemptyads})
and the gravitational background is equipped with the metric \eqref{sch-ads-brane-metric},
like e.g. the constant time slice of the Schwarzschild AdS$_{d+2}$ black brane
and of the Poincar\'e AdS$_{d+2}$ (see \eqref{fAdS-metric}),
let us denote the two branches of the integral lines by $x_{\textrm{\tiny$\lessgtr$}}(z; z_m)$. 
In this instance, as emphasized in Sec.\,\ref{sec-Sch-AdS-strip}, it suffices to concentrate on the subspace defined by the coordinates $(z, x)$
because the remaining coordinates primarily act as spectators during this analysis.
The metric \eqref{metricaortogonale} orthogonal to the integral lines reads
\begin{equation}
	ds^2_{\perp}= \left( \sqrt{h_{xx}} \, \rd x \pm  \sqrt{h_{zz}}\,  \rd z \right)^2+\frac{1}{z^2} \; \rd\boldsymbol{x}_\perp^2 
	\label{dsortogonale}
\end{equation}
where $h_{xx}$ and $h_{zz}$ can depend on $z$ and $z_m$. 
The choice of the sign in \eqref{dsortogonale} is determined by the branch of the bit thread  $x_{\textrm{\tiny$\lessgtr$}}(z; z_m)$.  Changing coordinates from $(z,x)$ to the adapted coordinates $(z,z_m)$,
the metric  \eqref{dsortogonale} can be written only in terms of the transverse element $\rd z_m$ on the orthogonal plane as follows
\begin{equation}
	ds^2_{\perp}= \left(\sqrt{h_{xx}} \ \partial_{z_m} x(z;z_m) \right)^2 \rd z_m^2 + \frac{1}{z^2} \,\rd \boldsymbol{x}_{\perp}^2\, .  
	\label{metricaortogonalefinale}
\end{equation} 
In our examples,
the factor $h_{xx}$ depends on the type of bit threads  under investigation
but it does not feel the difference between pure AdS$_{d+2}$ 
and  Schwarzschild AdS$_{d+2}$  black brane. 
For the geodesic bit threads (see Sec.\,\ref{stripemptyads} and Sec.\,\ref{sec-gbt-strip-ddim})
and the minimal hypersurface inspired bit threads (see Sec.\,\ref{sec5} and Sec.\,\ref{sec-strip-ddim-mhbt}),
we find respectively 
\be
h_{xx}
= \dfrac{z_m^{2} \,z_*^{2d} -z^2 \left(z_*^{2d}-z_m^{2d} \right)}{z^2 \,z_*^{2d} \, z_m^2}
\;\;\;\qquad\;\;\;
h_{xx}
=\dfrac{z_m^{2d} \,z_*^{2d} -z^{2d} \left(z_*^{2d}-z_m^{2d} \right)}{z^2 \, z_*^{2d} \,z_m^{2d}} 
\ee
where the notation of Sec.\,\ref{Schwarzschild AdS black brane} has been adopted 
and the  expressions for pure AdS$_{d+2}$ are obtained by replacing $x$ with $y$, $z$ with $w$ and $z_*$ with $w_*$.
By applying the general expression \eqref{formulageneralemodulo} for the determinant obtained from \eqref{metricaortogonalefinale}, we find
\begin{equation}
\label{mod-V-appA-strip}
	\big|\boldsymbol{V}\big|
	= \left| \, \left(\frac{z}{z_m} \right)^{d-1} \, 
	\frac{\sqrt{h_{xx}} \,\big|_{z=z_m}}{\sqrt{h_{xx}}}  \; \frac{\big(\partial_{z_m} x(z;z_m)\big)\big|_{z=z_m}}{\partial_{z_m} x(z;z_m)} \, \right| 
\end{equation}
where the absolute value is needed because the sign of \eqref{dsortogonale} is not well defined.

The analysis described above for the strip can be easily adapted to the case where 
the subsystem $A$ is a sphere,
and therefore, it is more convenient to adopt the spherical coordinates 
in the boundary of \eqref{sch-ads-brane-metric}.
From the integral lines $r(z,z_m)$ and their unit tangent vector $\boldsymbol{\tau}$,
we can write the metric of the hypersurfaces orthogonal to the bit threads
as follows
\begin{equation}
	ds^2_{\perp}= \left(\sqrt{h_{rr}} \ \partial_{z_m} r(z;z_m) \right)^2 \rd z_m^2 + \frac{r^2}{z^2} \; \rd \boldsymbol{\Omega}^2 \,.
	\label{metricaortogonalefinalesfera}
\end{equation} 
%
For the geodesic bit threads 
in pure AdS$_{d+2}$ and in Schwarzschild AdS$_{d+2}$ black brane
(see Sec.\,\ref{sec-ads-gbt-sphere} and Sec.\,\ref{sec-Sch-AdS-sphere}), 
we find that $h_{rr}$ is given respectively  by 
\begin{equation}
h_{rr}= \dfrac{{b}_0^2-w^2}{{b}_0^2 w^2}
\;\;\;\;\qquad\;\;\;
h_{rr}= \dfrac{{C}^2-z^2}{{C}^2 z^2} 
\end{equation}
where $b_0$ and $C$ have been defined in \eqref{c0-b-gbt-adsd} and \eqref{z*bhsphere} respectively.
Finally, the magnitude of the vector field is obtained by applying \eqref{formulageneralemodulo}
with the determinant of \eqref{metricaortogonalefinalesfera}, finding 
\begin{equation}
\label{mod-V-appA-sphere}
	\big|\boldsymbol{V} \big|
	= \left| \, 
	\bigg(\frac{z \, r_m}{z_m \,r} \bigg)^{d-1} 
	\frac{\sqrt{h_{rr}}\,\big|_{z=z_m}}{\sqrt{h_{rr}}} \; 
	\frac{\big(\partial_{z_m} r(z;z_m) \big)\big|_{z=z_m}}{\partial_{z_m} r(z;z_m)} 
	\, \right| 
\end{equation}
where, again, the absolute value is necessary because we have taken the square root of the coefficient of $\rd z_m^2$.

\section{Holographic entanglement entropy of $B$ in BTZ black brane }
\label{app-entropy-complementary-btz}

In this appendix
we discuss the holographic entanglement entropy of the region $B = \RR \setminus A$ complementary to the interval $A=(-b,b)$ 
when the gravitational background is the time slice of the BTZ black brane, which is equipped with the metric \eqref{btz-brane-metric}. 

\begin{figure}[t!]
	\vspace{-.5cm}
	\hspace{.6cm}
	\includegraphics[width=.9\textwidth]{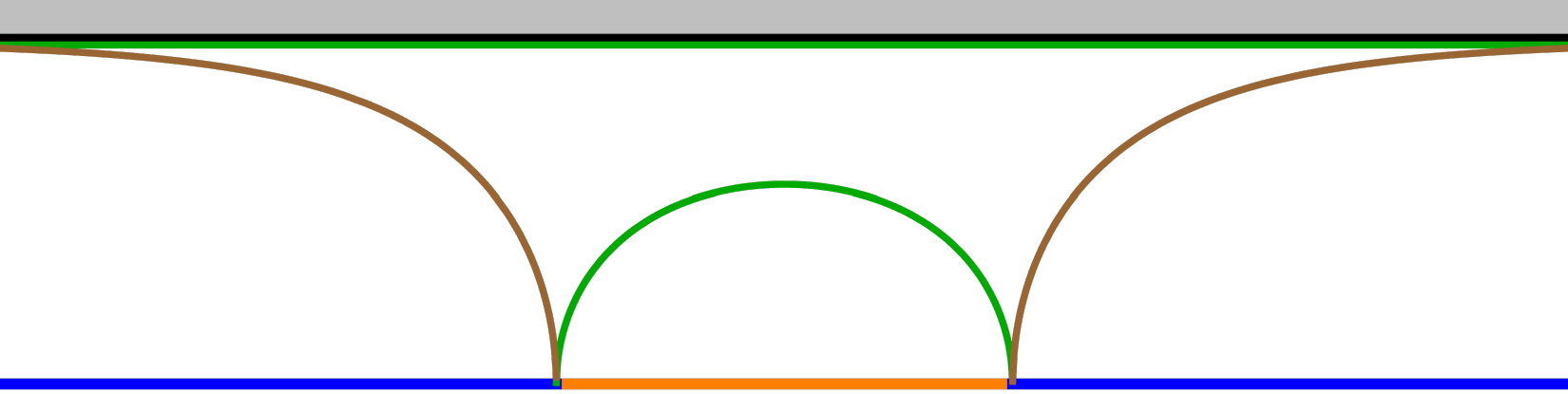}
	\vspace{.3cm}
	\caption{\small
		The configurations $\gamma_{B,1}$ (brown curves) and $\gamma_{B,2}$ (green curves) providing 
		the RT curve $\gamma_B$ of $B$ (infinite blue domain) in the BTZ black brane,
		whose regularized length determines (\ref{btz-hee-B-complementary}).}
	\label{fig-planar-btz-RT-compl}
\end{figure}

Since the RT curve $\gamma_B$ must satisfy both the minimal length condition and the homology constraint, 
we have to consider two qualitatively different candidates:
the configuration $\gamma_{B,1}$ given by the union of the two geodesics of the form \eqref{geod-btz-gen-bdy-inf} 
anchored to the endpoints of $B$ (which coincide with the endpoints of $A$), i.e. 
$\gamma_{B,1} \equiv \big\{ z = z_{\textrm{\tiny I/II}}^+ (x)|_{s=b} \big\} \cup \big\{ z = z_{\textrm{\tiny I/II}}^- (x)|_{s=-b}\big \}$,
and the configuration $\gamma_{B,2}$
defined as the union of the RT surface of $\gamma_A$ in \eqref{RT-curve-btz-brane} for $A$ and the horizon, i.e.
$\gamma_{B,2} \equiv  \big\{ z = z_m (x_m)   \} \cup \{ z = z_h \big\}$.
These two configurations are shown in Fig.\,\ref{fig-planar-btz-RT-compl}.

Since $\gamma_B$ must be homologous to $B$, which has infinite size, 
beside the usual holographic UV cutoff $z= \varepsilon_{\textrm{\tiny BTZ}} \ll 1$
we must also introduce in IR cutoff $L_{\textrm{\tiny IR}} \gg 1$ 
such that $|x| < L_{\textrm{\tiny IR}}$.

Computing the areas of $\gamma_{B,1}$ and $\gamma_{B,2}$, 
one observes that the minimal length curve is $\gamma_{B,1}$  if $b/\beta < \log (2)/4 \pi$, 
otherwise it is $\gamma_{B,2}$. 
Thus, the holographic entanglement entropy of $B$, 
which is given by 
$S_B= \tfrac{1}{4 G_{\textrm{\tiny N}}} \min \! \big\{ \textrm{Area}( \gamma_{B,1}), \textrm{Area}(\gamma_{B,2}) \big\} $,
reads
\be
\label{btz-hee-B-complementary}
S_B 
\,=\,
\left\{
\begin{array}{ll}
	\displaystyle
	\frac{2 \pi  \, c_{\textrm{\tiny BH}}\, (L_{\textrm{\tiny IR}}-b)}{3 \beta }
	+ \frac{ c_{\textrm{\tiny BH}}}{3} \, \log \!\left(\frac{\beta }{2 \pi \, \varepsilon_{\textrm{\tiny BTZ}}  }\right)
	&
	\displaystyle b/\beta < \frac{\log(2)}{4\pi} 
	\\
	\rule{0pt}{.9cm}
	\displaystyle
	\frac{2 \pi   \,c_{\textrm{\tiny BH}} \, L_{\textrm{\tiny IR}} }{3 \beta } + 
	\frac{ c_{\textrm{\tiny BH}} }{3} \, \log \!\left(\frac{\beta }{\pi  \,\varepsilon_{\textrm{\tiny BTZ}} } \, \sinh ( 2 \pi  b / \beta ) \right)
	\hspace{1cm}
	&
	\displaystyle b/\beta > \frac{\log(2)}{4\pi} 
\end{array}
\right.
\ee
where $c_{\textrm{\tiny BH}}$ is the Brown-Henneaux central charge \eqref{BH-central-charge}. 
Notice that $S_B$ in \eqref{btz-hee-B-complementary} is different from $\widetilde{S}_B $ in \eqref{SB-tilde}.

It is worth verifying the Araki-Lieb inequality \cite{Araki:1970ba} in this holographic setup. 
Since the whole system is in a thermal state, 
this inequality becomes
\be 
\label{AL-ineq}
\big|S_A -S_B\big|  \leqslant  S_{\textrm{\tiny th}}
\ee 
where, from (\ref{SBthermalentropy}), 
$S_{\textrm{\tiny th}} =  2 \pi   c_{\textrm{\tiny BH}} L_{\textrm{\tiny IR}} /( 3 \beta )$
is the holographic thermal entropy of the system. 
Thus, in our case, 
from \eqref{hee-thermal-from-C} and \eqref{btz-hee-B-complementary} we find that
\be 
\label{AL-check-planarBTZ}
\frac{ \big|S_A -S_B\big| }{ S_{\textrm{\tiny th}}}  
\,=\,
\left\{
\begin{array}{ll}
	\displaystyle 
	1 - \frac{b}{L_{\textrm{\tiny IR}} }
	-  \frac{\beta }{2 \pi L_{\textrm{\tiny IR}} } \,
	\log\!\big(2 \sinh(2 \pi b / \beta ) \big)
	\hspace{1.cm}
	&
	b/\beta < \log (2)/ (4 \pi)  
	\\
	\rule{0pt}{.5cm}
	1 
	\quad
	&
	b/\beta >  \log (2)/(4 \pi) 
\end{array}
\right.
\ee 
which is always positive and less or equal to $1$ when $L_{\textrm{\tiny IR}} \gg 1$;
hence the Araki-Lieb inequality (\ref{AL-ineq}) is satisfied. 
Notice that also in this case one observes the holographic entanglement plateaux \cite{Hubeny:2013gta},
mainly discussed for a dual CFT at finite volume.

\section{Geodesic bit threads in AdS$_4$ for the strip}
\label{GeodAdS4}

In Sec.\,\ref{stripemptyads}, we have shown that the geodesic bit threads in AdS$_4$ for the infinite strip can be defined
and in this appendix we discuss the vector field $\boldsymbol{V}$ characterizing them.

\begin{figure}[t!]
	\vspace{-.5cm}
	\hspace{-.9cm}
	\includegraphics[width=0.55\textwidth]{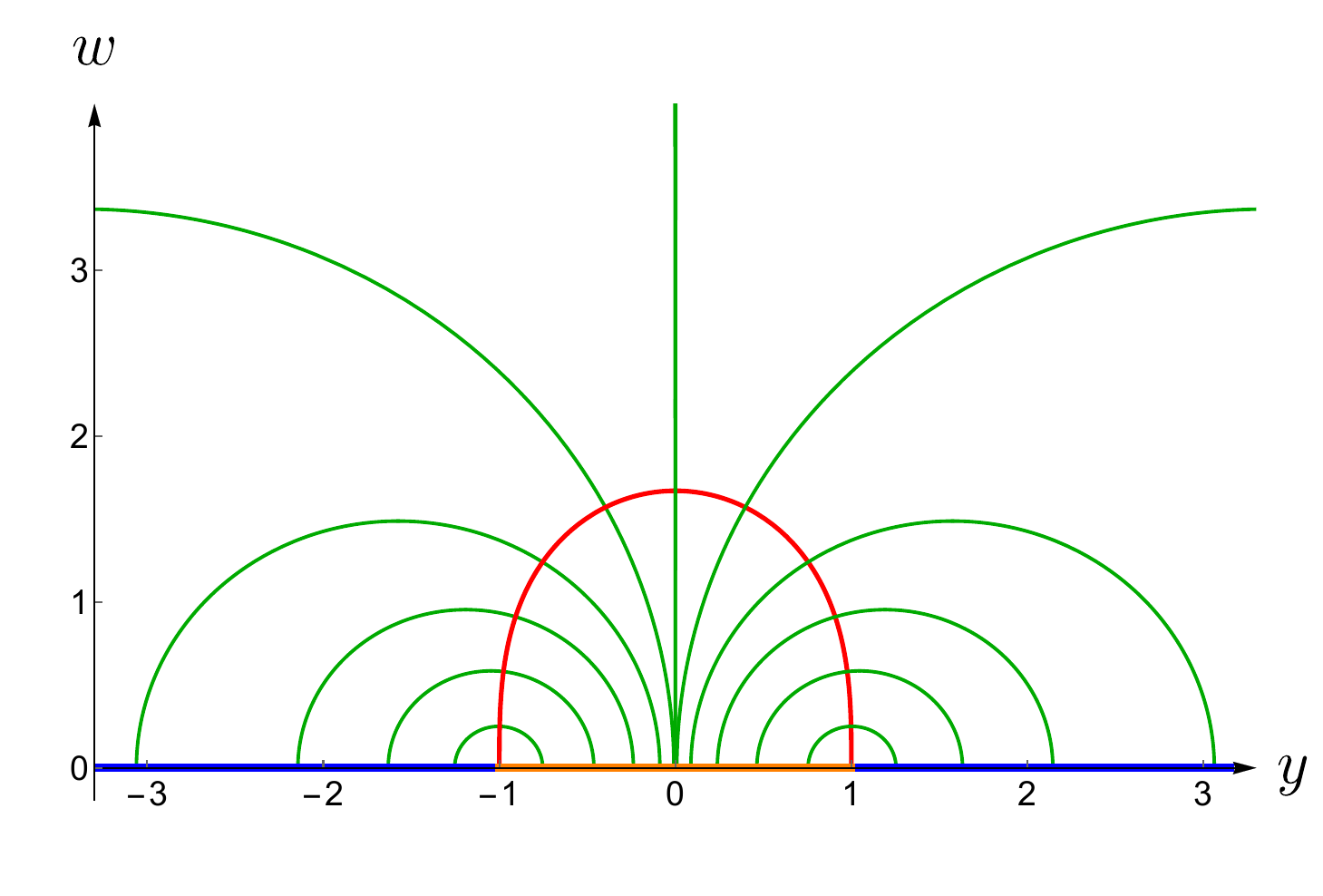}
	\vspace{-.3cm}
	\includegraphics[width=0.55\textwidth]{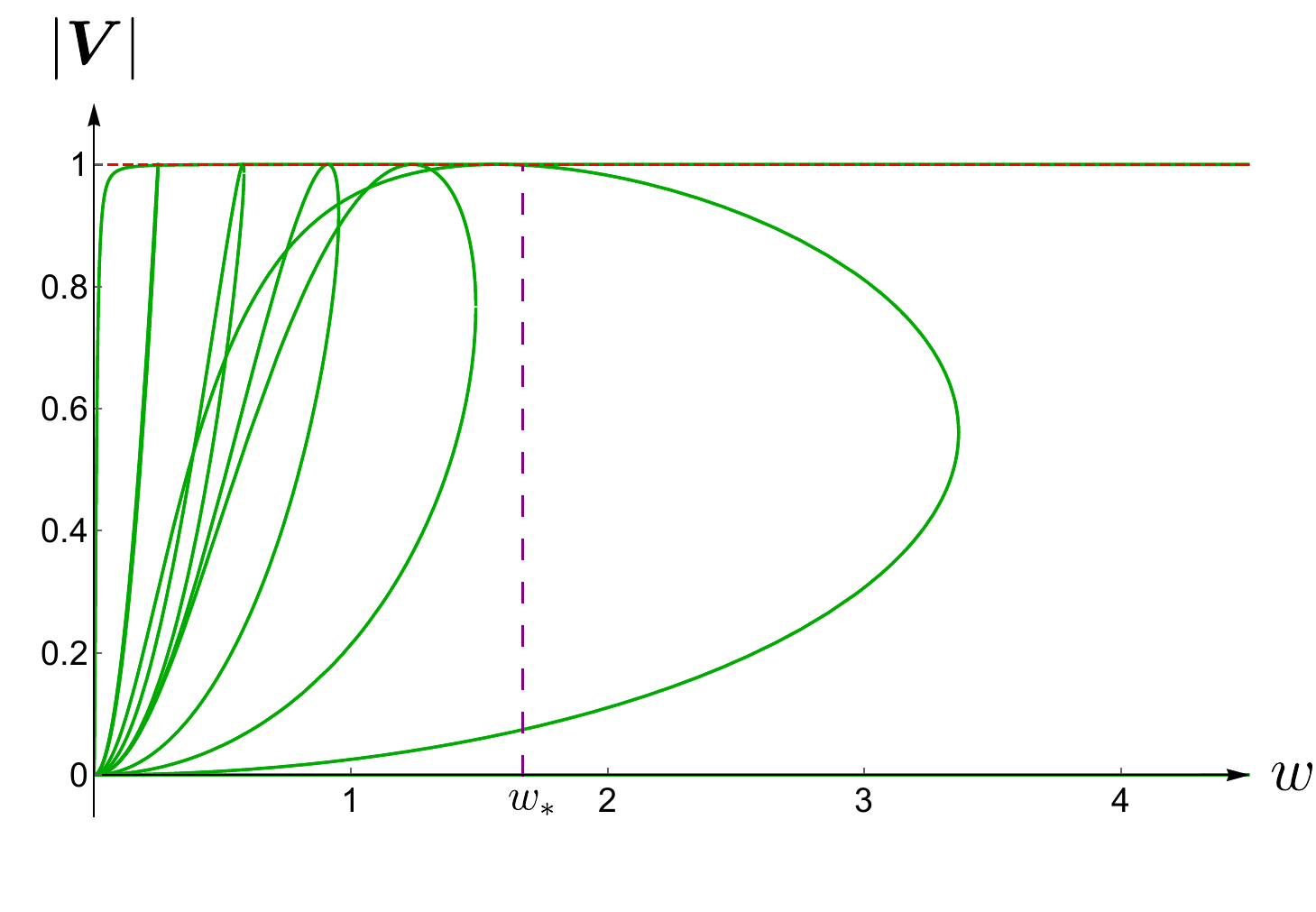}
	\caption{\label{AdS4vuoto}
		\small 
			Geodesic bit threads of the infinite strip, in AdS$_4$:
			integral curves (left) and modulus of the vector field (right)
			as the coordinate $w$ varies along the integral lines in the left panel. 
	}
	\label{fig:moduligeodeticivuoto}
\end{figure}

Following the procedure outlined in Appendix\;\ref{app-modulus}, 
we can calculate $\boldsymbol{V}$  in terms of $w$ and $w_m$.  
We distinguish two branches along any geodesic bit thread, 
depending on whether $w$ is increasing ($<$) or decreasing ($>$). 
After some algebraic manipulations,
for the magnitude of $\boldsymbol{V}$ we obtain 
\begin{equation}
	\label{sss}
	\big|\boldsymbol{V}_{\textrm{\tiny$\lessgtr$}}\big|
	= 
	\frac{w^2 \,(w_*^4-w_m^4) }{
		\left(w_m^3\mp \sqrt{w^2   \left(w_m^4-w_*^4\right)+w_*^4 w_m^2} \,\right)^2+w^2 \big(w_*^4-w_m^4\big)}
\end{equation}
where the expression under the square root is real along the particular bit thread
we are considering because the value of $w$ cannot exceed the maximum height of the bit thread, given by (\ref{zstarstrisciavuoto}).
We remark that, from (\ref{sss}), it is evident that $|\boldsymbol{V}_{\textrm{\tiny$\lessgtr$}} | \leqslant 1$.
In the right panel of Fig.\,\ref {AdS4vuoto} we show
$|\boldsymbol{V} |$ along the geodesic bit threads in AdS$_4$ for the strip 
whose integral lines are displayed in the left panel of the same figure. 

The unit tangent vector on the two branches is given by 
\begin{equation}
	{ \boldsymbol{\tau}}_{\textrm{\tiny$\lessgtr$}}
	=
	\big( \, { \tau}_{\textrm{\tiny$\lessgtr$}}^w \,, { \tau}_{\textrm{\tiny$\lessgtr$}}^y \, \big)
	=
	\frac{1}{L_{\textrm{\tiny AdS}}}
	\left(\pm\,\frac{w \, \sqrt{w^2 w_m^{4}+w_*^{4} \left(w_m^2-w^2\right)}}{w_m \, w_*^{2}}\, , \,\frac{w^2 \,\sqrt{w_*^{4}-w_m^{4}}}{w_m \,w_*^{2}} \, \right) .
\end{equation}

\begin{figure}[t!]
		\vspace{-.5cm}
		\hspace{1.3cm}
	\includegraphics[width=0.8\textwidth]{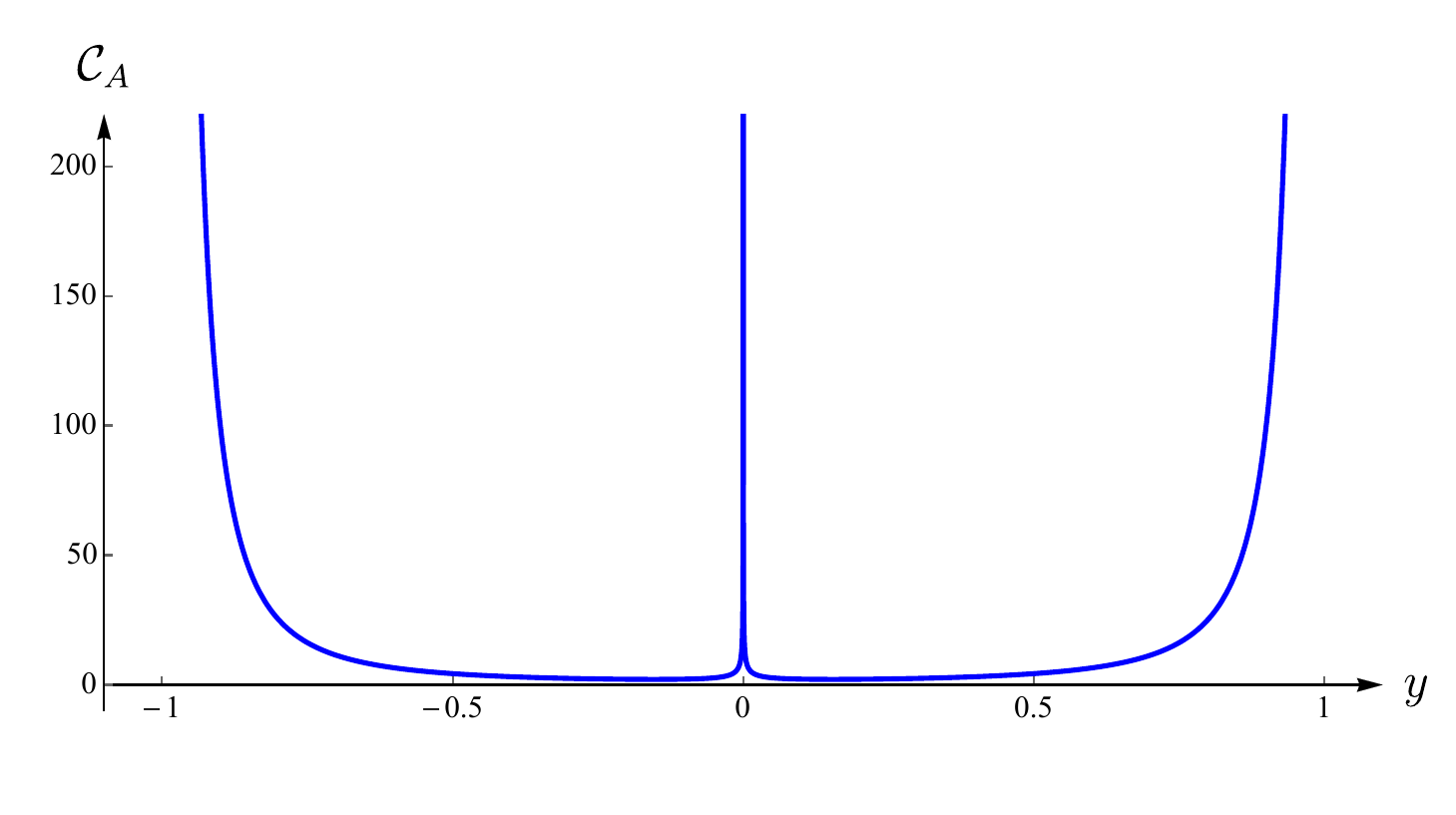}
		\vspace{-.5cm}
	\caption{
		\small   
		Holographic contour function induced in $A$ 
		by the geodesic bit threads of the infinite strip, in AdS$_4$
		(see (\ref{contourstrisciavuotoooA})).
	}
	\label{contourAdS4}
	\label{fig:HigherDstripcontoura}
\end{figure}

The explicit dependence on $w_m$ can be eliminated 
by expressing this quantity in terms of the points $(w, y)$ 
belonging to the geodesic bit thread corresponding to $w_m$. 
We can remove this dependence, at least numerically, 
and, consequently,  determine e.g. the holographic contour function in $A$, which reads
\be
\mathcal{C}_A(\boldsymbol{y})
=
\lim_{w \to 0^+}\left( \frac{1}{4 G_{\textrm{\tiny N}} } 
\,\big |\boldsymbol{V}_{\textrm{\tiny$<$}} \big| 
\, \tau_a  \, n^a   
\left(\frac{L_{\textrm{\tiny AdS}}}{w}\right)^2 \,\right)
=  
\frac{L^2_{\textrm{\tiny AdS}}}{4 G_{\textrm{\tiny N}} }
\; \frac{ w_*^2+  w^2_m(y_{\textrm{\tiny $A$}}) }{ w^2_m ( y_{\textrm{\tiny $A$}}) \big[w_*^2- w^2_m(y_{\textrm{\tiny $A$}}) \big] }
\label{contourstrisciavuotoooA}
\ee
where $\boldsymbol{y} \in A$
and $w_m(y_{\textrm{\tiny $A$}})$ is obtained by solving for $w_m$ the first of the two transcendental equations  in \eqref{es1-2}. 
The holographic contour function (\ref{contourstrisciavuotoooA}) is shown in Fig.\,\ref{fig:HigherDstripcontoura}.
The contour in the complementary region $B$ is given by
\be
\mathcal{C}_B(\boldsymbol{y})
= 
\lim_{w \to 0^+}
\left( -\frac{1}{4 G_{\textrm{\tiny N}} } \,
\big|\boldsymbol{V}_{\textrm{\tiny$>$}} \big| \,
\tau_a \, n^a   
\left(\frac{L_{\textrm{\tiny AdS}}}{w}\right)^2 \right)
=  
\frac{L^2_{\textrm{\tiny AdS}}}{4 G_{\textrm{\tiny N}} } \;
\frac{ w_*^2-w^2_m(y_{\textrm{\tiny $B$}}) }{w^2_m(y_{\textrm{\tiny $B$}}) \big[ w_*^2+w^2_m(y_{\textrm{\tiny $B$}})\big] }
\label{contourstrisciavuotoooB}
\ee
where $\boldsymbol{y} \in B$
and we have instead to solve the second of the two transcendental equations in \eqref{es1-2}  to remove the dependence on $w_m$.

The contour  \eqref{contourstrisciavuotoooA} is singular when  $y_{\textrm{\tiny $A$}} \to 0$ because $w_m(0)=w_*$.
Thus, the flux of $\boldsymbol{V}$ is not smooth
and, strictly speaking, it does not define a proper family of bit threads. 
However, the singularity of the holographic contour function (\ref{contourstrisciavuotoooA})
is  integrable; indeed 
\be
w_m(y_{\textrm{\tiny $A$}})=w_*-3^{2/3} \sqrt[3]{w_*} \; y_{\textrm{\tiny $A$}}^{2/3} + \cdots 
\;\;\;\qquad\;\;\;
y_{\textrm{\tiny $A$}} \to 0\,.
\ee
This implies that the integral of \eqref{contourstrisciavuotoooA} over a strip regularized through the UV cutoff $\varepsilon^{\textrm{\tiny $A$}}_{\textrm{\tiny bdy}}$, 
implicitly defined by $w_m(\varepsilon_{\textrm{\tiny AdS}})$,
gives the holographic entanglement entropy of $A$ as follows
\bea
S_A
&=&
\int_{A}  \mathcal{C}_A(\boldsymbol{y}) \,\rd^2 y
\,=\,
\frac{L_{\textrm{\tiny AdS}}^2 \, b_\perp}{G_{\textrm{\tiny N}} } 
\int^{b-\varepsilon^{\textrm{\tiny $A$}}_{\textrm{\tiny bdy}}}_{0} 
\frac{ w^2_m(y)+w_*^2 }{ w^2_m(y) \big[ w_*^2- w^2_m(y) \big] } \; \rd y 
\nonumber
\\
\rule{0pt}{.8cm}
&=&
\frac{L_{\textrm{\tiny AdS}}^2 \, b_\perp}{G_{\textrm{\tiny N}} } 
\int^{w_*}_{\varepsilon_{\textrm{\tiny AdS}}}
\frac{ \big| \partial_{w_m} y_{\textrm{\tiny{A}}} \big|  \left(w^2_m+w_*^2\right)}{w^2_m \big(w_*^2- w^2_m\big)} \; \rd w_m
=
\frac{L_{\textrm{\tiny AdS}}^2 \, b_\perp}{G_{\textrm{\tiny N}} } 
\int^{w_*}_{\varepsilon_{\textrm{\tiny AdS}}}  \frac{ w_*^2}{w_m^2\sqrt{w_*^{4}-w_m^{4}}} \;\rd w_m 
\nonumber
\\
\rule{0pt}{.8cm}
&=&
\frac{L_{\textrm{\tiny AdS}}^2 }{G_{\textrm{\tiny N}} } 
\left[\,   \frac{b_\perp}{\varepsilon_{\textrm{\tiny AdS}}} 
\;_2F_1 \! \left(-\frac{1}{4},\frac{1}{2};\frac{3}{4};\frac{\varepsilon_{\textrm{\tiny AdS}} ^4}{w_*^4}\right)
-
\frac{\sqrt{\pi } \;\Gamma(3/4)\, b_\perp}{ \Gamma (1/4) \, w_*}  \,\right]
\eea
which is the expression found in \cite{Ryu:2006ef}.

\section{Schwarzschild AdS$_{d+2}$ black brane: Nesting for the strip}
\label{Nestingtotal}

\noindent


In this appendix,
considering the infinite strip $A$ and the gravitational background 
given by the constant time slice of the Schwarzschild AdS$_{d+2}$ black brane,
we discuss the nesting property 
for the geodesic bit threads 
and for the minimal hypersurface inspired bit threads.

\begin{figure}[t!]
	\vspace{-.5cm}
	\hspace{-.9cm}
	\includegraphics[width=0.55\textwidth]{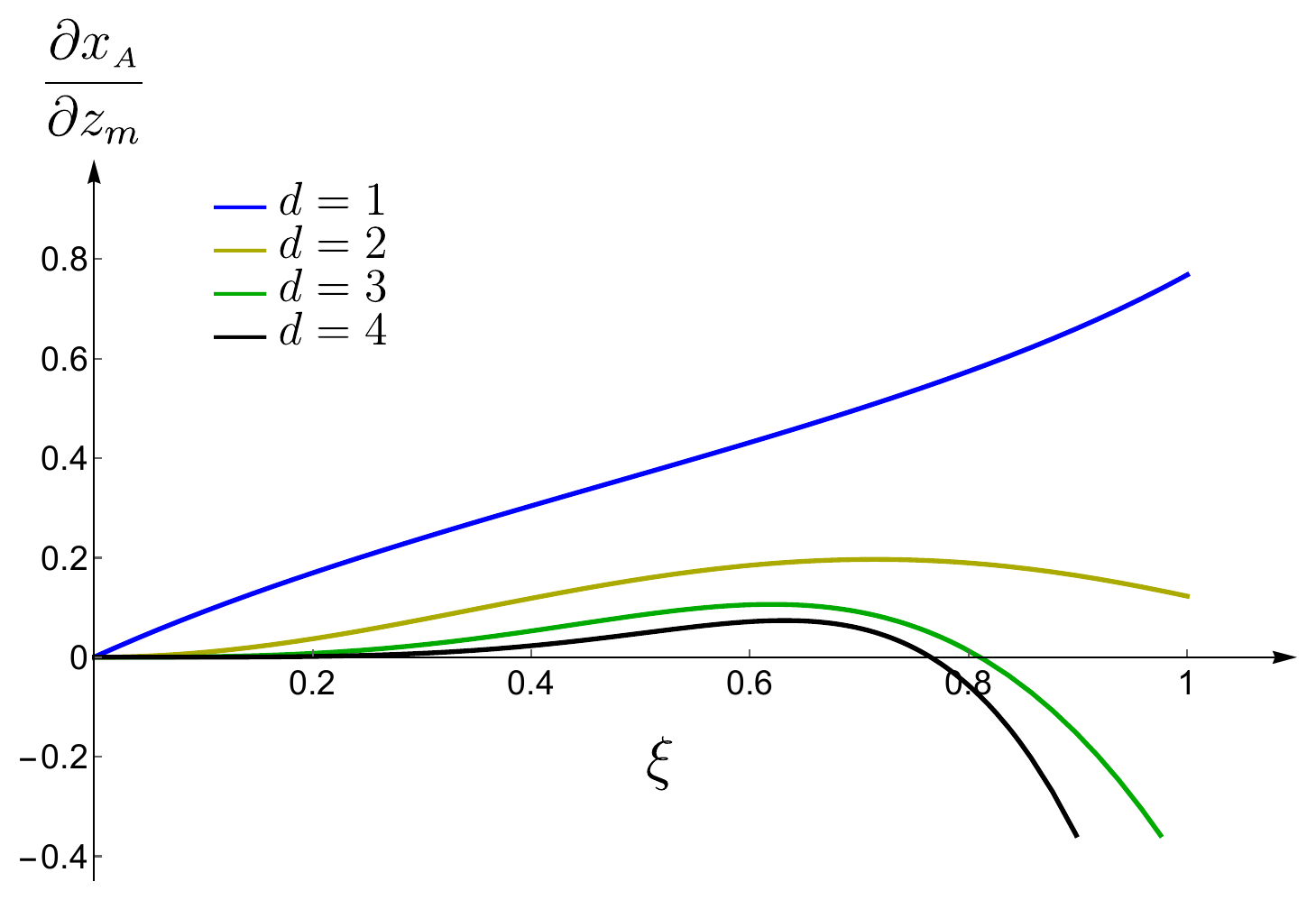}
	\vspace{.1cm}
	\includegraphics[width=0.55\textwidth]{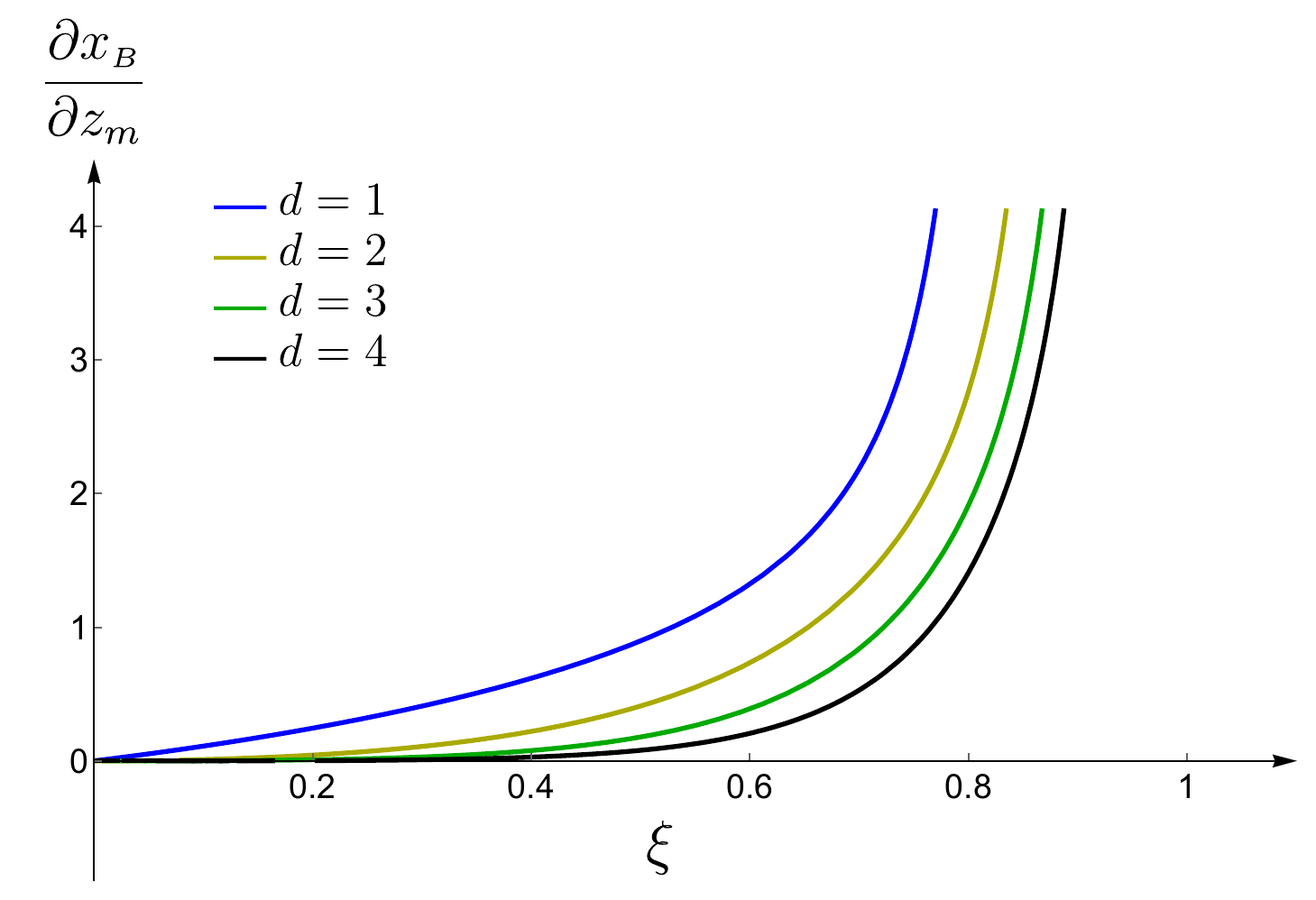}
	\caption{\label{fig:NestingstripBH}
		{\small  
		Nesting of the geodesic bit threads for the strip, in Schwarzschild AdS$_{d+2}$ black brane:
		$\partial_{z_m} x_{\textrm{\tiny{$A$}}}$  and $\partial_{z_m} x_{\textrm{\tiny{$B$}}}$
		(see (\ref{dyABHstrip}) and (\ref{dyBBHstrip}) respectively)
		for $\rho=2/3$ and different values of  $d$.
		The nesting property is satisfied only for $d=1$ and $d=2$.
		}
	}
\end{figure}

Let us focus first on the geodesic bit threads (see Sec.\,\ref{sec-gbt-strip-ddim}).
The nesting property can be studied 
from the sign of the derivative of the endpoint $x_{\textrm{\tiny{$A$}}}  \in A$ of the geodesic bit thread
w.r.t. $z_m$, 
which must be non-positive to have such a property satisfied. 
This derivative reads
\bea
\frac{\partial x_{\textrm{\tiny{$A$}}} }{\partial z_m} 
&=&
\frac{z_*^{2 d} \, z_h^{\frac{d+1}{2}}}{\sqrt{z_*^{2 d}-z_m^{2 d}}}
\left(
-\frac{ z_m^{-d} }{\sqrt{z_h^{d+1}-z_m^{d+1}} }
+
\int_0^{z_m}   \!\!
\frac{
	z \, z_m  \big[ (d-1) z_m^{2 d}+z_*^{2 d}\,\big] 
}{
	\sqrt{z_h^{d+1}-z^{d+1}}  \; \big[ z^2 z_m^{2 d}+z_*^{2 d} ( z_m^2-z^2 )\big]^{3/2}
}\; \rd z
\right) 
\nonumber
\\
& = &
-\frac{1}{\xi^{d}\,\sqrt{1-\xi ^{2 d}}} \; {\cal D}_{\textrm{\tiny $A$}}(\xi,\rho)
\label{dyABHstrip}
\eea
where in the last step we have introduced the dimensionless variables 
$\xi=z_m/z_*$, $\rho=z_*/z_h$, $t=z/z_m$
and also 
\be
{\cal D}_{\textrm{\tiny $A$}}(\xi,\rho)
\,\equiv \,
-\frac{1}{\xi^{d}\,\sqrt{1-\xi ^{2 d}}}
\left(
\frac{1}{ \sqrt{1-(\xi  \, \rho)^{d+1}}}
-
\int_0^1
\frac{\xi^d 
\big[(d-1)  \xi ^{2 d}+1\big] t
}{ 
\big[ t^2 \left(\xi ^{2 d}-1\right)+1\big]^{3/2} \,\sqrt{1-(\xi \, \rho \, t)^{d+1}}} 
   \;\rd t
   \right).
\ee
This expression remains consistently positive for $d\leqslant 2$ and any value of $\rho$; indeed
\be
{\cal D}_{\textrm{\tiny $A$}}(\xi,\rho)
\geqslant 
\frac{1}{\sqrt{1-(\xi \, \rho) ^{d+1}}}
\left( 1-
\int_0^1 \frac{\xi^d\,  \big[(d-1)  \xi ^{2 d}+1\big]\; t }{ \big[ (\xi ^{2d}-1)  t^2  +1 \big]^{3/2} }\; \rd t
\right)
=
\frac{\xi ^d \,\big[ 1+(1-d) \, \xi ^d \, \big] }{ \sqrt{1-(\xi \, \rho) ^{d+1}}\;(\xi ^d+1)}
\ee
whose r.h.s is non-negative both for $d=1$ and $d=2$.
However, for $d\geqslant 3$  the sign of ${\cal D}_{\textrm{\tiny $A$}}(\xi,\rho)$ is not always positive for arbitrary values of  $\rho$, 
as shown e.g. in the left panel of Fig.\,\ref{fig:NestingstripBH} for the special case of $\rho=2/3$. 
Hence, since for $d\geqslant 3$ the nesting property is not satisfied, 
the geodesics do not always provide well defined bit threads when $d\geqslant 3$.

The analysis of the nesting property requires to consider also 
the derivative of the endpoint $x_{\textrm{\tiny{$B$}}}  \in B$ of the geodesic bit thread w.r.t. $z_m$,
which must be always  positive. 
After a tedious but straightforward computation, we find
\be
\label{dyBBHstrip}
\frac{\partial x_{\textrm{\tiny{$B$}}}}{\partial z_m} 
\,=\,
\frac{1}{\xi^{d} \big(1-\xi^{2 d} \big)^{3/2}} \; {\cal D}_{\textrm{\tiny $B$}}(\rho,\xi)
\ee
where
\bea
\label{RB-def-app}
{\cal D}_{\textrm{\tiny $B$}}(\rho,\xi)
&\equiv &
\frac{ (2 d-1)\xi ^{2 d}+1 }{\sqrt{1-(\xi \,\rho)^{d+1}}} 
- 
\int_0^{1}\! \frac{t \,\xi ^{d} \big[ (d-1) \xi ^{2 d}+1\big] \big(1-\xi ^{2 d}\big) 
}{ 
	\big[ 1-t^2 (1-\xi ^{2 d})\big]^{3/2} \sqrt{1-(\xi \, \rho\, t)^{d+1}}
} \;\rd t 
\\
\rule{0pt}{.9cm}
& &
\hspace{3cm}
+ 
\int_{1}^{ \frac{1 }{ \sqrt{1 - \xi^{2d} }}}  
\frac{(d+1) \, \xi ^{2d+1} \big[ (d-1) \xi ^{2 d}+1\big]  \rho ^{d+1} \, t^d
}{ 
	\sqrt{1-t^2 \left(1-\xi ^{2 d}\right)} \, \big[ 1-(\xi \,\rho \,t)^{d+1}\big]^{3/2}
} \; \rd t
\nn
\eea
and $t$, $\rho$ and $\xi$ denote the same dimensionless variables introduced in (\ref{dyABHstrip}).
Since the last integral in the r.h.s. of (\ref{RB-def-app}) is positive, 
we can write the following inequality
\bea
\label{RB-ineq-app}
& &\hspace{-.4cm}
{\cal D}_{\textrm{\tiny $B$}}(\rho,\xi)
\geqslant 
\frac{(2 d-1) \xi ^{2 d}+1}{\sqrt{1-(\xi \, \rho)^{d+1}}} 
- 
\int_0^{1}  
\frac{t \, \xi ^{d} \big[(d-1) \xi ^{2d}+1\big] (1-\xi ^{2 d})}{ \big[ 1-t^2 \left(1-\xi ^{2 d}\right)\big]^{3/2}\sqrt{1-(\xi \,\rho\, t)^{d+1}}}  
\, \rd t
\\
\rule{0pt}{.8cm}
& &
=
\frac{\xi ^{2 d} \big[(d-1) \xi ^{2 d}+d+1\big] }{\sqrt{1-\xi ^{d+1} \rho ^{d+1}}}+\int_0^1 
\frac{t \,\xi ^d (1-\xi ^{2 d}) \big[ (d-1) \xi ^{2 d}+1\big] \big[ (d-1) (\xi  \rho  t)^{d+1}+2\big] }{
	2 \sqrt{t^2 \left(\xi ^{2 d}-1\right)+1} \; \big[1-(\xi  \rho  t)^{d+1}\big]^{3/2}} \; \rd t 
\, \geqslant \,
0
\nonumber
\eea
where the expression in the second line is obtained 
through an integration by parts 
based on the factor $1/ \big[ 1-t^2 \left(1-\xi ^{2d}\right)\big]^{3/2}$. 
The last inequality in (\ref{RB-ineq-app}) holds for any $d\geqslant 1$ because both terms are positive. 
A numerical analysis shown in the right panel of Fig.\,\ref{fig:NestingstripBH}, where $\rho=2/3$,
confirms that ${\cal D}_{\textrm{\tiny $B$}}(\rho,\xi)$ is consistently positive for the values of $d$ considered.


In the remaining part of this Appendix, we focus on the nesting property for the 
minimal hypersurface inspired bit threads 
considered in  Sec.\,\ref{sec-strip-ddim-mhbt}.

We find that $\partial_{z_m} x_{\textrm{\tiny{$A$}}}$ can be expressed as 
\be
\label{dyAMSbit}
\frac{\partial x_{\textrm{\tiny{$A$}}} }{\partial z_m}
=
-\frac{1}{\xi^d \,\sqrt{1-\xi ^{2 d}} \, \sqrt{1-(\xi \,\rho) ^{d+1}}}
+
\int_0^1\!
\frac{ d\;  t^d}{
	\sqrt{1-\xi ^{2 d}} \, \big[ 1- (1-\xi ^{2 d}) \,t^{2 d}\big]^{3/2} \sqrt{1-(\xi \,\rho\, t)^{d+1}}
} 
\; \rd t\,. 
\ee
This expression is derived in two steps. 
Firstly, we calculate the derivative of $x_{\textrm{\tiny{$A$}}}=x_{\textrm{\tiny$<$}}(0)$ with respect to $z_m$, 
where $x_{\textrm{\tiny$<$}}(0)$ is evaluated by using \eqref{msbitbuconero} and \eqref{zsandc0}. 
Secondly, we express the outcome in terms of the dimensionless variables $t$, $\rho$, and $\xi$.

To demonstrate that the above derivative remains non-positive for any value of $d$ and of the parameters $t$, $\rho$ and $\xi$, 
let us first observe that
\begin{equation}
	\label{ineqMS1}
	\frac{t^d}{\sqrt{1-(\xi \,\rho \,t) ^{d+1}}} \leqslant \frac{t^{d-1}}{\sqrt{1-(\xi \,\rho) ^{d+1}}}
\end{equation}
which holds because all the parameters belong to the interval $[0,1]$. 
By leveraging the inequality \eqref{ineqMS1} and the positivity of the integral in \eqref{dyAMSbit}, we can write
\begin{equation}
	\label{C7}
	\frac{\partial x_{\textrm{\tiny{$A$}}} }{\partial z_m}
	\leqslant
	-\frac{1}{\sqrt{1-\xi ^{2 d}} \, \sqrt{1-(\xi \rho) ^{d+1}}}
	\left[\, 
	\frac{1}{\xi^d}-\int_0^1  \frac{d \; t^{d-1} }{\big[ 1- (1-\xi ^{2 d} ) \,t^{2 d}\big]^{3/2}
	}\;\rd t
	\,\right]=0
\end{equation}
which shows that the derivative is indeed nonpositive for all values of $d$, $t$, $\rho$ and $\xi$. 
Next, we consider the derivative of the other endpoint $x_{\textrm{\tiny{$B$}}}=x_{\textrm{\tiny$>$}}(0) \in B$ of the bit thread, 
as provided in \eqref{msbitbuconero}, with respect to $z_m$.
We find 
\bea
\frac{\partial x_{\textrm{\tiny{$B$}}} }{\partial z_m} 
&=&
\frac{ 1+\xi ^{2 d} }{ \xi^d\left(1-\xi ^{2 d}\right)^{3/2} \sqrt{1-(\xi \,\rho) ^{d+1}}}
-
\int_0^1  \frac{d\; t^d }{
	\sqrt{1-\xi ^{2 d}} \, \big[1-\left(1-\xi ^{2 d}\right) t^{2 d}\big]^{3/2}
	\sqrt{1-(\xi \,\rho\, t)^{d+1}}} \; \rd t
\nonumber\\
\rule{0pt}{.8cm}
& &
+
\int ^1_{\frac{1}{\sqrt[2d]{1+\xi ^{2 d}}}} 
\frac{ (\xi  \, \rho \,  t)^{d+1}(3 d-1)-2 (d-1) }{ t^d\,\big(1-\xi ^{2 d}\big)^{3/2} \sqrt{1-\left(1-\xi ^{2
			d}\right) t^{2 d}} \; \big(1-(\xi  \,\rho \, t)^{d+1}\big)^{3/2}} \; \rd t \,.
\eea

To demonstrate the positivity of this expression, 
we can omit the second integral in the r.h.s., 
which  is evidently positive 
because its integrand is positive within the integration interval.
By applying the same bounding technique 
discussed in the previous inequality to the first integral,
we obtain
\begin{equation}
	\label{C9}
	\frac{\partial x_{\textrm{\tiny{$B$}}} }{\partial z_m} 
	\, \geqslant \, 
	\frac{2\xi^d }{\left(1-\xi ^{2 d}\right)^{3/2} \sqrt{1-(\xi \, \rho )^{d+1}}} \, \geqslant \, 0 \,.
\end{equation}
The results \eqref{C7} and \eqref{C9} tell us that the underlying curves 
are good candidates for being trajectories of proper bit threads.


\section{Translated and dilated bit threads in BTZ black brane} 
\label{app-traslatipedraza}

At the beginning of Sec.\,\ref{sec5} we mentioned 
the equivalence of the 
minimal hypersurface inspired bit threads \eqref{msgeneralcurves}
and the translated and dilated bit threads of \cite{Agon:2018lwq} 
in AdS$_{d+2}$.
In this appendix, we show that this equivalence does not hold 
in the constant time slice of the Schwarzschild AdS$_{d+2}$ black brane (see (\ref{sch-ads-brane-metric})).
For the sake of simplicity, we focus on the $d=1$ case
where the subsystem $A$ is an interval of length $2b$.

The RT curve $\gamma_A$ in the BTZ black brane geometry (\ref{btz-brane-metric}) 
with depth $z_*$ and centered in $x=0$ can be written for positive values of $x_m(z_m)$ as follows
\begin{equation}
	x_m(z_m)
	=
	z_h \; \mathrm{arcsinh} \Bigg( \sqrt{\frac{ z_*^2-z_m^2}{z_h^2-z_*^2} } \; \Bigg) \,.
	\label{supminimatraslatidilatati}
\end{equation}
This expression is essentially the inverse of \eqref{RT-curve-btz-brane} for positive $x$, but parameterized in terms of the maximal depth rather than the length of the interval $2b$. The second branch of this curve, valid for negative values of $x_m(z_m)$, is obtained through reflection symmetry around the axis $x=0$.
The expression \eqref{supminimatraslatidilatati} provides  
the building block to construct the integral curves of the translated and dilated bit threads,
which are written as follows
 (see Eq.\,(2.26) of \cite{Agon:2018lwq})
\begin{equation}
	x_{\textrm{\tiny$\gtrless$}}(z)=c_0 \pm \alpha \, x_m (z /\alpha) 
	\label{bittraslatidilatati}
\end{equation}
where $c_0$ is a term that shifts $\gamma_A$ along the $x$-axis
and $\alpha$ parameterizes the maximum height $\tilde{z}_* \equiv \alpha z_* $ of the curve.

This construction has suggested the one for 
minimal hypersurface inspired bit threads (see (\ref{msgeneralcurves})), 
with the crucial difference that the parameter $\alpha$ occurs in (\ref{bittraslatidilatati}).
The putative bit threads in \eqref{bittraslatidilatati} consist of two branches that share an endpoint at $z=\tilde{z}_\ast$;
hence, it is convenient to describe them separately.
Let us denote by $x_{\mbox{\tiny$<$}}(z)$ the branch corresponding to the minus sign, originating from the interval $A$, 
and by $x_{\mbox{\tiny$>$}}(z)$ the branch associated with the plus sign, originating from $\tilde z_\ast$ and extending to the complementary region $B$.

The parameters $c_0$ and $\tilde{z}_*$ are obtained through the standard procedure, i.e.
by imposing that the integral curves \eqref{bittraslatidilatati} intersect $\gamma_A$ orthogonally at the point $\big( z_m(x_m) , x_m\big)$. 
This gives
\begin{equation}
	\left\{ \begin{array}{l}
		z(x_m) = z_{m}(x_m)
		\\ 
		\rule{0pt}{.5cm}
		\big[ \,g_{zz} +  g_{xx} \, x’_m(z) \,x’_{m}(z) \,\big] \big|_{(z,x)=(z_m(x_m),x_m)} = 0
	\end{array}\right.
	\label{ortostrisciatraslati}
\end{equation}
where $g_{xx} = L^2_{\textrm{\tiny AdS}} / z^2$ 
and $g_{zz}= L^2_{\textrm{\tiny AdS}} /[ z^2 f(z) ]$ are  the diagonal components of the metric \eqref{sch-ads-brane-metric}. 
Solving \eqref{ortostrisciatraslati} gives
the depth $\tilde{z}_*$ of each integral curve
\begin{equation}
	\tilde{z}_*=  \frac{z_m \,\sqrt{z_h^2 \,z_*^2-z_m ^4-z_m ^2\, z_*^2+z_*^4 + \sqrt{\big(z_*^2 \,z_h^2-z_m^4-z_*^2 \,z_m^2+z_*^4\big)^2 - 4 z_*^2 \,z_h^2 \big(z_m^2-z_*^2\big)^2}}}{z_h \sqrt{2 \big(z_*^2-z_m ^2\big)}}
	\label{ztildetraslati}
\end{equation}  
and the center 
\begin{equation}
	c_0 = x_m(z_m)+ \alpha \, x_m (z_m / \alpha )\,.
\end{equation}
\begin{figure}[t!]
		\vspace{-.5cm}
		\hspace{.7cm}
	\includegraphics[width=0.9\textwidth]{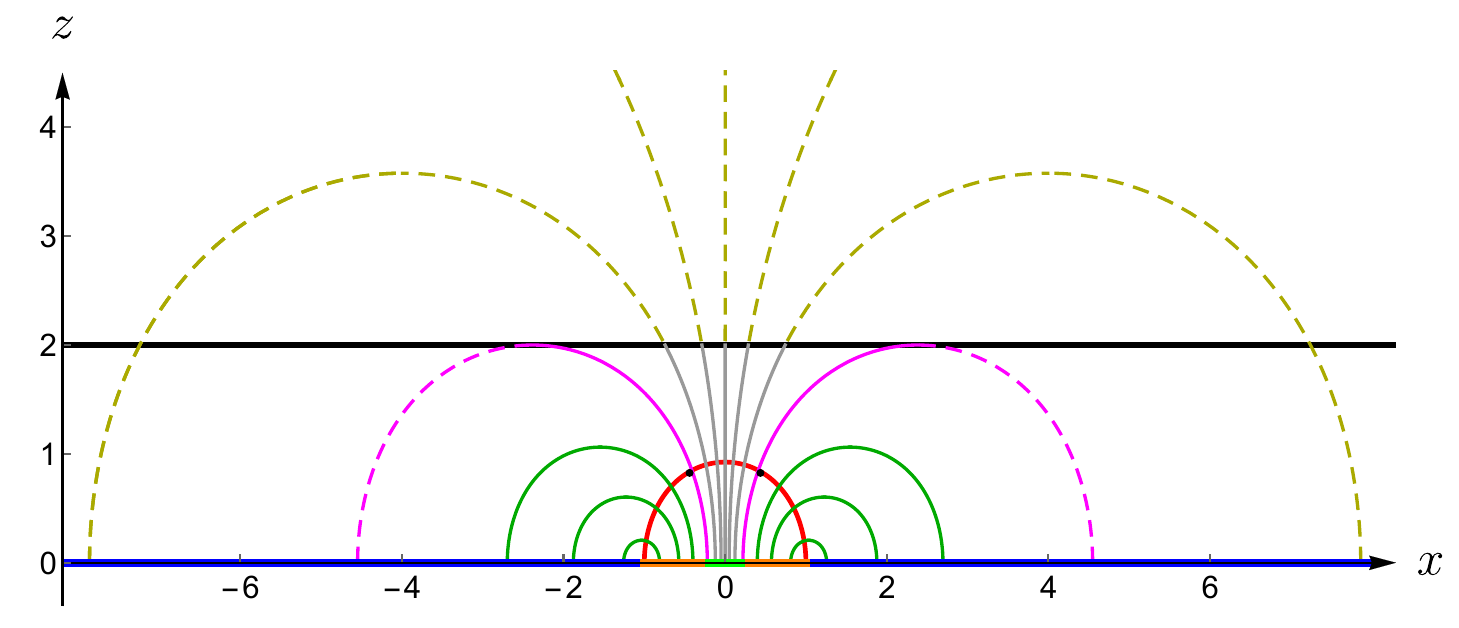}	\vspace{-.0cm}
	\caption{\small
	Integral lines for the 
	translated and dilated bit threads of an interval $A$ in the BTZ black hole. 
	The RT curve $\gamma_A$ corresponds to the solid red line. 
	The solid grey curves hit the horizon while the solid green curves connect $A$ and its complement $B$.
	The critical trajectories (magenta curves) 
	reach the horizon for a finite value of $x$,
	differently from the corresponding curves 
	for the geodesic bit threads 
	and the minimal hypersurface inspired bit threads (see Fig.\,\ref{fig:BTZ-brane-main}, bottom left panel).
	The black dots identify the portion $\gamma_{A,\beta} \subsetneq \gamma_A$ whose area is equal to $S_{A,\textrm{\tiny th}}$.
	The dashed dark yellow lines correspond to the extension of the solid grey curves.
	}
	\label{fig:TDbitintegrallinesbh}
\end{figure}
Thus, the integral curves in \eqref{bittraslatidilatati} can be written as follows
\begin{equation}
	x_{\textrm{\tiny$\gtrless$}}(z)
	=
	x_m(z_m)
	+ 
	z_h \, \frac{\tilde{z}_*}{z_*} \, 
	\arcsinh\! \left( \frac{z_* }{\tilde{z}_* } \, \sqrt{\frac{\tilde{z}_*^2-z_m^2}{z_h^2-z_*^2}} \;\right)
	\pm 
	z_h \,\frac{\tilde{z}_*}{z_*}  \, 
	\arcsinh\! \left( \frac{z_* }{\tilde{z}_* } \, \sqrt{\frac{\tilde{z}_*^2-z^2}{z_h^2-z_*^2}} \;\right)
	\label{bittraslativersionefinale}
\end{equation}
and some representatives are shown in Fig.\,\ref{fig:TDbitintegrallinesbh}. 
The first integral curve that reaches the horizon is obtained by imposing that 
the depth of the curves given in \eqref{ztildetraslati} equals $z_h$. 
Denoting by $z_{m,\beta}$ the value of $z_m$ defined by this condition, we find
\begin{equation}
	z_{m,\beta}= \sqrt{\frac{\sqrt{z_h^8+4 \,z_h^6 \, z_*^2 - 2 \, z_h^4 \, z_*^4+z_*^8}-z_h^4-z_*^4}{2( z_h^2- z_*^2)}} \;.
\end{equation}

These putative bit threads are different from 
the minimal hypersurface inspired bit threads and the geodesic bit threads.
Indeed, e.g. we have that their integral lines cross the horizon
(see the grey curves in Fig.\,\ref{fig:TDbitintegrallinesbh}).
Moreover, the first curves reaching the horizon 
(see the solid magenta curves in Fig.\,\ref{fig:TDbitintegrallinesbh})
arrive at the horizon at a finite value of the $x$-coordinate. 
This implies that the threads connecting $A$ and $B$ cover only a finite proper subset of $B$
(in Fig.\,\ref{fig:TDbitintegrallinesbh} it corresponds to the domain complementary to $A$
in the segment identified by the endpoints of the dashed magenta curves on the boundary).
The black dots in Fig.\,\ref{fig:TDbitintegrallinesbh} 
single out a portion $\gamma_{A,\beta} \subsetneq \gamma_A$ of finite length, 
whose area provides the holographic thermal entropy $S_{A,\textrm{\tiny th}}$. 
Since  the magenta geodesics do not intersect $\gamma_A$ at these points,
we have that $S_{A,\textrm{\tiny th}} \neq \widetilde S_{A,\textrm{\tiny th}}$,
being $\widetilde S_{A,\textrm{\tiny th}}$ defined as
the flux through the entire horizon.
In Fig.\,\ref{fig:TDbitintegrallinesbh},
the dashed dark yellow lines are the extension of the solid grey curves beyond the horizon. 
They are the analogue of the auxiliary geodesics displayed in
Fig.\,\ref{fig:BTZ-brane-main}, Fig.\,\ref{fig:bbetasphere}, 
Fig.\,\ref{GeoandModulusBHD} and Fig.\,\ref{fig:msbitintegrallinesbh},
and, in contrast with them, 
these curves probe the interior of the black hole before arriving to the boundary.

\begin{figure}[t!]
		\vspace{-.5cm}
		\hspace{.7cm}
	\includegraphics[width=0.9\textwidth]{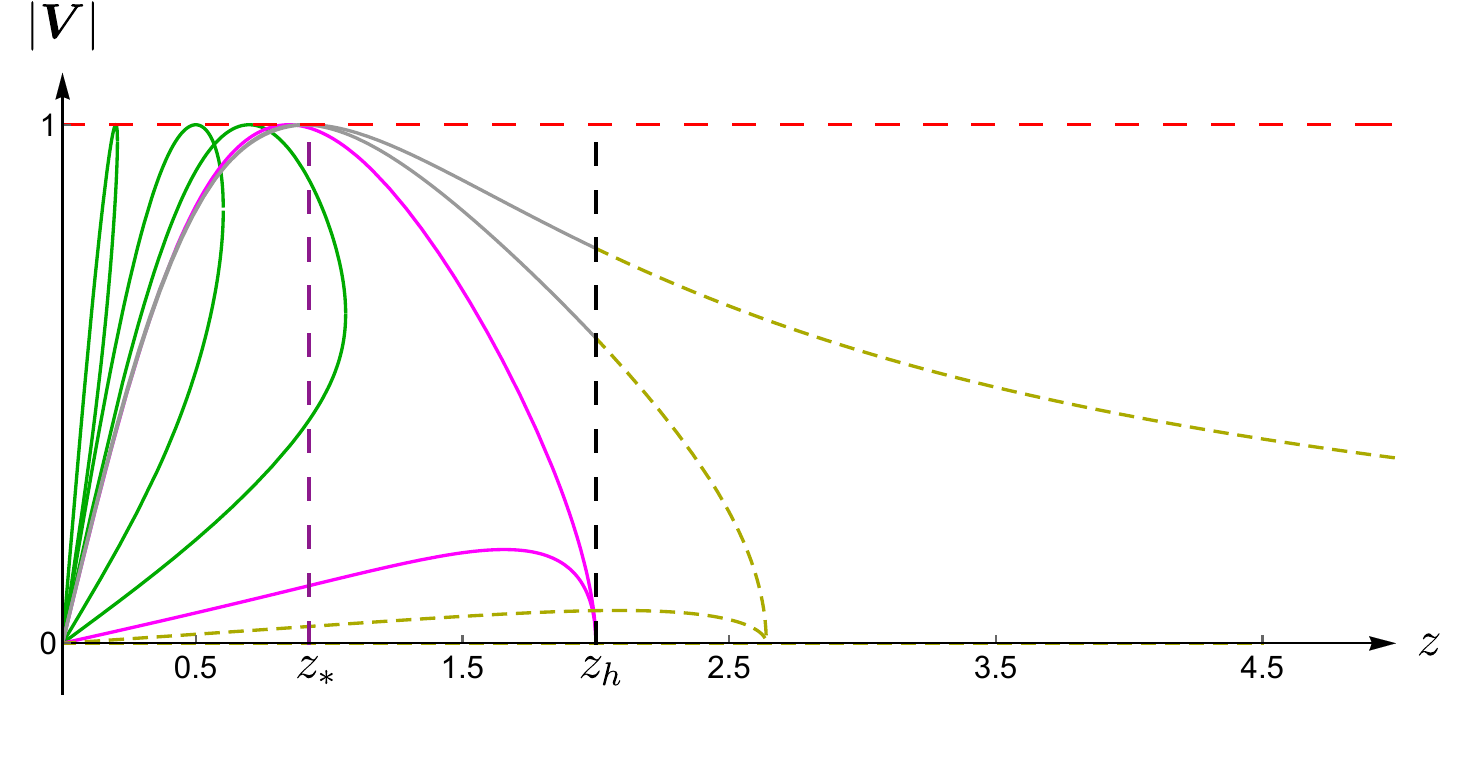}
			\vspace{-.3cm}
	\caption{\small Magnitude of the vector field $\boldsymbol{V}$ 
	along various translated and dilated bit threads of an interval $A$ in the BTZ black hole
	(see \eqref{modulotraslatidilatati}).
	}
	\label{fig:TDbitmodulo}
\end{figure}

Following the procedure described in Appendix\;\ref{app-modulus}, we can compute the modulus of the integral curves \eqref{bittraslativersionefinale}, finding 
\begin{equation}
	{|\boldsymbol{V}_{\!\!\textrm{\tiny$\lessgtr$}}|}
	=
	\left|\, 
	\frac{z}{z_m}\;
	\sqrt{
		\frac{\big( z_m^2-\tilde{z}_*^2 \big)\, \big( z_m^2 z_*^2-z_h^2 \tilde{z}_*^2 \big)\, \big( z_h^2 \tilde{z}_*^4+z^4 z_*^2-z^4 \tilde{z}_*^2-z^2 z_*^2 \tilde{z}_*^2 \big) 
		}{ 
			\big( z^2-\tilde{z}_*^2 \big)\, \big( z^2 z_*^2-z_h^2 \tilde{z}_*^2 \big)\, \big( z_h^2 \tilde{z}_*^4+z_m^4 z_*^2-z_m^4 \tilde{z}_*^2-z_m^2 z_*^2 \tilde{z}_*^2 \big)  }
	}\;
	\frac{\big(\partial_{z_m} x_{\textrm{\tiny$<$}}\big)\big|_{z=z_m} }{ \partial_{z_m} x_{\textrm{\tiny$\lessgtr$}}} 
	\,\right|  
	\label{modulotraslatidilatati}
\end{equation}
where $|\boldsymbol{V}_{\!\!\!\textrm{\tiny$<$}}|$ and $|\boldsymbol{V}_{\!\!\!\textrm{\tiny$>$}}|$
refer to the magnitude corresponding to the minus and plus branches, respectively. 
In Fig.\,\ref{fig:TDbitmodulo} we show the magnitude \eqref{modulotraslatidilatati} for some choices of $z_m$ as  $z$ varies,
verifying numerically that $|\boldsymbol{V}| \leqslant 1$ holds, as expected for well defined bit threads.

The holographic thermal entropy is $S_{A,\textrm{\tiny th}}= s_{\textrm{\tiny th} }V_1= \tfrac{L_{\textrm{\tiny AdS}}}{4 G_{\textrm{\tiny N}}} \, \tfrac{2b}{z_h} $
(see (\ref{HigherDStefanBoltzmann}) for $d=1$)
in the case that we are considering,
where we remind that $2b$ is the width of the interval $A$.
It is worth comparing this holographic thermal entropy
with the flux of the vector field discussed above through the region on the boundary 
(see the green interval in Fig.\,\ref{fig:TDbitintegrallinesbh})
identified by the threads reaching the horizon 
(see the solid grey and magenta curves in Fig.\,\ref{fig:TDbitintegrallinesbh}),
denoted by $\widetilde S_{A,\textrm{\tiny th}}$ in the following. 
The most straightforward approach to this computation consists in 
evaluating the area of $\tilde{\gamma}_{A,\beta}$ enclosed by $z_*$ and $z_{m,\beta}$.
This gives
\begin{equation}
\widetilde S_{A,\textrm{\tiny th}}
	=
	\frac{L_{\textrm{\tiny AdS}}}{4 G_{\textrm{\tiny N}}} \; 
	2 \int_{z_{m,\beta}}^{z_*} \frac{1}{z} \,\sqrt{\frac{1}{z^{\prime }(x)^2} + \frac{1}{f(z)}} \;\rd z
	\,=\,
	\frac{L_{\textrm{\tiny AdS}}}{4 G_{\textrm{\tiny N}}} \;  
	2 \int_{z_{m\beta}}^{z_*}   \frac{z_*}{z} \;\frac{1}{\sqrt{f(z) \left(z_*^2-z^2 \right)}} \;\rd z 
	\label{quasiflussoTDbit}
\end{equation}
which can be found  analytically in terms of $\zeta\equiv b /z_h$ and reads
\begin{equation}
\widetilde S_{A,\textrm{\tiny th}}
	= \frac{L_{\textrm{\tiny AdS}}}{2 G_{\textrm{\tiny N}}} \;  
	\arcoth \left(\tanh (\zeta ) \,\sqrt{\frac{ 1+2 \cosh ^4(\zeta )-\sqrt{1+\sinh ^2(2 \zeta ) \left(1+\cosh ^4(\zeta )\right)}}{2
			\cosh ^4(\zeta )-\sqrt{1+\sinh ^2(2 \zeta ) \left(1+\cosh ^4(\zeta )\right)}}} \; \right) .
	\label{flussotermicoTDbit}
\end{equation}
Since this complicated function of $\zeta$ is definitely not a straight line with slope $L_{\textrm{\tiny AdS}} /(2 G_{\textrm{\tiny N}}) $,
we conclude that $\widetilde S_{A,\textrm{\tiny th}} \neq S_{A,\textrm{\tiny th}}$ 
for this class of bit threads.

\section{Relating Poincar\'e AdS$_3$ and BTZ black hole} 
\label{app-BTZ-global-CTmap}

\noindent
In this appendix, we explore the connection between the geodesic bit threads in the constant time slice of AdS$_3$ (see Sec.\,\ref{sec-AdS3})
and the maximal set of geodesic in the constant time slice of the BTZ black hole discussed in Sec.\,\ref{sec-BTZ-black-hole}
by employing a map reported in \cite{Carlip:1994gc},
which relates a region of AdS$_3$ to the exterior of the BTZ black hole. 

\begin{figure}[t!]
	\vspace{-.5cm}
	\hspace{-.1cm}
	\begin{minipage}{0.5\textwidth}
		\centering
		\includegraphics[width=.9\textwidth]{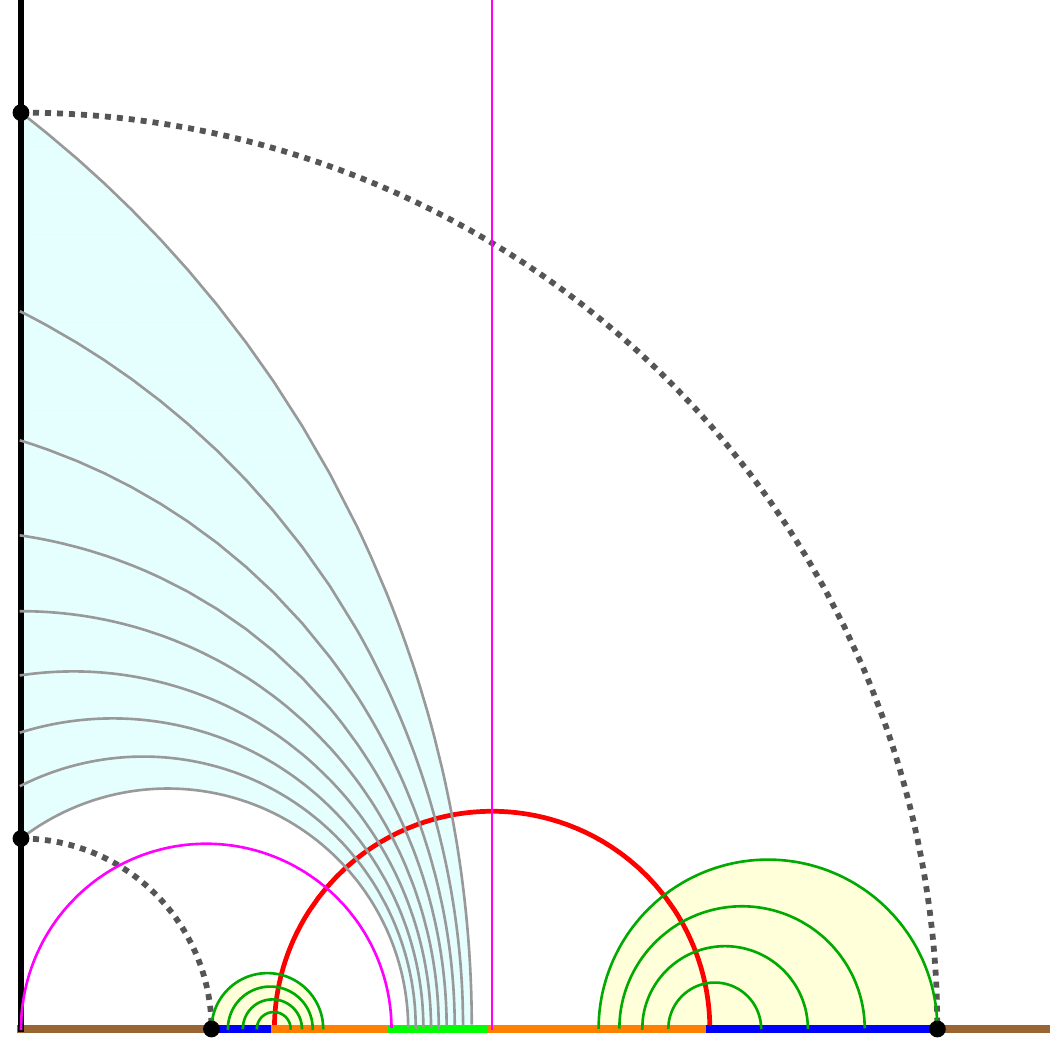}
	\end{minipage}
	\hfill 
	\begin{minipage}{0.5\textwidth}
		\centering
		\includegraphics[width=.9\textwidth]{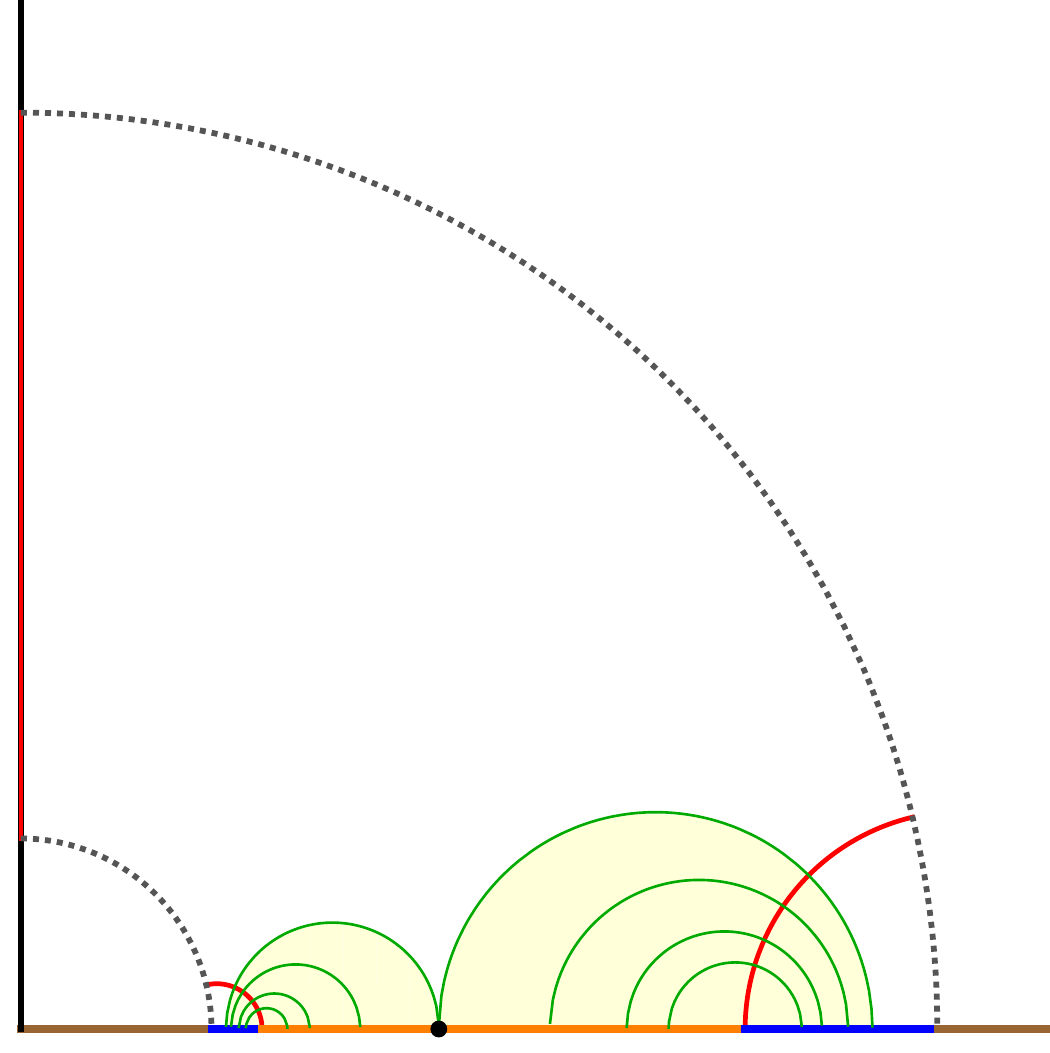}
	\end{minipage}
	\vspace{.2cm}
	\caption{\small
		Images through (\ref{CTmap-btz-global}) in the domain $\mathcal{D}$ 
		(contained in the constant time slice of Poincar\'e AdS$_3$)
		of the maximal sets of non-intersecting geodesics 
		shown
		in the left panel of Fig.\,\ref{fig-global-btz-threads-before} (left)
		and in the left panel of Fig.\,\ref{fig-global-btz-threads-after} (right).
	}
	\label{fig-global-to-ads-CT} 
\end{figure}

In the constant time slice of AdS$_3$ (see (\ref{H2-metric})),
let us consider the vertical half line $\mathcal{R}_0 $, whose points have $y=0$,
and the domain $\mathcal{D}$ defined by the points whose coordinates $(w,y)$
are such that $y>0$ and 
$\sqrt{\e^{-2 r_h \pi / L_{\textrm{\tiny AdS}} } -y^2} \leqslant  w \leqslant \sqrt{\e^{ 2 r_h \pi / L_{\textrm{\tiny AdS}} } -y^2}$
and by the identification of the two intersection points 
of the concentric arcs of circumferences $w = \sqrt{\e^{\pm 2 r_h \pi / L_{\textrm{\tiny AdS}} } -y^2}$ 
with a half line starting from their common center.
The following change of coordinates \cite{Carlip:1994gc} 
\be
y =  \sqrt{1- (r_h/r )^2 } \;  \e^{ r_h \phi / L_{\textrm{\tiny AdS}} }
\;\;\;\qquad\;\;\;
w = \frac{r_h}{r} \;  \e^{ r_h \phi / L_{\textrm{\tiny AdS}} }
\label{CTmap-btz-global}
\ee 
sends $\mathcal{D}$ equipped with (\ref{H2-metric})
onto the constant time slice of the BTZ black hole geometry \eqref{btz-global-metric}.
Notice that (\ref{CTmap-btz-global}) maps $\mathcal{R}_0 \cap \partial \mathcal{D}$  
into the horizon of the BTZ black hole.
In Fig.\,\ref{fig-global-to-ads-CT}, 
the concentric arcs of circumferences given by $w = \sqrt{\e^{\pm 2 r_h \pi / L_{\textrm{\tiny AdS}} } -y^2}$ 
provide the dotted grey arcs,
which correspond also to the grey dotted curves in left panel of  Fig.\,\ref{fig-global-btz-threads-before} 
and in both panels of Fig.\,\ref{fig-global-btz-threads-after}.

The arc $A = ( -\phi_b, \phi_b) $ on the boundary of the BTZ black hole geometry \eqref{btz-global-metric}
is sent by \eqref{CTmap-btz-global} onto the interval of $\partial \mathcal{D}$ on the boundary of AdS$_3$,
namely $\tilde{A} \equiv \big(  \e^{- r_h \phi_b/ L_{\textrm{\tiny AdS}} },  \e^{ r_h \phi_b/ L_{\textrm{\tiny AdS}} } \big) $.
Moreover, all the geodesics of the BTZ black hole introduced in Sec.\,\ref{sec-BTZ-black-hole}
are mapped by \eqref{CTmap-btz-global}  into arcs of geodesics in the constant time slice of AdS$_3$.
This includes the curves in $\gamma_A$ and geodesics of the corresponding maximal sets.
Thus, \eqref{CTmap-btz-global} allows us to describe through geodesics in $\mathcal{D}$
all the features of the geodesics in the BTZ black hole
discussed in Sec.\,\ref{sec-BTZ-black-hole}.
In particular, the homology constraint plays a crucial role to find 
the RT curve  for $\tilde{A}$ in $\mathcal{D}$.
When $\phi_b \leqslant \phi_b^*$ with $\phi_b^*$ defined in \eqref{crit-angle-RT-global-btz}, 
$\gamma_{\tilde{A}}$ is the half circumference with diameter $\tilde{A}$
(see the red curve in the left panel of Fig.\,\ref{fig-global-to-ads-CT}),
while for $\phi_b \geqslant \phi_b^*$ it is the union of two disconnected curves
given by the segment $\mathcal{R}_0 \cap \partial \mathcal{D}$ 
and by the curve made by the two arcs starting at the endpoints of $\tilde{A}$
and ending orthogonally on the arcs corresponding to  $w = \sqrt{\e^{\pm 2 r_h \pi / L_{\textrm{\tiny AdS}} } -y^2}$ 
(see the red curves in the right panel of Fig.\,\ref{fig-global-to-ads-CT}).

As for the maximal sets of geodesics in the BTZ black hole discussed in Sec.\,\ref{sec-BTZ-black-hole},
for $\phi_b \leqslant \phi_b^*$ and $\phi_b \geqslant \phi_b^*$
in the left and right panel of Fig.\,\ref{fig-global-to-ads-CT} we show 
the images through (\ref{CTmap-btz-global}) of the curves 
in the left panel of Fig.\,\ref{fig-global-btz-threads-before} and Fig.\,\ref{fig-global-btz-threads-after}
respectively, by adopting the same color code.
Thus, it is straightforward to repeat in $\mathcal{D}$ the considerations made in Sec.\,\ref{sec-BTZ-black-hole}.	
\section{Geodesics winding around the horizon in BTZ black hole}
\label{app-LongGeodesics-GlobalBTZ}

In this appendix, we focus on a particular set of geodesics in the constant time slice of the BTZ black hole 
(see \eqref{btz-global-metric}). 
These geodesics extremize the length functional but are not, in general, global minima (see e.g. \cite{Balasubramanian:2014sra}). 
Specifically, we first describe 
the subset containing the geodesics with both endpoints on the boundary 
and then the subset containing geodesics with one endpoint on the boundary and the other one on the horizon.

\begin{figure}[t!]
	\begin{minipage}{0.3\textwidth}
		\centering
		\includegraphics[width=.9\textwidth]{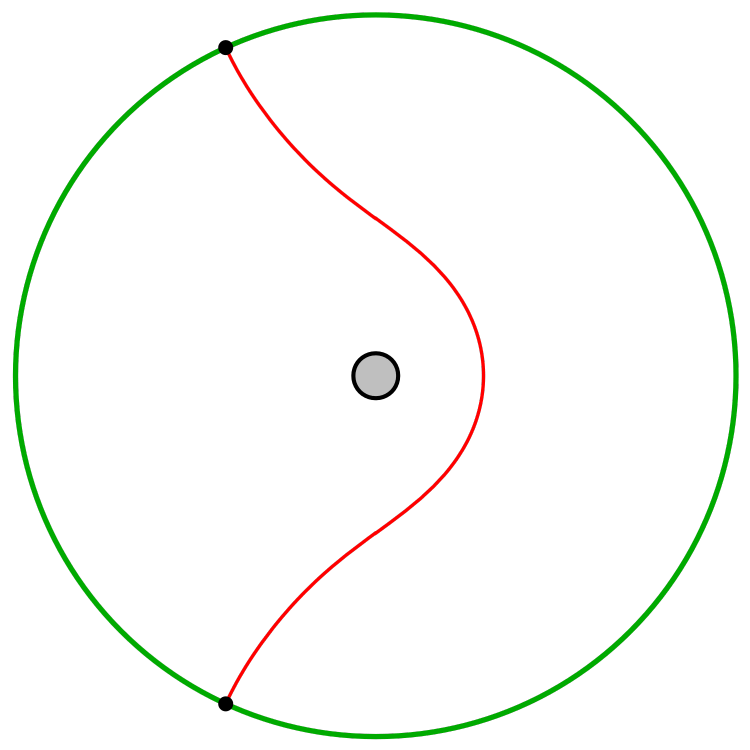}
	\end{minipage}\hfill
	\begin{minipage}{0.3\textwidth}
		\centering
		\includegraphics[width=.9\linewidth]{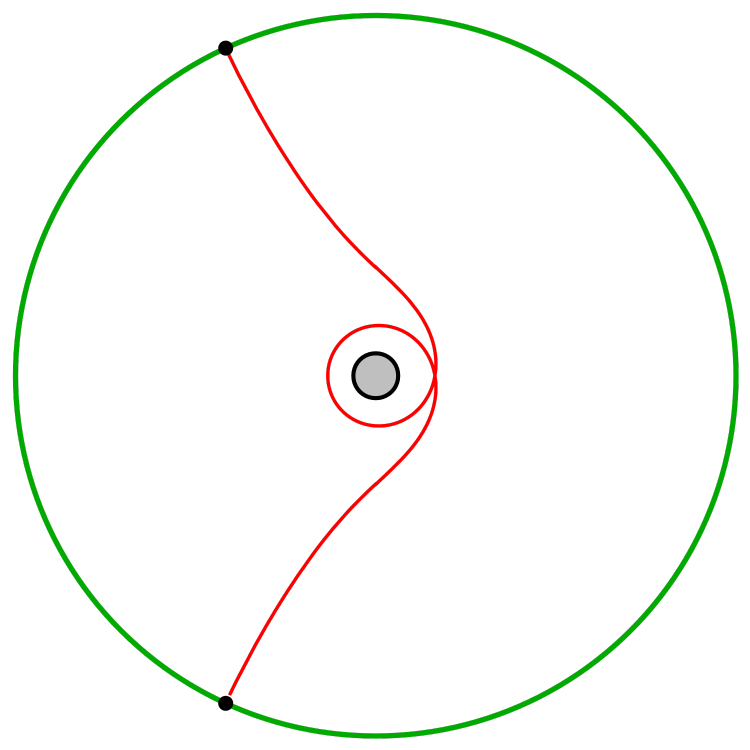}
	\end{minipage}
	\hfill
	\begin{minipage}{0.3\textwidth}
		\centering
		\includegraphics[width=.9\linewidth]{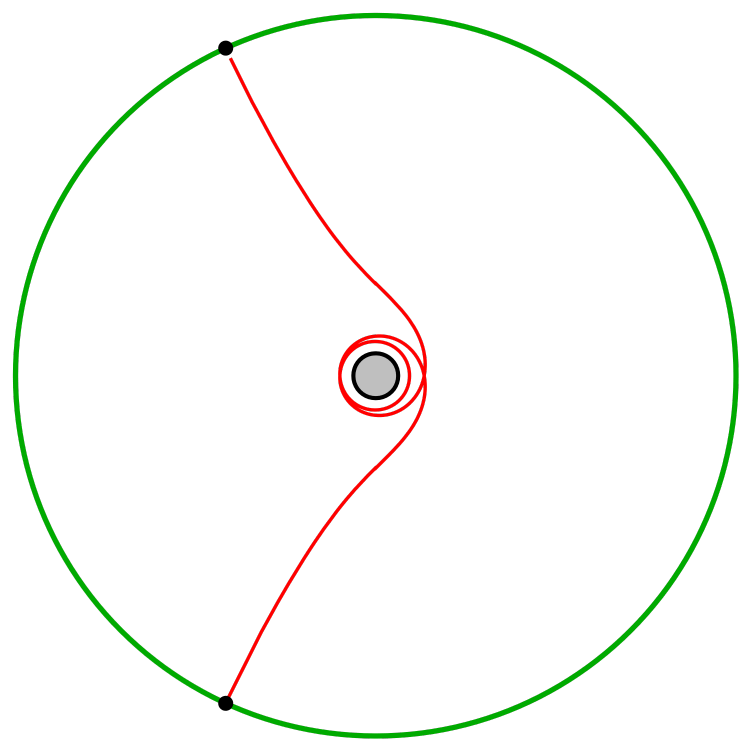}
	\end{minipage}
	\\ 
	\rule{0pt}{2.7cm}
	\begin{minipage}{0.3\textwidth}
		\centering
		\includegraphics[width=.9\textwidth]{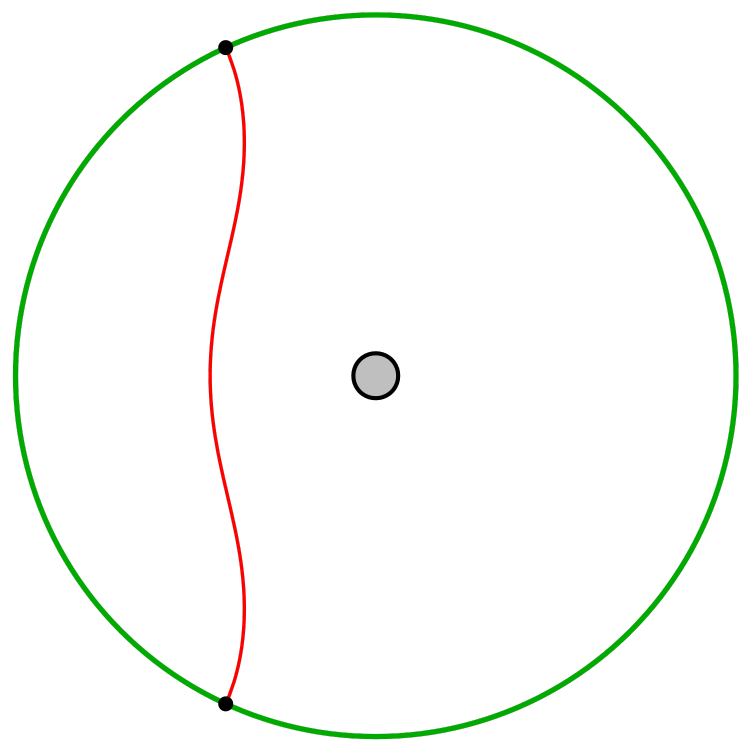}
	\end{minipage}\hfill
	\begin{minipage}{0.3\textwidth}
		\centering
		\includegraphics[width=.9\linewidth]{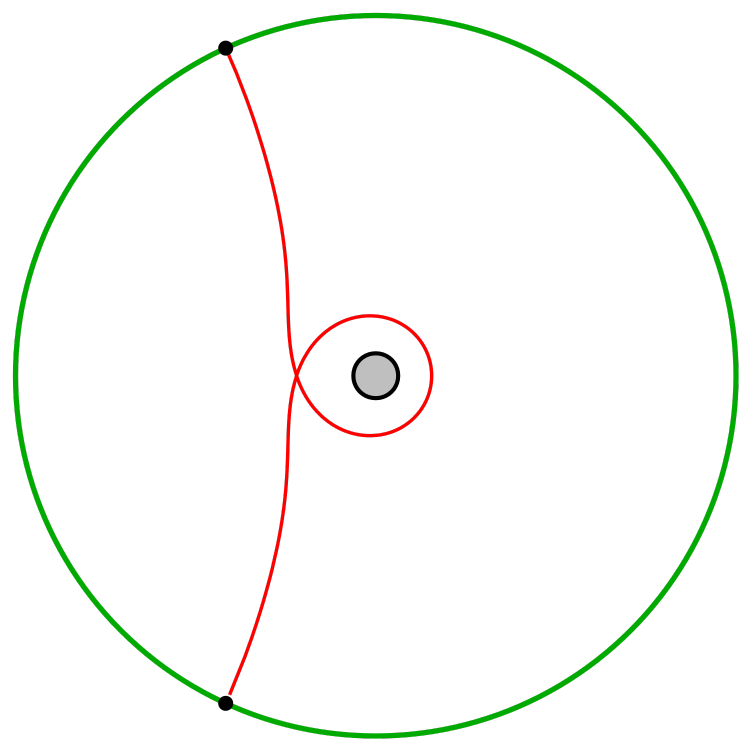}
	\end{minipage}
	\hfill
	\begin{minipage}{0.3\textwidth}
		\centering
		\includegraphics[width=.9\linewidth]{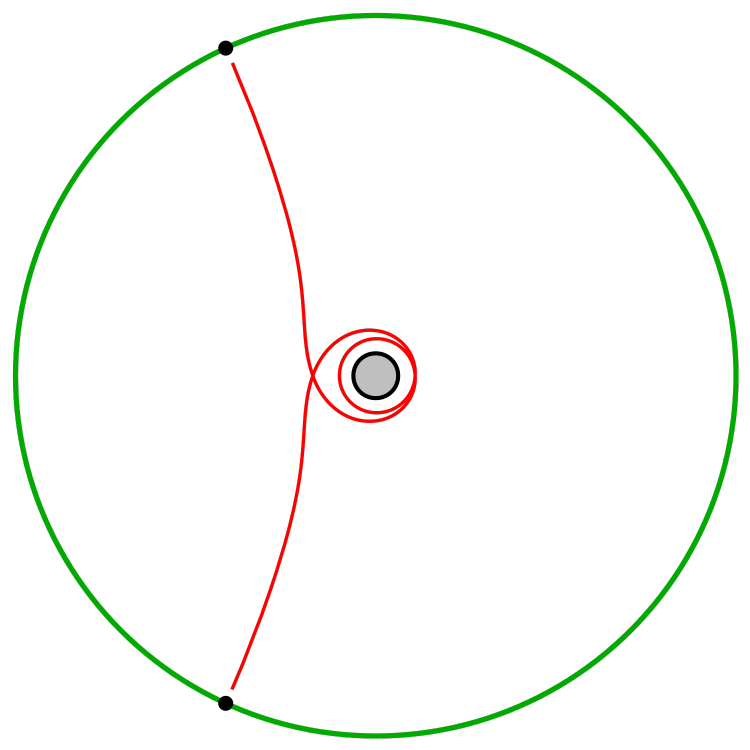}
	\end{minipage}
	\\
	\vspace{.3cm}
	\caption{\small
		Geodesics in the BTZ black hole \eqref{btz-global-metric} with both the endpoints on the boundary (black dots)
		winding either zero times (left) or one time (middle) or two times (right) around the horizon.}
	\label{fig-global-btz-long-geod-bdy-to-bdy}
\end{figure}

Concerning the geodesics whose endpoints are on the boundary,
with angular coordinates $\phi_1$ and $\phi_2$ belonging to the interval $(0, 2 \pi)$,
these curves  can be written in the form  \eqref{geo-btz-global-bh-2bdy}, 
where the integration constants $\theta_0$ and $\phi_0$ are  chosen to be 
\be
\label{long-geo-BTZ-1}
\theta_0 =  \frac{ |\phi_1 - \phi_2|}{2}  +   k  \pi  
\;\;\qquad\;\;
\phi_0  =  \frac{\phi_1 + \phi_2}{2}  +   k  \pi  
\;\;\qquad\;\;
k  \in \mathbb{Z}\,.
\ee
Some of these curves are shown in Fig.\,\ref{fig-global-btz-long-geod-bdy-to-bdy}.
Setting either $k =0$ or $k =-1$ in (\ref{long-geo-BTZ-1}),
one obtains the curves in the top left and bottom left panels of Fig.\,\ref{fig-global-btz-long-geod-bdy-to-bdy} respectively.
These geodesics do not wind around the horizon (i.e. their winding number is zero)
and occur in the evaluation of the holographic entanglement entropy
for the configurations displayed 
in Fig.\,\ref{fig-global-btz-threads-before} (red curve)
and Fig.\,\ref{fig-global-btz-threads-after} (red curve anchored to the boundary)
respectively.

Local extrema of the length functionals with nontrivial winding number 
are obtained by choosing 
other values of $k$  in (\ref{long-geo-BTZ-1}).
For instance, choosing either $k =1$ or $k= -2$  in (\ref{long-geo-BTZ-1}), 
we find the geodesics winding one time 
around the horizon. 
They are displayed respectively in the upper and lower middle panels of Fig.\,\ref{fig-global-btz-long-geod-bdy-to-bdy}.
While for either $k =2$ or $k =-3$ in (\ref{long-geo-BTZ-1})
we get the geodesics winding two times 
around the horizon. They are drawn respectively in the top right and bottom right panel of Fig.\,\ref{fig-global-btz-long-geod-bdy-to-bdy}.
%
 Counting how many times one of these geodesics wraps the horizon is straightforward. 
 Given the length  $2|\theta_0|$ of the interval covered by
$\phi$ in  \eqref{geo-btz-global-bh-2bdy}, 
when we move from one endpoint to the other,
which are reached when $\phi$ is equal either to $\theta_0+\phi_0$ or to $\phi_0-\theta_0$,
the number of times $w_{\text{\tiny I}}$ that a geodesic winds the horizon is simply the number of times that the interval $2\pi$ is strictly contained in $2|\theta_0|$,
namely
\be
w_{\text{\tiny I}}
=\left\lfloor\frac{2 |\theta_0|}{2\pi}\right\rfloor
= 
\left\lfloor \, \left |\frac{ |\phi_1-\phi_2|}{2\pi}+k\right| \, \right\rfloor
=
\left\{\begin{array}{cll} 
k & \hspace{.7cm} & k\geqslant 0
\\
\rule{0pt}{.5cm}
\!\! -\!(k+1) & & k<0
\end{array}\right.
\ee
where $\lfloor \cdots \rfloor$ denotes the integer part of a number.

\begin{figure}[t!]
	\begin{minipage}{0.3\textwidth}
		\centering
		\includegraphics[width=.9\textwidth]{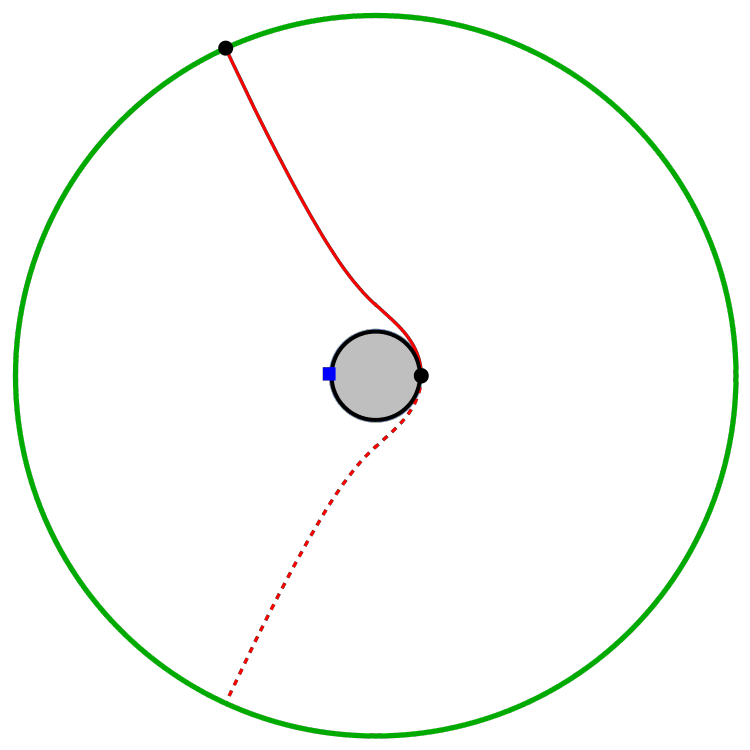}
	\end{minipage}\hfill
	\begin{minipage}{0.3\textwidth}
		\centering
		\includegraphics[width=.9\linewidth]{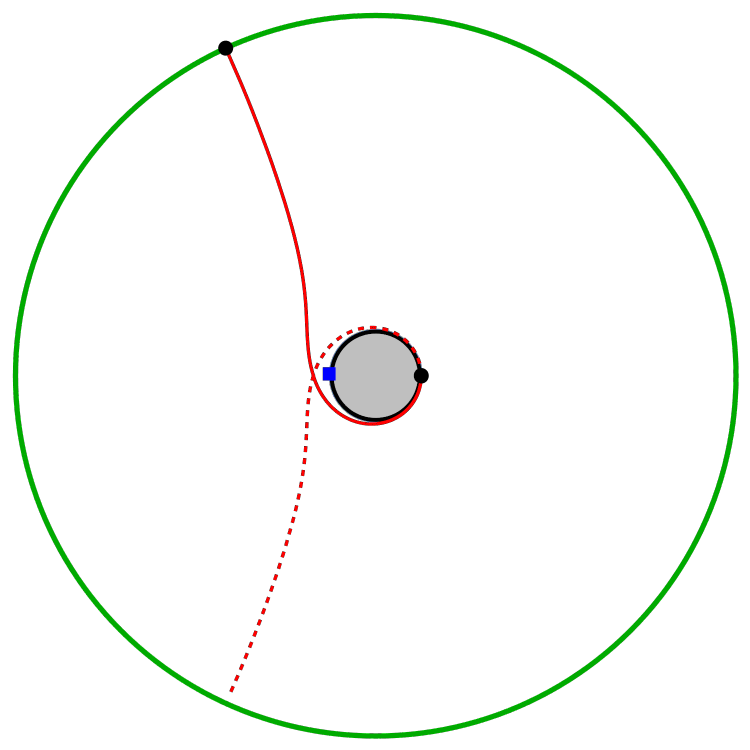}
	\end{minipage}
	\hfill
	\begin{minipage}{0.3\textwidth}
		\centering
		\includegraphics[width=.9\linewidth]{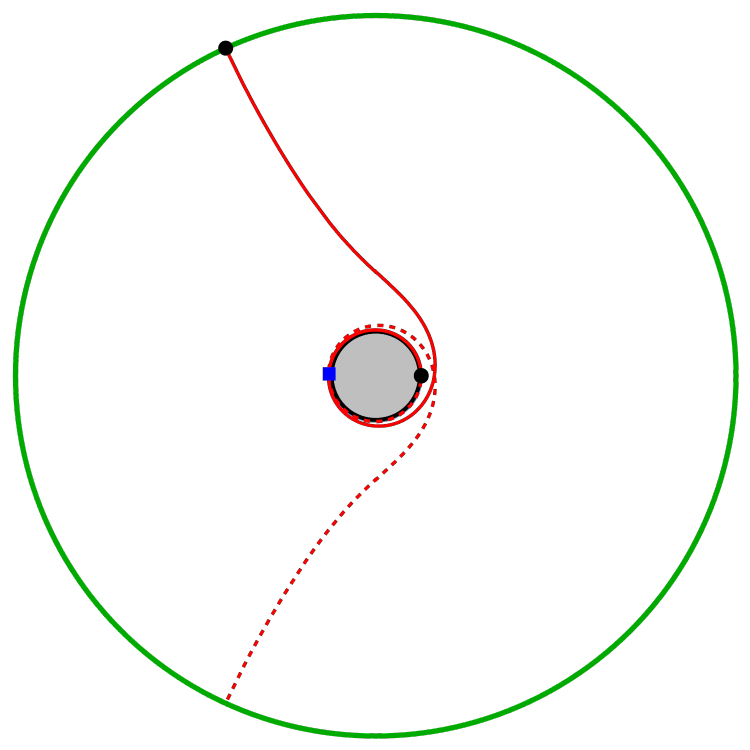}
	\end{minipage}
	\\
	\vspace{.3cm}
	\caption{\small
		Geodesics (red solid lines)  in the BTZ black hole \eqref{btz-global-metric} with one endpoint on the boundary 
		and the other endpoint on the horizon (black dots) and the corresponding auxiliaries (red dashed lines).	
		Their winding number is either zero (left) or one (middle) or two (right).
		}
	\label{fig-global-btz-long-geod-bdy-to-hor1}
\end{figure}

Next, we focus on the second set of geodesics with one endpoint on the boundary, at $\phi = \phi_b  \in [0, 2 \pi) $,
and the other one on the horizon, at $\phi = \phi_h  \in [0, 2 \pi) $. 
They are, for instance,  the grey geodesics covering the light blue region of  Fig.\,\ref{fig-global-btz-threads-before}. 
These curves are described by  \eqref{geo-btz-global-bh-hor}, 
where the parameter $\theta_0$ and $\phi_0$ are chosen to be
\be
\label{long-geo-BTZ-3}
\theta_0  = |\phi_b - \phi_h| + 2 n \pi 
\hspace{1.3cm}
\phi_0 = \phi_h
\hspace{1.3cm}
n \in \mathbb{Z}\,.
\ee  
Some of these curves are shown in Fig.\,\ref{fig-global-btz-long-geod-bdy-to-hor1} as solid red lines.
These geodesics can extend beyond the point where they intersect the horizon. 
Their extension includes an auxiliary branch that retraces from the horizon back to the boundary,
as indicated by the red dashed curves  in Fig.\,\ref{fig-global-btz-long-geod-bdy-to-hor1}. 
The two endpoints on the boundary of the maximal extension of these geodesics have coordinates $\theta_0+\phi_0$ and $\theta_0-\phi_0$ modulo $2\pi$.
In Fig.\,\ref{fig-global-btz-long-geod-bdy-to-hor1}, the solid red curves are geodesics  
with $n= 0 $ (left panel), $n= -1 $ (middle panel) and  $n= 1 $ (right panel).
We can easily calculate, for a generic value of $n$,  the number of times  $w_{\text{\tiny II}} $ that the extended geodesic (solid and dashed red line in 
Fig.\,\ref{fig-global-btz-long-geod-bdy-to-hor1}) wraps the horizon. This computation is identical to the previous case 
because the extended geodesic possesses two endpoints on the boundary,
with the only difference is that $k$ is replaced by $2n$. Thus, we find
\be
w_{\text{\tiny II}}=\left\{\! \!
\begin{array}{cll} 
2n & \hspace{.7cm} & n\geqslant 0
\\
\rule{0pt}{.5cm}
-\!(2n+1) & & n<0\,.
\end{array}
\right.
\ee
As a final remark, let us observe that,
given the analogy between \eqref{long-geo-BTZ-1} and \eqref{long-geo-BTZ-3} if we set $k = 2n$, 
one might be tempted to ask what happens when $\theta_0$ is such that $2n$ is replaced by $2n +1$ in \eqref{long-geo-BTZ-3}. 
This choice is incompatible with our initial data;
indeed, to complete the analogy with \eqref{long-geo-BTZ-1},  
the horizon hitting point has to be taken at $\phi_h + \pi$ and not at $\phi_h$, 
which correspond respectively to the blue square and to the black dot on the horizon 
in Fig.\,\ref{fig-global-btz-long-geod-bdy-to-hor1}.   

\bibliographystyle{nb}
\bibliography{refsBT}
	
\end{document}